\UseRawInputEncoding
\documentclass[%
superscriptaddress,
preprint,
 amsmath,amssymb,
 aps,
]{revtex4-2}

\usepackage{graphicx}% Include figure files
\usepackage{xtab}
\usepackage{dcolumn}% Align table columns on decimal point
\usepackage{bm}% bold math
\usepackage[utf8]{inputenc} % non-english names and addresses
\usepackage{hyperref}% add hypertext capabilities
\usepackage{isotope}
\usepackage{xcolor}
\usepackage[normalem]{ulem}
\usepackage{orcidlink}

\usepackage{float} % Nicole V added to allow figure to be forced to be exactly where ask with H

\newcommand{\nsat}{n_\mathrm{sat}}
\newcommand{\obs}{\boldsymbol{D}}
\newcommand{\params}{\boldsymbol{\alpha}}

\usepackage{subcaption}

\usepackage{titlesec}

\titleformat{\paragraph}
{\normalfont\normalsize\bfseries}{\theparagraph}{1em}{}
\titlespacing*{\paragraph}
{0pt}{3.25ex plus 1ex minus .2ex}{1.5ex plus .2ex}

\begin{document}

\preprint{FRIB theory whitepaper}

\title{Motivations for Early High-Profile FRIB Experiments}% Force line breaks with \\

% -------------- organizers
\author{B.~Alex~Brown\,\orcidlink{0000-0002-6111-1906}}
\affiliation{
Department of Physics and Astronomy and the Facility for Rare Isotope Beams, Michigan State University,
East Lansing, Michigan 48824-1321, USA
}

\author{Alexandra~Gade\,\orcidlink{0000-0001-8825-0976}}
\affiliation{
Department of Physics and Astronomy and the Facility for Rare Isotope Beams, Michigan State University,
East Lansing, Michigan 48824-1321, USA
}

\author{S.~Ragnar~Stroberg\,\orcidlink{0000-0002-0635-776X}}
\affiliation{
Department of Physics and Astronomy, University of Notre Dame, Notre Dame, Indiana 46556-5670, USA
}

%----------------- section leaders

\author{Jutta~Escher\,\orcidlink{0000-0002-0829-9153}}
\affiliation{
Nuclear and Chemical Sciences Division, Lawrence Livermore National Laboratory, Livermore, California 94551, USA
}

\author{Kevin~Fossez\,\orcidlink{0000-0002-5823-4446}}
\affiliation{
Department of Physics, Florida State University, Tallahassee, Florida 32306, USA
}

\author{Pablo~Giuliani\,\orcidlink{0000-0002-8145-0745}}
\affiliation{
Facility for Rare Isotope Beams, Michigan State University,
East Lansing, Michigan 48824-1321, USA
}
\affiliation{Department of Statistics and Probability, Michigan State University, East Lansing, Michigan 48824, USA}

\author{Calem~R.~Hoffman\,\orcidlink{0000-0001-7141-9827}}
\affiliation{
Physics Division, Argonne National Laboratory, 9700 S. Cass Ave., Argonne, Illinois 60439, USA
}

\author{Witold~Nazarewicz\,\orcidlink{0000-0002-8084-7425}}
\affiliation{
Department of Physics and Astronomy and the Facility for Rare Isotope Beams, Michigan State University,
East Lansing, Michigan 48824-1321, USA
}

\author{Chien-Yeah~Seng\,\orcidlink{0000-0002-3062-0118}}
\affiliation{
Facility for Rare Isotope Beams, Michigan State University,
East Lansing, Michigan 48824-1321, USA
}
\affiliation{
Department of Physics, University of Washington, Seattle, Washington 98195-1560, USA
}

\author{Agnieszka~Sorensen\,\orcidlink{0000-0002-1984-8023}} 
\affiliation{
Institute for Nuclear Theory, University of Washington, 
%Box 351550, 
Seattle, Washington 98195, USA
}

\author{Nicole~Vassh\,\orcidlink{0000-0002-3305-4326}}
\affiliation{
TRIUMF, 4004 Wesbrook Mall, Vancouver, British Columbia V6T 2A3, Canada
}

%------------------  contributors

\author{Daniel~Bazin\,\orcidlink{0000-}}
\affiliation{
Department of Physics and Astronomy and the Facility for Rare Isotope Beams, Michigan State University,
East Lansing, Michigan 48824-1321, USA
}

\author{Kyle~W.~Brown\,\orcidlink{0000-0003-1923-3595}}
\affiliation{
Department of Chemistry and the Facility for Rare Isotope Beams, Michigan State University, 
East Lansing, Michigan 48824-1321, USA
}

\author{Mark~A.~Caprio\,\orcidlink{0000-0001-5138-3740}}
\affiliation{
Department of Physics and Astronomy, University of Notre Dame, Notre Dame, Indiana 46556-5670, USA
}

\author{Heather~Crawford\,\orcidlink{0000-0002-7765-4235}}
\affiliation{
Nuclear Science Division, Lawrence Berkeley National Laboratory, Berkeley, California 94720, USA
}

\author{Pawel~Danielewicz\,\orcidlink{0000-0002-1989-5241}}
\affiliation{
Department of Physics and Astronomy and the Facility for Rare Isotope Beams, Michigan State University,
East Lansing, Michigan 48824-1321, USA
}

\author{Christian~Drischler\,\orcidlink{0000-0003-1534-6285}}
\affiliation{
Department of Physics and Astronomy and Institute of Nuclear and Particle Physics, Ohio University, Athens, Ohio 45701, USA
}

\author{Ronald~F.~Garcia~Ruiz\,\orcidlink{0000-0002-2926-5569}}
\affiliation{
Massachusetts Institute of Technology, Cambridge, Massachusetts 02139, USA
} 

\author{Kyle~Godbey\,\orcidlink{0000-0003-0622-3646}}
\affiliation{
Facility for Rare Isotope Beams, Michigan State University,
East Lansing, Michigan 48824-1321, USA
}

\author{Robert~Grzywacz\,\orcidlink{0000-0002-0920-2587}}
\affiliation{
Department of Physics and Astronomy, University of Tennessee, Knoxville, Tennessee 37996, USA
}

\author{Linda~Hlophe\,\orcidlink{0000-0001-6675-6132}} 
\affiliation{Nuclear and Chemical Sciences Division, Lawrence Livermore National Laboratory, Livermore, California 94551, USA;
Facility for Rare Isotope Beams, Michigan State University, East Lansing, Michigan 48824-1321, USA;Los Alamos National Laboratory, Los Alamos, New Mexico 87545, USA
}

\author{Jeremy~W.~Holt\,\orcidlink{0000-0003-4373-3856}}
\affiliation{
Cyclotron Institute, Texas A\&M University, College Station, Texas 77843, USA;
Department of Physics and Astronomy, Texas A\&M University, College Station, Texas 77843, USA
}

\author{Hiro~Iwasaki\,\orcidlink{0000-0003-0750-380X}} 
\affiliation{
Department of Physics and Astronomy and the Facility for Rare Isotope Beams, Michigan State University,
East Lansing, Michigan 48824-1321, USA
}

\author{Dean~Lee\,\orcidlink{0000-0002-3630-567X}}
\affiliation{
Department of Physics and Astronomy, and the Facility for Rare Isotope Beams, Michigan State University,
East Lansing, Michigan 48824-1321, USA
}

\author{Silvia~M.~Lenzi\,\orcidlink{0000-0002-4666-8431}}
\affiliation{
Dipartimento di Fisica e Astronomia dell’Universit\`a and INFN, Sezione di Padova, I-35131 Padova, Italy
} 
\author{Sean~Liddick\,\orcidlink{0000-0003-1589-891X}}
\affiliation{
Department of Chemistry and the Facility for Rare Isotope Beams, Michigan State University,
East Lansing, Michigan 48824-1321, USA
}

\author{Rebeka~Lubna\,\orcidlink{0000-0002-3144-1285}} 
\affiliation{
Facility for Rare Isotope Beams, Michigan State University,
East Lansing, Michigan 48824-1321, USA
}

\author{Augusto~O.~Macchiavelli\,\orcidlink{0000-0002-3144-1285}} 
\affiliation{
Physics Division, Oak Ridge National Laboratory, Oak Ridge, Tennessee 37831, USA
}

\author{Gabriel~Martínez~Pinedo\,\orcidlink{0000-0002-3825-0131}}
\affiliation{
GSI Helmholtzzentrum für Schwerionenforschung, Planckstraße 1, 64291 Darmstadt, Germany}
\affiliation{Institut fur Kernphysik (Theoriezentrum), Fachbereich Physik, 
Technische Universität Darmstadt, Schlossgartenstraße 2, 64289 Darmstadt, Germany}
\affiliation{Helmholtz Forschungsakademie Hessen fur FAIR, GSI Helmholtzzentrum für 
Schwerionenforschung, Planckstraße 1, 64291 Darmstadt, Germany}

\author{Anna~McCoy\,\orcidlink{0000-}} 
\affiliation{
Facility for Rare Isotope Beams, Michigan State University,
East Lansing, Michigan 48824-1321, USA
}
\affiliation{
Department of Physics, Washington University in Saint Louis, Saint Louis, Missouri 63130, USA
}

\author{Alexis Mercenne\,\orcidlink{0000-0002-2624-3911}}
\affiliation{Department of Physics and Astronomy, Louisiana State University, Baton Rouge, LA 70803, USA}

\author{Kei~Minamisono\,\orcidlink{0000-0003-2315-5032}}  
\affiliation{
Department of Physics and Astronomy and the Facility for Rare Isotope Beams, Michigan State University,
East Lansing, Michigan 48824-1321, USA
}

\author{Belen~Monteagudo}
\affiliation{
Department of Physics, Hope College, Holland, Michigan 49423-9000
}

\author{Petr~Navratil\,\orcidlink{0000-0002-6535-2141}}
\affiliation{
TRIUMF, 4004 Wesbrook Mall, Vancouver, British Columbia V6T 2A3, Canada
}

%\author{Filomena~Nunes\,\orcidlink{0000-0001-8765-3693}}\affiliation{Department of Physics and Astronomy, and the Facility for Rare Isotope Beams, Michigan State University,East Lansing, Michigan 48824-1321, USA}

\author{Ryan~Ringle\,\orcidlink{0000-0002-7478-259X}}
\affiliation{
Department of Physics and Astronomy and the Facility for Rare Isotope Beams, Michigan State University, 
East Lansing, Michigan 48824-1321, USA
}

\author{Grigor~H.~Sargsyan\,\orcidlink{0000-0002-3589-2315}}
\affiliation{
Nuclear and Chemical Sciences Division, Lawrence Livermore National Laboratory, Livermore, California 94551, USA
}
\affiliation{
Facility for Rare Isotope Beams, Michigan State University,
East Lansing, Michigan 48824-1321, USA
}

\author{Hendrik~Schatz\,\orcidlink{0000-0003-1674-4859}}
\affiliation{
Department of Physics and Astronomy and the Facility for Rare Isotope Beams, Michigan State University,
East Lansing, Michigan 48824-1321, USA
}

\author{Mark-Christoph~Spieker\,\orcidlink{0000-0002-7214-7656}}
\affiliation{
Department of Physics, Florida State University, Tallahassee, Florida 32306, USA
}

\author{Alexander~Volya\,\orcidlink{0000-0002-1765-6466}}
\affiliation{
Department of Physics, Florida State University, Tallahassee, Florida 32306, USA
} 

\author{Remco~G.~T.~Zegers\,\orcidlink{0000-0001-6076-5898}}  
\affiliation{
Department of Physics and Astronomy and the Facility for Rare Isotope Beams, Michigan State University,
East Lansing, Michigan 48824-1321, USA
}

\author{Vladimir~Zelevinsky\,\orcidlink{0000-0002-1705-5660}}
\affiliation{
Department of Physics and Astronomy and the Facility for Rare Isotope Beams, Michigan State University,
East Lansing, Michigan 48824-1321, USA
}

\author{Xilin~Zhang\,\orcidlink{0000-0001-9278-5359}}
\affiliation{
Facility for Rare Isotope Beams, Michigan State University,
East Lansing, Michigan 48824-1321, USA
}

\date{\today}% It is always \today, today,

\begin{abstract}
This white paper is the result of a collaboration by those that attended a workshop at the
Facility for Rare Isotope Beams (FRIB), organized by the FRIB Theory Alliance (FRIB-TA), on ``Theoretical Justifications and Motivations for Early High-Profile FRIB Experiments''.
It covers a wide range of topics related to
the science that will be explored at FRIB.
After a brief introduction, the sections address: (II) Overview of
theoretical methods,
(III) Experimental capabilities, 
(IV) Structure, (V) Near-threshold Physics, 
(VI) Reaction mechanisms,
(VII) Nuclear equations of state,
(VIII) Nuclear astrophysics, 
(IX) Fundamental symmetries, and
(X) Experimental design and uncertainty quantification.
\end{abstract}

\maketitle

\tableofcontents

\section{Introduction}

In 2008, the DOE Office of Science awarded a project to Michigan State University (MSU)
to build the Facility for Rare Isotope Beams (FRIB). In May of 2022
there was a ribbon cutting ceremony for its completion,
and the experimental program was begun.
Since then, over 200 proposals have been submitted to the 
Program Advisory Committee (PAC). Those selected to run are
chosen for their scientific impact and feasibility. 
The scientific impact of the proposals as well as the
publication of the results relies 
on a strong collaboration
of experiment with nuclear theory and astrophysical modeling.

In May of 2023, an FRIB Theory Alliance (FRIB-TA) workshop
%organized by Alex Brown, Alexandra Gade and Ragnar Stroberg,
was held at FRIB on 
``Theoretical Justifications and Motivations for Early High-Profile FRIB Experiments''.
This workshop brought together researchers with a broad range of interests in FRIB science. 

\begin{figure}
\includegraphics[width=\textwidth]{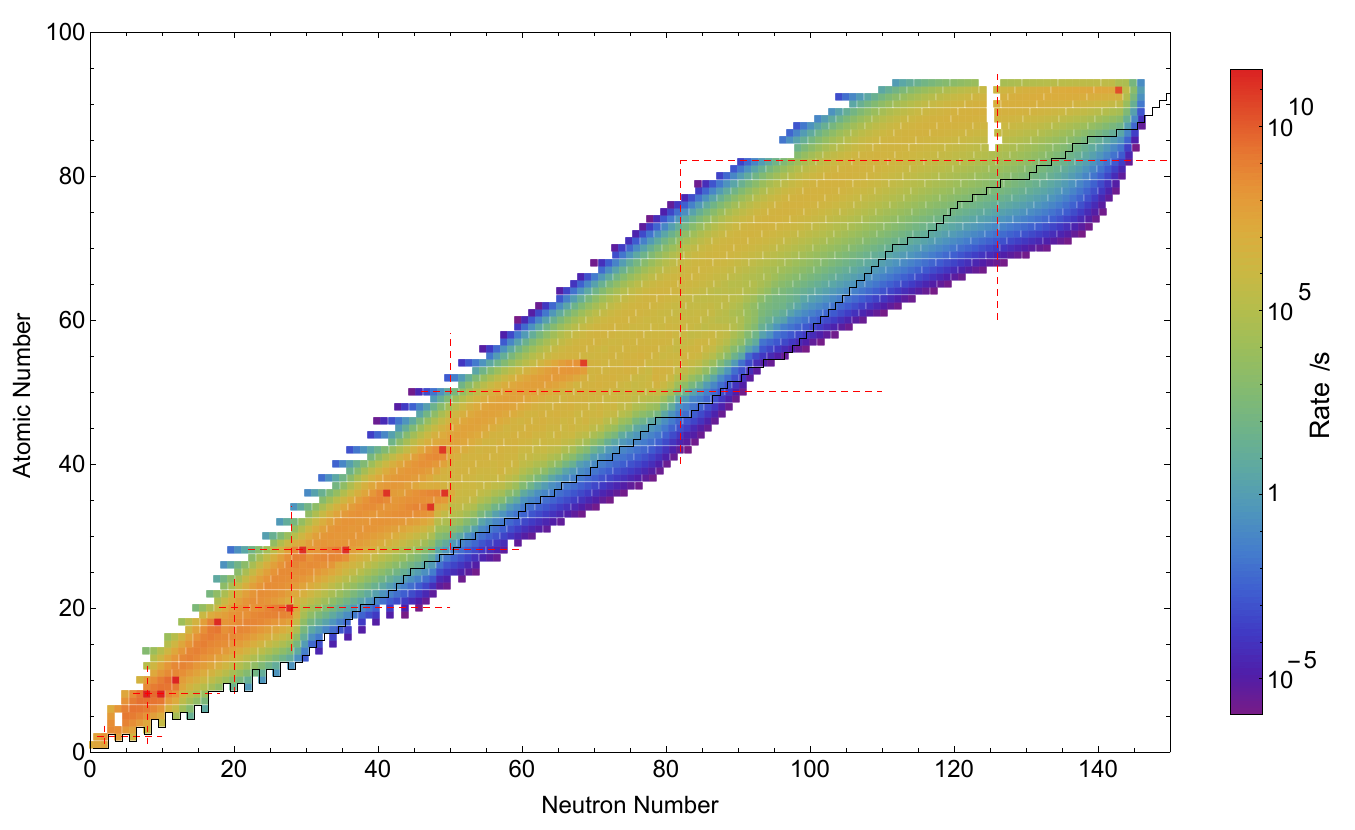}
\caption[]{\small 
FRIB beam rates based on primary beam intensities of 10 kW
for nuclei out to the proton and neutron drip lines
predicted by calculations with the UNEDF1 energy-density-functional \cite{UNEDF2013}.
The thin black line on the neutron-rich side is the approximate boundary of known nuclei.
The red squares are the primary beams produced by FRIB.
These are fragmented to produce the other nuclei shown at the rates indicated on the right-hand side.
The nuclei of interest for a given experiment are selected by the fragment separator.
}
\label{YearTwoPlot}

\end{figure}

The workshop started with a presentation 
%by the FRIB Scientific Director Brad Sherrill
on the capabilities for fast, stopped, and
reaccelerated beams at FRIB. 
Opportunities for experiments over a wide range  
of the nuclear chart, out to the limits of
nuclear stability~\cite{thoe2004}, will be opened up as the
beam power is ramped up from 10 kW at present to
400 kW over the next decade~\cite{ sher2018}.
The calculated beam intensities,
assuming a beam power of 10 kW, are shown in 
Fig.~\ref{YearTwoPlot}.
Proposals for the third Program Advisory Committee (PAC3)
were submitted in October of 2024.
%(https://frib.msu.edu/users/pac/call.html).

The feasibility of a proposal for an FRIB experiment is associated with the 
production of specific nuclei
together with the instrumentation available.
At the workshop, experimentalists gave talks on a
wide range of instrumentation and
types of observables that will be measured.
Proposals are most often based on collaborations with
theorists to understand and articulate the connection 
between the measured observables and the scientific
impact.
At the FRIB-TA workshop, theorists gave talks on
their present results and future plans. This
was followed by discussions between theory and experiment
to consider how the impact of the early FRIB experiments can be maximized by a 
close experiment-theory collaboration where results from both sides advance each other. 

This white paper is the product of collaboration by those that attended the workshop.
It covers a wide range of topics related
to the science at FRIB.
The sections are: (II) Overview of modern
theoretical methods,
(III) Experimental capabilities, 
(IV) Structure, (V) Near-threshold Physics, 
(VI) Reaction mechanisms,
(VII) Nuclear equations of state,
(VIII) Nuclear astrophysics, 
(IX) Fundamental symmetries, and
(X) Experimental design and uncertainty quantification,
followed by a brief summary.

The author list starts with the workshop organizers, 
followed by the list of section leaders and then
the list of contributors. 
Abbreviations used in the text are given in an Appendix.

%In addition we have included a section on experimental methods, an appendix for a review of the notations for 
%shell model orbitals, model spaces and Hamiltonians.

%The coauthor list starts with the workshop and whitepaper organizers,followed by the section leaders, followed by the many contributors.

%A review of nuclear shell-model properties, model spaces, and Hamiltonians is given in Appendix A.1. Appendix A.2 provides a list of acronyms used.

%For each topic the writers were asked to consider:\\
%What are the best high-impact programs to carry out?\\
%What theoretical input is needed to achieve this?\\
%What theoretical methods need to be improved or developed?\\
%What do experiments need from theory?\\
%What do theorists need from experiment?

%______________%

%______________%
%\section{Methods}
\section{Overview of modern theoretical methods\label{sec:TheoryMethods}}
%%% Methods section
%%% Coordinator: Witek Nazarewicz

%\subsection{Overview of modern theoretical methods}
\label{sec:overview_theo_methods}
%% Brief overviews of shell model, DFT, ab initio, reaction theory, transport theory,...

%% Witek
The goal of low-energy nuclear structure theory is to build a unified microscopic framework in which nuclei, nuclear matter, and nuclear reactions can all be explained, described, and predicted. While this goal is ambitious, it is not beyond the realm of possibility. Hand in hand with breakthroughs  in radioactive ion beam (RIB) experimentation, a rapid transition in nuclear modeling is taking place. Due to  novel theoretical concepts, interdisciplinary collaborations, and high performance computing, nuclear theorists have been quite successful in cracking the nuclear many-body problem.

These are exciting times for theoretical nuclear structure
research \cite{LRP2015,Nazarewicz2016,Nunes2021,Exascale}.  
In the area of quantum chromodynamics (QCD), the underlying theory of strong interactions that governs the  properties and dynamics of quarks and gluons that form baryons and mesons,  significant progress is being made by computing properties of 
 nuclear interactions, operators, and the lightest nuclei \cite{Savage2016,Drischler2021,Cirigliano2022,Amarasinghe2023}. 
 
The  nuclear many-body problem with protons and neutrons is an effective low-energy approximation to QCD. Nuclear
effective field
theory (EFT) has enabled theorists to construct high-quality two- and three-body inter-nucleon
interactions consistent with the 
% chiral symmetry
low-energy symmetries 
of QCD
\cite{Entem2020,Epelbaum2020,Piarulli2020,Jiang2020}. The  low-energy coupling constants (LECs) that incorporate unresolved short-distance (i.e., high-energy) physics, must be determined from experiment \cite{Tews2020,Ekstrom2020} or eventually from QCD \cite{McIlroy2018,Tews2020}.

%%% Ragnar Stroberg and Witek
Ab initio methods, i.e., systematically improvable 
 $A$-body approaches based on inter-nucleon interactions \cite{Ekstrom2023}, have seen dramatic progress over the past decade~\cite{hergert2020guided,Hu2022}.
Open-shell systems, excited states, heavy nuclei, and nuclear reactions are now within reach, and electroweak currents consistent with the strong interaction are available. 
Broadly, we can divide ab initio many-body methods into quasi-exact solutions (no-core shell model~\cite{Barrett2013}, quantum Monte Carlo~\cite{Carlson2015}, lattice effective field theory~\cite{Lee2020}), and more approximate but more widely applicable polynomial-scaling methods (coupled cluster~\cite{Hagen2014}, self-consistent Green's function~\cite{Soma2020}, in-medium similarity renormalization group~\cite{Hergert2016}, many-body perturbation theory~\cite{Tichai2020}).
In an ab initio framework, comparison to experimental data can reveal shortcomings either in the input (nuclear interaction, operators) or in the approximations made in the many-body solution.
A proper understanding of both of these sources of error will then enable predictions with quantified uncertainties.

%% Witek
For light and medium-mass  systems,   configuration-interaction (CI) methods employing effective shell-model interactions optimized to various nuclear structure data offer detailed descriptions of nuclear excitations
and  decays \cite{Caurier2005,Shimizu2018,Magilligan2020}. In the area of weakly bound or unbound nuclear states, and reactions at near-threshold energies,
modern continuum shell-model approaches  unify
nuclear bound states with resonances and scattering continuum within one consistent framework \cite{Michel2008,Johnson2020}.

For heavy complex nuclei the tool of choice is nuclear density functional theory (DFT)  \cite{Bender2003,Drut2010}; the  validated global  energy density functionals  often provide a level of accuracy typical of phenomenological approaches based on parameters locally optimized to experiment, and enable extrapolations into nuclear {\it terra incognita} \cite{Neufcourt2020a}. The time-independent and time dependent  extensions of DFT \cite{Nakatsukasa2016,Sheikh2021} have provided quantitative descriptions of one of the toughest problems of nuclear structure: nuclear large amplitude collective  motion, which includes such phenomena as shape coexistence, fission, and heavy-ion fusion \cite{Garrett2019,Fission2020,Godbey2020prc}.

%% Jutta on nuclear reactions - revised 2023-08-31
These are exciting times also for nuclear reaction theory. 
For light nuclei, the combination of the no-core shell model (NCSM) with the resonating group method (RGM) enables ab initio descriptions of reactions with both nucleons and light composite projectiles.
The method uses modern interactions rooted in chiral EFT, can include 3-nucleon forces and continuum effects, and has successfully predicted scattering, transfer, and capture reactions
%~\cite{Navratil:16,
~\cite{navr2016,
Quaglioni:20, Hupin:19, Hebborn:22}.
The use of symmetry-adapted bases provides a viable path to extend the approach to medium-mass nuclei~\cite{Launey:21,Mercenne:22}. 
%% Xilin added the following sentence. 
In addition, methods are being explored to extract nuclear scattering and reactions from computing eigen-energies of the nuclear systems in external traps~\cite{Zhang:2019cai, Zhang:2020rhz, Bagnarol:2023crb}, or from continuum states at complex energies~\cite{XilinZhangTRIUMF2023,Zhang:2024ril}.  
%%(Move to reaction section) Since the wave functions in these calculations are spatially localized, these approaches could enable microscopic continuum calculations wherever the structure calculations are feasible. 

For medium-mass and heavy nuclei, we distinguish between direct and compound nuclear reactions.
Traditional descriptions of direct reactions, such as the Distorted-Wave Born Approximation (DWBA) and Adiabatic Distorted Wave Approximation (ADWA) are increasingly combined with structure information from microscopic theories, in order to improve their predictive power~\cite{Brida:11, Grinyer:12, Nobre:10, Nobre:11, Dupuis:19}. 
Extensions of standard DWBA and eikonal approaches have been introduced to account for the dynamics between the transferred nucleon and the nucleus in descriptions of one-nucleon transfer and knockout reactions~\cite{Potel:17, Hebborn:23}. 
The full coupled-channels (CC) approach accommodates multistep reaction mechanisms and the Continuum-Discretized Coupled Channels (CDCC) and Faddeev methods provide improved treatments of the breakup contributions~\cite{Johnson2020, Hagino:22, Alt:67, Deltuva:19}. Extensions to include more complicated reaction mechanisms, such as core excitations and four-body channels, have been the focus of recent work~\cite{Timofeyuk:20, Descouvemont:18, Watanabe:21}.

Compound-nuclear reactions are treated in a phenomenological R-Matrix approach (for isolated resonances) or statistical Hauser-Feshbach (HF) theory (for overlapping resonances)~\cite{Descouvemont:10, Thompson:09book}.  
Efforts have been devoted to compiling recommendations for default inputs for HF calculations, which enable data evaluations as well as initial calculations for thousands of isotopes for astrophysics application~\cite{Capote:09, Koning:2023wz}.  At the same time, significant progress has been in replacing the phenomenological models used in HF codes by microscopically-calculated quantities~\cite{hebborn2023optical, Koning:2023wz}.
% End - Jutta on reactions ---

From a theoretical point of view, short-lived exotic nuclei far from stability, studied at RIB facilities, offer a unique test of those aspects of the many-body theory that depend on the isospin degrees of freedom and the coupling to the continuum space. The challenge is to develop methods to reliably calculate and understand the properties of new physical systems, identify the impact of new observables on theoretical models, quantify correlations between predicted observables, and assess uncertainties of theoretical predictions.  The use of advanced  tools of uncertainty quantification and machine learning will  help to speed  up the cycle ``observation-theory-prediction--experiment-'' of the scientific method \cite{Boehnlein2022}.   These  new tools are expected to provide meaningful input for planned measurements at FRIB and will be used to interpret and use new nuclear and astrophysical information obtained at FRIB.

%% Agnieszka
Using rare isotopes in heavy-ion collisions (HICs) produces dense nuclear matter within a large range of isospin asymmetry~\cite{FRIB400} and provides a connection to dense neutron-rich matter present in neutron stars and neutron star mergers~\cite{Sorensen:2023zkk,Huth:2021bsp,Lovato:2022vgq}. In particular, comparisons of HIC observables to transport model simulations put constraints on the nuclear matter equation of state at densities well above the saturation density~\cite{Danielewicz:2002pu,Fuchs:2003pc,Lynch:2009vc,LeFevre:2015paj,Oliinychenko:2022uvy,Lynch:2021xkq}, currently inaccessible to ab initio approaches~\cite{Drischler:2020hwi}. Simulations are also used to identify promising observables~\cite{Kruse:1985hy,Aichelin:1985rbt,Li:2002qx,Li:2000bj,Li:2014oda,Danielewicz:2021vqq,Berkowitz:2022pgz}, informing detector and experimental designs. An outstanding challenge in these studies is a quantitative understanding of model uncertainties~\cite{Sorensen:2023zkk,TMEP:2022xjg}, addressed by investigations of different transport models in well-controlled simulation setups as well as through Bayesian analyses~\cite{Morfouace:2019jky,Liyanage:2022byj,Oliinychenko:2022uvy} and Bayesian model mixing
%~\cite{Phillips:2020dmw}. 
~\cite{BANDmanifesto}.
The availability of high-performance computing has only recently made such studies feasible, providing a path to a comprehensive understanding of HIC dynamics~\cite{Sorensen:2023zkk}.

New theoretical ideas in describing atomic nuclei, progress in RIB
experimentation, the  arrival of exascale computing platforms \cite{Exascale}, and
increased collaboration between nuclear theorists with their computer
science and applied mathematics partners \cite{UNEDF2013,BANDmanifesto} -- all are paving the way for
today's progress in theoretical nuclear structure studies. The important challenge for the field is to connect different many-body approaches, describing the nucleus  at different resolution scales,  in the regions of the nuclear landscape where they overlap. By bridging the gaps, one is aiming at developing one comprehensive  picture of the atomic nucleus, from the single nucleon  all the way to the superheavy species. This is an exciting  prospect.
On the journey to the comprehensive nuclear theory framework, important milestones will be marked by designer nuclei with characteristics adjusted to specific research needs
\cite{Sherrill2008,Jones2010,Nazarewicz2016}. Those rare isotopes are the key to answering questions in many areas of science and they will also provide society with numerous 
opportunities.

%______________%

\section{Experimental Capabilities\label{sec:ExpMethods}}
\subsection{Mass measurements}
Expanding the reach of precision mass measurements at a level of $\delta$m/m $\approx$ 10$^{-6}$ to 10$^{-8}$ \cite{Lunney.2003} for rare isotopes to the most exotic regions of the nuclear chart is of utmost importance, enabling the direct observation of nuclear structure phenomena, such as (sub-)shell closures, pairing effects, and the onset of deformation. FRIB is set to revolutionize precision mass measurements by opening a fresh territory within the nuclear chart, over 1300 isotopes with sufficient production rates and long enough half-lives with mass uncertainties greater than 50 keV/c$^2$, thus, in collaboration with theory, paving the way for groundbreaking discoveries.

The two primary techniques employed in performing direct mass measurements of rare isotopes at FRIB are the TOF-B$\rho$ technique~\cite{Matos2012,Meisel.2013} and using Penning traps at the low-energy beam and ion trap facility (LEBIT) facility~\cite{Ringle.2013}.  The TOF-B$\rho$ technique uses the fast beams produced directly by FRIB and can perform measurements on rare isotopes with half lives down to 100's of nanoseconds with a mass precision down to 100 keV/c$^2$.  
% Maybe add similar metrics for the Penning trap measurements

\subsection{Charge radii and nuclear moment measurements}

In the last several decades, laser spectroscopy techniques have been extensively used to determine 
mean-square charge radius, $R_{\rm ch}$, and moments 
of ground and isomeric states \cite{otte1989, bill1995, klug2003, chea2010, blau2013,camp2016,neug2017, yang2023}. Especially, the bunched-beam collinear laser spectroscopy (CLS) has been almost exclusively used for charge radius measurements of long chains of radioactive isotopes and nuclei occurring toward nucleon drip lines. This is thanks to the bunched-beam CLS technique's high resolution due to the Doppler compression \cite{kauf1976}, and its high sensitivity due to the large background suppression by bunching the beam in fluorescence measurements \cite{niem2002, camp2002, barq2017} or by the virtual background free ion detection in resonant ionization measurements \cite{flan2013}. 

At FRIB, such CLS experiments are performed at the BEam COoling and LAser spectroscopy (BECOLA) facility \cite{mina2013, ross2014p093503}, which is designed to accept low-energy beams typically 30 keV/u from the gas stopping systems \cite{sumi2020, lund2020}, the offline batch mode ion source (BMIS) \cite{sumi2023} for long-lived isotopes, and a dedicated local offline ion source \cite{ryde2015} at BECOLA. Time-resolved laser resonant fluorescence and resonant ionization spectroscopy measurements with bunched beams \cite{ross2014p093503, barq2017, flan2013} are performed for highly sensitive measurements of hyperfine spectra and their isotope shifts. A beam rate of as low as $\sim$10/s has been achieved for $^{36}$Ca ions \cite{mill2019} and $^{54}$Ni atoms \cite{pine2021}. BECOLA has been extensively used to determine charge radius \cite{ross2015, mina2016, mill2019, pine2021, somm2022, koni2023} and electromagnetic moments \cite{mill2017, klos2019, powe2022} of radioactive nuclei. The Resonance Ionization Spectroscopy Experiment (RISE) instrument was recently integrated into the BECOLA facility in collaboration with MIT to realize $\sim$1 ion/s sensitivity for laser spectroscopy. RISE has been commissioned with offline beams and is planning to take online beams in 2024. Similar to LEBIT, BECOLA is the only CLS facility worldwide making use of rare isotopes from projectile fragmentation, complementing capabilities at ISOL-type facilities.

The extraction of  $R_{\rm ch}$ from the isotope shift depends on atomic factors that must
be calculated \cite{yang2023,maas2019}
or measured using the King plot \cite{king1984} involving three or more isotopes
whose $R_{\rm ch}$ are well determined through electron scattering and/or  $\mu$-capture measurements.
The uncertainties involve two terms, one for the uncertainty of the hyperfine spectra, and
the other related to the uncertainty of the atomic factors.

\subsection{Charge-exchange reactions}

At the NSCL, the ($t$,$^{3}$He) reaction \cite{Osterfeld:1992,ICHIMURA2006446,Fujita2011549,RevModPhys.75.819, FRE18,HAR01,Langanke2021,zegers2020}, used for extracting GT strengths from stable nuclei at the S800 spectrometer \cite{bazin03}, has proven to be a versatile probe. For early operation at FRIB, tritium beam intensities that are more than 30 times higher can be achieved. At these intensities, detailed studies of Gamow-Teller strength distributions in the $\beta^{+}$ direction in heavier stable nuclei can be investigated, which was previously very difficult. It will become possible to use the Summing NaI(Tl) gamma-ray detector (SuN) 
%\cite{SIMON201316}  
\cite{simo2013}  
in coincidence with the S800, and use the Oslo method for extracting level densities and $\gamma$-ray strength functions, similar to the $\beta$-Oslo method %\cite{LARSEN201969}.
\cite{Lar19a}.

The isolation of $\Delta L=0$ contributions from the measured excitation energy spectrum is facilitated through a multipole decomposition analysis ~\cite{bon84,Moinester:1989,WAK97}. The proportionality between strength and differential cross section, represented by the so-called unit cross section ($\hat{\sigma}$) is facilitated through a calibration using a pair of states for which the transition strength is known from  beta decay or electron capture ~\cite{Taddeucci1987125}.  The Gamow-Teller (GT) or Fermi (F) transition strength is proportional to the charge-exchange cross section at small momentum transfer ($q\approx 0)$ ~\cite{Taddeucci1987125,Perdikakis:2011}. If such calibrations are not available, phenomenological mass-dependent relations for $\hat{\sigma}$ are employed (\cite{zegers07,Sasano:2009}. 

Recently, the ($d$,$^{2}$He) reaction in inverse kinematics was developed for probing GT strengths in the $\beta^{+}$ direction on unstable nuclei \cite{PhysRevLett.130.232301}, by using the Active Target Time Projection Chamber (AT-TPC) 
%\cite{AYYAD2020161341} 
\cite{ayya2020} 
in combination of the S800 spectrometer \cite{bazin03}. At FRIB, it will be possible to probe GT strengths on neutron-rich nuclei that are important for astrophysics. The early science program will likely focus on light and medium-heavy systems, to gain experience with the injection of highly charged ions into the AT-TPC.    

The ($p$,$n$) reaction in inverse kinematics for probing GT strengths in the $\beta^{-}$ direction was developed some time ago. By combining the measurement of the heavy residual particle  in a spectrometer \cite{bazin03} with the detection of the low-energy recoil neutron from the ($p$,$n$) reaction \cite{PhysRevLett.107.202501,PhysRevC.106.054323} in the Low-Energy Neutron Detector Array (LENDA), isotopes across the chart of nuclei can be studied. An attractive opportunity arises from the coincident study of the ($p$,$n$) reaction and the decay-in-flight by $\gamma$ emission of the residual nucleus by using GRETINA \cite{WEISSHAAR2017187}. Such measurements will be of impact for nuclear structure and decay studies, and enable the application of the charge-exchange Oslo method with unstable nuclei. 

At FRIB, information about isovector giant resonances that are excited in unstable nuclei was obtained by using the above-mentioned ($p$,$n$) \cite{PhysRevLett.107.202501,PhysRevLett.121.132501} and ($d$,$^2$He) \cite{PhysRevLett.130.232301} reactions in inverse kinematics. To reliably extract resonance parameters, the ($p$,$n$) reaction in inverse kinematics is the more likely candidate for early FRIB experiments due to the high luminosity that can be achieved even with modest beam intensities, see e.g., the extracted GT resonance strength distribution from a $^{132}$Sn($p$,$n$) experiment at RIBF. \cite{PhysRevLett.121.132501}.

Another opportunity for studying isovector giant resonances is by using rare-isotope beams as novel reaction probes for stable targets. By cleverly making use of rare-isotope beams, it is possible to isolate resonances that are otherwise very difficult to study. A good example is the ($^{10}$Be,$^{10}$B) reaction, in which a secondary $^{10}$Be beam is used. The measurement of the $\gamma$ decay in GRETINA from $^{10}$B excited states provides separate $\Delta S=0$, $\Delta T=1$ and $\Delta S=1$, $\Delta T=1$ filters in one experiment. The $^{10}$B detection in the S800 spectrometer is used to extract the excitation energy in the target nucleus. This probe was first used successfully to extract information about the Isovector Giant Monopole Resonance (IVGMR) in $^{28}$Al by using the $^{28}$Si($^{10}$Be,$^{10}$B$^*$(IAS)) reaction \cite{PhysRevLett.118.172501}. With the much higher beam intensities available at FRIB, studies of heavy nuclei will be possible. The ($^{10}$C,$^{10}$B) reaction can be used to investigate the $\beta^-$ direction \cite{Sasamoto2011,Sasamoto2012}, based on the same principle. 

Ultimately, the High Rigidity Spectrometer (HRS) \cite{HRS} at FRIB will be a game changer for charge-exchange experiments in inverse kinematics. Until the HRS is available, improvements to the detector systems used in combination with the S800 spectrometer will enhance experiments. Increased solid angle coverage of GRETINA and adding the ability to perform pulse-shape discrimination in LENDA to remove $\gamma$ background in ($p$,$n$) experiments are important enhancements. 

\subsection{In-beam gamma-ray spectroscopy}

In-beam gamma-ray spectroscopy is a work-horse capability for radioactive ion beam facilities across the globe.  In-beam spectroscopy enables studies of both nuclear structure and reactions from stability to the most exotic nuclear systems, leveraging techniques from direct reactions to Coulomb excitation to compound nuclear reactions to explore a broad range of nuclear properties.  At FRIB, in-beam $\gamma$-ray spectroscopy is enabled by a number of advanced detector systems.

The Segmented Germanium Array (SeGA) is an array of 18 single-crystal HPGe crystals each segmented with 32 outer surface contacts~\cite{SeGA}.  While limited in solid-angle coverage and thus efficiency, SeGA offers flexibility for varied installation configurations including close-packed geometries surrounding a target or configurations at larger distances with varied angles, allowing optimization of any given experiment.  The segmentation of the SeGA crystals allows good Doppler correction for in-beam spectroscopy, reducing the finite opening angle of the detectors.  SeGA is a versatile detection system which will continue to play an important role at FRIB in the next years.

For experiments which require high detection efficiency and can tolerate lower energy resolution (of order 10\% FWHM at 1~MeV), the scintillator array CAESAR (CAESium iodide ARray) is another important tool~\cite{WeisshaarNIM2010}.  CAESAR offers an efficiency of approximately 40\% at 1~MeV with 192 CsI(Na) scintillation crystals in a close-packed array with very high solid angle coverage.  For exotic light systems, where level density is low and experiments may only expect to populate a handful of excited states, CAESAR provides a powerful capability to push to very low production rates.

Finally, the Gamma-Ray Energy Tracking Array (GRETA) will be completed in 2026 with a full complement of 120 large-volume HPGe detectors housed in 30 Quad modules and covering approximately 80\% of the solid angle around a target position.  GRETA takes advantage of the 36-fold segmentation of each HPGe crystal, combined with digital signal processing and advanced computational capabilities to reconstruct the position of each gamma-ray Compton scatter or photoelectric interaction.  With interaction positions reconstructed with several mm accuracy, $\gamma$-ray tracking algorithms can cluster these together and identify the most likely scattering sequences, thus allowing rejection of background events which either did not originate at the target position or did not deposit their full energy.  GRETA will provide a world-unique capability at FRIB, with high-resolution ($\sim$0.2\% FWHM), high efficiency ($\geq 30$\% at 1.3~MeV) spectroscopic capability with excellent peak-to-total ($\geq 50$\% at 1.3~MeV) for experiments with fast and with re-accelerated beams.  Combined with advanced targets and a broad range of auxiliary devices, GRETA will be a key device for a large part of the FRIB science program.

\subsection{Charged-particle, beta and gamma decays}

%The Facility for Rare Isotope Beams (FRIB) will provide unprecedented access to exotic nuclei; approximately 80% of the isotopes predicted to exist up to uranium (Z = 92) will be produced \cite{Erl12,Afa13}. The FRIB Decay Station, which will be used to study the decay properties and structure of these isotopes and will impact our understanding of nuclear structure, nuclear astrophysics, fundamental symmetries, and isotopes of importance to applications.  The high sensitivity and relatively low beam-rate requirements of the various decay spectroscopy techniques, enable decay measurements at the very limits of the production capabilities at FRIB. In addition, for nuclei produced at higher rates, the FDSi will enable high-precision measurements for thorough characterization of emergent phenomena, which can be used to benchmark and differentiate between leading theoretical models.

The FRIB Decay Station Initiator (FDSi) is an efficient, granular, and modular multi-detector system designed under a common infrastructure \cite{FDS}. The FDSi brings multiple complementary detection modes together in a framework capable of performing decay spectroscopy with multiple radiation types over a range of beam production rates spanning ten orders of magnitude.  At the core of the FDSi is a system for stopping incoming radioactive ions and detecting their subsequent charged-particle decay emissions. Stopping systems include inorganic scintillators \cite{yoko2019} or large area segmented silicon detectors.  The choice between the two detector options depends the needs of the experimental investigation.  Additional detector arrays surround the central implantation detector to monitor delayed radiation in the form of photons and/or neutrons. Again, the specific configuration is adjustable to match the science goals of each experiment. The FDSi can be instrumented with either a 2$\pi$ or 4$\pi$ array of HPGe Clover detectors combined with LaBr/CeBr scintillators for precision timing measurements.  Neutron detection is accomplished with the NEXTi/VANDLE \cite{pete2016} neutron time-of-flight detection system. Thermal neutron detection is also possible with $^3$He gaseous counters.  Measuring the strength function following beta-decay can be accomplished using a total absorption spectrometer.  As part of the FDSi the Modular Total Absorption Spectrometer can be used.  MTAS is a one-ton volume of NaI scintillator detectors segmented parallel to the beam axis.  Separate from the FDSi, the Summing NaI (SuN) total absorption spectrometer is also available \cite{simo2013}. SuN is a 16 inch right cylinder of NaI segmented in eight separate detectors segmented perpendicular to the beam axis.  Both MTAS \cite{cox2024} and SuN \cite{ong2020} have been used with fast fragment beam to extract the strength distribution of exotic isotopes. The early science program of the FDSi will predominately focus on the structure of exotic nuclear systems in medium mass nuclei using a system with 2$\pi$ of HPGe Clover detectors combined with 2$\pi$ of time-of-flight neutron detection \cite{Crawford2022PRL,gray2023} before proceeding to heavier systems.  The development of the FDS will occur concurrently with the operation of the FDSi.

%{\color{red} experimental methods for beta decay} These need to be explained in a little more detail. Summing NaI detector (SUN) and the FRIB Decay Station initiator (FDSi) currently at FRIB. The full FRIB Decay Station (FDS) is under construction \cite{FDS}.

\subsection{Neutron decay}

The experimental study of neutron-unbound systems near and beyond the neutron drip line is generally based on invariant-mass spectroscopy, as the direct measurement of the core+xn system is not possible due to its extremely short lifetime. Invariant-mass measurements allow for the reconstruction of the relative energy spectrum from the energy and momenta of the coincident decay particles (charged fragment, neutron, and $\gamma$s). A key strength of this approach lies in the ability to probe correlations between decay particles, including neutrons. This makes invariant-mass spectroscopy a valuable tool to study exotic phenomena such as neutron halos or multi-neutron decays. 

At FRIB, the standard Sweeper-Mona setup is dedicated to invariant-mass studies. The separation of the charged fragments from the neutrons is provided by the 5.3 Tm dipole Sweeper magnet placed at 30 degrees from the beam axis. The Particle Identification (PID) of the fragment of interest is then determined from energy loss (Z selection), position tracking, and time-of-flight (A selection) measurements using a suite of charged particle detectors positioned at the focal plane of the magnet. The neutron is detected by the MoNA-LISA position-sensitive neutron detector~\cite{BaumannNIM2005}, placed at around 8 meters from the target. The momentum of the neutron is reconstructed from the hit position and the time-of-flight information between the MoNA-LISA array and a plastic scintillator positioned upstream of the target. In addition, the $\gamma$ detector CAESAR~\cite{WeisshaarNIM2010} can be used to identify decays to excited bound states in the charged fragment.

One of the main advantages of the kinematic complete measurement is the study of the correlations of the decay particles, including the correlations between the decay neutrons. An important limiting factor is, however, the neutron detection efficiency and resolution. Given that a neutron only loses a part of its energy in a single interaction, the same neutron can be therefore detected multiple times in the array, a phenomenon known as cross-talk. The current MoNA-LISA efficiency for single neutron detection is approximately 70\%. Yet, the efficiency drops significantly for multi-neutron events, primarily due to challenges in identifying and rejecting cross-talk events.

The dominating factor for the overall resolution of the reconstructed relative energy in the MoNA-Sweeper setup is the position resolution of the neutron detector. In order to overcome the constraints of the current design, the MoNA collaboration is actively developing a next-generation neutron detector. This new design aims to significantly enhance the precision in determining the vertex of neutron interactions by replacing the Photomultiplier Tubes with Silicon Photomultiplier arrays as readout technology. Additionally, the new neutron array will be arranged as a tile-like configuration, offering more flexibility in adjusting the active area to specific experiment requirements. 

% Kyle Brown
\subsection{Heavy-ion collisions}
\label{heavy-ion_collisions_exp_capability}

Heavy-ion collision experiments at FRIB will proceed in distinct phases, reflecting the availability of detector technology and beams. 

The first experiments at FRIB will rely on arrays of charged-particle and neutron detectors. The construction of a %$\sim$
nearly-4$\pi$ array of thin, fast plastic scintillators to infer the plane of the reaction is underway at MSU. This new detector will be added to the High Resolution Array (HiRA) \cite{HIRA} and upgraded Large Area Neutron Array (LANA) \cite{LANA}. HiRA is a small array of Si-Si-CsI(Tl) telescopes for detecting protons and other light, charged particles. Two versions of HiRA exist with 4 and 10 cm CsI(Tl) crystals, capable of stopping 115 and 200 MeV protons, respectively. The LANA neutron walls are comprised of liquid scintillator bars for measuring neutron energies by time-of-flight. These walls utilize pulse-shape discrimination to separate neutrons from gamma-rays, and have a thin, fast-plastic scintillator wall in front for vetoing of charged particles. These detector arrays will be utilized to measure neutron-to-proton ratios, directed and elliptic flow, and other observables sensitive to the nuclear equation of state.

The next phase of heavy-ion collision experiments at FRIB will focus more on ratios of charged pions, similarly to the S$\pi$RIT experiment at RIKEN ~\cite{SRIT:2021gcy,SpRIT:2020blg}. These experiments will require the development of a new Time Projection Chamber~(TPC). The most straightforward development path in this case is to construct a TPC similar to the Active-Target Time Projection Chamber and SOLARIS solenoid magnet~\cite{ayya2020} to run with 200 MeV/nucleon beams. A major task here will be the construction of an inner field cage with the goal of reducing the space charge effects, arising from the column of positive ions that builds up in the beam region from the ionized gas. Likely, a higher density pad plane will also be required to manage the relatively high charged-particle multiplicities in heavy-ion collisions. The modified time projection chamber can be used to measure both flow as well as the pion ratios.

The final phase of development currently planned for heavy-ion collision experiments at FRIB will utilize the High Resolution Spectrometer (HRS) and the upgrade of available beam energies to 400~MeV/nucleon with FRIB400. These experiments will require the development of a TPC that will fit into the gap of the first dipole magnet of the HRS. The planned, usable gap of the HRS dipole is 15 cm smaller than that of the SAMURAI magnet at RIKEN, which means that the S$\pi$RIT TPC cannot be utilized in this case without major modifications to the design, including designing the field cage, readout, and other detector components anew. Therefore, it will be likely more efficient to construct an entirely new detector. While there is an overlap between the science enabled by the HRS TPC and the solenoid TPC, described in the previous paragraph, the HRS TPC will have a higher efficiency for pion identification and will allow for coincident neutron detection.

% Daniel Bazin
\subsection{Re-accelerated radioactive beams}
\label{ReA_beams_exp_capability}

A unique capability at FRIB is the possibility to post accelerate radioactive beams produced via fragmentation. This scheme enables the production of high quality beams at lower energies than the typical fragmentation regime, regardless of the chemical configuration of the selected ions. The main scientific goals of this capability involve the use of reactions to explore the properties of radioactive nuclei from structural, astrophysical and reaction point of views.

In its present status the ReA linear accelerator is capable to accelerate charged radioactive ions at energies ranging from 0.3 to 10 MeV/u with emittance qualities typical of primary beams. An upgrade to higher energies is planned to augment the energy reach to about 20 MeV/u \cite{ReAWhitePaper}, depending on the mass-to-charge ratio. The instruments making use of these beams are the Solenoidal Spectrometer Apparatus for Reaction Studies (SOLARIS) \cite{SOLARIS, CHEN2024138678}, the Separator for Capture Reactions (SECAR) \cite{SECAR}, and the JANUS setup \cite{Lun18a}. In addition, general purpose beam lines are available to accommodate other instruments.

SOLARIS is a dual mode solenoidal spectrometer that can be used either with a Silicon array detection setup similar to the HELIOS spectrometer \cite{HELIOS}, or with the AT-TPC \cite{ayya2020}. The main purpose of this instrument is to gain access to the structure of radioactive isotopes via simple, one-step and momentum-matched reactions in inverse kinematics. The Silicon array mode is suitable for beam intensities of 10$^4$ pps and above, and provides excellent missing mass resolution approaching 100 keV and a large solid angle coverage. SOLARIS provides arrays both backward and forward relative to the target location, hence covering multiple reaction channels simultaneously. The AT-TPC mode extends the reach to use similar reactions at very low intensities, down to 100 pps. This is possible thanks to the 2 to 3 orders of magnitude gains provided by the large thickness and solid angle coverage of the AT-TPC, without degradation of the energy and angular resolutions. In addition, the AT-TPC can be used with the high quality beams of ReA to measure excitation functions in a very efficient manner, simply by using the energy loss of the beam traveling through the gas and the recorded vertex location of each reaction. This opens the door to resonance scattering experiments on radioactive nuclei. Another capability of the AT-TPC is possibility to detect and measure the particle decay of unbound resonances and reconstruct their excitation spectrum using invariant mass methods. This is particularly interesting for the study of threshold resonances.

The SECAR separator is primarily dedicated to the study of (p,$\gamma$) and ($\alpha$,$\gamma$) capture reactions in inverse kinematics on radioactive nuclei at low energy, directly related to the study of hydrogen and helium burning in explosive stellar environments. It is equipped with a windowless target system and is designed with acceptances large enough to transmit the heavy recoils without significant loss. The primary beam rejection design goal of this instrument is 10$^{-13}$, with an additional 10$^{-4}$ rejection provided by the focal plane detection system. This extremely high rejection should allow the measurement of resonance strength down to 1 $\mu$eV. 

The JANUS setup is dedicated to the measurement of Coulomb cross sections induced by a heavy target on radioactive beams. In the low energy domain provided by ReA, multiple excitations can be used to reach higher excited states, and characterize their degree of collectivity from the deduction of quadrupole moments. It is composed of a set of silicon strip detectors that detect the scattered beam particles, while their Doppler-shifted $\gamma$-ray decays are detected by the surrounding Segmented Ge Array (SeGA). Once it is complete, the GRETA array \cite{greta} can be used instead, and will provide greatly enhanced sensitivity.

%______________%
\section{Structure\label{sec:Structure}}
 %% Please remember to add a comment with your name/initials
%% above any content that you add. -SRS

Experiments at FRIB will provide unprecedented access to properties of neutron-rich nuclei out to the neutron dripline. Connections between theory and experiment have been discussed in recent reviews \cite{Otsuka2020,nowa2021,brow2022p525,busk2024}. The following sections discuss open theoretical  questions in nuclear structure and how they relate to  the variety of experimental data.

In the study of nuclear structure, the goal is to understand the properties of nuclei by identifying the relevant degrees of freedom, and by connecting these properties to the underlying interaction between nucleons.
The relevant degrees of freedom could be collective modes like clustering, superfluidity, vibrations and deformations; they could be related to a set of active orbitals defined by shell gaps; or they could be related to the proximity of the continuum.
The main new information provided by FRIB will relate to how these features are intertwined and modified in very neutron-rich nuclei, and how these various degrees of freedom interplay with the continuum.
Historically, the various degrees of freedom have been understood phenomenologically, \textit{e.g.}, by putting in a single-particle spin-orbit potential ``by hand”; however these features must arise from the underlying nuclear forces. Studying neutron-rich systems will help to lift the degeneracy of various explanations of how this occurs. For example, from an \textit{ab initio} standpoint, the one-body spin-orbit potential gets contributions from both two-body and three-body interactions. Due to the Pauli principle, the three-neutron force is well understood, hence, changes in the effective spin-orbit potential as neutrons are added can illuminate its microscopic origin.

First, we examine the evolution of shell structure
with examples from several regions of the nuclear
chart. Following this, there are separate sections
on collectivity, isospin symmetry, spin-isospin response,
and nuclear pairing.
Notations and abbreviations for the shell-model orbitals and commonly used 
truncations (model spaces) are described in the Appendix.

\subsection{Evolution of shell structure}

%\textcolor{red}{Add 1-2 sentences about what is meant by ``shell structure''.}
In nuclear physics, ``shell structure'' refers to the independent-particle picture in which nucleons move in an effective mean-field potential and groups of degenerate or quasi-degenerate single-particle orbitals are separated by energy gaps.
The number of nucleons required to fill all orbitals below one of these gaps is called a ``magic number''.
Many of the regularities of nuclear data can be qualitatively understood in terms of these magic numbers.
In addition, quantitative configuration-interaction (CI) calculations make use of magic numbers 
to define
the model space to be used for a given region 
of the nuclear chart.
For example, the $  0d_{5/2}, 1s_{1/2}, 0d_{3/2}  $ orbitals define
the $  sd  $ model space as a starting point for calculations
of the properties of nuclei between the
magic numbers 8 and 20.
The textbook magic numbers 2, 8, 20, 28, 50, 82, and 126 apply to nuclei in the valley of stability.
Moving away from stability, the mean field changes and shell gaps disappear or appear in new places, generating new magic numbers.
This phenomenon is called shell evolution, and it has been argued that the primary drivers of shell evolution are the monopole component of the tensor force~\cite{otsu2005p232502,Otsuka2020} and continuum effects~\cite{Doba07,Mich10}.
FRIB is especially well-suited to explore the latter. 
%Hamiltonians used for configurations of the orbitals inside the model space
%are renormalized to take into account
%interactions with the configurations involving orbitals
%outside of the model space.

\begin{figure}[th]
\includegraphics[scale=0.7]{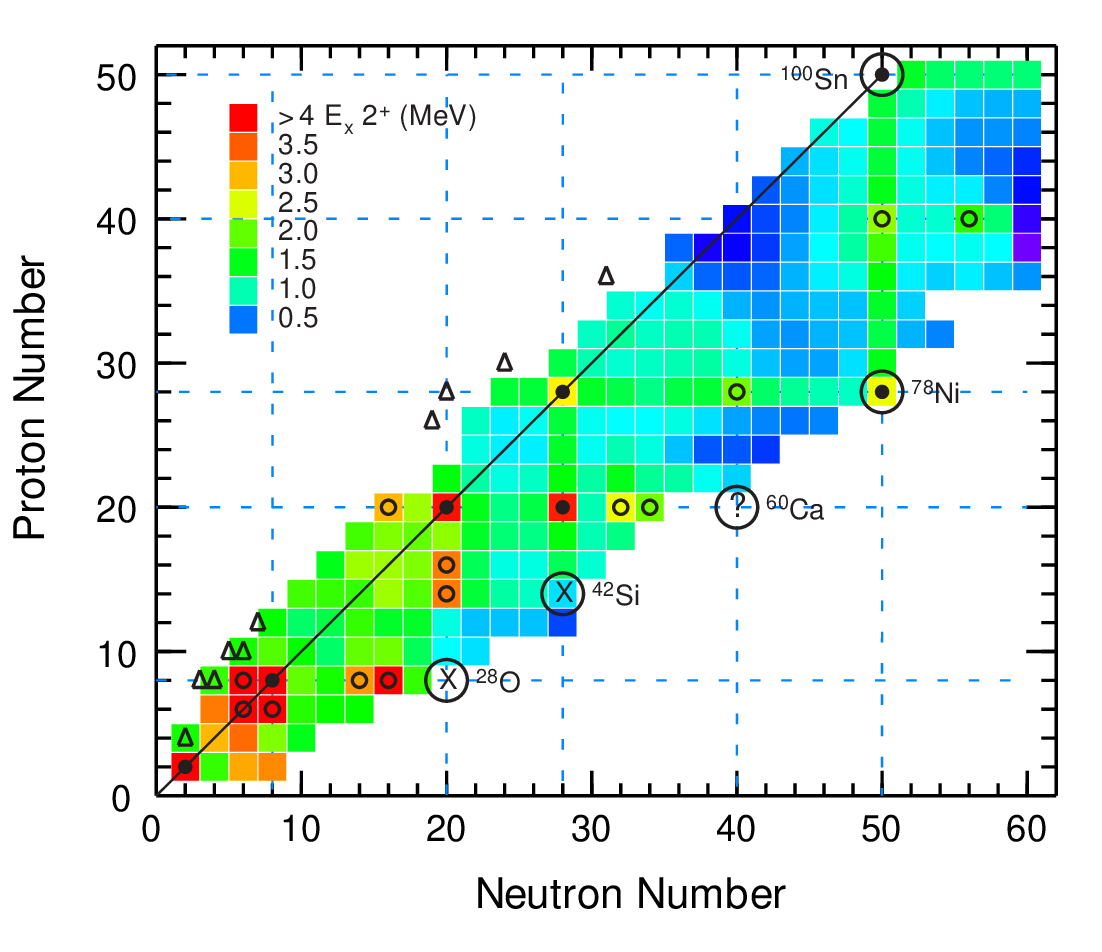}
\caption{Even-even isotopes in the lower region of the nuclear chart.
The colors indicate the energy of the first 2$^{ + }$ state.
The filled black circles show the doubly-magic nuclei
associated with the magic numbers 8, 20, 28  and 50.
The small open circles show the doubly-magic nuclei associated with
the $  j  $-filled numbers 14, 16, 32, 34 and 40.
The large open circles indicate the nuclei near the
driplines that are the focus of this introduction.
Two open circles with crosses indicate that these nuclei
are observed to be in islands of inversion.
The open circle with ``?'' for $^{60}$Ca
indicates that its properties are not yet known.
The triangles are those nuclei observed to decay by two-proton emission.
Adapted from ref.~\cite{brow2022}.
}
%\label{(1)}
\label{fig:E2+Chart}
\end{figure}

%There are many properties of nuclei that 
%are connected shell structure and how it evolves
%from the valley of stability to the neutron drip line.
Shell structure is not a directly observable property; it is a framework in which to interpret and understand experimental data.
Consequently, various nuclear properties should be considered in order to develop a coherent picture of evolving shell structure.
Important signals of a shell gap include a relatively high energy of the first excited $2^+$ state (for even-even nuclei); a relatively small electric quadrupole transition strength $B(E2;2^+ \! \to\! 0^+)$; a gap in the effective single-particle energies obtained from a combination of energy levels and spectroscopic factors; and relatively large jumps in differential quantities derived from binding energies and radii.

%Fig. \ref{fig:E2+Chart} shows the energies of 2$^{ + }_{1}$ states in the region $N,Z\lesssim 50$. A relatively high 2$^{ + }$ energy is indicative of an underlying magic number (a large energy gap).
%(see Appendix I).

Fig. \ref{fig:E2+Chart} shows the energies of 2$^{ + }_{1}$ states in the region $N,Z\lesssim 50$.
%As observed from the 2$^{ + }$ energies in  Fig. \ref{fig:E2+Chart},
It is evident that
the $  N=20  $ magic number disappears below $  Z=14  $.
This region has been called an ``island of inversion'' \cite{warb1990}.
Most nuclei with proton and neutron numbers between 8 and 20
have ground and low-lying states that are described by
configurations within the  $  sd  $ model space.
All of these nuclei also have excited ``intruder'' states
that are described by configurations involving
the excitation of nucleons from $  p  $ into $  sd  $ or $  sd  $ into $  fp  $ orbitals.
For nuclei inside the island of inversion, this order
is inverted (the intruder configuration is the ground state).
Other islands of inversion have been identified in the regions around 
$  N=8  $ for $Z<5$,
$  N=28  $  below $^{48}$Ca, and $  N=40  $  below $^{68}$Ni
\cite{brow2010p104}.
Based on CI calculations, a 5$^{th}$ island of inversion is predicted 
below $^{78}$Ni \cite{nowa2016}.
In general, whether a nucleus is inside the island of inversion depends on the competition between shell structure and correlation energy~\cite{Poves2018}.
The connection to shape-coexistence has been recently reviewed in ref.~\cite{bona2023}.
%The change from normal to inverted energy order comes
%from the enhanced proton-neutron interaction driven
%deformation energy
%below $  Z=14  $ \cite{pove1987}, \cite{nowa2021}, as well as a narrowing of the
%of the $  N=20  $ energy gap between the 
%effective single-particle energies (ESPE) of the
%$  0d_{3/2}  $ and the $  (0f_{7/2},1p_{3/2})  $ neutron orbitals.
All of these islands of inversion lie in neutron-rich regions of nuclei that will
be studied at FRIB.
%\textcolor{red}{Tie this back to FRIB.}

\begin{figure}
\includegraphics[height=7cm]{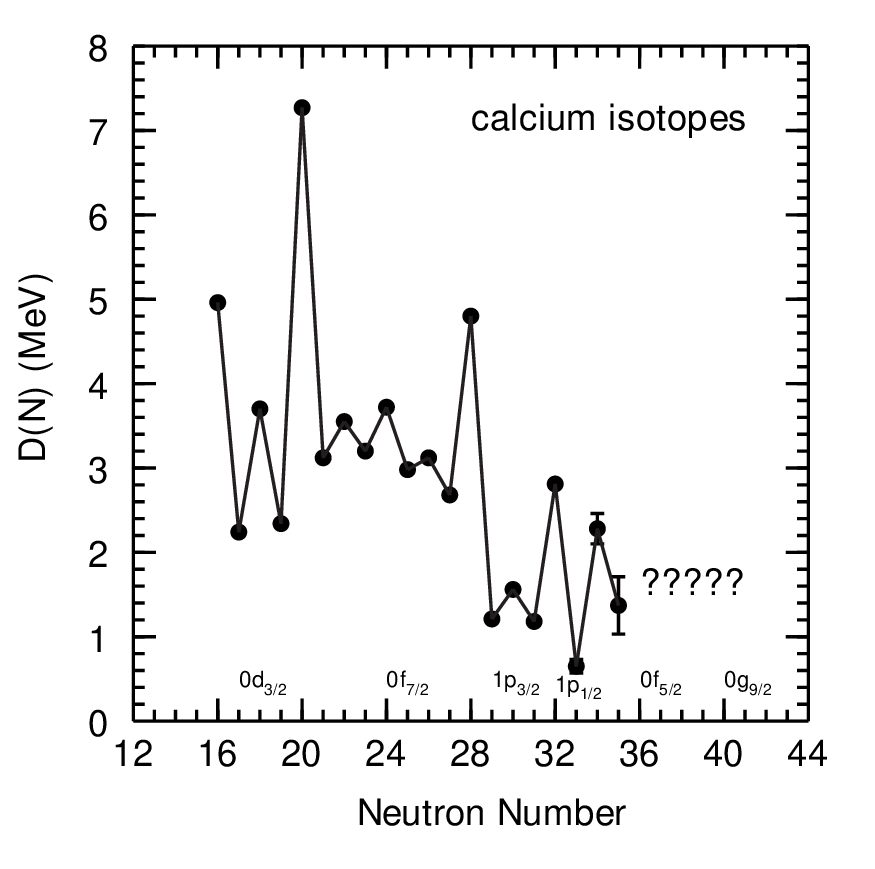}
\includegraphics[height=7cm]{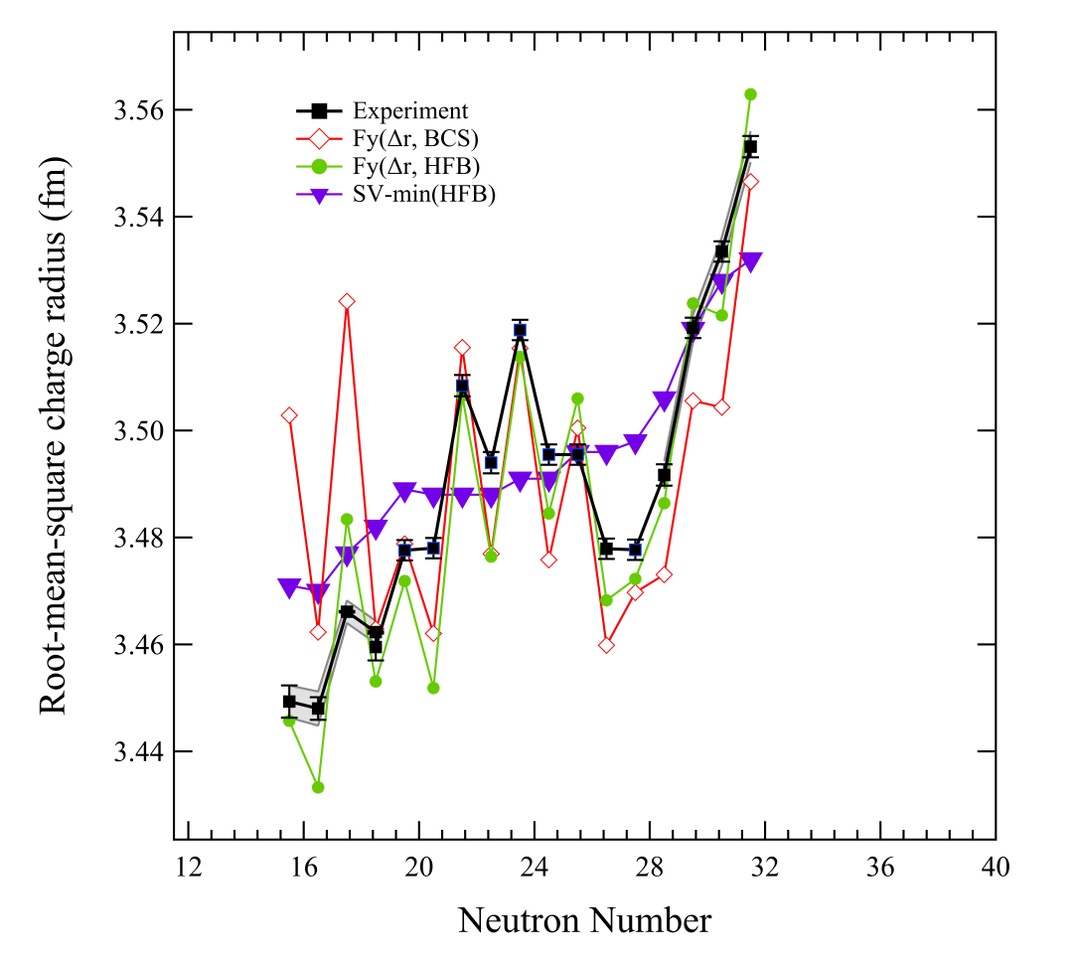}
\caption{  Left: Experimental $D(N)  $ for $  Z=20  $ given by Eq. (1). The black dots with error bars
are the experimental data. The orbitals that
are being filled are shown (adapted from ref.~\cite{brow2022}).
Right: Experimental rms charge radii for the calcium isotopes, compared to theoretical calculations~\cite{mill2019}.
}
%\label{(2)}
\label{fig:CaDN}
\end{figure}

Another experimental signature of magic numbers is given by the double difference in the binding energies (BE) defined by 
\begin{equation}\label{eq:Dq}
D(q) = (-1)^{q} [2 \, \, {\rm BE}(q) - {\rm BE}(q+1) + {\rm BE}(q-1)],
\end{equation}
for isotopes ($  q=N  $ with $  Z  $ held fixed) or isotones ($  q=Z  $ with $  N  $ held fixed). The example of $  D(N)  $ for the calcium isotopes is shown in Fig. \ref{fig:CaDN}, 
revealing peaks at various (sub)shell closures.
%where peaks at the magic numbers 16, 20, 28, 32 and 34 are observed.
The corresponding shell-model orbitals are indicated at the bottom of 
 Fig. \ref{fig:CaDN}.
 %The most recent datum related to these points in Fig. \ref{(2)} is for the mass of $^{35}$Ca \cite{lala2023}.
% It is clear from the $  E_{x}(2^{+})  $ data and from the $  D(q)  $ data that energy gaps associated with some magic numbers are smaller than others.
 %In \cite{brow2022p525} it was shown that the height of the $  D(N)  $ peaks at the magic number is close to the shell gap at those magic numbers.
 %The average values of $  D(N)  $ between the magic numbers are the pairing energies.
 %It was noted \cite{brow2022p525} that the drop of the pairing energy at $  N=33  $ was related to the filling of the low-$  j  $ $  1p_{1/2}  $ orbital at $  N=33  $.
 The row of ``?'' beyond $  N=36  $ indicates one of the regions of nuclei to be explored by FRIB. One will learn how pairing depends on the neutron excess, whether or not $^{60}$Ca
 is a doubly-magic nucleus, and to what extent the nuclear landscape extends beyond $^{60}$Ca.
 %\textcolor{red}{Specifically what questions will this answer?}

Likewise, the changes in nuclear charge radii contain important 
experimental signatures of shell structure. As an example, the charge radii of the calcium isotopes are shown in Fig. \ref{fig:CaDN},
displaying several important features.
In particular, there is an obvious odd-even oscillation which has been variously attributed to pairing effects~\cite{rein2017} and to quadrupole deformation~\cite{caur2001,brow2022}.
In addition, there are changes in the slope of the charge radii at $N=20$ and $N=28$
that are associated with the change in the 
nominally  filled neutron valence
orbitals in the sequence  $0d_{3/2}$, $0f_{7/2}$ and $1p_{3/2}$.
These slope changes in the evolution of charge radii,  
which appear in all isotopes 
at the magic numbers, provide 
a challenging test for nuclear structure models.
%\cite{brow2022}.
Experimental data for the charge and matter radii for the
heavier calcium isotopes will provide information on the neutron skin
with its linkage to the nuclear equation of state.

\begin{figure}
\includegraphics[height=7.5cm]{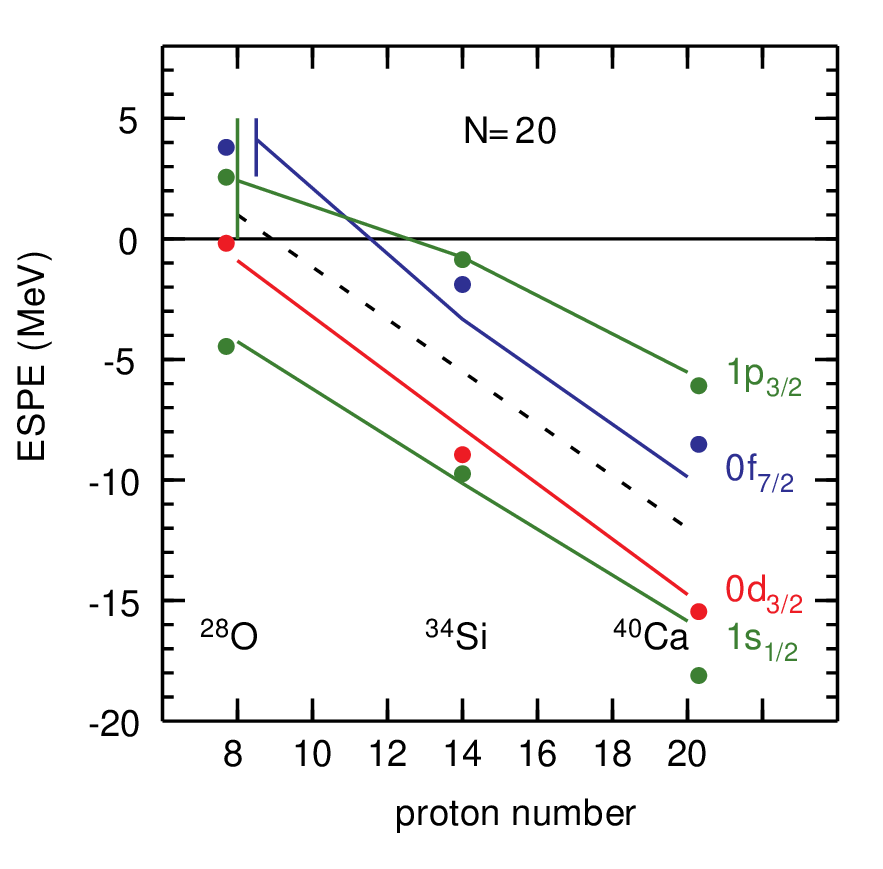}
\includegraphics[height=7.5cm]{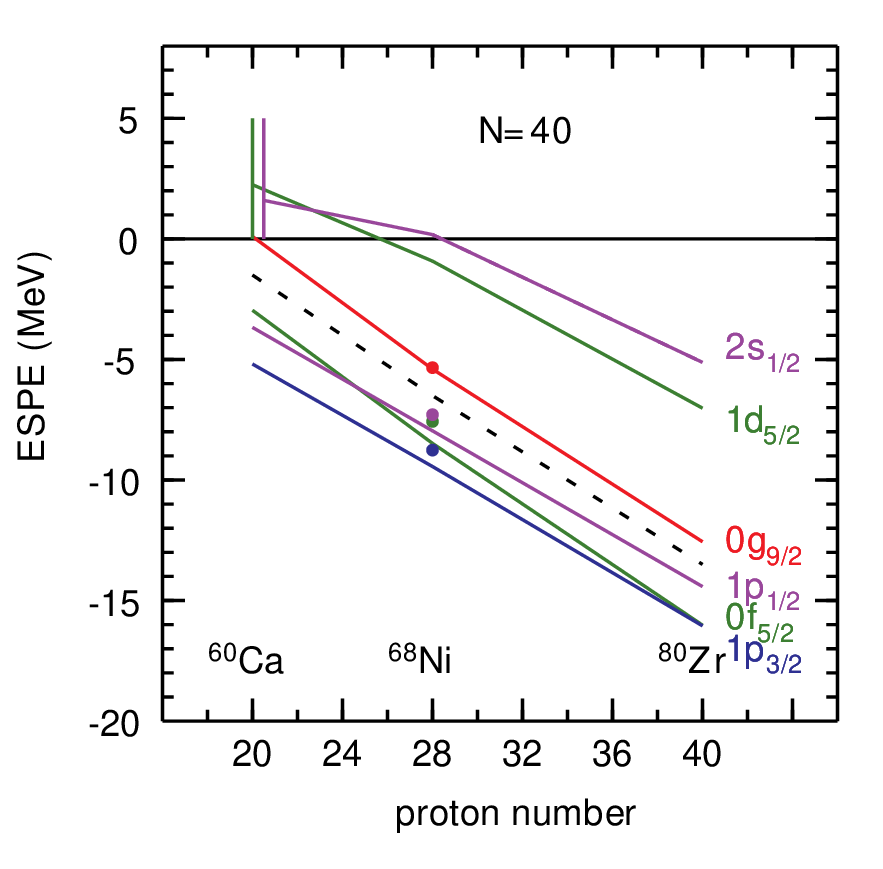}
\caption{Left: $N=20$ neutron effective single-particle energies as a function of proton  number, computed with the Skx functional~\cite{brow1998} (solid lines) and the FSU shell-model interaction\cite{lubn2019,lubn2020} (filled circles).
The neutron orbitals below the dashed line are filled.
The vertical lines at $N=8$ indicate orbitals
in the continuum.
Right: Same, but for $N=40$, using Skx (lines) and the jj44a Hamiltonian~\cite{lise2004} (circles).}
\label{fig:ESPE_N20_N40}
\end{figure}

On the theory side, shell evolution can be visualized by  calculating ESPE as a function of $N$ or $Z$.
Neutron ESPE around $N=20$ and $N=40$ as a function of proton number are shown in Fig.~\ref{fig:ESPE_N20_N40}.
%Calculations for the neutron ESPE below and above $  N=20  $ as a function of proton number are shown in Fig. \ref{fig:ESPE_N20_N40}.
Representative CI results are obtained with the FSU~\cite{lubn2019,lubn2020} and jj44a~\cite{lise2004} Hamiltonians for $N=20,40$, respectively.
These are compared with the results from a representative energy-density functional using the Skx interaction~\cite{brow1998}.
%CI results are represented 
%by those obtained with the FSU Hamiltonian.
%The FSU ESPE are compared to those from a "typical"
%energy-density functional calculation obtained with 
%the Skx Skyrme-type interaction \cite{brow1998}.
The difference in the $N=20$ gap predicted by FSU and Skx can be traced partly to the strong tensor component of the FSU interaction, which has the effect of repelling the $0d_{3/2}$ and $0f_{7/2}$ ESPE~\cite{otsu2020}.
%The $N=20$ FSU shell gap at $^{34}$Si is larger than
%that obtained with Skx partly because the
%contribution from the tensor interaction contained in
%the FSU Hamiltonian is maximized and has the effect of repelling the $0d_{3/2}$ and $0f_{7/2}$ ESPE
%\cite{otsu2020}. 
%
Moving from  $^{34}$Si 
to $^{28}$O, the $0p_{3/2}$ ESPE bends over and moves into the continuum (the vertical line), which generally tends to reduce the energy of low-$\ell$ orbits. The combination of tensor
and continuum effects reduce the 
$N=20$ gap for nuclei from about 6 MeV in $^{34}$Si
to about 3 MeV in $^{28}$O,
resulting in the island of inversion below $^{34}$Si.
A similar story plays out at $N=40$, shown on the right panel of Fig.~\ref{fig:ESPE_N20_N40}, but there are interesting differences related to the different $\ell$ values involved. Comparing $  0g_{9/2}  $ with $  0f_{7/2}  $, the smaller gap at $  N=Z=40  $ relative to $  N=Z=20  $ results in a deformed $^{80}$Zr ground state compared to a spherical $^{40}$Ca ground state ($^{40}$Ca has low-lying deformed excited states).
The $  0g_{9/2}-1d_{5/2}  $ ESPE gap is larger than that for $  0f_{7/2}-1p_{3/2}  $, which is why $^{100}$Sn is a better doubly-closed-shell nucleus than  $^{56}$Ni. Despite this relatively large gap, CI calculations need to explicitly include the  $  1d_{5/2}  $ in the model space to obtain the large proton-neutron deformation energy in the region of $^{80}$Zr and in the $  N=40  $ island of inversion below $^{68}$Ni.
Continuing towards  $^{60}$Ca, the low-$\ell$ $2s_{1/2}$ and $1d_{5/2}$ orbits enter the continuum, closing the $N=50$ gap.
For $^{60}$Ca, these Skyrme
ESPE are very different from those obtained with the
LNPS Hamiltonian~\cite{lenz2010} where the $  0g_{9/2}-1d_{5/2}  $
gap is close to zero (see Fig. 1 in \cite{lenz2010}).

The calculations shown in Fig.~\ref{fig:ESPE_N20_N40} do not explicitly contain continuum components,
but the ESPE implicitly contain effects of the continuum because the interactions are fit to data.
In the regions around $^{42}$Si, $^{60}$Ca, and $^{78}$Ni, the experimental data required to constrain empirical Hamiltonians (or EDFs) is limited.
This will be filled in by new FRIB experiments.
Single-particle models of
nucleons in a finite potential well can be used as a qualitative
guide to energies and neutron decay widths,
but ultimately,
\textit{ab initio} Hamiltonians must be constructed in a basis
that explicitly contains the continuum.
%A separate (section) of this whitepaper is devoted to theoretical models for the single-particle and many-body states in the continuum.
Section~\ref{sec:Continuum} of this whitepaper is devoted to the physics of the single-particle and many-body states in the continuum.

Experimentally, shell structure can also 
be illuminated by reactions probing the single-particle structure of nuclei.
Such reactions include light-particle reactions such as $(d,p)$, performed in inverse kinematics at FRIB, and knockout reactions at fragmentation-beam energies~\cite{Hansen2003}.
The cross sections for these reactions can be used to infer spectroscopic factors and experimental ESPEs, though some challenges remain in the interpretation of these reactions (see section~\ref{sec:Reactions}).

Below, we highlight several ``outpost'' nuclei, indicated by circles in Fig.~\ref{fig:E2+Chart}.
These nuclei involve relatively simple configurations that can be connected with
experiments and used to assess the theoretical models.
Some specific questions of interest for each of these
regions are also discussed.
%The nuclei in the large open circles in Fig. \ref{fig:E2+Chart} serve as ``outposts''
%for theoretical calculations over this region because they involve
%relatively simple configurations that can be connected with
%experiments and used to asses the theoretical models.
%Some specific questions of interest for each of these
%regions are discussed in the following subsections.

\subsubsection{The region around \texorpdfstring{$^{28}$O}{28O}: Neutron clustering}

Due to the closed proton shell, the dominant correlations in the neutron-rich oxygen isotopes are among the neutrons.
$^{28}$O lies beyond the neutron dripline, and has recently
been observed to decay by emission of four neutrons with a total
energy of 0.46(5) MeV \cite{kond2023}.
The phenomenological interpretation \cite{kond2023} is that the
the dominant configuration for the $^{28}$O ground state
has two neutrons excited from the $  sd  $
to the $  fp  $ shell, placing it in the $N=20$ island of inversion.

The charge rms radii, $R_{\rm ch}$, for oxygen and fluorine isotopes
have been calculated with \textit{ab initio} \cite{lapo2016} and relativistic mean-field theory \cite{shuk2011}, among others. However, experimental data on the ground-state $R_{\rm ch}$ and electromagnetic moments for these isotopes are not yet available. Laser spectroscopy measurements for oxygen and fluorine isotopes are planned to be performed at the BECOLA facility at FRIB. 
The obtained data will be critical for an understanding of structure and continuum effects around $^{24}$O and to benchmark theory in a continuum-imacted region of nuclei most accessible to \textit{ab initio} type calculations. 
%\textcolor{red}{[SRS: This seems vague.]}

\subsubsection{The region around \texorpdfstring{$^{42}$Si}{42Si}: Quadrupole plus continuum}

%Experimental information on nuclei near $^{42}$Si
%is very incomplete.
The nucleus $^{42}$Si lies in the chain of the
$  jj  $ magic nuclei $^{132}$Sn and $^{78}$Ni,
but the ESPE gaps at $  Z=14  $ and $  N=28  $
are not large enough to provide magicity.
As with the $N=20$ island of inversion, both proton and neutron gaps are small, so quadrupole correlations become important.
There
are rapid shape changes in these nuclei \cite{tsun2020};
CI calculations predict a positive $2^{ + }$ $^{42}$Si quadrupole
moment, indicative of a prolate shape \cite{brow2022p525}, and a negative $2^{ + }$ quadrupole moment for 
 $^{40}$Mg. 
%Below $^{42}$Si, $^{40}$Mg has been observed \cite{craw2019}.

Due to  the lack of experimental information,
the empirical Hamiltonians for the region of $^{42}$Si
are not well established. For the two
most widely used Hamiltonians,
SDPF-MU  \cite{utsu2012} and SDPF-U-Si \cite{nowa2009},
the predictions  for excited states above the 2$^{ + }$ in $^{42}$Si
are very different \cite{tost2013,gade2019}.
In the region of $^{42}$Si near the neutron dripline, the
ESPE for the $  1p_{3/2}  $ and $  1p_{1/2}  $ orbitals
will be influenced by the proximity to the continuum,
%The importance of transfer reactions to determine
%the ESPE are discussed in (section).
which is also important for describing 
two-neutron halos around $N=28$ \cite{sing2024}.
%Ab-initio Hamiltonians that explicitly take into account the
%continuum for need to be further developed.

The first published FRIB experiment established
new $\beta$-decay lifetimes for the nuclei in the region
around $  N=28  $ towards the neutron dripline
\cite{Crawford2022PRL}.
The results were in reasonable agreement with calculations
\cite{yosh2018p054321}
based on the SDPF-MU  \cite{utsu2012} Hamiltonian.
These $\beta$ decays involve first-forbidden $\beta$
transitions to low-lying states and Gamow-Teller transitions
to excited states that cross the $Z=20$ shell gap starting around 3 MeV in excitation
(see Fig. 3 in \cite{yosh2018p054321}).
The second FRIB experiment with the FDSi
studied the detailed $\gamma$ and neutron decay spectra of $^{45}$Cl to establish the $\beta$ branching 
to states up to 8 MeV in $^{45}$Ar \cite{cox2024}.
Compared to theoretical results with the SDPF-MU
Hamiltonian, it was found that the $Z=20$ shell
gap needed to be reduced by about 1~MeV to describe the
$\beta$-decay strength function.
This type of interplay between experiment and theory will be important for future FRIB experiments and advances in the modeling of nuclei.

\subsubsection{The region around \texorpdfstring{$^{60}$Ca}{60Ca}: Pairing in the continuum}

%\textcolor{red}{Need a clear 1-2 sentence explanation of why it is interesting whether $^{60}$Ca is magic. This is a flagship nucleus for FRIB.}
Like $^{28}$O, the nucleus $^{60}$Ca has a nominally closed-shell proton configuration, and so the dominant collective mode should be pairing.
One difference from $^{28}$O is that, primarily due to the mean-field spin-orbit potential, the $N=40$ gap is weaker to begin with and there is a higher density of states near the Fermi surface.
This could lead to a strong interplay between collective pairing and continuum effects (see section~\ref{sec:pairing} below).

The nucleus $^{60}$Ca has been observed to be stable against neutron decay in its ground state \cite{tara2018},
but the properties required to determine whether or not
$  N=40  $ is a magic number for $^{60}$Ca
are not yet known. In the region of $^{60}$Ca, there are many theoretical extrapolations and predictions
for the location of the neutron dripline \cite{erler12_1297,ravl2023,neufcourt2019neutron,stro2021p022501}
and the properties of excited states of nuclei in the region of $^{60}$Ca \cite{lenz2010,hage2012,hage2013,herg2014,hage2016,holt2014,magi2021,chen2023}.
The ESPE gap between the $  0f_{5/2}  $ and $  0g_{9/2}  $
orbitals is the most important unknown quantity.
Lenzi {\it et al.} \cite{lenz2010} have used the LNPS Hamiltonian to extrapolate the neutron
ESPE from $  Z=28  $ down to $  Z=20  $. Their $  0f_{5/2}-0g_{9/2}  $ ESPE gap for $^{60}$Ca is close to zero
(see Fig. 1 in Ref. \cite{lenz2010}) with the implication that
$^{60}$Ca will not exhibit $  N=40  $ doubly-magic features. The recent calculations of Li for the $N=40$ and $N=50$ islands of inversion \cite{li2023} result in $^{60}$Ca and $^{68}$Ni having similar doubly-magic properties,
but the ESPE near used $^{60}$Ca in \cite{li2023} do 
not include the continuum properties observed in Fig.~\ref{fig:ESPE_N20_N40}.

%\begin{figure}
%\includegraphics[scale=0.8]{babn40.eps}
%\caption{$N=40$ neutron effective single-particle energies (those obtained with closed-shell configurations for $^{60}$Ca, $^{68}$Ni and $^{80}$Zr)  as a function of proton  number.
%The lines are based on calculations with the the Skx EDF \cite{brow1998}.
%The filled circles are based on calculations with the 
%jj44a Hamiltonian \cite{lise2004}.
%The neutron orbitals below the dashed line are filled.
%The vertical lines at $N=20$ indicate  orbitals
%in the continuum.}
%\label{(4)}
%\end{figure}

The excitation energy of 2$^{ + }$ states in $^{56,58}$Ca
have recently been reported \cite{chen2023}. If $^{60}$Ca had a closed shell for $  N=40  $,
one would expect the 2$^{ + }$ states of $^{56,58}$Ca, coming from the
relatively isolated $  (0f_{5/2})^{n}  $ configurations, to have about the same
excitation energy \cite{magi2021}. The measured energies \cite{chen2023} fall from
1456(12) keV in $^{56}$Ca to 1115(34) keV in $^{58}$Ca. It
was concluded \cite{chen2023} that
this was caused by a small $  0f_{5/2}-0g_{9/2}  $ ESPE gap.
The $\gamma$-ray transition for $^{58}$Ca only has about four counts above the background in the spectrum 
(Fig. 1 in \cite{chen2023}) and the experimental evidence for this 2$^{ + }$ state needs to
be strengthened. $^{61,62,63}$Ti are observed \cite{wimm2019} to lie in the
$  N=40  $ island of inversion  below $^{68}$Ni.
%
%
%
%Fig. \ref{fig:ESPE_N20_N40} shows results of the Skx Skyrme EDF \cite{brow1998} calculations for the effective single-particale energies (ESPE) below and above $  N=40  $ analogous to those shown in Fig. \ref{fig:ESPE_N20_N40} for $  N=20  $. There are interesting differences related to the different $\ell$ values involved. Comparing $  0g_{9/2}  $ with $  0f_{7/2}  $, the smaller gap at $  N=Z=40  $ relative to $  N=Z=20  $ results in a deformed $^{80}$Zr ground state compared to a spherical $^{40}$Ca ground state ($^{40}$Ca has low-lying deformed excited states.) The $  0g_{9/2}-1d_{5/2}  $ ESPE gap is larger than that for $  0f_{7/2}-1p_{3/2}  $. This is reason why $^{100}$Sn is a better doubly closed-shell nucleus than  $^{56}$Ni. Although there is a relatively large $  0g_{9/2}-1d_{5/2}  $ gap, the inclusion of $  1d_{5/2}  $ in the model space is needed to obtain the large proton-neutron deformation energy in the region of $^{80}$Zr and in the $  N=40  $ island of inversion below $^{68}$Ni see (section) appendix.
%Due to the higher $\ell$ value, the $  1d_{5/2}  $ ESPE near $^{60}$Ca,
%(Fig. \ref{(4)})
%does not bend down as much as the $  1p_{3/2}  $ ESPE does near $^{28}$O
%(Fig. \ref{(3)}). Near $^{60}$Ca the $  1d_{5/2}  $ 
%$  2s_{1/2}  $ orbitals
%are in the
%continuum.
%For $^{60}$Ca, these Skyrme
%ESPE are very different from those obtained with the
%LPNS Hamiltonian \cite{lenz2010} where the $  0g_{9/2}-1d_{5/2}  $
%gap is close to zero (see Fig. 1 in \cite{lenz2010}).
A 9/2$^{ + }$ isomeric state has been reported in
$^{63}$Ti \cite{wimm2019}. One needs much more experimental spectroscopic
information related to the $  0f_{5/2}  $, $  0g_{9/2}  $
and $  1d_{5/2}  $ orbitals for $  N>36  $ near $^{60}$Ca
to develop a clearer picture of the structure in this region of rapid shell evolution.
 %to test and improve theoretical models for this region.

\subsubsection{The region around \texorpdfstring{$^{78}$Ni}{78Ni}: A \texorpdfstring{5$^{th}$} island of inversion?}

The doubly-magic properties of $^{78}$Ni have been established by the observation of the 2$^{ + }$ state at 2.60 MeV \cite{tani2019}. Energy spacings of proton ESPE have been deduced from the $\gamma$-ray decays of low-lying states in $^{79}$Cu \cite{oliv2017}. Based on the excited-state energy systematics for $  Z=29  $, one expects the ground state of $^{79}$Cu to have $  J^{\pi }  $ = 5/2$^{-}$ with an excited state of $  J^{\pi }  $ = 3/2$^{-}$ at 0.66 MeV. This establishes the approximate spacing between the  $  0f_{5/2}  $ and $  1p_{3/2}  $ proton ESPE. The second excited state observed at 1.51 MeV would then be expected to have $  J^{\pi }  $ = 1/2$^{-}$ and be related to the ESPE for the $  1p_{1/2}  $ orbital. The absolute SPE for these states and other SPE around $^{78}$Ni require mass measurements for $^{77,78,79}$Ni, $^{79}$Cu and $^{77}$Co. Below $^{78}$Ni, CI calculations predict $^{76}$Fe to be the start of a 5$^{th}$ island of inversion with the breaking of the $  N=50  $ magic number below $  Z=28  $ \cite{nowa2016}.
The structure physics involves
configurations built on 2p-2h (and higher) neutron excitations
across $N=50$. This configuration is predicted to
start as an excited state band below 3 MeV $^{78}$Ni,
and then is lowered in energy to become
deformed  ground-state bands (with $\beta \approx 0.3$) in nuclei below $^{78}$Ni. Measurements of the radii, electric quadrupole moment, and magnetic dipole moments of isotopes in the neighborhood of $^{78}$Ni will provide important guidance to understand the structure of this region \cite{deGroote2017,deGroote2019}.
The $N=50$ gap depends upon the neutron single-particle
energies for $0g_{7/2}$, $1d_{5/2}$ and $2s_{1/2}$
which are connected to energies of low-lying states in $^{79}$Ni.

As discussed in section \ref{sec:Astro}, the magic number $N=82$ for
isotopes below $^{132}$Sn is responsible
for the binding-energy discontinuity at $N=82$ which leads
to the r-process abundance peak for $A \approx 130$.
How does the structure physics leading to the
5$^{th}$ island of inversion for $N=50$ influence
the models and predictions for the structure of nuclei below $^{132}$Sn?
The 2p-2h excitation across $N=82$ will lead to deformed excited states
(isomers) and influence the binding-energy trend across $N=82$.

%\textcolor{red}{So what if it is?}

\subsubsection{The region around \texorpdfstring{$^{100}$Sn}{100Sn}: The end of the \texorpdfstring{$N=Z$}{N=Z} line}

At the top end of the $  N=Z  $ line, the exploration of  nuclei around $^{100}$Sn will provide many opportunities for connections between theory and experiment. Quantitative ESPE need to be established from masses and spectra of $^{99,100,101}$Sn, $^{99}$In and $^{101}$Sb.
%The mass of the isomeric state of $^{99}$In has recently been measured \cite{nies2023}, providing a gap of 671(37) keV between the ESPE for the $  0g_{9/2}  $ and $  1p_{1/2}  $ proton-hole orbitals. The energies of the states associated with $  1p_{3/2}  $ and $  0f_{5/2}  $ need to found, perhaps by gammas following proton knockout from $^{100}$Sn. The doublet of states separated by 172(2) keV in $^{101}$Sn have been populated following the alpha decay of $^{105}$Te. Based on the alpha decay patterns, the tentative $  J^{\pi }  $ assignments are 7/2$^{ + }$ for the ground state and 5/2$^{ + }$ for the excited state. These would be associated with the neutron ESPE for the $  0g_{7/2}  $ and $  1d_{5/2}  $ orbitals. This order is in contradiction to that suggested by systematics of the energies of states with these $  J^{\pi }  $ in nuclei with $  N=51  $.  Various type of nucleon transfer experiment are required to confirm these $  J^{\pi }  $ assignments and to find states in $^{101}$Sn related to the $  1d_{3/2}  $, $  2s_{1/2}  $ and $  0h_{11/2}  $ orbitals. Some of these may lie above the proton separation energy of about 3.4 MeV. 
%
%
The nucleus $^{100}$Sn also provides a unique opportunity to study Gamow-Teller $\beta$ decay.
The observed transition probability $B(GT)$ is about a factor of four smaller than the closed-shell prediction.
%With a closed shell configuration, the filled $  0g_{9/2}  $ orbital $\beta^{ + }$ decays into the $  0g_{7/2}  $ orbital with $B(GT)$ = 17.8. But the observed value is about a factor of four smaller.
Details of theoretical results compared to experiment for this decay are discussed in section~\ref{sec:FundamentalSymmetries} on Fundamental Symmetries.
For other nuclei in this region, the large $Q_{\beta}$ values enable the study of Gamow-Teller strength up to and beyond the proton separation energy with total-absorption spectroscopy (TAS), presenting an interesting challenge for nuclear theory.
More details are presented in section~\ref{sec:beta}.
As described below in section~\ref{sec:pairing}, the region of the heaviest $N=Z$ nucleus will probe the evolution of isoscalar pairing approaching the continuum.
%Beta decay is more complicated in other nuclei where many final states can be populated.
%Experimentally, the details of final state decays can be used to establish the spin-parities for low-lying levels in the daughter nuclei. For the most proton ($  q=p  $) and neutron-rich ($  q=n  $) nuclei, the large $  Q_{\beta }  $ value enables the study Gamow-Teller strength up to and beyond the separation energies $  S_{q}  $ with total-energy-absorption (TAS) $\gamma$ detection (for example \cite{spyr2016}). Gamow-Teller strength to final states above $  S_{q}  $ also leads to $\beta$ delated nucleon emission in competition to $\gamma$ decay. It is a theoretical challenge to calculate the details of these complex decays involving up to thousands of final states. More examples are presented in section~\ref{sec:beta}.

Below $^{100}$Sn,
the nucleus of $^{80}$Zr \cite{hamaker2021precision} turns out to be substantially more bound and lighter than expected from systematic trends of nuclear masses, suggesting $^{80}$Zr to have a deformed double shell closure.
The isotopes $^{90,96}$Zr are near spherical with a very rapid onset of strong deformation
beyond  $^{96}$Zr within the isotopic chain \cite{chei1970, roub2017}. 
Measurement of the charge radii for the Zr isotopes, combined with theoretical models,
will help clarify the rapidly changing structural evolution taking place in these isotopes.
For the odd-mass Zr isotopes the nuclear spin, magnetic dipole, and electric quadrupole moments can also be determined from the hyperfine structure via laser spectroscopy. These nuclear moments
will provide an excellent testing ground for nuclear models \cite{vern2022,karthein2023electromagnetic}.

\subsection{Collectivity}

%% Anna McCoy, Mark Caprio, and Ronald Garcia Ruiz
\subsubsection{Collective degrees of freedom and connections to microscopic theory}
\label{sec:structure:collective}

Collectivity gives rise to simple patterns 
in the spectra and decay modes reflecting the emergence of new
degrees of freedom  related to the coherent motion of many nucleons.
FRIB experiments
in conjunction with theory have the potential to offer new insight into a wide
range of questions associated with collectivity, especially in nuclei with
extreme proton-neutron imbalance. How do deformation, shape
coexistence~\cite{heyd2011},
clusterization~\cite{free2018}, and
pairing~\cite{free2018} influence nuclear structure and
observables?  What are the dominant collective excitation modes giving rise to
the low-lying structure?  How is the structure of halo states influenced by
collective correlations? What are the relevant approximate symmetries arising in
the many-body system, \textit{e.g.},
$\mathrm{Sp}(3,R)$~\cite{rowe1985,dytr2007,mcco2020, dytrych:PRL2020},
$\mathrm{SU}(3)$~\cite{harv1968,dytr2013}, and
$\mathrm{SU}(4)$~\cite{wign1937}, and how do they help us to
understand the relevant degrees of freedom?

\textit{Ab initio} theory in combination with experiment can help to identify
the relevant degrees of freedom for collective structure. This, in turn, can lead to
a microscopically motivated simple picture that accurately describes collective
structure. While descriptions of collective structure were traditionally the
domain of phenomenological models, \textit{ab initio} approaches are placing
nuclear structure theory on a predictive footing and give insight into how
collective phenomena emerge. Signatures of collective phenomena, including deformation \cite{dytr2007,dytrych:PRL2020},
rotation~\cite{capr2013,strob2016,jans2016,capr2020,yao2020,hage2022,hu2024}
and
clustering~\cite{piep2004,neff2004,mari2012,yosh2013,rome2016,navr2016,free2018,Dreyfuss:20,Shen:2022bak},
emerge in \textit{ab initio} predictions.  However, computational
challenges limit the ability of \textit{ab initio} calculations to resolve
certain aspects of collective correlations, notably the overall scale of
quadrupole correlations.  Furthermore, structure can be highly sensitive
to details of the choice of internucleon
interaction~(Sec.~\ref{sec:overview_theo_methods}), while electromagnetic (and
weak interaction) observables are also sensitive to in-medium effects arising
from the fact that the nucleus does not consist of point
particles~\cite{past2013,gysb2019}. Therefore it is critical to have experimental validation to
give us confidence in the collective picture emerging from \textit{ab initio}
theory.

Traditional signatures of collective behavior~\cite{cast2000} provide a
first point of contact between theory and experiment.  These include binding
energies, $2^+$ excitation energies, $E2$ transitions, and quadrupole moments.
FRIB experiments can provide access to regions of the nuclear chart near,
\textit{e.g.}, islands of inversion and near proton and neutron driplines, where new
collective degrees of freedom appear in the low-lying spectrum.  FRIB
experiments can also provide measurements of relevant observables across larger
swaths of the nuclear chart, thereby enabling a systematic study of the
evolution of these observables along isotopic chains, to more rigorously test
theoretical predictions.  Notably, laser-spectroscopy experiments at BECOLA-FRIB
will enable precision measurements of nuclear electromagnetic properties
such as charge radii, nuclear quadrupole moments, and nuclear dipole magnetic
moments~\cite{yang2023}. Recent experimental
developments, such as the Resonance Ionization Spectroscopy Experiment (RISE) at
BECOLA, will enable these measurements across the nuclear chart. For light
nuclei, \textit{e.g.}, in the $pf$ shell, measurements can be achieved from the
proton dripline to the neutron dripline. In such light nuclei, where radii do
not simply follow the global empirical formula \cite{Ruiz2016,mill2019}, both the radius and quadrupole
moment must be taken into consideration in determining the nuclear deformation.

However, these basic observables are not sufficient to test the more detailed
picture arising in \textit{ab initio} calculations, \textit{e.g.}, rotational
bands, proton-neutron triaxiality~\cite{kana1997},
shape coexistence~\cite{capr2022:10be}, and halo structure.
Further spectroscopic information on excited states and transitions is crucial.
Even where \textit{ab initio} methods cannot presently resolve the overall scale
of the binding energy or $E2$ strengths, correlations among calculated
observables~\cite{calc2016,sarg2022}
permit robust predictions of relative properties (\textit{e.g.}, excitation
energies and $E2$
ratios)~\cite{capr2021,capr2022:8li-trans,capr2022:emnorm}, to be
tested against experiment.  In particular, data on electromagnetic transitions
are sparse in the light nuclei where many of the \textit{ab initio} approaches
are most applicable. For $E2$ transition strengths, Coulex experiments are
critical, as lifetimes are often dominated by $M1$ decay.  These measurements
will require reaccelerated beams.  Because of the importance of proton-neutron
asymmetry in exotic light nuclei, neutron quadrupole matrix elements provide an
important probe of the collective structure complementary to the electromagnetic
measurements.  Ratios of neutron and proton matrix elements
($M_n/M_p$)~\cite{bern1983} can be obtained by combining
Coulex with inelastic scattering
experiments~\cite{furu2019}.  Recent
developments in precision laser spectroscopy experiments with radioactive
molecules~\cite{Gar20a,udre2021}
have the potential to enable access to yet-to-be-explored electroweak nuclear
properties across isotopic
chains~\cite{arro2023}. These experiments
promise to yield information on the weak charge distribution within the nucleus
from which information of the quadrupole distribution of the neutrons could be
extracted~\cite{arro2023}. FRIB, in
combination with the ongoing isotope harvesting
program~\cite{abel2019}, will provide various
opportunities for these studies.

\subsubsection{Octupole excitations}
%% Mark Spieker

Electric octupole, $E3$, excitations are observed throughout the nuclear chart \cite{Kib02a}. However, significantly enhanced $B(E3)$ strengths are only expected in regions of the nuclear chart where the Fermi surface lies between spherical single-particle levels differing by $\Delta j= \Delta l = 3$, i.e., at proton and neutron numbers of approximately 34, 56, 88, and 136\,\cite{But96a,But16a}. Possibly only a handful out of these nuclei with enhanced $B(E3)$ strengths are statically octupole deformed in their ground states\,\cite{But21a}.
A sudden increase of the $B(E3;3^- \rightarrow 0^+_1)$ strength to 30 W.u. and more was recently discussed around the $A=72$ prolate-oblate shape transitional point \cite{Spi22a}, \textit{i.e.}, on the neutron-deficient side of the nuclear chart where $Z \approx N \approx 34$. A clear theoretical interpretation of the $B(E3)$ strength increase is missing.
Systematic studies of such effects in the corresponding neutron-rich isotopes are also missing so far. Experimentally, many of the nuclei in the $A=68$ mass region and $A=90$ mass region, where $^{96}$Zr~\cite{Isk19a} is located, would already be accessible at FRIB with 20-kW beam power. Their octupole collectivity could be determined using, {\it e.g.}, inelastic proton scattering in inverse kinematics with fast beams.
Theoretically, octupole deformation has often been studied at the mean-field level \cite{Agb17a, Cao20a, Che21a}. 
From a shell-model perspective, one needs to understand the interplay between the $sd$ shell, $fp$ shell and the $sdg$ orbitals in generating enhanced $B(E3)$ strengths in the $A=68$ and $A=90$ mass regions. 

\subsubsection{Pygmy Dipole Resonance}
%% Mark Spieker

Properties of the giant dipole resonance (GDR) excited by isovector ($\Delta T=1$) probes for excited states with $J^{\pi}=1^{-}$
are well established for stable nuclei and are understood as the collective out-of-phase oscillation of all protons against all neutrons. 
In the shell model, these are coherent mixtures of 1$\hbar \omega$  1p-1h states with collective energies pushed up
above 1$\hbar \omega$.
In neutron-rich nuclei, the smaller isoscalar and isovector dipole strength observed below the GDR is called the pygmy dipole resonance (PDR).
It has traditionally been interpreted as a collective oscillation of the neutron skin against the core, though this interpretation has been questioned~\cite{Lan23a}.
Understanding the PDR in terms of 1p-1h and collective contributions remains a theoretical challenge~\cite{Roc12a, Sav13a, Vre12a, Bra19a, Lan23a}.

The isovector part of the PDR can influence neutron-capture rates when calculated with statistical Hauser-Feshbach approaches \cite{Lit09a, Tso15a, Ton17a, Lar19a}. 
To provide more robust predictions for $\gamma$-ray strength functions of rare isotopes involved in nucleosynthesis processes~\cite{Ang12a, Mar17a, Isa19a, Mar21a}, continued experimental and theoretical studies of the structure of the PDR are essential. 
Early FRIB experiments will be able to indirectly measure the $\gamma$-ray strength function and access its possible influence on $(n,\gamma)$ reactions by using experimental approaches such as the $\beta$-Oslo method \cite{Spy14a} or $(d,p\gamma)$ surrogate reactions 
%\cite{Rat19a}.
\cite{Ratkiewicz:19prl}.
$(d,p)$ one-neutron transfer reactions have already been used to explore the neutron 1p-1h structure of the PDR in more detail for $^{208}$Pb \cite{Spi20a}, $^{120}$Sn \cite{Wei21a}, and $^{62}$Ni \cite{Spi23b}. 
At FRIB, these types of experiments can be carried out in inverse kinematics at ReA using, for example, SOLARIS \cite{sol18a, sol23a}. 
Beyond-mean-field approaches are needed to understand such data and the structure of the PDR \cite{Sav11a,Avi20a,Lan23a}. 
The science case for direct studies of the complete electric dipole strength of rare isotopes with the energy upgrade of FRIB,  FRIB400, has been presented \cite{FRIB400}.

\subsection{Isospin symmetry}

%% Intro added by SRS
Isospin is an approximate symmetry of the nuclear force which can be traced back to the light masses of the up and down quarks as compared to typical QCD scales~\cite{vankolck1998,Miller2006}.
Isospin is also broken by the electromagnetic interaction.
Naively, because the Coulomb force is understood and significantly weaker than the strong nuclear force, isospin-breaking corrections should be trivial to calculate.
However, the interplay of correlations with Coulomb and other isospin-breaking forces leads to non-trivial effects in nuclei which remain a challenge for theory.
In particular, it is difficult to unambiguously disentangle the effects due to isospin-breaking strong interactions from those due to Coulomb combined with isospin-conserving strong interactions.
Beyond its intrinsic interest, a proper understanding of how isospin-breaking forces manifest in nuclei is an important component of the theoretical corrections for superallowed $\beta$ decays, described in section~\ref{sec:FS_CKM}.

Powerful probes of isospin breaking effects in nuclei are the masses within an isobaric  multiplet, labeled by isospin $T$.
In the limit of isospin symmetry, the $2T+1$ constituents of the multiplet would have identical masses (or equivalently, identical binding energies).
Isospin-breaking interactions lead to a shift of these masses, given in first-order perturbation theory by
the Isobaric Multiplet Mass Equation (IMME):
$BE(T_z) = a + bT_z + cT_z^2$, where $T_z = (N-Z)/2$, and $a,b,c$ are parameters unconstrained by isospin symmetry.
Higher-order terms like $dT_z^3$ and $eT_z^4$ can arise from second-order effects or many-body forces.
Experimental results for these coefficients have been compiled ~\cite{MacCormick.2014}.
CI calculations of these terms can be linked to the underlying Coulomb and 
isospin-dependent interactions~\cite{Ormand.1989,Ormand.1995,Lam.20138ai,Lam.2013},
while, so far, {\it ab initio} calculations have had difficulties predicting the IMME coefficients~\cite{martin2021}.
Charge radii of the $T=\pm T_z$ members of the multiplet that can be measured at FRIB
provide a further constraint to help disentangle the contributions of the various sources of isospin symmetry breaking (see also section~\ref{sec:FS_CKM}). 

In addition to the IMME,
%Another important property regards the isospin symmetry and independence of the nuclear interaction.
valuable information on
isospin-symmetry breaking in nuclei
%the validity or violation of these symmetries
can be obtained from the study of excitation energy differences of analog states in mirror nuclei (Mirror energy differences, MED) and isobaric triplets (Triplet Energy Difference, TED), as well as from the measurement of transition probabilities between analog states in these nuclei. 
From the theoretical side, approaches based on the interacting shell model have been shown to reproduce the MED and TED with good accuracy, provided an isospin-breaking term is added to the effective isospin-conserving nuclear interaction once Coulomb effects have been taken into account~\cite{zuke2002,bent2007}. While a single correction term seems to hold for the MEDs measured so far, its origin has not yet been fully understood. Tests are needed very far from stability. 
Large MEDs have obtained between isobaric states of non-natural parity due to the different energy gaps for protons and neutrons caused mainly by the electromagnetic spin-orbit interaction~\cite{ekma2004p132502, lala2022}. Moreover, the development of accurate effective interactions that involve more than one major shell are needed to describe non-natural-parity MEDs.

On the experimental side, these studies become quite demanding for the most proton-rich members of isobaric multiplets because of their low production cross sections towards the proton dripline and challenges in beam purity. Using proton-rich rare-isotope beams from FRIB, these studies can be extended to the proton dripline. The high-resolution, high-efficiency GRETA $\gamma$-ray spectrometer will enable extending the MED and TED studies and probe the validity of our understanding of isospin symmetry breaking at largest values of isospin.

The investigation of MEDs has also put into evidence the role of the continuum (see also section~\ref{sec:Continuum});
specifically, the role of the larger radius of low-{\em l} orbits with respect to the other orbits in a major shell \cite{lenz2001,bonn2016} in the evolution of the MED as a function of the spin or excitation energy.  More recently, MED studies in the $sd$ shell indicate that, when fractionally occupied, the radius of the $s_{1/2}$ orbital is about 1.7 fm larger than the $d$ orbitals. This difference suddenly decreases, to about 0.6 fm, when occupied by at least one nucleon \cite{bonn2018}. A similar behavior is observed between the $p$ and the $f$ orbitals in the $fp$ shell~\cite{fern2021,enci2022}. 
The measurement of the charge radii and isotope shifts in different regions of the nuclear chart, and in particular in the proton-rich side, together with an extension of MEDs into the upper $pf$ shell, will be essential to understand this phenomenon.

\subsection{Spin-isospin response}

For a given initial state one can study final states reached by processes associated with one-body operators characterized by their orbital $\Delta L$, spin $\Delta S$, and isospin $\Delta T$ tensor structure, that have $\Delta T_{z} = -1,0,+1$. There are experiments at FRIB focused on charge-exchange processes  $\Delta T_{z} = -1,+1$ associated with $\beta$ decay and charge-exchange reactions. Simple examples are Fermi $\beta$ decay with ($\Delta L$, $\Delta S$, $\Delta T$ =(0,0,1), and Gamow-Teller transitions  with ($\Delta L$, $\Delta S$, $\Delta T$ = (0,1,1) induced by $\beta$ decay and  charge-exchange reactions such as ($t$,$^{3}$He) ($E_{\rm beam}\gtrsim 100$ MeV/$u$). Experimental methods and instrumentation used for these is discussed in section~\ref{sec:ExpMethods}. The purpose is to study properties of individual low-lying final states, and to asses strength functions over a large excitation-energy range of final states \cite{Fujita2011549,HAR01,zegers2020}. Experiments at FRIB will study the $\beta$ decay of very neutron-rich nuclei, and charge exchange on targets away from stability via inverse kinematics. 

Exploration of the spin-isospin response of nuclei offers a unique view at the single-particle degree of freedom as well as bulk properties of the nuclear medium. Characterizations of the allowed Gamow-Teller spin-isospin response of nuclei uniquely probe the validity of nuclear structure models up to high excitation energy. In hydrodynamical models of the nucleus, isovector giant resonances are interpreted as density oscillations of the neutron vs. the proton fluid and continue to provide information on macroscopic nuclear properties associated with isovector fields.

\subsubsection{Beta decay\label{sec:beta}}
%redo of robert grzywacz

FRIB beams will provide an unprecedented opportunity to access very neutron-rich nuclei with large $\beta$-decay energy windows. Possible at the lowest beam rates, one can obtain essential information on the half-lives and the decay properties of very neutron-rich nuclei that can be used to probe theoretical models at the extremes. The spin and parity selection rules inherent to $\beta$ decay enable a characterization of the structure of low-lying states in the decay daughter, often complementing the structure information obtained from in-beam $\gamma$-ray spectroscopy. The studies of $\beta$ decay near neutron-rich magic nuclei, such as $^{54}$Ca, $^{60}$Ca, $^{78}$Ni, and $^{132}$Sn  \cite{xu23prl}, are particularly important in this regard to benchmark the dominant configurations of the respective model spaces defined by the magic nuclei. 

When the $\beta$ $Q$ value is larger than the neutron separation energy of the daughter nucleus, neutron unbound states will be populated.   
Calculations for the $\gamma$-particle decay competition for these states requires configurations in the continuum \cite{Oko03}. 
The neutron emission hindrance has been linked to the mismatch of the wavefunctions between the precursor and decay daughter \cite{spyr2016}.
The other aspect of $\beta$-delayed neutron spectroscopy is that it enables selective
studies of neutron unbound states and provides information about the location and
widths of nuclear resonances in nuclei far from stability. This kind of effort is
expected to mainly focus on low-Z nuclei where the expected widths of tens of keV \cite{Ohm76}
are measurable with current experimental techniques. Here the goal is
to connect data and theoretical models, including continuum coupling effects \cite{Oko03}, which will reveal themselves through modifications of widths and excitation energies. Neutron-rich nuclei can have multi-neutron emission \cite{Azuma79}. Interesting questions as to the simultaneous or sequential character of $\beta$-delayed multi-neutron emission remained unanswered in the first attempt of a correlation measurement \cite{Folch18} and pose an interesting challenge for experiment and theory in the future.

When combined with neutron detection and total absorption methods, the detailed spectroscopy of $\beta$ decays will provide access to a broad spectrum of excited states in very proton-neutron asymmetric systems. The same experimental methods employed in deformed regions will help establish how the strength distribution evolves with deformation in very neutron-rich systems \cite{Sarriguren}. New-generation experiments with polarized $\beta$-delayed emitters can produce spin-resolved strength distributions. The decay studies performed on island-of-inversion nuclei will provide another set of constraints for nuclear models that set out to describe such quantities at the intersection of statistical and discrete descriptions of nuclear physics. 
Non-statistical aspects of neutron emitting states are discussed in (section \ref{sec:reactions_statistical}).

Neutron emission probabilities and $\beta$ decay half-lives are important for modeling the r-process path (see section~\ref{sec:AstroNeutronRich}).
For applications to nuclear astrophysics, experiments will only provide some information on a limited number of nuclei. All other information needed relies on calculations provided by nuclear theory. Beta decay for lighter nuclei mainly involves Gamow-Teller type transitions. But for heavier nuclei, one must include first-forbidden type transitions.  One must asses the renormalization (quenching) needed for these operators in neutron-rich nuclei either 
%\cite{Gysbers19} 
\cite{gysb2019}
empirically or with the explicit addition of two-body currents. The latter was also found to be essential for improving the predictions of nuclear magnetic moments \cite{Miy24}.

Theoretical calculations provide detailed information on the $\beta$ strength distributions to final states from which neutron-decay probabilities can be obtained. Experimental data related to these final-state distributions and neutron decay modes will be obtained with total-energy $\gamma$ absorption spectrometers such as the Summing NaI detector (SUN) and the Modular Total Absorption Spectrometer (MTAS) that is part of the FRIB Decay Station Initiator (FDSi). 
One of the early FDSi experiments at FRIB reported a rapid increase in the $\beta$-decay strength distribution above the neutron separation energy in $^{45}$Ar in the decay of $^{45}$Cl, interpreted to be caused by the transitioning of neutrons into protons excited across the $Z=20$ shell gap \cite{cox2024}. This demonstrates the potential of integrated quantities such as strength functions to probe the proton single particle degree in neutron-rich nuclei. Advancing decay spectroscopy in general, the full FRIB Decay Station (FDS) is aspired by the community \cite{FDS}.

% (moved to reactions section) It is often assumed that $\beta$ decay, and neutron emission, proceed through the compound nucleus stage, which is needed for the implementation of Hauser-Feshbach codes in $\beta$-n models \cite{Kaw08} and decouples the nuclear structure effects from the neutron emission. Recently, evidence for non-statistical neutron emission was discovered, and the concept of doorway states was used to explain the data \cite{Hei23}. Mounting evidence \cite{SLAUGHTER197222, Piersa19} shows that the neutron emitting states in medium and heavy nuclei are very narrow.  For all of these situations, a better microscopic description is needed for the description of $\beta$-delayed neutron emission. 

%FRIB and other rare-isotope facilities provide ample access to $\beta$-delayed neutron emitters in different regions of the nuclear chart, offering up various bound and unbound single-particle configurations in various $Q$-value windows.

\subsubsection{Charge-exchange reactions}

Charge-exchange reactions at intermediate beam energies ($E_{beam}\gtrsim 100$ MeV/$u$) provide excellent ways to probe the isovector response of nuclei. 
\cite{Osterfeld:1992,ICHIMURA2006446,Fujita2011549,RevModPhys.75.819, FRE18,HAR01,Langanke2021,zegers2020}. 
Charge-exchange reactions can be used to extract weak-interaction strengths that cannot be obtained from $\beta$ decay.  Even though the charge-exchange reaction is mediated via the strong nuclear force, it can be used to extract information about allowed weak transitions associated with the transfer of zero angular momentum ($\Delta L=0$), with the transfer of spin ($\Delta S=1$, Gamow-Teller type) or without the transfer of spin ($\Delta S=0$, Fermi type)
(see the experimental methods in section \ref{sec:ExpMethods})

%\textcolor{red}{$<$-This paragraph should go in the methods section. AG: I am fine with this going elsewhere - no structure or science question in general}

Charge-exchange reactions provide a unique experimental method to obtain Gamow-Teller strengths in the EC $\beta^{+}$/EC direction in neutron-rich nuclei. 
From a structure viewpoint this part of the Gamow-Teller strength is
blocked by the filled orbitals in the neutron excess. Thus, any strength 
observed provides a test for theoretical models of correlations among the neutrons. 
These theoretical models provide EC rates needed for astrophysical
applications (see section \ref{sec:Astro}). 

%Charge-exchange reactions are used for extracting transition strengths that are important in astrophysical phenomena, as well as for studies related to double-$\beta$ decay ~\cite{RevModPhys.75.819,Langanke2021,FRE18,Fujita2011549}. For example, charge-exchange reactions in the $\beta^{+}$/EC direction are used to benchmark astrophysical EC rates that are used in stellar evolution models and inspire the development of improved theoretical models \cite{RevModPhys.75.819,PhysRevC.83.064318,sull2018,PhysRevC.105.055801,PhysRevC.101.025805}. Experiments are carried out for a selected set of targets. Collaboration with theory is needed to help choose the optimal targets, and then to provide systematic results for all other nuclei needed for astrophysics.  \textcolor{red}{If it's relevant for fundamental symmetries or astro, it should be described in those sections. AG: I cannot defend this paragraph's presence in the structure section}

There is significant interest to study isovector giant resonances, which, in a macroscopic picture, are out-of-phase density oscillations of the neutron and proton matter inside the nucleus. The characteristics of these isovector giant resonances provide information about bulk properties of nuclei and the nuclear equation of state that is complementary to that obtained from isoscalar resonances ~\cite{GARG201855} and helpful for extrapolating to systems with small or large neutron-to-proton ratios \cite{PhysRevC.52.R1175,PhysRevLett.121.132501}.  Microscopically, isovector giant resonances are described as collective one-particle-one-hole ($1p-1h$) excitations in combination with the transfer of isospin. A systematic study of the excitation energies and collectivity of isovector giant resonances 
can be directly compared to \textit{ab initio} ~\cite{PhysRevC.106.054323,PhysRevLett.130.232301}, configuration-interaction \cite{Fujita2011549, PhysRevC.74.034333, Langanke2021}, and density-functional \cite{HAR01,PhysRevC.72.064310,zegers2020}
models to shed light on remaining ambiguities in the isovector part of the nucleon-nucleon effective interactions \cite{PhysRevC.72.064310,PhysRevC.98.051301}.

\subsection{Pairing effects\label{sec:pairing}}

\begin{figure}
\centering
\includegraphics[trim=80 200 100 120, clip,width=0.8\columnwidth, angle=0]{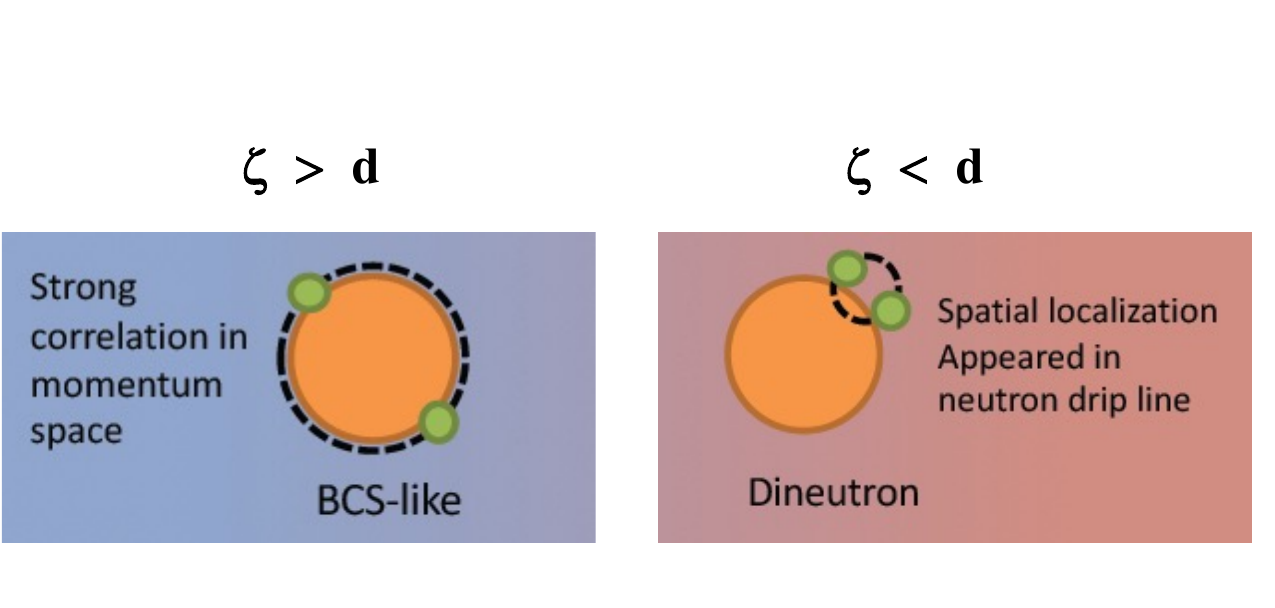}
\caption{Schematic illustration of the expected change in behavior between Cooper pairs and di-neutrons is indicated. Figure adapted from Ref.~\cite{Matsuo1}.}
\label{fig:fig4}
\end{figure}

%{\color{red}SRS cut-paste.}
%\subsection{Pairing Correlations in neutron rich nuclei}
%ADDED By Augusto Macchiavelli on 8/26/23
In 1958, Bohr, Mottelson and Pines \cite{bmp} suggested a pairing mechanism in the atomic nucleus 
analogous to that observed in superconductors~\cite{BCS}.  Since the publication of that seminal paper, 
 a wealth of experimental data have been accumulated, supporting the important role played by 
 neutron-neutron and proton-proton ``Cooper pairs'' in modifying many nuclear properties such as 
 deformation, moments of inertia, alignments, etc.~\cite{BrogliaBrink,50yearsBCS,DeanMorten}.  Driven by advances in experimental techniques, the development of sensitive and highly 
efficient instruments  and the availability of rare-isotope beams, the pairing correlations 
can now be studied in exotic nuclei out to and beyond the neutron dripline.  Of particular interest is the role 
of neutron-neutron pairing in neutron-rich isotopes, where the effects of weak binding and continuum coupling are important. Illustrative results for nuclear matter~\cite{Matsuo1} show that the correlation length, $\zeta$, of an $nn$ pair and their separation distance, $d$, depend on the nuclear matter density. At normal density $\zeta >d$, typical of Cooper pairs.  At lower densities that could be relevant in the surface of weakly bound nuclei, $\zeta < d$,  and a localization of di-neutrons appears, signaling a possible BCS to BEC crossover transition as schematically shown in  Fig.~\ref{fig:fig4}.

An indicator for weak binding effects could be considered from the following general arguments.  The neutron separation energy is given by $S_n \approx \lambda +\Delta$, with $\lambda$ the Fermi level and $\Delta$ the pairing gap. Near the valley of stability,  the quasi-particle binding is dominated by the mean field and  $S_n \approx \lambda$. As we increase the number of neutrons, approaching the drip line, $\lambda \rightarrow 0$ and $S_n \approx \Delta$, \textit{i.e.}, correlations dominate. Thus, it is natural to expect that the transition between the two regimes will start to take place when $S_n \approx \Delta$.

Two-neutron transfer reactions  have provided a unique tool to understand neutron pairing correlations in nuclei~\cite{Yos62,Bayman68,Broglia73}.  In formal analogy with the case of quadrupole shape fluctuations~\cite{Broglia73,BTG}, where an important measure of collective effects is provided by the reduced transition probabilities ({\it i.e.}, $B(E2)$'s), one can associate a similar role to the transition operators $\langle f|a^\dagger a^\dagger|i\rangle$ and $\langle f|a a|i\rangle$ in the two-particle transfer mechanism between the initial $|i\rangle$ and
final $|f\rangle$ states.

The criteria discussed above can also be related to the asymptotic behaviour of the Cooper pairs $\rightarrow e^{-Kr}$.  For strongly bound nuclei,  $K \approx 2\kappa$, the tail of the particle density. For weakly bound nuclei, $K <  2\kappa$ and the pair condensate extends further outside the surface.  While a reaction such as ($^{18}$O,$^{16}$O) might be of interest, it seems clear that $(t,p)$ reactions will continue  to play a crucial role to study pairing correlations in neutron rich nuclei since they are particularly suited to directly probe the $2n$ pair density.

\begin{figure}
\centering
\includegraphics[trim=100 100 120 80, clip,width=0.6\columnwidth, angle=270]{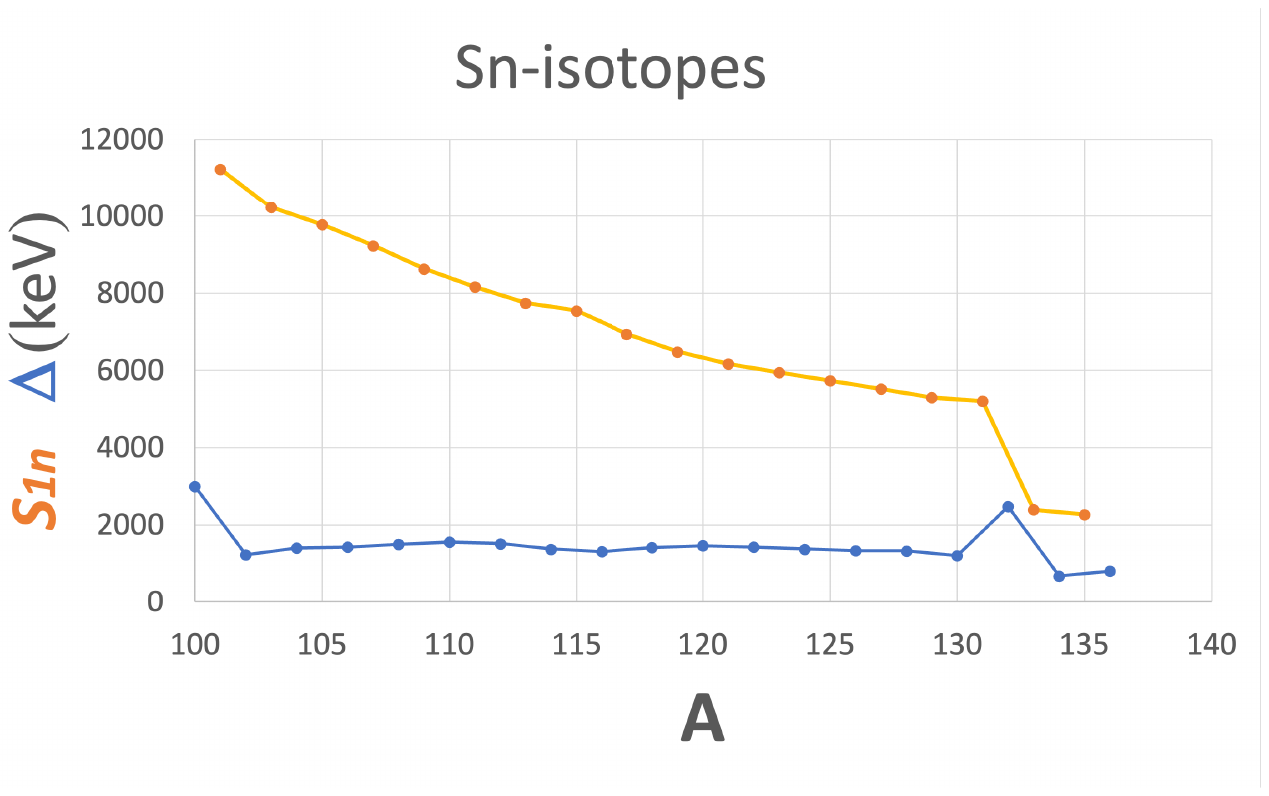}
\caption{Comparison of one neutron separation energy, $S_{1n}$ and pairing gap, $\Delta$, for the tin isotopic chain.}
\label{fig:Sns}
\end{figure}

The Sn isotopes are the classic example of superfluidity in atomic nuclei and their ground states are well described as a BCS neutron condensate.  Following from the criteria introduced earlier, in Fig.~\ref{fig:Sns} the $S_{1n}$ and $\Delta$ are compared across the isotopic chain, showing an indication that just above the $N=82$ shell closure weak binding effects may start to influence the low-lying pairing properties of $^{134}$Sn and heavier isotopes.

In fact, theoretical calculations based on Skyrme-Hartree-Fock mean field
and continuum RPA predict a significant increase in the neutron pair-transfer strength to
low-lying excited $0^+$ pairing vibrational states. For very neutron-rich
Sn nuclei with $A > 140$, a large increase in the pairing gap is expected, which results
in an increased ground state to ground state pair-transfer strength~\cite{Matsuo2,Matsuo3}. This behavior
is attributed to the loosely bound $2p_{1/2}$  and $2p_{3/2}$ orbitals, extending far beyond the
nuclear surface. Currently, it is not possible to study Sn nuclei with $A > 140$,  but the region where strong transitions to pairing vibrational states  are predicted, just above $^{132}$Sn, will be  within reach at FRIB once it reaches full power and even more so with the envisioned 400 MeV/u energy upgrade. Having this in mind, it is worth considering some other cases that would be of much interest to study. An inspection of the condition $S_{1n} \approx \Delta$ suggests that weak-binding  effects might start to show beyond $^{56}$Ca and $^{78}$Ni.

Crucial theoretical input for these studies will be systematic state-of-the art calculations of two-neutron transfer strength for a given isotopic chain including the effects of weak binding and continuum coupling. These will guide the experimental program by suggesting the best cases to study and provide  input to the reaction codes needed to calculate the reaction cross sections measured in the experiment.
%\begin{figure}
%\centering
%\includegraphics[trim=20 90 40 60, clip,width=1.0\columnwidth, angle=0]%{Structure/Presentation1.jpg}
%\caption{As in Fig.~\ref{fig:Sns} for the Calcium and Nickel isotopic chains}
%\label{fig:CaNi}
%\end{figure}

%SRS maybe remove this, since it doesn't provide any physics motivation.
%We have focused our discussions on heavier systems for which pairing collective effects are anticipated. Nevertheless, studies of $(t,p)$ reactions in light nuclei, for example $^{11}$Li$(t,p)$,  $^{22}$O$(t,p)$, and $^{40}$Mg$(t,p)$ will certainly receive much attention. 

%{\color{red}SRS cut-paste. Need to incorporate with the rest of the text.}
The odd-even staggering in $BE(Z,N)$ is linked to the pairing interactions
in nuclei. Theoretically, the pairing is determined directly from the neutron-neutron 
$J=0$, $T=1$ interaction within the valence space,  
and indirectly from the proton-neutron interaction between neutrons in the valence space
and protons in the core. Thus, $BE$ for neutron-rich nuclei will provide new insight
into the role of these two mechanisms.

%\color{red}SRS: cut-pasted from Lenzi's section. Need to incorporate with the rest of the text.}
A very interesting region in the Segr\`e chart is the one around the $N=Z$ line. Here, protons and neutrons occupy the same single-particle orbitals and, therefore, their interaction is maximized. These nuclei open the possibility of studying different fundamental properties of the nuclear interaction as for example, the proton-neutron pairing, that presents two different isospin channels, isoscalar $T=0$ and isovector $T=1$. While the latter is present also in the proton-proton and neutron-neutron well-known isovector pairing interaction, the former is exclusive of the proton-neutron interaction and the $T=0$ pairing should manifest mainly in $N=Z$ nuclei~\cite{Frau2014,yosh2014}.

%{\color{red}SRS cut-paste. Need to incorporate with the rest of the text.}
%The odd-even staggering in $BE(Z,N)$ is linked to the pairing interactions
%in nuclei. Theoretically, the pairing is determined directly from the neutron-neutron 
%$J=0$, $T=1$ interaction within the valence space,  
%and indirectly from the proton-neutron interaction between neutrons in the valence space
%and protons in the core. Thus, $BE$ for neutron-rich nuclei will provide new insight
%into the role of these two mechanisms.  

The $BE$ of nuclei around $Z=N$ can be used to deduce  $T=0$ proton-neutron pairing energies,
often denoted as V$_{np}$.
The V$_{np}$ extracted for nuclei up to $^{74}$Rb have placed an essential constraint on
microscopic models, and have led to the construction of models with SU(4) symmetry~\cite{Isacker.1995}.
High-precision mass measurement for nuclei above $^{78}$Rb will provide the information 
needed to address the question of how V$_{np}$ evolves~\cite{Isacker.1995,Chartier.1996} from
the strongly deformed $^{80}$Zr~\cite{hamaker2021precision} nucleus to the doubly-magic  $^{100}$Sn nucleus.

\bigskip

%{\color{red}Move this to Experimental methods}

\bigskip

\section{Continuum Physics\label{sec:Continuum}}

% Confirmed: 
% W. Nazarewicz: two-proton decay [done]
% P. Navratil: ab initio description of resonances in light nuclei? [general motivation done]
% A. Volya: near-threshold clustered states? [done]
% H. Iwasaki: bound state spectroscopy and transition strengths measurements of weakly bound nuclei. [done]
% C. R. Hoffman: neutron-drip line in F isotopes? [done]
% B. Monteagudo Godoy: p-wave halos in medium-mass nuclei [done]
% A. Macchiavelli: pairing correlations in exotic nuclei?
% A. Brown: widths estimates in the SM (e.g. for (p,g) rates)

%A Volya

Continuum physics is the study of how the structure of quantum systems and the nearby presence of the continuum of scattering states and decay channels affect each other. 
Systems in which continuum couplings are important are called open quantum systems. 
In low-energy nuclear physics, continuum physics naturally lies at the intersection between structure and reaction theory, which are historically concerned with how nucleons self-organize into bound states (Sec.~\ref{sec:Structure}), and how nuclei interact over time in scattering processes (Sec.~\ref{sec:Reactions}), respectively. 
Continuum physics is thus important to understand properties of nuclei that: i) are excited near or beyond a given particle emission threshold, or ii) have a large $N/Z$ imbalance and lie near or beyond the edges of nuclear stability. 

The phenomenology of continuum physics is broad and encompasses halo and universal $S$-wave physics (Sec.~\ref{sec:InterplayContEmergent}, ~\ref{sec:AbInitioContinuum}), decay and notably exotic decay modes (Sec.~\ref{sec:2NucleonDecay}), continuum-mediated phenomena such as superradiance (Sec.~\ref{sec:NearThreshold}), as well as complex physics involving an interplay between continuum couplings and standard emergent phenomena such as pairing, deformation, or clustering (Sec.~\ref{sec:InterplayContEmergent}). 
Some of the most pressing open problems in the description of nuclei as open quantum systems are, for instance, understanding the origin of near-threshold clustering, developing accurate and scalable theoretical methods to describe many-body resonances and broad resonances, and unifying nuclear structure and reactions in an approach where all decay channels are open.
Progress on any of these problems will help move forward nuclear astrophysics (Sec.~\ref{sec:Astro}) and better constrain nuclear forces in new extreme conditions.

In the coming decades, FRIB will dramatically push forward the exploration of the drip lines and lead to the discovery of thousands of new isotopes. 
Continuum physics will thus be critical for the success of the FRIB mission.
Below, we first provide a more detailed account of near-threshold physics, and then selected cases of recent developments and studies that illustrate the status of the field.

\subsection{Near-threshold physics}\label{sec:NearThreshold}

The study of exotic nuclei  provides an unparalleled opportunity to explore the physics of open quantum many-body systems. 
Quantum systems near thresholds exhibit a rich array of features, reflecting the coupling between discrete and continuum spaces. 
In particular, halo phenomena and the associated threshold discontinuities have been recognized for a long time \cite{barker1964, baz1958Energy, baz1959Resonance, baz1961Energy, breit1957Energy, inglis1962Nuclear, wigner1948Behavior}, but only recently have their direct manifestations been observed in experiments. 
Single-particle states with low Coulomb and centrifugal barriers are particularly affected by the continuum threshold resulting, for instance, in discontinuities in various observables reflecting structural modifications of states such as spectroscopic factors 
%\cite{michel09_2,
\cite{Michel2008,
volya:2006a:art,volya:2003:art}, and in a modification of single-particle energies \cite{kay:2017,hoffman:2014,sorlin2020Reduced}.
The theoretical treatment of threshold phenomena is  difficult. 
Even for bound systems typical basis expansion techniques become impractical near thresholds, and alternative methods for including continuum couplings must be considered \cite{michel:2021,baz1959Resonance}. 

Above the threshold, for states embedded in the continuum, additional coupling via the continuum emerges. 
Interactions between overlapping resonances leads to collectivization and distribution of decay widths \cite{volya:2003:art}. 
The phenomenon of superradiance \cite{auerbach:2011} in which overlapping resonances naturally segregate into, on the one hand superradiant states strongly coupled to the continuum and, on the other hand, trapped or decoupled resonances, is an interesting phenomenon to investigate in future experiments. 
The complete understanding of resonances and their widths will require understanding the time dependence of formation and decay, background and non-resonant components.
Aspects of non-Breit-Wigner spectral functions and non-exponential decay are explored in \cite{Wang2023}.
% KF: reference?
% Cases such as the two nearby $3/2^+$ states in $^{13}$C and $^{13}$N offer remarkable opportunities for understanding the physics of unstable systems. 

The well-known near-threshold alpha clustering phenomenon \cite{ikeda:1968} is another excellent arena for exploring the effects of continuum couplings. Numerous broad alpha resonances have been experimentally observed, and there are indications of the isospin symmetry being broken due to superradiant alignment \cite{volya:2022a}.
The conservation of probability and the associated unitarity of the scattering matrix \cite{volya:2006a:art,auerbach:2011} are responsible for non-resonant near-threshold features such as threshold cusps \cite{wigner1948Behavior}. An examination of the experimental and theoretical implications and perspectives on clustering effects in exotic nuclei near threshold energies, resulting from a recent FRIB Topical Program
on “Few-Body Clusters in Exotic Nuclei and Their Role
in FRIB Experiments”, provides additional clues on how to study these effects \cite{bazin:2023}

A remarkable example showing the full complexity of the near-threshold physics is the case of $\beta$-delayed proton decay of $^{11}$Be. 
The excessive decay rate observed for the beta-delayed proton decay in $^{11}$Be \cite{ayyad:2019}
prompted speculations on the nature of the decay including exotic processes beyond the standard model \cite{volya:2020a:art,pfutzner:2018}. The $1/2^{+}$ ground state  of $^{11}$Be nucleus  is by itself a remarkable example of a one-neutron $s$-wave halo state bound by only 0.5 MeV just below the expected $p$-wave state. 
The sequential decay process occurs via a proton resonance in $^{11}$B \cite{ayyad:2022,lopez-saavedra:2022:art} conveniently located near the threshold and thus strongly enhancing proton decay instead of alpha decay, despite the latter channel being nearly 3.0 MeV above the $\alpha+^{7}$Li threshold \cite{volya:2020a:art,okolowicz:2020,atkinson:2022}. The proximity of the proton resonance to the neutron decay threshold is also noteworthy. Near-threshold resonances strongly coupled to their corresponding channels are expected to play an important role in astrophysical processes.

Theoretical advancements in time-dependent techniques \cite{volya:2009a:art,wang:2021} have opened up new possibilities for studying unstable nuclei and exploring open quantum systems more broadly. The dynamics of decay provides valuable insights on the interplay between internal dynamics and decay processes, and more specifically on the formation of the decaying state, exponential decay, and post-exponential processes driven by non-resonant components. In decays involving a three-body final state, where energy, momentum, and angular momentum conservation laws are insufficient to fully constrain the dynamics, correlations between observables may offer experimental methods for tracking the evolution of the wave function 
%\cite{Wang:2023aa}.
\cite{Wang2023}.

%Witek
\subsection{Two-nucleon emission}\label{sec:2NucleonDecay}

Due to the presence of the Coulomb barrier having a confining effect on the proton density, the one- and two-proton (2p) drip lines lie relatively close to the valley of $\beta$-stability. 
As a result, relatively long-lived, proton-unstable nuclei can exist beyond the drip line \cite{Pfutzner2012,Pfutzner2023}. The vast territory of proton-unstable nuclides contains rich and unique information on nuclear structure and dynamics in the presence of the low-lying proton continuum.

The phenomenon of ground state 2p radioactivity has generated much excitement as the experimental data on lifetimes and correlations between emitted protons provide us with unique structural information. 
Direct 2p decays have been
 found in a handful of even-$Z$ isotopes, in which single-proton decay is energetically blocked. 
 Currently, 2p radioactivity has been detected in: $^{19}$Mg,
$^{45}$Fe, $^{48}$Ni, $^{54}$Zn, and $^{67}$Kr. In addition, several broad resonances associated with prompt 2p decay were reported in, e.g., $^{6}$Be and $^{11,12}$O.

In Ref.~\cite{Neufcourt2020b}, the position of the two-proton drip line has been determined with the help of a Bayesian model averaging technique by using several global mass models. The most promising new candidates for the  two-proton radioactivity searches at FRIB,
with the predicted lifetimes between $10^{-7}$\,s and $10^{-1}$\,s
turned out to be: $^{30}$Ar, $^{34}$Ca, $^{39}$Ti, $^{42}$Cr, $^{58}$Ge, $^{62}$Se, $^{66}$Kr, $^{70}$Sr, $^{74}$Zr, $^{78}$Mo, $^{82}$Ru, $^{86}$Pd, $^{90}$Cd, and $^{103}$Te. 
In some heavy nuclei, such as $^{103}$Te and $^{145}$Hf, a competition between alpha and two-proton decay is expected.
\begin{figure}[htb]
    \includegraphics[scale=0.6]{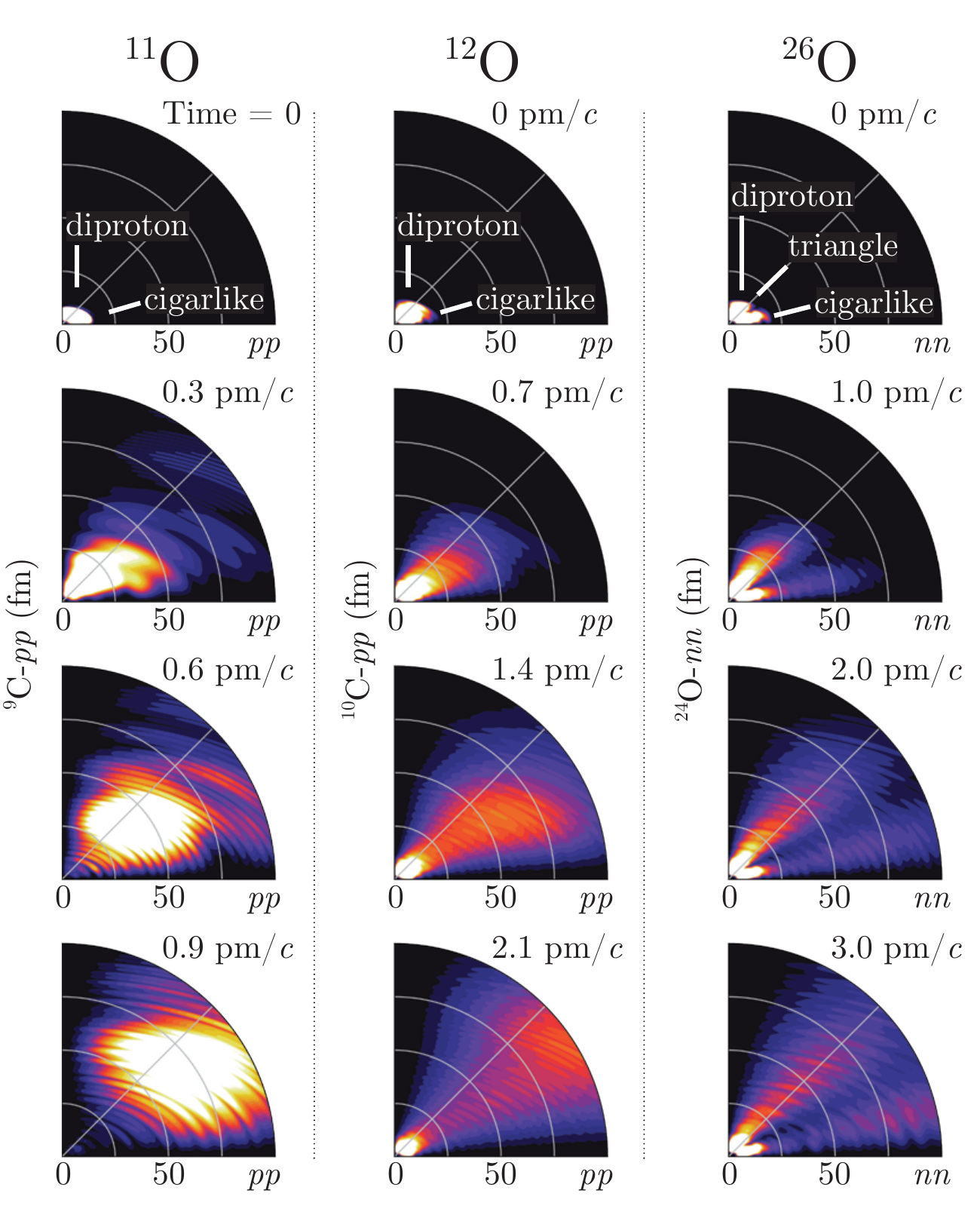}
    \caption{The density distributions of two-nucleon decays from the ground states of $^{11,12}$O (2p emitters) and  $^{26}$O (2n emitter)
isotopes for four different time slices. From Ref.\,\cite{Wang2022}}
    \label{fig:figO}
\end{figure}

For decaying proton pairs, long-range
%finite-state  %SRS change
final-state
Coulomb interaction is essential. Consequently, to study the mechanism of 2p decay,
theoretical models must fully control the behavior of the decaying system at large distances and long propagation times. Recently,  a
realistic time-dependent framework has been developed \cite{wang:2021} providing precise solutions with correct three-body asymptotics.  This study demonstrated that the energy and angular correlations in the Jacobi-Y angle between emitted nucleons strongly depend on the initial-state structure. Hence, the inter-nucleon correlations  provide invaluable information on the dinucleon structure in the initial state. An illustrative example of time-dependent calculations of two-nucleon decay is shown on Fig.~\ref{fig:figO} for the exotic isotopes
of oxygen.
Such results indicate that the anticipated high resolution 
data from FRIB on energy and angular nucleon-nucleon
correlations  will provide unique
insights into the structure of proton and neutron
pairs in rare isotopes.

\subsection{Experiments near the neutron dripline between N=20 and N=28}\label{sec:InterplayContEmergent}

\begin{figure}
    \centering
    \includegraphics[scale=0.75]{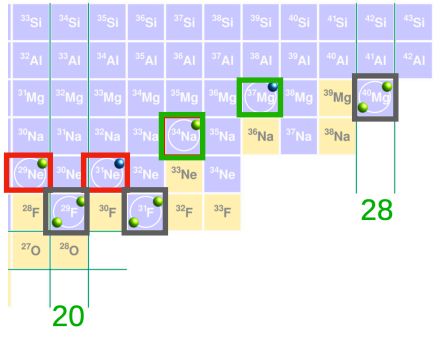}
    \caption{Subset of the chart of nuclides ranging from the neutron-rich O ($Z=8$) to the Si ($Z=14$) isotopes. The red, grey, and green squares denote known $p$-wave neutron halo ground-state systems, anticipated ground-state halo nuclei, and single-neutron $p$-wave halo ground-state systems that will be the focus of early FRIB experimental work, respectively.}
    \label{fig:fig1}
\end{figure}

The nuclear region suitable for early investigation at FRIB includes neutron-rich nuclei spanning across the neutron magic numbers $N=20$ and $N=28$. Fig.~\ref{fig:fig1}.
In this region, the dominance of low-angular momentum components in the ground-state wave functions is becoming apparent
%~\cite{PhysRevC.89.061305,
~\cite{hoffman:2014,
PhysRevC.99.024319}. 
One can 
investigate the interplay among the evolution of the single-particle structure, deformation, effects due to weak binding and continuum,
which may lead to unexpected emergent phenomena. The interpretation of data will rely on
collaborations with theory (discussed in the next section) that takes into account such rich aspects.

The interplay between emergent phenomena and near-threshold physics presents new interesting ways to study nuclear structure. 
Halo states, characterized by weak binding and a spatially extended wave function of valence nucleons, are illustrative. 
Experimental signatures for the halo formation in light dripline nuclei have been obtained from studies of the ground-state properties through measurements on masses, charge and matter radii, and spectroscopic information. 
However, effects of weak binding on their excitation properties remain largely unexplored~\cite{Macchiavelli2022}. 
Therefore, important open questions remain with respect to the existence of various soft or decoupled excitation modes~\cite{WARNER1997145}, the role of core excitations and deformation in halo formation and its properties, and the influence of the spatially extended wave functions on collective modes such as rotations and vibrations. A detailed understanding of nuclear structure properties of halo nuclei or near-threshold states in general is also important to predict new halo candidates in heavier mass regions at and beyond $N$~=~50 where mechanisms for the halo formation are expected to be diverse~\cite{MISU199744,PhysRevC.99.024319,FRIB400}.

In general, the formation of halos is favored in weakly-bound states with typical neutron separation energies $S_n<1$~MeV and a low centrifugal barrier ($\ell=0,1$). Therefore, the location of the neutron orbitals $0f_{7/2}$ and $1p_{3/2}$ in the $N=20$-$28$ region is critical. The near degeneracy of these orbitals leads to deformation driven via the Elliott-Jahn-Teller effect, which tends to lower the $p$-wave over the $f$-wave states. 
%Recent studies have shown that halo formation towards the neutron drip line is not solely a characteristic feature of the $sd$-shell. Heavier $p$-wave neutron halos may as well develop where deformation is prominent.
\newcommand{\nuc}[2]{$^{#1}$#2}
Halo aspects have been experimentally observed in  (\nuc{29}{F}~\cite{BagchiPRL20}, \nuc{29}{Ne}~\cite{KobayashiPRC2016}, \nuc{31}{Ne}~\cite{NakamuraPRL2014}, \nuc{37}{Mg}~\cite{KobayashiPRC2014}) and more are predicted by theory: \nuc{34}{Na} as a one-neutron halo, \nuc{31}{F} and \nuc{40}{Mg} as two-neutron halos. An overview of observed and predicted $p$-wave halo nuclei in the region of interest is presented in Fig.~\ref{fig:fig1}. 

These nuclei will be explored at FRIB with a variety of experimental techniques.
$\beta$ decay properties are among first to be studied as in 
the FRIB Day-One experiment of Crawford et al., \cite{Crawford2022PRL}.
One- and two-proton removal are used to populate more neutron-rich nuclei. For bound states, one can carry out excited-state lifetime measurements \cite{PhysRevLett.121.262501} aimed at transition strengths and collective modes.
Invariant-mass spectroscopy based on neutron decay followed by the detection of $\gamma$-ray decays
provides binding and excited-state energies for unbound systems.
Inelastic scattering on targets such as the ($d$,$d'$) or ($p$,$p'$) provides selectivity in the population and characterization of collective states.

For Coulomb breakup reactions, the amplitude and shape of the $dB(E1)/dE_{rel}$ distribution is sensitive to the state configuration (spectroscopic factors, $\ell$ value), the neutron separation energy, and perhaps deformation. The expected integrated cross sections and $B(E1)$ strength are exceptionally large in halo nuclei as a result of the dominance of the electric dipole strength at low excitation energies (soft $E1$ excitation).  
An increase in the $dB(E1)/dE_{rel}$ distribution at small $E_{rel}$ values will be indicative of a halo character related to $p$-wave contributions.

One particular region of interest is the F ($Z=9$) isotopic chain where a two-neutron halo structure has recently been suggested in $^{29}$F~\cite{BagchiPRL20}.    
FRIB experiments out to $^{33}$F will provide key pieces of new empirical information pertaining to the competition between continuum effects, deformation, and coherent correlations, and other central or many-body effects, on ground and excited states
%~\cite{ref:Fossez2016,
~\cite{Fossez16B,
ref:Fossez2022,
ref:Luo2021}. Of particular interest is the role played by the occupancy of the $\nu 1p_{3/2}$ orbital.

The ground state of $^{40}$Mg presents an intriguing case where an interplay between weak binding and collectivity might explain the curious low-lying structure observed. The first spectroscopy studies of $^{40}$Mg~\cite{Crawford14,Crawford19}, carried out at RIKEN/RIBF, revealed some intriguing differences  between the observed $\gamma$-ray spectrum and those of the neighboring $^{36,38}$Mg isotopes. The transition energies deviate from 
the smooth trend seen in lighter isotopes and cannot be reproduced by state-of-the-art shell-model calculations that have been very successful in this region.  Thus, one might be tempted to speculate that this could be due to the effects of weak binding. In fact, $^{40}$Mg lies close to the neutron dripline and is a potential candidate for a halo nucleus with two neutrons occupying the  $p_{3/2}$ orbital~\cite{Caurier14,Hamamoto07,Nakada18}, 
Of particular interest are the effects of the continuum on nuclear rotational motion, that have been studied in Ref.~\cite{Fossez16A,
Fossez16B} 
in the framework of the particle-plus-core problem using a non-adiabatic coupled-channel formalism.  The subtle interplay between deformation, shell structure, and continuum coupling can result in a variety of excitations in an unbound nucleus just outside the neutron dripline, as predicted for the low-energy structure of $^{39}$Mg. 

In a recent work~\cite{Macchiavelli2022}, the coupling of a two-neutron halo to a core was studied in a phenomenological approach. From the known properties of $^{38,40}$Mg, possible particle-core coupling schemes and their impact on the low-lying excitation spectrum were presented.   It is natural to expect that effects of weak binding on  excited states will show when the energy scales of the two degrees of freedom become comparable: 
\begin{equation}
 E_{core}(2^+) \approx  E_{2n}(2^+)
\end{equation}
The picture that emerges is one where the low-lying $2^+$ excitations in $^{40}$Mg result from a strong competition of the core and the $2n$ subsystems. 
While this scenario seems appealing, it is clear that further experimental and theoretical works will be required to fully elucidate the structure of $^{40}$Mg. 
%In that regards, two experiments have been approved by the FRIB PAC which are anticipated to run in the near future. One proposal~\cite{FRIBTOF} is to measure the mass of $^{40}$Mg to determine the two-neutron separation energy, a key ingredient to signal the presence of a halo, currently evaluated to be $667 \pm 706$ keV ~\cite{AME2021B,AME2021C}.  The other proposal~\cite{FRIBCrossSection} aims to measure the total reaction cross-section directly related to the nuclear matter radius.  The extended matter radius exhibited by a two-neutron halo nucleus~\cite{Hansen87, Ozawa01} can be expressed in terms of the separation energy of the weakly bound neutrons ($S_{2n}$) through the tunneling parameter $X = r_{c} \sqrt{2\mu S_{2n}}/\hbar $ derived from the exponential nature of the asymptotic wavefunction, with $r_c$ being the core radius.

%An important benchmark could be available soon from the results of the FRIB Day-one experiment, where the new $\beta$ decay lifetimes of a number of nuclei near $N=28$ were measured, including $^{36,38}$Mg.  
%With the accelerator now reaching 10 kW of power, it is anticipated that, in a second run, the statistics will be improved by a factor 10 and data for the decay of $^{40}$Mg will be obtained.

It will be important to push the early FRIB $\beta$ decay experiments
in this mass region \cite{Crawford2022PRL} out to  $^{40}$Mg and beyond.
Beta decay of halo nuclei has recently been reviewed in Ref.~\cite{Riisager2022}. 
%In this particular case, the situation is schematically shown in Fig.~\ref{fig:betadecay}.
%\begin{figure}[htbp]
%\centering
%  \includegraphics[trim=200 120 100 100, clip,width=\columnwidth,angle=0]{Continuum/Beta1.pdf}
%  \caption{Schematic representation of the $\beta$-decay of a two-neutron halo $^{40}$Mg (bottom) %as compared to that of the more bound $^{36,38}$Mg (top).}
%  \label{fig:betadecay}
%\end{figure}
The large available $Q_{\beta}$ value 
%(see Fig.~\ref{fig:betadecay2}) 
means that a large fraction of the decays will proceed via delayed one- and two-neutron emissions.
How is the final state distribution effected by the two-neutron halo structure?
%For that reason, it is important to obtain a relative comparison of not only the ground state lifetimes, but also of the strength distribution with energy between $^{40}$Mg and lighter isotopes. 
%Qualitatively, one would expect that the contribution of both paths in the halo scenario will result in a redistribution of strength to higher energies since the dineutron configuration will favor the overlap with excited states in $^{40}$Al. 

%Without a doubt, a systematic study of the evolution of these observables is likely to serve as an indicator for potential weak binding effects and at the same time provide a stringent test for theory.  Predictions from large-scale shell model calculations  will be essential for providing a needed reference to highlight any deviation, even if relative across the chain. 

%\begin{figure}[htbp]
%\centering
%  \includegraphics[trim=100 100 100 100, clip,width=\columnwidth,angle=0]{Continuum/Beta2.pdf}
%  \caption{  $\beta$-decay $Q$-values for the isotopes of interest.}
%  \label{fig:betadecay2}
%\end{figure}

%K. Fossez, P. Navratil
\subsection{Ab initio calculations in light nuclei and FRIB physics}\label{sec:AbInitioContinuum}

The \textit{ab initio} framework plays an important role in the FRIB scientific mission by providing a feedback between models of nuclear forces and observations. 
However, while the reach of \textit{ab initio} calculations has increased dramatically in the past few decades 
%\cite{hergert20_2412}, 
\cite{hergert2020guided}, 
there remains many challenges to provide reliable predictions in the medium-mass region and beyond, 
notably due to uncertain constraints in some sectors of nuclear forces \cite{calci16_2128}. 
Light nuclei ($A \leq 16$) offer interesting opportunities to test our understanding of nuclear forces within the \textit{ab initio} framework, 
not only because systematic quasi-exact calculations appear computationally feasible, 
but also due to the presence of near-threshold phenomena and their interplays with emergent features such as clustering or deformation, 
on which nuclear forces have seldomly been tested. 

Many light nuclei have been studied that are particle unbound in their ground state, and all of them have unbound excited states. Neutron-halo states, where the last few neutrons
have a large ground-state spatial extension are be found in $^{6,8}$He, $^8$B, $^{11}$Li, and $^{11}$Be.  
Bound states are characterized by their spin-parity $J^\pi$ and energy position $E$, 
unbound states can manifest themselves as decaying resonances defined by their width $\Gamma$, inversely related to their lifetime $\tau = \hbar/\Gamma$. There are also $\ell=0$  
antibound states or virtual resonances affecting the low-energy scattering cross section. 

The description of these inherently time-dependent states starting from high-precision nuclear forces thus requires a unified approach to nuclear structure and reactions.
Extending \textit{ab initio} methods to weakly bound and unbound systems thus presents a considerable challenge. 
Some of the state-of-the-art works in this region include for instance  
the Faddeev-Yakubovsky calculations with the uniform complex-scaling method \cite{lazauskas18_2032}, 
the no-core shell model (NCSM) and its extensions (NCSM-RGM and NCSMC) based on the resonating group method \cite{baroni13_753,vorabbi18_1977}, 
and the Gamow density matrix renormalization group (G-DMRG) method \cite{papadimitriou13_441,fossez17_1916} 
and the no-core Gamow shell model (NCGSM) \cite{fossez17_1916,li21_2395} based on the Berggren basis \cite{berggren68_32,berggren93_481}. 

Providing stronger constraints on nuclear forces using near-threshold phenomena will hopefully improve \textit{ab initio} predictions in regions of interest for FRIB, 
and perhaps offer valuable insight into the the many-body dynamic of near-threshold states and their interplays with emergent features, 
which in turn will help better understand the formation of the drip lines \cite{erler12_1297,
%neufcourt20_2388}, 
Neufcourt2020a}, 
notably around islands of inversion. 

%\subsubsection{Recent NCSMC investigations of near threshold resonances and halo states}

As an illustrative example, the NCSMC approach has been applied to investigate near threshold resonances in several systems. The $^3$H(d,n)$^4$He and $^3$He(d,p)$^4$He reactions are leading processes in the primordial formation of the very light elements (mass number, $A\le7$), affecting the predictions of Big Bang Nuleosynthesis (BBN) for light nucleus abundances~\cite{1475-7516-2004-12-010}. With its low activation energy and high yield, $^3$H(d,n)$^4$He is also the easiest reaction to achieve on Earth, and is pursued by research facilities directed toward developing fusion power by either magnetic ({\em e.g.}\ ITER) or inertial ({\em e.g.}\ NIF) confinement. At low energies, the fusion reaction rate is significantly enhanced by the $3/2^+$ resonance appearing just about 50 keV above the DT (D$^3$He) threshold; compared to 210 keV in the $^3$He(d,p)$^4$He reaction. Using a chiral NN+3N interaction, the NCSMC calculations~\cite{Navratil2012,
%Hupin2019} 
Hupin:19} 
were able to reproduce this resonance quite well showing its $S$-wave nature in the  DT channel. The outgoing $D$-wave $n{-}^4$He (or $p{-}^4$He) channel is characterized by a strongly varying diagonal phase shift that does not reach $90^\circ$. The NCSMC calculations were also able to predict the enhancement of the fusion cross section when the reacting nuclei are polarized
%~\cite{Hupin2019}. 
~\cite{Hupin:19}.
It should be noted that this resonance is responsible for the production of 99\% of $^4$He in BBN with the remaining 1\% of primordial $^4$He coming from the $^3$He(d,p)$^4$He reaction, which benefits from the same mirror $3/2^+$ resonance~\cite{Smith93,Chadwick:2023xni}.

The NCSMC was also used to investigate resonances in $^7$Li and $^7$Be~\cite{DohetEraly2016,Vorabbi2019}. In Ref.~\cite{Vorabbi2019}, an $S$-wave $1/2^+$ resonance just above the threshold of $^6$He+p was reported. For technical reasons, the calculations had been performed by considering the $^3$H+$^4$He, $^6$Li+n, and $^6$He+p mass partitions separately without taking into account their coupling. Apparently the same $S$-wave $1/2^+$ resonance was observed in the $^6$Li+n mass partition just below the $^6$Li($J^\pi{=}0^+,T{=}1$)+n threshold. In a separate ongoing investigation, a broader $1/2^+$ resonance seem to play a role in the $^3$H+$^4$He scattering at $\sim$7 MeV. An experimental search for the predicted $1/2^+$ resonance above the $^6$He+p has been performed without success~\cite{PhysRevC.107.L061303} although an isospin anti-analog $(3/2^-,1/2)$ resonance just above the $(3/2^-,3/2)$ state also predicted in Ref.~\cite{Vorabbi2019} was found. It is likely that the discussed $1/2^+$ resonance exist between the $^6$Li+n and $^6$He+p thresholds in $^7$Li and a more realistic prediction of its position and width requires proper coupling of all the mass partitions. Calculations in this direction are under way.

Recently the NCSMC capability was extended to calculate Fermi and Gamow-Teller $\beta$-decays with the final state bound or in the continuum~\cite{atkinson:2022}. Motivated by the TRIUMF experiment~\cite{ayyad:2022}, calculations were performed for the $^{11}$Be $\beta$-decay with the proton emission. Focusing on the $1/2^+$ final states, a resonances in $^{11}$B have been identified with a substantial Gamow-Teller strength. By applying a phenomenological adjustment (dubbed NCSMC-pheno) of the {\it ab-initio} calculation, the resonance was shifted to the position observed in the experiment~\cite{ayyad:2022}. The NCSMC-pheno predicted width of the resonance matched closely the experimentally found width~\cite{atkinson:2022}. The calculated Gamow-Teller branching ratio was about an order of magnitude smaller than that reported in the TRIUMF experiment, although still two orders of magnitude higher than non-resonant value~\cite{Baye:2011}. It should be noted that the chiral NN+3N interaction used in that work reproduced the parity inversion in the $^{11}$Be ground state as well as in the isospin-analog $(1/2^\pm,3/2)$ resonances in $^{11}$B.

The parity inversion in $^{11}$Be and its extended halo ground state was investigated within NCSMC in Ref~\cite{calci16_2128}. It was found that only a chiral Hamiltonian with a non-local 3N interaction reproduces the inversion. We note that a non-local chiral 3N interaction was utilized in Ref.~\cite{atkinson:2022}. A near-threshold enhancement of the $E1$ strength in the photo-dissociation of $^{11}$Be was investigated in Ref.~\cite{calci16_2128} with obtained strength in agreement with experiment. The calculations confirm the non-resonant nature of this enhancement. The ground-state asymptotic normalization coefficient (ANC) predicted in these calculations have been later confirmed in phenomenological calculations investigating the breakup of $^{11}$Be and $^{10}\mathrm{Be}(d,p)^{11}\mathrm{Be}$ transfer reaction~\cite{PhysRevC.98.054602}. 

In a related unpublished study, the $S$-wave halo $1/2^+$ ground state and the $D$-wave extended excited $5/2^+$ state of $^{15}$C have been investigated utilizing the NCSMC with the $^{14}$C+n continuum channels. The calculated ANCs, $C_{1/2^+} {=} 1.282$ fm$^{-1/2}$  and $C_{5/2^+} {=} 0.048$ fm$^{-1/2}$, have been later validated in the experimental data analysis in Ref.~\cite{PhysRevC.100.044615}.

%______________%

%______________%
\section{Reaction mechanisms\label{sec:Reactions}}
\label{sec:reactions}

%{\em Contributors to this section: Jutta Escher, Alexis Mercenne, Linda Hlope, Dean Lee, Xilin Zhang, Daniel Bazin, Alex Gade, Grigor Sargsyan.

Describing the dynamics of interacting many-body systems produced at radioactive beam facilities provides unique challenges and broad opportunities. 
Integrating structure and reaction theory into a unified framework forms the basis for microscopic descriptions of interacting nuclei and provides critical links to observables. 
FRIB experiments will play an important role in testing current reaction theories on nuclei that have so far been out of reach.
This will highlight areas in need of improvement and increase the reliability of the nuclear reaction inputs used in applications. 
Accurate reaction calculations are needed for applications:  Reaction cross sections are important ingredients for multi-physics simulations used in the areas of astrophysics, national security, medicine, and nuclear energy.
In addition, a robust understanding of reaction mechanisms aids in planning and interpreting the experiments that are used to probe structure aspects of nuclei.

A significant challenge for nuclear reaction theories is the integration of reliable structure information in the calculations.  To quantitatively describe and predict nuclear reaction outcomes, both the structure information and the reaction mechanisms included in the theory have to be accurate.  In the past decade, efforts have focused on treating structure and reactions on the same footing in {\em ab initio} approaches, which have primarily been applied to light nuclei, and to integrate more advanced structure descriptions into both direct and statistical reaction theories, which are used for medium-mass and heavy nuclei.  FRIB experiments will have the opportunity to provide data for chains of isotopes, to study the impact of the evolution of shell structure and collectivity on reaction observables, and to probe changes in the relative importance of the reaction mechanisms as one moves away from stability to nuclei with lower level densities and less binding.  

%To reproduce measurements and to formulate predictions for reaction observables

\subsection{\textit{Ab initio} approaches to reactions} \label{sec:abinitio_reactions}

An important goal of nuclear theory is to predict both static and dynamic properties of nuclei in a consistent framework.  
Remarkable progress has been made in recent years in the development of many-body approaches from first principles to scattering and nuclear reactions (see Refs. \cite{Johnson2020, 
%Navratil:16, 
navr2016, 
hebborn2023optical} for reviews).
Truly microscopic approaches take into account nucleon degrees of freedom along with their correlations within and between the reaction fragments. Coupled with realistic inter-nucleon interactions, these approaches provide {\em ab initio} predictions of reaction observables, including phase shifts and reaction and scattering cross sections.

For light nuclei, the combination of the no-core shell model (NCSM) with the resonating group method (RGM), currently applicable for reactions with both nucleons and light composite projectiles, treats nuclear structure and reactions on equal footing and has successfully predicted scattering, transfer, and capture reactions~\cite{navr2016,Hupin:19, Quaglioni:20, Kravvaris:24}.

%%Navratil:16, 
%navr2016,
%Hupin:19, Quaglioni:20, Kravvaris:24}.
%JE(KK):This integrated treatment allows for novel interpretations of experimental measurements but also presents with new data needs. For example, systematic measurements of energy spectra along entire isotopic chains of light nuclei can serve as unique constraints on the nuclear force. Furthermore, high-precision measurements of decay widths and transition rates in highly collective nuclei will also provide stringent benchmarks that will reveal possible theoretical weaknesses. The enhanced predictive power that comes from employing such first-principle theories will lead to a better understanding of astrophysical reaction rates and of the emergence of collectivity from the fundamental nuclear interaction.
%

The use of symmetry-adapted (SA) bases, within the SA-NCSM~\cite{launey:PPNP2016, dytrych:PRL2020}, provides a viable path to extend the NCSM/RGM approach to describe reactions with medium-mass nuclei~\cite{Launey:21,Mercenne:22}. 
%The SA-NCSM \cite{launey:PPNP2016, dytrych:PRL2020} is an {\em ab initio} framework that exploits SU(3) and Sp(3,$\mathbb{R}$) symmetries, found to be approximately valid in many nuclei up to the calcium region \cite{dytrych:PRL2007, dytrych:PRL2013},  to resolve the scale explosion problem in nuclear structure calculations. 
The SA-NCSM can accommodate larger model spaces and reach heavier nuclei than traditional NCSM methods, including \nuc{20}{Ne} \cite{dytrych:PRL2020}, \nuc{21}{Mg} \cite{ruotsalainen:PRC2019}, ${}^{22}$Mg \cite{henderson:PLB2018}, ${}^{28}$Mg \cite{williams:PRC2019}, ${}^{32}$Ne and ${}^{48}$Ti \cite{launey:AIPCP2018}.  
These larger model spaces make it possible to predict collective observables, such as $E2$ transition strengths and quadrupole moments, without effective charges, as well as $\alpha$-decay partial widths and asymptotic normalization constants (ANCs)~\cite{henderson:PLB2018,Dreyfuss:20,Sargsyan:23}.  Measurements of these quantities serve as stringent tests of the predictive power of these {\em ab initio} approaches, which aim at describing both structural nuclear properties and reaction outcomes, and help identify potential weaknesses in the theory.

RGM-based methods will provide theoretical support for planned and upcoming FRIB experiments for neutron-rich isotopes, especially for understanding deformation and the role of intruder states in the proximity of the drip line (e.g., for He, C, Ne, Mg, and Ca isotopes).  Comparisons between predictions and FRIB data will also be critical for understanding the impact of these phenomena on reaction outcomes. Collective correlations have been shown to play a key role in measurements of nuclei across the nuclear chart, and experiments are now starting to explore such properties for exotic nuclei, see, e.g. Refs.~\cite{Holl:21, Kanungo:20handbook, Chen:22}. Charged-particle inelastic scattering is particularly sensitive to collective effects and can~--- with some additional development~--- be performed in inverse kinematics. 
FRIB scattering measurements will not only shed light on the evolution of collectivity away from stability; they will also test to what extent our present theoretical approaches capture the relevant correlations~\cite{Launey:21,Mercenne:22,navratil:22handbook,bazin:2023}.  

Nuclear lattice simulations provide a very different, but complementary, approach to computing {\it ab initio} nuclear structure and reactions using chiral effective field theory \cite{Elhatisari:2015iga}.  Recent progress has allowed calculations to be performed using high-fidelity chiral interactions to produce accurate predictions for the properties of light and medium-mass nuclei \cite{Elhatisari:2022zrb} and measure $A$-body correlations between nucleons to probe intrinsic shapes, deformation, and clustering \cite{Elhatisari:2017eno,Shen:2022bak}.  Efforts are now in progress to develop efficient algorithms to compute the energies and decay widths of nuclear resonances.   Of particular interest are the properties of neutron-rich nuclei.  This includes both excited states of bound nuclei as well as unbound nuclei past the dripline such as $^{28}$O 
%\cite{Kondo:2023lty} 
\cite{kond2023} 
that decay into multi-neutron channels.  There is much interest in understanding continuum scattering features for such multi-neutron systems \cite{Kisamori:2016jie,Duer:2022ehf,Lazauskas:2022mvq}.

In addition, two other methods are being explored to extract nuclear scattering, reactions, and response functions from spatially localized wave functions: (1) eigenenergies of the nuclear systems in external traps~\cite{Zhang:2019cai, Zhang:2020rhz, Bagnarol:2023crb}, or (2) continuum states at complex energies~\cite{XilinZhangTRIUMF2023,Zhang:2024ril}.  
Because these methods use localized wave functions, they could potentially take advantage of the recent progress of \textit{ab initio} structure methods in computing continuum state observables in the nuclear chart wherever the structure methods are feasible, thus expanding the applicability of \textit{ab initio} continuum calculations. In addition, for both methods, fast emulations for the input parameters are being developed, which will enable uncertainty quantification of the continuum predictions and allow for more meaningful comparisons with experiments.
The anticipated FRIB detections of new drip-line nuclei, either shallow bound or resonant states, and their binding energy and/or width measurements, can provide valuable benchmarks for these methods and the nucleon interaction theories on which these methods are based. These continuum methods can then provide inputs for other physics areas,  such as capture reaction cross-sections for nuclear astrophysics and neutrino-nucleus scattering cross-sections for neutrino oscillation measurements.
%\textcolor{red}{What physics questions will be addressed with these methods?}
%
%The first method generalizes the so-called L\"{u}scher method~\cite{Luscher:1990ux}, a method extensively used in Lattice QCD to infer hadronic scatterings from hadrons' eigenenergies in finite spatial boxes~\cite{Briceno:2017max}, to extract nuclear scatterings from the eigenenergies of the scattering systems trapped in harmonic potential wells~\cite{Zhang:2019cai, Zhang:2020rhz,Bagnarol:2023crb}. 
%The second method extrapolates continuum state calculations at complex scattering energies above the real energy axis to the other regions of the complex plane (e.g., real energies) by using projection-based emulations~\cite{XilinZhangTRIUMF2023}. In both approaches, the nuclear many-body wave function solutions are spatially localized, and therefore could potentially be computed using current nuclear structure codes. They allow us to take advantage of the progress of \textit{ab initio} structure methods to compute continuum state observables in the nuclear chart wherever the structure methods are feasible. The coming FRIB detections of new drip-line nuclei, either shallow bound or resonant states, and their binding energy and/or width measurements, can provide valuable benchmarks for these methods and the nucleon interaction theories that these methods are based on. 

\subsection{Direct reactions: Transfers, inelastic scattering, breakup, and knockout}

%\subsection{Direct reactions - transfers, inelastic scattering, and breakup}

%For medium-mass and heavy nuclei, we distinguish between direct and compound nuclear reactions.
Direct reactions change few degrees of freedom in the reaction partners and serve as excellent probes of nuclear structure.
Of interest are both individual single-particle and collective degrees of freedom. Particle transfer reactions are useful for determining the former, and inelastic scattering is used to probe the latter.  
Scattering experiments can also be used to study resonant states in light exotic nuclei. 
%For instance, $^1$H and $^4$He gas active target devices can be used to populate proton and alpha resonances in inverse kinematics, such as a recently measured $^{11}$B resonance observed in p+$^{10}$Be and alpha+$^{7}$Li experiments~\cite{Ayyad:22prl}.
Both transfer reactions and inelastic scattering have been used extensively to study properties of stable nuclei~--- and both reaction types can be used to explore exotic nuclei in inverse-kinematics experiments.  
In confronting experimental observables, theory needs to improve the underlying structure description that is assumed in the reaction model, e.g., moving from simplistic collective or single-particle models to more sophisticated structure models. It also needs to account for reaction mechanisms that have been studied, but that are not regularly included when describing experimental observables, e.g., including two-step reactions, couplings to excited states in deformed nuclei, breakup contributions, etc. 
This will be relevant not only to better extract the structure properties of exotic nuclei from FRIB reaction experiments; it will also shed light on the reaction mechanisms at work.

An important goal of nuclear theory is to obtain accurate descriptions of the nuclear reaction dynamics, while taking into account the underlying nuclear structure. While {\em ab initio} descriptions of direct nuclear reactions have seen great advances for light systems, a broader application across the nuclear chart, for a wide range of beam energies, is presently not feasible. 
However, in many cases the complicated many-body problem can be reduced to a few-body problem. 
For elastic and inelastic scattering, single and multichannel optical potentials are typically used to solve an effective two-body Schr\"odinger equation. For single-particle transfer reactions, one considers an effective three-body problem whose exact solution can be obtained using Faddeev methods~\cite{Faddeev1961}. The approach can be generalized to two-particle transfer processes leading to a four-body scattering problem that is described within the Faddeev-AGS equations in momentum space~\cite{Deltuva2007} and Faddeev-Yakubovsky equations in coordinate space~\cite{Lazauskas2005}. The numerical cost associated with the exact few-body methods has led to various approximate techniques that have validity within specific kinematic conditions. Commonly utilized techniques include the distorted-wave Born approximation (DWBA), the adiabatic distorted-wave approximation ADWA, the continuum-discretized coupled channels (CDCC), and variants thereof~\cite{Thompson:09book, Johnson2020, Hagino:22, Alt:67, Deltuva:19, Timofeyuk:20, Descouvemont:18, Watanabe:21}.

Complementing these developments are efforts to use structure information from microscopic theories in reaction calculations. One-body overlap functions from the Green's function Monte Carlo and variational Monte Carlo methods are used in descriptions of transfer and knockout reactions~\cite{Brida:11, Grinyer:12} and transition densities from the (quasiparticle) random phase approximation enter calculations of inelastic scattering and reaction cross sections~\cite{Nobre:10, Nobre:11, Dupuis:19, Chimanski:23preprint}.

To better understand the interplay of nuclear structure effects and reaction mechanisms, multiple types of reaction data are needed.  Elastic and inelastic scattering, in particular angular distributions, provide stringent checks of the optical model and structure models used.  Inelastic scattering data can be used to study the effect of couplings between ground and excited states, as well as re-arrangement effects (due to density differences between the ground and excited states) on the scattering observables~\cite{Dupuis:19}.  
Experimental efforts that study a chain of isotopes, or at least multiple nuclei with large isospin differences, will be particularly valuable, as they can shed light on the evolution of nuclear properties with neutron excess and/or as a function of deformation.
Exclusive and inclusive cross sections are needed to better understand the relative importance of elastic and inelastic breakup mechanisms, as well as complete and incomplete fusion in transfer reactions~\cite{Lei:15a, Lei:15b, Potel:15, Carlson:16, Potel:17, Lei:19a, Lei:19b, Nunes:20rev, GomezRamos:22, bazin:2023, Lei:23, Torabi:23}.

A combination of elastic scattering and breakup measurements can potentially provide information on the structure of one-neutron halo nuclei.  This is true if the prediction of the recoil excitation and breakup (REB) model holds, which finds that the elastic scattering pattern of one-neutron halo nuclei from a heavier target is very similar to the angular pattern obtained when the halo nucleus breaks up into its core and halo neutron~\cite{Capel:10}. Within the REB model, the ratio of the angular distributions for elastic scattering and breakup is only a function of the projectile nuclear wave function and can thus provide sensitive information on the halo structure of the projectile. This ratio method has been studied theoretically for both one-neutron and one-proton halo nuclei~\cite{Capel:11, Capel:13, Capel:20}.  Given the potential impact of this approach, it is important to experimentally test the predictions of the underlying reaction model.
In particular, it would be valuable to confirm that this new reaction observable is independent of the reaction process and to establish under which experimental conditions this independence holds. 

%Advances in this area enable the community to better understand reaction mechanisms \textcolor{red}{(examples?)(e.g. the role of multi-step reaction contributions) } for stable and exotic nuclei, to extract information on shell closures from transfer reactions, and to study collective modes of excitation in very neutron-rich and proton-rich isotopes.
%%Experimental efforts that study a chain of isotopes, or at least multiple nuclei with large isospin differences, will be particularly valuable, as they can shed light on the evolution of nuclear properties with neutron excess and/or as a function of deformation.

%\subsection{Direct reactions - knockout}
The status of the theoretical treatment of nucleon knockout reactions, 
heavy-ion as well as proton-induced, was reviewed recently in great 
detail~\cite{Aumann2021}. The broad conclusions are that (i) no 
consistency is achieved yet between different ways of modeling $(p,2p)$ 
and $(p,pn)$ reactions and that (ii) the reduction factor dependence as 
function of the asymmetry at the Fermi surface reported for Be- and 
C-induced knockout reactions has not been found to manifest itself in 
the handful of quasi-free knockout and transfer reactions at the 
extremes of $\Delta S$ published so far \cite{Aumann2021, Kay2022}. For Be- and C-induced knockout, novel approaches 
are being developed to explore inclusion of processes that hinder core 
survival when a minority nucleon is removed and the majority nucleon 
separation energy is low 
%\cite{Hebborn2023a}. 
\cite{Hebborn:23}.
First valuable attempts to 
quantify the uncertainty in knockout reactions evaluate a single aspect, such as optical model potentials that 
carry their own sizable uncertainty \cite{Hebborn2023b}, but do not yet interrogate, as a whole, something akin to the consistent formalism used over decades to analyze knockout data \cite{Tostevin2021,Gade2008}. For example, as shown in \cite{Gade2008}, the geometry of the bound-state potential is crucial and aspects such as the optical model potentials should not be viewed in isolation. In general, the use of direct reactions for the extraction of absolute spectroscopic 
factors remains difficult due to uncertainties in reactions models 
which, at present, are not on the same footing with the nuclear structure 
calculations that provided needed input~--- this is not different for 
knockout reactions and remains a challenge and opportunity for nuclear 
theory that could advance the field tremendously. Quasi-free knockout of the $(p,2p)$ or $(p,pn)$ type becomes only 
competitive at FRIB after the energy upgrade to FRIB400 \cite{FRIB400} so that the nucleons in the exit channel have at least 
200 MeV of energy. This energy regime for the exit-channel proton or neutron inside the nucleus starts to minimize distortions encountered in the energy-dependent nucleon-nucleon ($NN$) cross sections that are typically input to models used to describe quasi-free knockout reactions~\cite{FRIB400,Aumann2021}.

\subsection{Statistical descriptions of compound-nuclear reactions}
\label{sec:reactions_statistical}

Nuclear astrophysics and other applications require cross sections for compound-nuclear reactions, in particular for neutron-induced reactions~\cite{Arcones:17, Arnould:20, Hayes:17}.  Application needs drive many measurements and nuclear reaction data evaluations~\cite{Brown:18endf8, Koning:19tendl}, which culminate in evaluated data being broadly made available in databases at the NNDC, IAEA, and other data centers around the globe. To a large extent, the reaction evaluations rely on reaction descriptions that contain phenomenological components and thus require nuclear data for parameter adjustments.  For reactions on isotopes that have not been measured, or for unknown reaction channels, the calculations often have to rely on regional, or even global, systematics.  Experiments that can provide constraints on the models used have potentially very high impact, not only because a specific reaction involving a particular nucleus will be better known, but also because the measurements can affect the regional systematics developed and the results can be propagated through simulations relevant to an application.  Moreover, experiments that probe the limits of validity of the reaction descriptions can provide impetus and guidance for developing the next generation of reaction theories.

Compound-nuclear reactions are traditionally treated in a phenomenological $R$-matrix approach (for isolated resonances) or statistical Hauser-Feshbach (HF) theory (for overlapping resonances)~\cite{Azuma:10, Descouvemont:10, Thompson:09book, Capote:09, Carlson:14, Koning:2023wz}.  Practical HF calculations require optical-model potentials and nuclear-structure information in form of level densities and gamma-ray strength functions, plus fission barriers where decay by fission has to be considered.  Spins and parities of low-lying nuclear levels, and their decay branchings, are needed as well.

Optical-model potentials, which are crucial for both direct and statistical nuclear reaction calculations, have been the focus of recent research that seeks to ground the potentials in microscopic calculations, to extend our knowledge of these potentials to unstable targets, and to quantify associated uncertainties. 
Some aspects are discussed in the next subsection and a more in-depth review of recent optical-model developments is given in Ref.~\cite{hebborn2023optical}. 

Phenomenological level density models and gamma-ray strength functions have become much more accurate in the last decade, and significant progress has been made in obtaining these quantities from microscopic approaches. Both shell model and DFT-based approaches have been employed to predict level densities~\cite{Koning:08ld, Goriely:08ld, Hilaire:12ld, alhassid:15, shim2016, Karampagia:20, Ormand:20, Goriely:22ld}.   While Hartree-Fock(-Boguliubov) methods, combined with a combinatorial approach, can be used across the isotopic chart, correlations due to many-particle many-hole configurations and/or rotational excitations, which do affect the level densities, have to be accounted for through (adjustable) correction factors~\cite{Dossing:19ld}.  Very recent work has laid out a path to go beyond the mean-field approximation by exploiting boson-expansion methods~\cite{Hilaire:2023ux}.  
Approaches based on the shell model, on the other hand, are limited by the model space sizes and interactions that are available to shell-model calculations and therefore are typically restricted to light and medium-mass nuclei, or heavier nuclei near closed shells.  
The shell model Monte Carlo (SMMC) method addresses the challenge of the model space size by reformulating the shell model in the framework of auxiliary field Monte Carlo \cite{alhassid_2017_book, alhassid_PRL_2008, gilbreth_CPC_2015}. Where suitable interactions are available, SMMC can provide microscopic predictions of nuclear properties across a wide range of nuclear systems, extending from medium-sized to rare earth nuclei~\cite{alhassid_PRC_2015,alhassid_PRL_2017,fanto_arxiv_2021}.
Shell-model approaches do include correlations beyond those contained in mean-field models, and innovative methods for efficiently calculating moments and smart model-space truncations~\cite{shim2016, Karampagia:20, Ormand:20, Johnson:23}, coupled with high-performance computers, are pushing the existing boundaries.

Microscopic calculations of gamma-ray strength functions are challenging, as theory needs to cover a wide range of transition energies, several multipole-parity combinations ($E1$, $M1$, $E2$, etc), and contain the collective correlations that shape the strength functions.  This is difficult for the shell model~--- again for reasons related to model-space size and availability of suitable interactions (in particular for cross-shell excitations).  DFT-based methods, in particular the QRPA, have been used to calculate strength functions for a large number of nuclei~\cite{Goriely:19, Goriely:19dipole, Goriely:19gSF}. Since these calculations construct the strength function from excitations built on the ground state, they cannot provide reliable predictions for the low-energy portion of the strength function. Work is underway to push the limits of the shell model~\cite{Johnson:23} and to account for additional correlations in the DFT-based approaches~\cite{Tsoneva:16, Tsoneva:18, Peru:23pc}.
Such theory extensions are expected to yield predictions of phenomena such as pygmy resonances, toroidal resonances, and the shape of the low-energy tail of the gamma-ray strength functions (``low-energy enhancement" \cite{lars2014}).  Since these features cannot be directly observed in experiments, it is important to integrate the structure into (direct) reaction descriptions and plan experiments that are sensitive to the predictions. Shedding light on these features is not only interesting from a nuclear structure perspective, it is also important for developing more reliable statistical reaction calculations of neutron capture cross sections.

Experimental quantities, such the average spacings between $s$-wave and $p$-wave neutron resonances at the neutron separation energy ($D_0$ and $D_1$, respectively), and the average radiative width ($\langle \Gamma_{\gamma} \rangle$), are known to provide strong constraints for level density models and calculated gamma-ray strength functions, but these quantities are not available for unstable isotopes. 
Early work showed some promise in getting information on resonance properties from beta-delayed neutron emission~\cite{Raman:83, Raman:84}.  However, many questions remain.  In particular, it is an open question whether beta-decay populates states in the daughter nucleus that damp into a compound nucleus (which then decays statistically via gamma or neutron emission) or produces states that decay via direct or semi-direct decay mechanisms~\cite{Tain:15, Madurga:16, Spyrou:16, Gottardo:17, 
%Heideman:23, 
Hei23, 
Xu:24prl}. In the former case, one may infer constraints on level densities and strength functions from observing the beta-delayed neutron and/or gamma emission.

Similarly, if beta decay, with subsequent neutron emission, proceeds through the compound nucleus stage, one can  decouple the nuclear structure effects from the neutron emission, which is a precondition for using HF codes to describe $\beta$-$n$ processes \cite{Kaw08}. Recently, however, evidence for non-statistical neutron emission was discovered, and the concept of doorway states was used to explain the data \cite{Hei23, Xu:24prl}. Mounting evidence \cite{SLAUGHTER197222, Piersa19} shows that the neutron emitting states in medium and heavy nuclei are very narrow.  For all of these situations, a better microscopic description is needed for the description of $\beta$-delayed neutron emission. 
FRIB will provide ample access to $\beta$-delayed neutron emitters in different regions of the nuclear chart, offering up various bound and unbound single-particle configurations in various $Q$-value windows.

Multiple indirect methods have been introduced to obtain information on level densities and gamma-ray strength functions~\cite{Escher:12rmp, Loher:13, Escher:16a, 
%Larsen:19}. 
Lar19a}.  
One challenge for experiments aiming to obtain information on level densities and gamma-ray strength functions is that the observables reflect convolutions of these quantities. It is therefore important to explore multiple experimental approaches which can complement each other.  One should also consider forward modeling of the reaction (using theory inputs where needed) and the experimental set-up, with uncertainty propagation, in order to identify how specific aspects of the level densities and gamma-ray strength functions are reflected in the observables of a planned measurement. 
For level densities, experiments that can shed light on the impact of deformation and shell closures, the role of correlations, and the dependence on neutron excess would be particularly valuable.  Typical approximations for level density models, such as their dependence on spin and parity, need to be further investigated.
For gamma-ray strength functions, one needs to better understand the microscopic origin of the observable features of the strength function including the behavior at low gamma energies, the role of deformation and collectivity, and the dependence of these properties on neutron excess.

The use of the HF statistical reaction description is only justified when the fusion of the projectile with the target nucleus produces a compound nucleus in an energy regime of strongly-overlapping resonances. For light ($sd$-shell and below) nuclei and nuclei near the dripline, characterized by low level densities, it is not \textit{a priori} justified to use the HF description.  Nuclear data evaluators routinely connect $R$-matrix descriptions at very low projectile energies with HF descriptions at higher energies, using experimental data as guide for the evaluations~\cite{Rochman:17,Rochman:20}. No information is available to guide calculations for very neutron-rich nuclei near the dripline.  In addition, at low level densities one expects contributions from direct and semi-direct reactions~\cite{Rauscher:98, Goriely:12direct, Xu:14, Thompson:14, Sieja:2021a, Saito:23, Wang:24pCap}.  Experimental guidance and reaction theory developments are needed, e.g., for providing neutron capture rates for astrophysical simulations.

\subsection{Surrogate reactions and reaction mechanisms}
\label{sec:reactions_surrogate}

In the absence of theoretical predictions for the major ingredients to HF calculations (optical model potentials, level densities, gamma-ray strength functions), it is important to obtain constraints for cross section calculations.  The surrogate reaction method provides such constraints for compound nucleus (CN) reactions through a combination of theory and experiment.  The CN designation does not simply refer to a composite system made of two nuclei~--- it indicates that the reaction partners fuse and produce a system in statistical equilibrium.  This system remembers the conserved energy, angular momentum, and parity, but otherwise forgets how it was produced.  In the second stage of the reaction, the CN decays~--- by a combination of fission, gamma emission, and particle evaporation.  Many reactions of interest to nuclear astrophysics and societal application are CN reactions.

The surrogate reaction method uses a direct reaction~--- typically charged-particle scattering or a transfer reaction - to produce a CN (more precisely, it produces a doorway state that evolves into a CN).  The subsequent decay of the nucleus is observed, and a quantity that is characteristic of the exit channel of interest (e.g., a fission fragment or a $\gamma$-ray characteristic of a reaction product) is measured in coincidence with the outgoing surrogate-reaction particle.  By combining a theoretical description of the direct-reaction mechanism used to produce the CN with modeling the decay, and comparing with the coincidence measurement, one can extract constraints on the CN decay models.  This then allows one to produce a data-constrained calculation of the desired CN reaction.
The method has been reviewed in Ref. \cite{Escher:12rmp}, a pedagogical introduction into the underlying ideas was given in Ref. \cite{Escher:16a}, and more recent developments for $(d,p)$ reactions have been discussed in Ref.~\cite{Potel:17}.

Proof-of-principle applications to $^{90}$Zr$(n,\gamma)$, $^{95}$Mo$(n,\gamma)$, utilizing $(p,d)$ and $(d,p)$ reactions, and the determination of $^{87}$Y$(n,\gamma)$ for the unstable $^{87}$Y isotope have demonstrated the application to neutron-capture reactions~\cite{Escher:18prl, Ratkiewicz:19prl}. Inelastic alpha particle scattering was utilized to simultaneously determine neutron-induced fission and radiative capture cross sections for $^{239}$Pu~\cite{PerezSanchez:20}. More recent work has focused on obtaining $(n,\gamma)$, $(n,n')$, and $(n,2n)$ reactions from inelastic scattering with protons and light ions~\cite{Gorton:23, Escher:23dnp} and on measurements in inverse-kinematics. Also considered are $(n,p)$ reactions~\cite{Sharma:22}. Fission reactions, which were the focus of the earliest surrogate reaction measurements, are being revisited~\cite{Casperson:14, Hughes:14}, including in inverse-kinematics experiments~\cite{Bennett:23}. 

The theoretical description of the reaction that produces the doorway state is non-trivial, as the excitation energies of the system produced are in the range of a few MeV to 20--30~MeV, two-step reaction mechanisms have been found to contribute, and the role of width fluctuations needs to be considered in some cases~\cite{Carlson:16, Escher:18prl, Bouland:19, Bouland:20, Escher:23dnp, Escher:23}.  Modeling of the CN decay, to connect with the experimental observable, benefits from knowledge of the structure of the decaying nucleus.  Finally, a better understanding of the evolution from a doorway state to a CN is desirable.  

In addition to investigating the reaction mechanisms in more detail, it is important to expand, improve, and test the experimental techniques to be used with radioactive beams: Experimental conditions may make it more difficult to obtain and utilize the decay observables that have been used in the past, cover a more limited range of excitation energies in the compound nucleus, have different background challenges, and/or have lower count statistics. Validating the method by comparing the results from two different surrogate reaction mechanisms, e.g., $(d,p)$ and inelastic scattering, against each other and against other approaches (indirect measurements or predictive theory) will be valuable. 
Theoretical and experimental work that sheds light on these aspects will not only enable the broader applicability of the surrogate reaction method; it will also provide new insights into the interplay of direct and compound reaction mechanisms.

\subsection{Optical-model potentials for reaction calculations}
\label{sec:reactions_omp}

Optical-model potentials (OMPs) describe the effective interaction between a target nucleus and a projectile. They are essential ingredients in analyses of direct reactions, such as elastic and inelastic scattering, transfers, breakup and knockout reactions, as well as for the description of compound nuclear reactions. 
The imaginary part of the OMP reproduces the loss of flux from the elastic scattering channel due to all other processes that are not explicitly accounted for in the calculation. The importance of such potentials and various techniques for obtaining OMPs are discussed in a recent review~\cite{hebborn2023optical}.
The role of (dispersive) optical-model potentials for constraining the equation of state (EOS) and for providing input to transport model simulations are discussed in Section~\ref{EOS_OMP}.

Phenomenological nucleon-nucleus optical models that have been adjusted to reproduce a large collection of nuclear data, such as the Koning-Delaroche (KD) and Chapel-Hill (CH) OMPs, have been tremendously successful in reproducing not only the data they have been trained on, but also newer measurements~\cite{Koning:03, Varner:91}, especially if uncertainties are incorporated in the study~\cite{Pruitt:23}.  The extensive collection of data on stable isotopes that was used for determining these models is unlikely to find a counterpart for unstable nuclei. Thus it becomes important to (a)~pursue approaches that can utilize theoretical predictions of nuclei and nuclear matter, (b)~identify experimental observables that constrain various components of the optical potentials needed, and (c)~strive to collect additional data for OMP training and testing whenever it is feasible in FRIB experiments.  Of particular interest are elastic scattering cross sections, charge-exchange reactions, and ratios of cross sections measured on different isotopes along an isotopic or isotonic chain. Bound-state properties can also be used to constrain and test dispersive optical models~\cite{Dickhoff:19, hebborn2023optical}.

\begin{figure}
    \centering
    \includegraphics[width=0.95\textwidth]{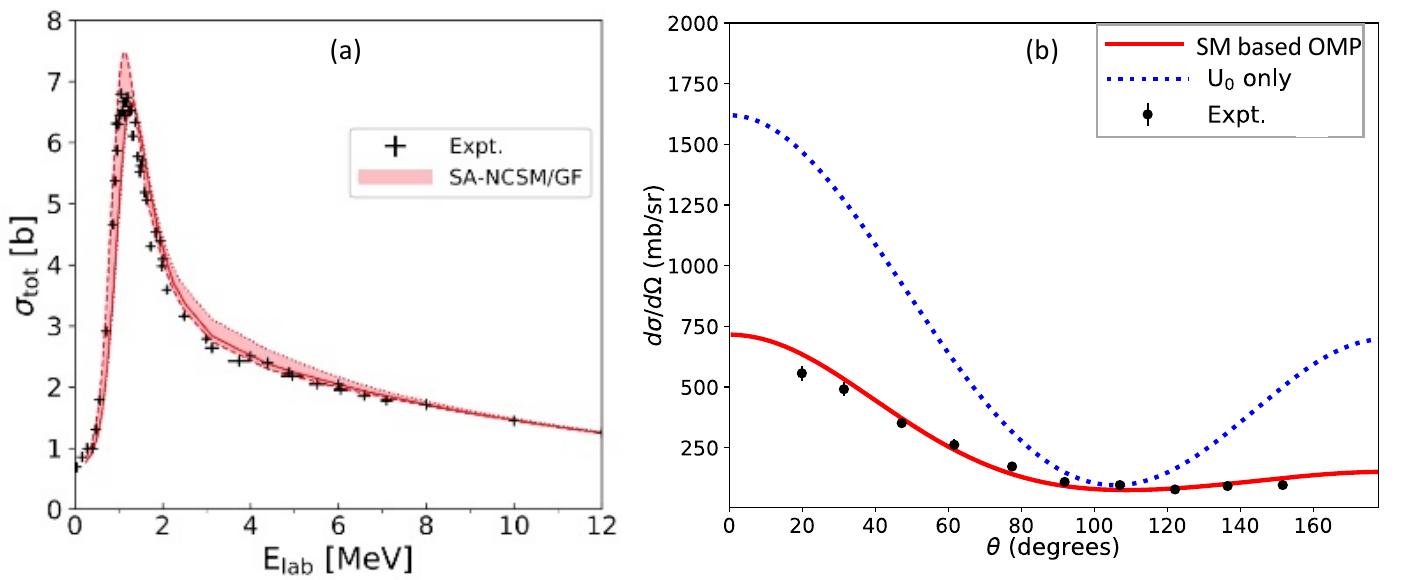}
    \caption{(a)~Total cross section for $n+^4$He \textit{vs.} the neutron kinetic energy in the laboratory frame calculated using the \textit{ab initio} SA-NCSM with Green’s function technique. The light red band shows the spread from calculations within a range of HO frequencies $\hbar\Omega$=12--20~MeV. Experimental data is from Refs.~\cite{HaesnerHKD1983, Coon1959total, BashkinMP1951}. (b) Differential cross section for neutron elastic scattering off $^{24}$Mg at 1.9~MeV bombarding energy. The solid red curve corresponds to predictions using the OMP constructed with around 600 intrinsic states from shell model calculations. The blue dotted line corresponds to calculations with the real Woods-Saxon potential only. Results with the full potential are in excellent agreement with experimental data from Ref.~\cite{ThomsonCL1962}, indicated by black circles. }
    \label{fig:24Mg_and_He}
\end{figure}

Recent theoretical approaches have employed realistic inter-nucleon interactions, typically derived within chiral effective field theory, without the need to fit interaction parameters in the nuclear medium. These models have evolved from earlier theoretical frameworks, such as the one introduced by Feshbach, which led to the Green's function formulation \cite{mahaux:86NPA, mahaux:91book, dickhoff:08book} and the successful dispersive optical model, as well as the pioneering work by Watson \cite{watson:53PR,kerman:AP59}, leading to the spectator expansion of the multiple scattering theory \cite{siciliano:PRC77}.
Notably successful recent applications include \textit{ab initio} nucleon-nucleus potentials for elastic scattering of closed-shell nuclei at low projectile energies ($\leq$ 20 MeV per nucleon) based on the Green's function technique with the coupled-cluster method \cite{rotureau:PRC17, rotureau:PRC18}, the SA-NCSM~\cite{Burrows:24},  and the self-consistent Green's function method \cite{idini:PRL19}. 
In these approaches a precise experimental threshold can be a vital input for achieving faster convergence and decreasing the uncertainty on the calculated reaction observables. Fig.~\ref{fig:24Mg_and_He}(a) shows an example of such a calculations for $^4$He$+n$.
Additionally, \textit{ab initio} approaches have been employed for light targets in the intermediate-energy regime ($\geq$ 65 MeV per nucleon) using the spectator expansion of the multiple scattering theory in conjunction with the \textit{ab initio} no-core shell model \cite{burrows:PRC20, vorabbi:PRC22, Johnson2020}.
Furthermore, optical potentials have been derived from two- and three-nucleon chiral forces in nuclear matter \cite{whitehead:PRC19}.
These potentials provide cross sections for elastic proton or neutron scattering and can be used as input for modeling $(d,p)$ and $(d,n)$ reactions \cite{rotureau:JPGNP20}.

It is possible to integrate microscopic nuclear structure information into the construction of OMPs by using the Feshbach formulation \cite{Feshbach1958}. This approach links the OMP to the underlying nucleon-nucleon interaction and allows one to build a nonlocal dispersive potential, where the nonlocality arises from the nonlocal nature of the underlying nucleon-nucleon interactions \cite{ArellanoB2018}. 
The general form of the OMP in Feshbach formulation is given by $V_{OP}=U_{0}+V_{\mathrm{pol}}$, where $U_0$ is a static energy independent mean field potential, while $V_{\mathrm{pol}}$ is an energy dependent non-local term constructed using the underlying structure of the many-body system.

Using these ingredients in the Feshbach formulation, it becomes possible to construct OMPs based solely on microscopic nuclear structure predicted by theory. An example is discussed in Ref.~\cite{SargsyanPKE2023}, where an OMP for $n+^{24}$Mg was constructed from intrinsic states provided by shell-model calculations.  The resulting neutron scattering predictions are in good agreement with the experimental scattering data, as shown in Fig.~\ref{fig:24Mg_and_He}(b). 
Approaches like this can provide a viable path for extending OMPs to very exotic nuclei, where measurements will remain scarce, but important for validating the predictions.
To use this approach to construct optical-model potentials for exotic nuclei, data is needed to develop and test shell-model interactions for nuclei in the region of interest, and to test the predictions of the newly-constructed potentials, e.g., through comparisons to differential proton elastic scattering data for selected nuclei.

\subsection{Fission}
\label{sec:reactions_fission}

%{\em This subsection to be completed by Kyle Godbey, Kyle Brown, and Witek Nazarewicz.}

%This intro is just a template, please revise and add
Nuclear fission, the process by which a heavy nucleus splits into two or more smaller nuclei along with the release of energy, has been a cornerstone of nuclear physics since its discovery.
Recent advancements and a heightened understanding of its intricacies have spurred renewed interest in the field, highlighting its significance in both basic and applied nuclear science.
Applications spanning the energy sciences, medical isotope production, and stockpile stewardship have the fission process as a central reaction channel and require high quality nuclear experimental and theoretical data.
In nuclear astrophysics, a comprehensive understanding of fission is vital to the pin down the endpoint of r-process nucleosynthesis and to understand the outgoing fragments' masses and energetics in fission recycling \cite{horowitz2018,Giuliani2020}.
This same understanding of the fission decay can also inform our knowledge on the structure and stability of superheavy elements and is at the heart of any analysis of superheavy searches \cite{GiulianiRMP,Smits2023}.

In recent years, advances in theory and computation have permitted more comprehensive microscopic studies of nuclear fission (see Refs.~\cite{schunck2016,Bender2020,Bulgac2020} for recent reviews).
This includes advances in the fidelity of the many-body methods as well as improved analysis techniques to perform systematic studies for a range of isotopes and fission modes~\cite{Flynn2022}.
While microscopic global studies are still a major computational challenge, continued methods development~\cite{sadh2022} and the adoption of machine learning into the pipeline~\cite{lovell2020quantifying,lay2023neural} promises to ease the burden of integrating microscopic approaches into r-process simulations.
The enormous progress in leadership class computing systems has enabled unprecedented fidelity in the simulation of the real-time dynamics of fissioning systems, though it remains incredibly challenging to properly quantify the uncertainties in a Bayesian manner.
Even more challenging is the inclusion of fission observables in the calibration of microscopic models \cite{Bertsch2015}, though such data promises to be very information-rich.
Observables that would be strong tests for theory include charge and mass yields, total kinetic energies, and their correlations. 

One method of producing isotopes of interest is via near-barrier fusion reactions with exotic projectiles.
Relatively high rates of reaccelerated beams in the tin and calcium region in particular will allow for systematic studies on entrance channel effects in heavy-ion fusion.
Microscopic methods have been successful in describing these reactions in recent years, though current state-of-the-art many-body methods suitable for large scale calculations are known to be under-constrained by experimental data~\cite{godbey2022} or otherwise deficient~\cite{desouza2023search}.
At FRIB one promising region for early experiments of fusion-fission reactions is in the vicinity of mercury and platinum.
This region has several nuclei that exhibit interesting asymmetric fission modes~\cite{Andreyev2010,Warda2012,TSEKHANOVICH2019583,Jhingan2022}, making them good candidates to investigate the impact of shell effects as one sweeps along the isotopic chains.
The same is true of heavier elements, though higher excitation energies and competing reaction channels such as quasifission may complicate direct fission studies.
While this does pose a challenge, the onset of quasifission provides an opportunity to test time-dependent theoretical simulations of the reaction dynamics at play in near-barrier heavy-ion reactions.
Recent theoretical studies have noted the similarities in fragment yields and deformed shell effects between fission and quasifission~\cite{godbey2019,godbey2022,Mcglynn2023}, a phenomenon that can be probed by exploring different entrance channels leading to the same compound at the same excitation energy. One of the first approved experiments at FRIB will tackle this with beams of $^{48,49}$Ca on $^{174,173}$Yb making $^{222}$Th.
Due to the varying shell effects and transfer dynamics~\cite{godbey2017} in these systems, some combinations of target and projectile will lead to substantially longer-lived pseudo-compound quasifission fragments.
If the build-up of particle-number fluctuations increases as suggested in~\cite{Simenel2020}, this implies that the quasifission mass yields should be broader for more asymmetric entrance channels.
This will also inform similar time-dependent studies of multinucleon transfer reactions where the fragment mass widths predicted by theory need stringent tests~\cite{Sekizawa2019} as well as the strong correlation between proton and neutron transfer noted in some  studies~\cite{Godbey2020}.

\section{Equation of State\label{sec:EOS}}
% Agnieszka Sorensen
Investigations of the equation of state (EOS) play an essential role in nuclear physics by pursuing the understanding of bulk properties of nuclear matter and by providing a crucial input for studies of collisions of heavy nuclei, cold neutron stars and neutron star mergers, and core-collapse supernovae. Since the EOS can be both meaningfully constrained by comparisons of phenomenological approaches with data~\cite{Danielewicz:2002pu,Zhang:2007hmv,SRIT:2021gcy,SpRIT:2020blg,Lynch:2021xkq,Huth:2021bsp,Oliinychenko:2022uvy,Zhang:2022bni,Tsang:2023vhh} as well as calculated from, e.g., \textit{ab initio} approaches~\cite{Hebeler:2015hla,Lynn:2019rdt,Drischler:2021kqh,Drischler:2021kxf}, it can provide important benchmarks for effective field theories (EFTs) of nuclear interactions.
Furthermore, constraints on the EOS can provide a unique handle on the dependence of elementary interactions on the surrounding medium~\cite{Gale:1987zz,Danielewicz:2002pu,Sammarruca:2021bpn,Li:2018lpy}, including nonperturbative regions currently inaccessible to \textit{ab initio} approaches (e.g., at densities over twice that of the saturation density of symmetric nuclear matter $\nsat$ and at high temperatures attained in neutron star mergers~\cite{Most:2018eaw,Radice:2020ddv} and energetic heavy-ion collisions~\cite{HADES:2019auv,Sorensen:2023zkk}).

The EOS can be written as a sum of isospin-symmetric and isospin-asymmetric contributions. In terms of the energy per particle $E$, one often uses the expansion around $\delta = 0$,
\begin{equation}
E(n, \delta) = E(n) + S(n) \delta^2 + \mathcal{O}(\delta^4)~,
\end{equation}
where $n$ is the baryon density, $S(n)$ (sometimes also denoted as $E_{\rm{sym}}$) is known as the symmetry energy, $\delta$ is the isospin asymmetry parameter defined as $\delta = (n_N - n_P)/(n_N + n_P)$, where $n_N$ and $n_P$ are the neutron and proton densities, respectively, and $\mathcal{O}(\delta^4)$ indicates higher-order terms which are often neglected (for a discussion of these terms, see, e.g., Ref.~\cite{Wen:2020nqs}). In the above equation, the first term represents the energy per particle of symmetric nuclear matter, while the remaining terms account for the additional contribution occurring when $\delta \neq 0$. Both the symmetric and the asymmetric contributions are often further expanded about the saturation density of symmetric nuclear matter $n_{\rm{sat}}$, where the properties of nuclear matter have been extensively studied. Thus, one can write
\begin{equation}
E(n) = E_0 + \frac{1}{2!} K_0 \left( \frac{n - n_{\rm{sat}}}{3n_{\rm{sat}}}\right)^2 + \frac{1}{3!} Q_0 \left( \frac{n - n_{\rm{sat}}}{3n_{\rm{sat}}}\right)^3 + \dots~,
\end{equation}
where $E_0$ is the binding energy of symmetric nuclear matter at the saturation density, $n=n_{\rm{sat}}$, and \mbox{$K_0 \equiv \left(9 n^2 \frac{d^2 E}{dn^2} \right)_{n=n_{\rm{sat}}} $} is the incompressibility of symmetric nuclear matter at $n=n_{\rm{sat}}$. Similarly, the symmetry energy can also be expanded as
\begin{equation}
S(n) = S_0 + L_{\rm{sym}} \left( \frac{n - n_{\rm{sat}}}{3n_{\rm{sat}}}\right) + \frac{1}{2!} K_{\rm{sym}} \left( \frac{n - n_{\rm{sat}}}{3n_{\rm{sat}}}\right)^2 + \dots~,
\end{equation}
where $S_0$ is the symmetry energy at $n = n_{\rm{sat}}$ and $L_{\rm{sym}}$ (sometimes also simply denoted as $L$) is the slope of the symmetry energy at $n = n_{\rm{sat}}$.

While there exist fairly well-established constraints on the EOS for conditions similar to those characteristic of normal nuclear matter (where $E_0$, $K_0$, and $S_0$ are relatively tightly constrained~\cite{Bethe:1971xm,Dutra:2012mb,
%Garg:2018uam,
GARG201855,
Lynch:2021xkq}), its behavior away from~$\nsat$ and for systems with large asymmetry in the isospin content remains poorly known~\cite{Sorensen:2023zkk}. It is, therefore, the subject of numerous theoretical and experimental efforts worldwide. Currently available and planned experiments at FRIB include studies of the giant monopole resonance in neutron-rich isotopes, nucleon elastic and inelastic scattering on medium-mass and heavy isotopes, and central reactions of heavy nuclei. These experiments will probe the EOS over the widest accessible range of isospin asymmetry, both around saturation and up to densities \mbox{$n \approx 1.5\nsat$~}\cite{Sorensen:2023zkk}, allowing for robust connections with the physics of neutron stars and their mergers. With the proposed FRIB400 upgrade of the available beam energies, it will become possible to study both the isospin- and the density-dependence of the EOS up to~\mbox{$n \approx 2\nsat$}~\cite{FRIB400,Sorensen:2023zkk}.

\subsection{Microscopic calculations of the EOS within chiral EFT}
\label{Microscopic_calculations_of_the_EOS_within_chiral_EFT}

% Christian Drischler
In the FRIB and multi-messenger astronomy era, the density regime $n \approx (1$--$2)\nsat$, which can be probed by nuclear theory, nuclear experiments, and neutron star observations, provides a ``golden window'' of opportunities for dense matter research~\cite{Drischler:2021bup}.
Chiral EFT, combined with algorithmic and computational advances in many-body theory (see Sec.~\ref{sec:overview_theo_methods}), has made tremendous progress in describing nuclear matter at low densities, $n \lesssim 2\nsat$, with quantified theoretical uncertainties; see, e.g., Refs.~\cite{Drischler:2021kxf,Sorensen:2023zkk,MUSES:2023hyz} for extensive review articles.

Theoretical uncertainties in chiral EFT calculations increase rapidly for densities above $n\approx \nsat$.
Robust statistical comparisons of nuclear theory predictions with empirical constraints in this regime will enable rigorous benchmarks of chiral interactions~\cite{Essick:2020flb,Rose:2023uui} and model selection between competing EFT implementations, leading to the construction of improved microscopic EOS models.
To this end, advances in chiral EFT are needed to improve its predictive power in the ``golden window'' and to resolve issues~\cite{Ekstrom:2015rta,Drischler:2017wtt,Dyhdalo:2016ygz,Stroberg:2019bch,Furnstahl:2021rfk} related to, e.g., regulator artifacts, modified power counting, and differing predictions in medium-mass to heavy nuclei.  
In particular, improving chiral three-nucleon forces for next-generation \textit{ab~initio} predictions~\cite{Stroberg:2019bch,Wesolowski:2021cni,Hu:2021trw} of neutron-rich nuclei, now increasingly available in experiments, including those at FRIB, will be crucial for confronting theoretical results with nuclear data.
An example of such a cross-cutting research direction is predicting and measuring mirror nuclei~\cite{Yang:2017vih,Reinhard2022M} as well as neutron-rich nuclei expected to have large neutron skins~\cite{FRIB400}, such as~${}^{86}$Ni.
Predictions for the structure and evolution of neutron stars based on the same nuclear interactions can then be confronted with astrophysical observations, covering physics across more than 18 orders of magnitude.
These advances will elucidate key questions in \textit{ab~initio} many-body theory, including:
(i)~Where does chiral EFT break down, and what is the underlying mechanism?;
(ii)~How should chiral two- and few-body forces be organized (i.e., the power counting) in a medium?; and
(iii)~What are the phases of neutron star matter at $n \gtrsim \nsat$?

\subsection{Isospin dependence of the EOS}
\label{Isospin_dependence_of_the_EOS}

% Christian Drischler
Enhancing our understanding of the dependence of the current EOS models on isospin asymmetry, including studies of nonquadratic and nonanalytic terms in the isospin expansion of the EOS~\cite{Wen:2020nqs,Somasundaram:2020chb}, will be an important task in the FRIB era. 
To this end, new experimental constraints on low-energy EOS parameters, such as the nuclear symmetry energy at $\nsat$ and nuclear incompressibility, are necessary. 
As an example, for densities around %the saturation 
$n_{\rm{sat}}$ the approved FRIB experiment ``The Isoscalar Giant Monopole Resonance in~${}^{132}$Sn: Implications on the Nuclear Incompressibility'' (21056) will shed light on the isospin expansion of the incompressibility of bulk nuclear matter $K = K_0 + K_\tau \, \delta^2$, where $K_{\tau}$ represents the deviation of the incompressibility of bulk nuclear matter away from the symmetric limit~($\delta = 0$) up to the second order in $\delta$ (note that $K_{\tau} \neq K_{\rm{sym}}$ due to the fact that for $\delta \neq 0$, the value of the nuclear saturation density changes, see Refs.~\cite{Piekarewicz:2008nh,Howard:2020nda}).

\begin{figure}
    \includegraphics[width=0.525\textwidth]{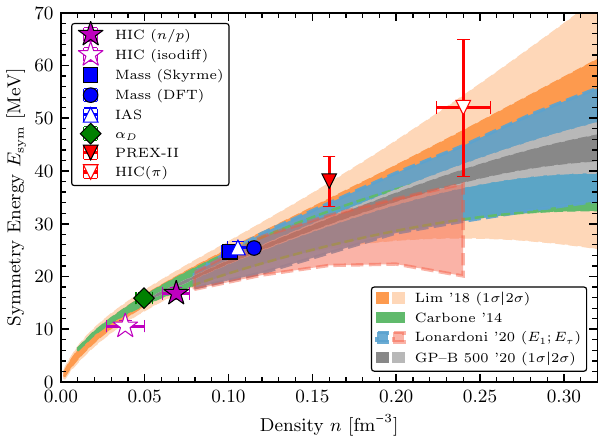}
    \includegraphics[width=0.465\textwidth]{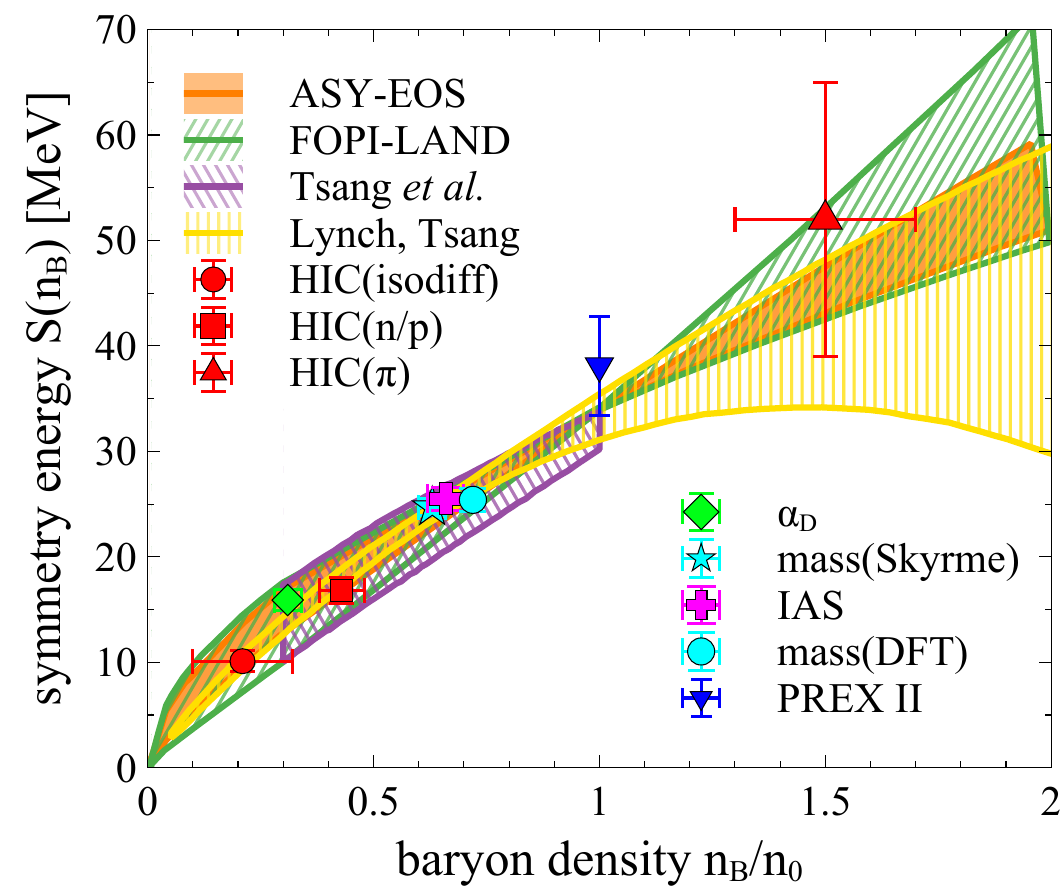}
    \caption{\textit{Left:} Symmetry energy as a function of the baryon density as predicted by several chiral EFT calculations (bands). 
    Various experimental constraints are depicted by symbols with error bars.
    Theoretical uncertainties are significant and increase rapidly in the density regime $n \approx (1$--$2)\nsat$.
    Figure modified from Ref.~\cite{Drischler:2021kxf}. \textit{Right:} Selected constraints on the symmetry energy obtained, among others, from comparisons of experimental data to results of transport simulations of heavy-ion collisions (bands labeled as ``ASY-EOS'', ``FOPI-LAND'', and ``Tsang \textit{et al.}'' and symbols labeled as ``HIC(isodiff)'', ``HIC(n/p)'', and ``HIC($\pi$)''). Figure modified from~\cite{Sorensen:2023zkk}.
    }
    \label{fig:esym}
\end{figure}

As previously discussed, the isospin-dependence of the EOS, which also depends on density, is not well-constrained at~\mbox{$n\gtrsim\nsat$}. This can be seen in the left panel of Fig.~\ref{fig:esym}, in which theoretical predictions for the symmetry energy (i.e., the contribution to the energy per particle due to the isospin asymmetry) are shown as a function of the baryon density up to $n \approx 2\nsat$.
Although different EFT predictions agree with one another within uncertainties (shown as bands), these uncertainties tend to be significant in the ``golden window''.
The uncertainties are further magnified when constructing, e.g., the neutron star EOS through physics-guided extrapolations toward high densities present at the centers of heavy neutron stars (see, e.g., Ref.~\cite{Drischler:2020fvz}).
The left panel of Fig.~\ref{fig:esym} also shows a selection of experimental constraints on the symmetry energy, depicted as symbols with error bars.
More precise experimental constraints on the symmetry energy, e.g., from extractions of neutron skins and measurements of isospin-dependent observables in collisions of heavy neutron-rich nuclei, will provide valuable guidance for microscopic models of the EOS for astrophysical simulations.
Here, density functional theory (DFT)~\cite{Bender2003,Drut2010} is critical for determining these low-density EOS parameters from nuclear experiments.
However, as PREX--II~\cite{PREX:2021umo} and CREX~\cite{CREX:2022kgg} emphasized, extrapolations to the neutron-rich regime of current DFT models need improvements, for which microscopic EOS calculations of neutron-rich matter may provide guidance.

EFT calculations of pure neutron matter are particularly relevant here because, in this case, chiral three-nucleon forces are simplified and have no unknown parameters through next-to-next-to-next-to-leading order (N$^3$LO). This results in a good agreement between different many-body approaches and overall relatively small uncertainties at $n \lesssim \nsat$. At densities away from $\nsat$, experimental constraints on the symmetry energy can be obtained from studies of central heavy-ion reactions, which we elaborate on in the next subsection.

%% SRS inserted this paragraph (cut from the structure section).
Differences in the root-mean-square charge radii of mirror nuclei are also sensitive to the symmetry energy, in particular to the slope parameter $L_{\rm sym}$~\cite{brow2017p122502, yang2018}.
The size of the difference depends on the product $DL_{\rm sym}$, where $D=|N-Z|$ is the difference between the number of neutrons and the number of protons in a nucleus.
Several cases have already been measured: 
$^{36}$Ca-$^{36}$S ($D=4$)~\cite{brow2020}, 
$^{54}$Ni-$^{54}$Fe ($D=2$)~\cite{pine2021}, and  
$^{32}$Ar-$^{32}$Si ($D=4$)~\cite{koni2023b}, with 
results for $L_{\rm sym}$ consistent with those extracted from the neutron skin of $^{48}$Ca and from
the analysis of the gravitational waveform of the binary neutron star merger GW170817~\cite{rait2019}.
These $L_{\rm sym}$ determinations are model dependent~\cite{brow2022, 
%rein2022p021301}, 
Reinhard2022M}, 
and more data are required to assess and reduce the model dependence.
FRIB will provide access to the proton-rich nucleus $^{52}$Ni in the mirror pair
$^{52}$Ni-$^{52}$Cr ($D=4$), and both nuclei in the mirror pair $^{22}$Si-$^{22}$O ($D=6$). 
%Recently the lattice Monte-Carlo simulations with the wave function matching method \cite{elha2023} has been applied as well as VS-IMSRG calculations to compute the $R_{\rm ch}$ difference  for the $A=32$ mirror pair. 
%Such lattice calculations are planned for other mirror pairs.
Ultimately, the mirror pair $^{48}$Ca-$^{48}$Ni is of great interest due to its large $D = 8$ and the connection to the neutron skin of $^{48}$Ca. However, current estimates suggest that the production rate of $^{48}$Ni, even with FRIB operating at full beam power, will remain too low for laser spectroscopy in the near future.

\subsection{EOS and in-medium properties of nuclear matter from heavy-ion collisions}
\label{EOS_and_in-medium_properties_of_nuclear_matter_from_heavy-ion_collisions}

% Pawel Danielewicz, Agnieszka Sorensen
Heavy-ion collisions (also referred to as central heavy-ion reactions) at beam energies from a few tens of MeV/nucleon to several GeV/nucleon in the fixed-target frame probe the broadest range of baryon densities (from a few tenths of to a few times the saturation density) and temperatures (from a few MeV to well above one hundred) available in terrestrial laboratories~\cite{Sorensen:2023zkk}. In particular, the near-term and future heavy-ion collision experiments at FRIB, given its world-leading range of accessible proton- and neutron-rich heavy isotopes, are uniquely posed to probe the symmetry energy component of the EOS at densities up to~$2n_{\rm{sat}}$~\cite{Sorensen:2023zkk, FRIB400}. 

At the same time, heavy-ion collision experiments can only measure final momentum-space distributions of particles, and any properties of matter created in the collisions must be inferred from their influence on the evolution of the collision. Moreover, systems created in central heavy-ion reactions are short-lived and out-of-equilibrium for significant portions of their evolution.  
Consequently, conclusions on the properties of nuclear matter based on experiments colliding heavy nuclei require comparisons to results of simulations obtained within the framework of transport models. 

The challenge before the transport theory lies in describing all stages of the reaction, from the initial nuclei to fragments reaching the detectors, while taking into account the interplay between numerous details of the underlying physics, often gaining prominence at varying times. This complex problem becomes more manageable for experimental designs in which $4\pi$ detectors register vast amounts of information on reaction events (primarily about charged products of the reaction, as neutral particle measurements prove difficult, albeit not impossible). The resulting data sets contain information on particles spanning broad kinematic ranges and, therefore, carry insights on various physics elements driving the dynamics of the collisions. The data are then analyzed over several years, if not decades, often yielding answers to questions that were not originally considered when the experiment was staged.

Numerous studies~\cite{Stoecker:1980vf,Hartnack:1994ce,Danielewicz:2002pu,LeFevre:2015paj,Nara:2021fuu,Li:2002qx,Li:2000bj,Li:2014oda} identify heavy-ion observables particularly sensitive to the nuclear matter EOS. As an example, the right panel of Fig.~\ref{fig:esym} shows selected constraints on the symmetry energy, including constraints from heavy-ion collisions obtained using proton and neutron anisotropic collective flows~\cite{Russotto:2011hq}, isospin diffusion~\cite{Zhang:2006vb,Zhang:2007hmv}, ratios of neutron to proton spectra~\cite{Zhang:2006vb,Zhang:2014sva}, neutron to charged fragments ratios~\cite{Russotto:2016ucm}, and ratios of pion spectra
%~\cite{SRIT:2021gcy,SpRIT:2020blg}. 
~\cite{SRIT:2021gcy,SpRIT:2020blg}. 
However, the complexity of heavy-ion reactions also means that different reaction stages and physics inputs often contribute to the same observable. Consequently, no single observable tests \textit{exclusively} the EOS. The impact of the EOS on a given EOS observable, such as anisotropic collective flows or subthreshold meson yields% in central collisions of heavy nuclei
, often competes with that of, e.g., the momentum dependence of particle mean fields~\cite{Danielewicz:1999zn,Pan:1992ef} or details of subthreshold particle production
%~\cite{SpiRIT:2021gtq,
~\cite{SRIT:2021gcy,
Cozma:2021tfu}. It is a joint task for the theorists and experimenters, aided by recent developments in statistical approaches and artificial intelligence methods, to determine and quantify the influence of particular physics inputs on chosen observables, thus enabling more stringent constraints.

This challenge presents a unique opportunity to shed light on in-medium properties of nuclear matter and address multiple outstanding problems in nuclear physics, including the momentum-dependence of the nuclear EOS at densities away from the saturation density, in-medium hadron cross-sections, off-shell dynamics, subthreshold particle production, and cluster production mechanisms. 
This opportunity must be met with a concerted program to further develop theoretical foundations of various elements of transport theory and phenomenological solutions used in simulations.
Notably, the dependence on the isospin asymmetry of many processes that constitute an input to transport models will be thoroughly studied in upcoming experiments at FRIB, creating an opportunity for collaborations across different subfields.
Similarly, the fact that \textit{ab~initio} calculations are beginning to be carried out at finite temperatures~\cite{Wellenhofer:2014hya,Keller:2022crb} can also become highly important for interpretations of experimental heavy-ion data by providing more relevant reference studies. Indeed, the extrapolation from a non-equilibrium system created in a heavy-ion collision to the nearest equilibrium state at a given density and temperature is expected to be more reliable than a similar extrapolation to such a state at the same density and zero temperature.

Below, we briefly overview several key properties of nuclear matter that significantly influence interpretations of heavy-ion collision data. For a complementary discussion of these and other inputs to simulations of heavy-ion collisions, see Ref.~\cite{Sorensen:2023zkk}.

\subsubsection{Momentum dependence of nucleon interactions}

The momentum dependence of nuclear potentials emerges from elementary nucleon-nucleon (NN) interactions (which are predominantly attractive at low NN energies but turn into predominantly repulsive at higher energies) and from exchange terms~\cite{Botermans:1990qi,Sammarruca:2021bpn}. The momentum dependence of mean fields in cold matter at normal and subnormal densities can be tested in elastic scattering of nucleons off nuclei. However, in heavy-ion collisions, the momentum dependence is probed over different densities and in hot rather than cold matter. Fortunately, while studying the EOS effects without the influence of the momentum dependence effects is impossible, the opposite is true. The EOS effects subside in collisions of lighter heavy-ion systems and at higher impact parameters. At the same time, the influence of momentum dependence is enhanced at high transverse momenta. These factors allow one to calibrate the momentum dependence (as a function of density and isospin imbalance) before attempting to extract the EOS from data. 
Moreover, microscopic many-body calculations of nucleonic potentials as a function of density, momentum, and neutron-proton imbalance (and recently even temperature~\cite{Wellenhofer:2014hya,Keller:2022crb}) make the momentum dependence at different nucleonic densities a subject of interest in its own right. Furthermore, these calculations can aid the modeling of the collisions by narrowing the ranges of interaction parameter values that need to be confronted with data. 

FRIB, naturally, offers a greater variation of the isospin composition of collided nuclei  than has been available at other facilities to date. At early stages of the collisions, the relative isospin imbalance in high-density matter primarily reflects that of the original nuclei. The different response of isospin positive and negative particles to the asymmetric component of the EOS influences their dynamics during the collision and is ultimately reflected in final-state observables, such as, e.g., charged pion ratios~\cite{Li:2002qx,
%SpiRIT:2021gtq,%SpiRIT:2020sfn}. 
SRIT:2021gcy,SpRIT:2020blg}.
Thus, comparisons of simulations with observables measured in FRIB experiments bear the promise to simultaneously constrain, e.g., both the density and the momentum dependence of the asymmetric contribution to the EOS, which would be equivalent to constraining the symmetry energy parameters and the effective mass splitting between neutrons and protons
%~\cite{SpiRIT:2021gtq,
~\cite{SRIT:2021gcy,
Cozma:2021tfu}. Here, careful evaluations of physics assumptions and simulation frameworks are crucial to meaningful extractions of the properties of nuclear matter; multiple such efforts have already been started~\cite{TMEP:2023ifw,TMEP:2022xjg}.

\subsubsection{Influence of a medium on particle cross-sections}

The inputs to transport simulations also include cross-sections for elementary collisions. Following Occam's Razor principle, early transport calculations adopted cross-sections obtained from measurements of elementary processes. 
However, these typically led to excessive nucleon stopping as compared to measurements from central heavy-ion collisions, resulting in a mismatch between the simulations and the data already on the level of simple rapidity distributions~\cite{Persram:2001dg,Zhang:2007gd}. Moreover, reduced in-medium cross-sections (i.e., cross-sections smaller than those in free space) not only seemed to be needed by phenomenology, but they also emerged from microscopic calculations~\cite{Muther:2000qx,Pandharipande:1992zz,Persram:2001dg,Li:2005jy,Chen:2013bwa, PhysRevC.73.014001}. As a result, parametrizations for the in-medium cross-section reductions were developed and employed in collision simulations covering broad ranges of incident energy and masses of colliding systems~\cite{PhysRevLett.71.1986, LI2022137019, PhysRevC.99.034607}. However, the consensus on the level of cross-section reduction was shaken by recent measurements by the S$\pi$RIT Collaboration of Sn + Sn systems at 270 MeV/nucleon~\cite{SpiRIT:2022sqt}. Virtually all transport codes~\cite{TMEP:2022xjg} need larger in-medium cross sections to reproduce rapidity distributions for central collisions as measured by the S$\pi$RIT experiment than in the case of other data in this general energy region~\cite{FOPI:2010xrt,FOPI:2003fyz,Ono:2020zsx}. Among the possible explanations for this puzzle, one can consider the fact that other systems in this energy regime were either lighter or heavier, or that the centrality selection was less stringent. Still, the issue must be clarified from both the theoretical and the experimental sides. FRIB will operate in the energy regime in question and can contribute more significant variations in the isospin composition of the colliding systems at a given mass than other accelerators. Notably, the isospin dependence of in-medium cross sections remains poorly understood~\cite{Zhang:2010th, PhysRevLett.86.975, PhysRevC.73.014001}.

\subsubsection{Development of off-shell properties and subthreshold production}

Subthreshold production in different isospin channels can be a sensitive tool for testing the symmetry energy around $2 \nsat$~\cite{SpiRIT:2016eac,FRIB400}. 
Indeed, during a heavy-ion collision, nucleons and other particles in the dense region of the system undergo frequent collisions with each other, building up their widths (or further augmenting them if any were present under vacuum conditions). Significant widths imply that energy conservation can admit additional elementary interactions, facilitating subthreshold particle production~\cite{PhysRevC.42.1564, PhysRevC.60.051901, 
%BERTSCH199655}. 
Bertsch:1995ig}, which will, in general, be characterized by different thresholds for different isospin states.
Even transport calculations with no in-medium width enhancements~\cite{Cozma:2021tfu} have already demonstrated a strong sensitivity of predicted near-threshold pion yields to relatively modest in-medium particle-energy variations. The full exploitation of subthreshold production as a tool for constraining the symmetry energy calls for using off-shell transport to reach meaningful conclusions from comparisons to data.
While off-shell transport with positive definite distribution functions, amenable to Monte-Carlo sampling, has been put forward in the past~\cite{Botermans:1990qi, cassing_semiclassical_2000, Buss:2011mx,Linnyk:2015rco,Moreau:2019vhw,Aichelin:2019tnk}, it has been sparsely used for simulations of lower-energy heavy-ion collisions.

\subsubsection{Nuclear clusters and correlations}

While nuclear clusters emerge as independent degrees of freedom at subnormal densities~\cite{PhysRevC.55.2109, bastian_light_2016, doi:10.1126/science.abe4688,Shen:2022bak,
%RevModPhys.90.035004}, 
free2018}, 
their formation in heavy-ion collisions is generally initiated at higher densities (at least only modestly subnormal) since several nucleons need to appear in the same phase-space region for the formation to occur~\cite{DORSO1995197, PhysRevC.54.R28}. Moreover, depending on the kinematic region where given collision products are considered (e.g., midrapidity vs.\ forward rapidity), in certain conditions the density region where cluster formation occurs might only emerge through decompression from high-density matter, with the relative neutron-proton imbalance retained from the high compression stage throughout the nearly adiabatic expansion~\cite{SpiRIT:2021och, PhysRevC.51.716}. In this context, the dependence of cluster yields on isospin in select kinematic regions becomes a probe of the symmetry energy at supranormal densities and, more generally, of isovector characteristics at those densities.  

In order to fully utilize observables related to nuclear clusters and extract the isospin properties of nuclear matter, models of cluster production need further development. Currently invoked production mechanisms~\cite{Ono:2019jxm} can be typically divided into cluster-finding algorithms based on coalescence, that is with criteria based on the proximity of particles in the coordinate and momentum space (sometimes also including criteria related to the binding energy of a cluster~\cite{LeFevre:2019wuj}), and into models where nuclear clusters are introduced as independent degrees of freedom that can be produced in elementary scatterings~\cite{Danielewicz:1991dh, Oliinychenko:2018ugs} (which in principle can accommodate the modification of cluster properties due to the surrounding medium). Both categories of approaches have their shortcomings: coalescence-based algorithms, typically applied toward the end of a heavy-ion collision evolution, do not account for the influence of nuclear clusters on the collision dynamics. On the other hand, approaches treating clusters as independent entities usually consider only the lightest clusters (often just deuterons) due to challenges in including the increasing number of relevant production channels for heavier clusters. Furthermore, on the theoretical level, cluster formation is related to intermediate-range correlations and strong quantum effects, extending beyond the classical mean-field picture utilized in transport. Therefore, the problem of cluster formation in heavy-ion collisions touches on fundamental properties of nuclear matter, and successful phenomenological descriptions have the potential to shed light on particular mechanisms involved. 

Notably, similar fundamental questions are apparent in short-range correlations (SRCs) that can be measured in, e.g., nucleon knock-out experiments~\cite{Arrington:2011xs,Hen:2014nza,Hen:2016kwk,Arrington:2022sov}. Through phenomenological descriptions such as inclusion of high-momentum tails in the initial nucleon distribution function or considerations of three- and many-body collisions, SRCs have been likewise shown to appreciably influence heavy-ion collision dynamics~\cite{Li:2014vua,Wang:2017odj,Bertsch:1995ig,Bonasera:1992kjm}. Moreover, SRCs are also shown to be strongly dependent on the isospin composition of the surrounding medium~\cite{Xu:2012hf,Rios:2013zqa}. Theoretical descriptions of heavy-ion collisions at FRIB and FRIB400, utilizing models accounting for effects due to SRCs, hold the promise to complement the current and planned programs investigating SRCs at the Jefferson Lab and the future Electron-Ion Collider by studying the dynamical evolution of SRCs in dense isospin-asymmetric nuclear matter. Here, sensitive observables may include both pion yields~\cite{Bertsch:1995ig} and bulk observables~\cite{Bonasera:1992kjm}. Such studies can also further advance the understanding of the role of tensor forces~\cite{Rios:2013zqa,Rios:2009gb} at high densities and the development of off-shell transport models.

\subsection{Optical potentials}
\label{EOS_OMP}

%Jeremy W.\ Holt:

The nuclear EOS can be accessed through the momentum- and energy-dependent nucleon single-particle propagator~$G(q,\varepsilon)$ in nuclear matter, which is related to the irreducible self-energy~$\Sigma(q, \varepsilon)$ through the Dyson equation~$G(q,\varepsilon) = G_0(q,\varepsilon) + G_0(q,\varepsilon)\Sigma(q, \varepsilon)G(q, \varepsilon)$, where $q$ and $\varepsilon$ are the single-particle momentum and energy, respectively. The nucleon self-energy at positive energies can be identified with the optical potential~\cite{bell1959}, which is widely used in the theoretical description of nucleon elastic and inelastic scattering on medium-mass and heavy isotopes, and is closely connected to experiments. For isospin-asymmetric target nuclei, one empirically observes in the optical potentials for proton and neutron projectiles a splitting of the Lane form~\cite{lane1962}: $U_{n,p}(E) = U_0(E) \pm U_I(E) \delta_{np}$, where~$\delta_{np} = (N-Z)/(N+Z)$ is the isospin asymmetry parameter expressed in terms of the number of neutrons~$N$ and protons~$Z$, $U_I(E)$ is called the isovector optical potential, and the $+$ ($-$) sign applies in the case of neutrons (protons). Quasi-elastic charge-exchange reactions~$(p,n)$ are especially sensitive to~$U_I(E)$~\cite{lane1962}. This has been exploited in Ref.~\cite{dani2017} to refine the isovector radius and diffuseness parameters of nucleon-nucleus optical potentials and study correlations between these geometry parameters and the symmetry energy slope parameter~$L$.

The optical potential at positive energies is related to the nuclear shell model potential at negative energies through a dispersion integral connecting the real and imaginary components of the self-energy, see, e.g., Ref.~\cite{char2007}. Therefore, with a sufficient amount of bound-state and scattering data, dispersive optical potentials can provide unique insights into nuclear structure and connections to the EOS. For instance, in Ref.~\cite{prui2020} the neutron-skin thicknesses $\Delta r_{np}$ of $^{16, 18}$O, $^{40,48}$Ca, $^{58,64}$Ni, $^{112,124}$Sn, and $^{208}$Pb were obtained from a dispersive optical model analysis that included data from elastic scattering angular distributions, total cross sections, reaction cross sections, and level properties. In particular, the obtained value of $\Delta r_{np}(^{208}\text{Pb}) = 0.12-0.25$\,fm is consistent with models predicting moderate values of the symmetry energy slope parameter $L = 30-90$\,MeV.

From the Hugenholtz-Van Hove theorem~\cite{huge1958}, the single-particle energy at the Fermi surface is directly related to the energy per particle $\bar E(k_f)$ of the medium through
\begin{align}
\frac{k_f^2}{2M_N} + \Sigma\big(k_f, \varepsilon(k_f)\big) = \bar E(k_f) + \frac{k_f}{3}\frac{\partial \bar E(k_f)}{\partial k_f}~,
\end{align}
where $k_f$ is the Fermi momentum and $M_N$ is the nucleon mass. This relationship has been used~\cite{xu2010} to derive an explicit relationship between the symmetry energy at saturation~$S_0$, the slope parameter~$L$, and the optical potential. Using the global set of nucleon-nucleus optical potentials available at the time, Ref.~\cite{xu2010} found~\mbox{$S_0 = 31.3 \pm 4.5$\,MeV} and~\mbox{$L = 52.7 \pm 22.5$\,MeV}. More generally, the self-energy defines the spectral function 
\begin{align}
A(q, \varepsilon) = \frac{-2\text{Im} \Sigma}{\big( \varepsilon - \frac{q^2}{2M} + \text{Re} \Sigma \big)^2 + \big(\text{Im} \Sigma\big)^2}~,
\end{align}
which can be integrated over the single-particle energy and momentum to obtain the bulk energy per particle through the Galitskii-Migdal-Koltun sum rule~\cite{gali1958,kolt1974}. This technique is commonly used in self-consistent Green's function theory to obtain the EOS of nuclear matter at zero and finite temperature~\cite{carb2018,rios2020}. In the future, consistent microscopic modeling~\cite{holt2017,whit2021} of the EOS and optical potentials based on the same underlying nuclear forces will provide an opportunity to establish correlations between the empirical parameters of the EOS and optical potentials.

Finally, proton and neutron optical potentials (self-energies) are an essential input for transport model simulations of medium-energy heavy-ion collisions~\cite{Sorensen:2023zkk}, which provide an additional experimental means of probing the (hot) dense matter EOS. In the past, phenomenological single-particle potentials have typically been used in transport simulation studies. However, more recently microscopic predictions from chiral EFT have been used to constrain single-particle potentials and study their impact on heavy-ion collision observables~\cite{xu2019,zhan2018}.

\subsection{Upcoming and proposed experiments at FRIB}
\label{Upcoming_and_proposed_experiments_at_FRIB}

% Kyle Brown
The experimental approach to studying the nuclear EOS away from saturation density at FRIB is split into several phases, reflecting the availability of beams and equipment (see Sec.~\ref{heavy-ion_collisions_exp_capability}). The first heavy-ion collision experiment at FRIB, ``Measuring the isospin dependence of the nucleon effective mass at supersaturation density'' (23058), has been approved in PAC2. That experiment will focus on constraining the momentum dependence of the nuclear potential around \mbox{(1--1.5)$n_{\rm{sat}}$} through measurements of spectral ratios of neutrons and protons as well as directed and elliptic flow in \mbox{$^{56,70}$Ni + $^{58,64}$Ni} reactions at projectile energies of 175 MeV/nucleon. While the sensitivity of the spectral ratio to the momentum dependence of the symmetry energy drops off with increasing density~\cite{Zhang:2014sva}, flow remains a sensitive observable for the symmetry energy up to 1 GeV/nucleon~\cite{wang2020135249}. The mid-mass, 200 MeV/nucleon beams available in the first few years at FRIB will probe a density window in which both of these observables can yield information about the symmetry energy. This is especially advantageous as model uncertainties from transport calculations will likely be different between the two observables. The obtained results can be combined with previous experimental constraints at low density (both from prior heavy-ion collision experiments at NSCL~\cite{Morfouace:2019jky} as well as with independent constraints from optical models~\cite{li2013276}), while at high density the reference is provided by the current state-of-the-art results from the S$\pi$RIT campaign at RIKEN
%~\cite{SpiRIT:2021gtq,SpiRIT:2020sfn}, 
~\cite{SRIT:2021gcy,SpRIT:2020blg}, 
studying \mbox{Sn + Sn} reactions at beam energies of 270 MeV/nucleon.

In later phases, the experiments will shift to focus on the flow and yield ratios of charged pions. These pion measurements will initially concentrate on reactions at lower energies than those studied at RIKEN. This will allow for the study of sub-threshold pion production, which is an important input to transport model calculations of the pion ratio~\cite{Hong:2014yva}. Subsequently, the increased intensities and larger isospin asymmetry of the FRIB400 beams will allow for unprecedented studies of the nuclear EOS using spectral ratios of charged pions. In particular, the increased beam energy will lead to higher overall yields of charged pions. This will allow for studies with lower beam intensities, which areexpected in the case of beams of rare isotopes with high isospin asymmetry.

\section{Nuclear Astrophysics\label{sec:Astro}}
\subsection{Motivation and needs for progress in Nuclear Astrophysics}

2022 not only marked the first ever FRIB experiment, but also coincided with the 65th anniversary of the work that established the first comprehensive theory of nucleosynthesis laying the foundation of the modern field of nuclear astrophysics \cite{B2FH}. It was evident from even the early days that studies of the origin of the elements are intimately tied to understanding the properties and reactions of both, stable and unstable nuclei (see also \cite{Cameron1957}). Therefore laboratory measurements have been central to driving progress in nuclear astrophysics from the beginning.

With FRIB now online, hundreds of species yet to be probed in terrestrial environments can become available to experiments. This is particularly timely since in the last few years observation has begun a revolution of its own, with the first gravitational waves from binary neutron star mergers detected \cite{AbbottGW170817}. Now, we have seen the light emitted from the first multi-messenger neutron star merger event (GW170817) \cite{Cowperthwaite2017,Villar}, which pointed to the presence of lanthanide elements, implying mergers to be a site of heavy element production. Such observational abilities will only be enhanced by future gravitational wave dectectors. For example, Cosmic Explorer \cite{CosmicExplorer, CosmicExplorer2} and the Einstein Telescope \cite{EinsteinTelescope} could present the opportunity to observe many more binary mergers than LIGO. Additionally, GW170817 and its associated kilonova optical transient being observed across the electromagnetic spectrum by over 70 observing teams well-demonstrates the incredible ability of the observational community to perform highly detailed follow-up of events that have been localized by early detection via various messengers such as gravitational waves, gamma-rays, optical transient surveys, or neutrinos. In addition to this new access to multi-messenger events, our bank of metal-poor star observations (which can be tied more directly to individual events due to their being enriched by fewer events than our Sun \cite{FarouqiThielemann}) is set to grow as well. Observational campaigns such as the 
R-Process Alliance \cite{r-process-alliance} will present a new wealth of abundance data that we must decipher and integrate into global analyses with other observables. 

Studies have shown conclusively that interpretation of observables requires nucleosynthesis calculations which account for the uncertainties in the nuclear physics properties of relevant species (e.g. \cite{CoteGW170817}).  This applies to all nucleosynthesis observables such as Solar and stellar abundances \cite{B2FH,Holmbeck2019}, meteorites \cite{CoteICm,XiluRadioisotopes}, light curves \cite{BarnesKN2016,Kasen,Zhu2021Kilonova,BarnesKN2020}, and MeV gamma emission \cite{WangVasshMeVgammafiss}. This intimate connection between observables and nuclear physics properties presents the opportunity for not only nuclear physics to inform astrophysics, but vice versa as well~\cite{Cowan.Sneden.ea:2021}. For instance, questions remain on the ultimate reach of the rapid neutron capture process ($r$ process) in neutron star mergers but should late time heating of the environment from fission or alpha decay take place, this could show itself in the light curve \cite{Cfpaper,Wu.Barnes.ea:2019}. This would not only conclusively point to the production of actinides at this site, but would constrain the fission barriers and branching ratios in the actinide regions \cite{VasshJPhysGFiss,
%Giuliani.Martinez-Pinedo.ea:2020}, 
Giuliani2020}, 
requiring them to be such that production of late time long-lived fission species is possible. Ultimately, it is fundamental to be able to connect astrophysical simulations that describe the properties and dynamics of the ejected material~\cite{Just.Vijayan.ea:2023} with radiative transfer modelling of the kilonova spectra~\cite{Shingles.Collins.ea:2023} to be able to determine which specific elements are produced and under which conditions. Similar efforts are critical to understand observables of core collapse and thermonuclear supernovae, novae, and X-ray bursts. 

Therefore the need for a back and forth between theoretical studies of nucleosynthesis and nuclear physics experiments is clear, with the FRIB era now providing new exciting prospects to tackle several of the fundamental questions in nuclear astrophysics (e.g. \cite{horowitz2018,CowanReview}). In order to study astrophysical processes, reaction network inputs are needed based on realistic astrophysical simulations \cite{MeyerReview,SkyNet}, as well as experimental information to constrain the relevant astrophysical reactions. The nuclear data needs of reaction networks vary among astrophysical scenarios, and even individual model efforts. As far as unstable nuclei are concerned, networks typically require capture rates (neutron capture, proton capture, and alpha capture rates being particularly crucial), other charged particle and neutron induced rates such as ($\alpha$,n), ($\alpha$,p), (n,p), (p,$\alpha$),  beta-decay rates with branching ratios for the emission of protons or neutrons, continuum electron capture rates, and fission. In neutron stars, heavy ion fusion rates with unstable nuclei may also play a role. Masses serve as both a direct network input in the form of separation energies as well as a required input to model theoretical capture and decay rates where measured rate values are unavailable. Additionally, the need for nuclear data is not isolated to solely the nucleosynthesis network. For example, post-processing of network outputs with additional pieces of nuclear data, such as the energy liberated in individual decays in the different products: electrons, gammas, alpha-particles, \ldots, is required to make predictions for electromagnetic signals from astrophysical events. This in addition requires knowledge of the atomic processes that determine the spectra formation at the photospheric and nebular phases~\cite{Floers.Silva.ea:2023,Pognan.Grumer.ea:2023}. Recent studies have also shown that considering isomers can affect the overall decay of synthesized species at late times and impact the light curve \cite{Fujimoto.Hashimoto:2020,AstromerMischApJL,AstromerFRIB}. The wealth of nuclear physics information needed highlights that nuclear astrophysics will be an FRIB customer for many years to come. Despite the fact that many important investigations will require longer timescales due to the large number of measurements and the required synchroneous advances in multiple fields, there are  fundamental questions that can start to be systematically addressed in the near term. We detail below some of the motivations and needs for some specific nearer term and longer timescale investigations.

\subsection{Nucleosynthesis on the neutron-rich side of stability\label{sec:AstroNeutronRich}}

\subsubsection{Investigations to constrain nuclear theory approaches and extrapolations}

Investigating the astrophysical production of the elements inevitably leads to a need to study neutron capture reactions since reaching the heaviest species is only possible through neutron capture nucleosynthesis. The astrophysical $s$-, $r$-, and $i$ processes (slow, rapid, and intermediate respectively) all proceed via a series of neutron captures to make heavier isotopes and beta-decays to reach the next highest element number. It is via this dance between neutron capture and beta-decay that elements all the way up to the actinides can be made in astrophysical environments. Thus global models for neutron capture, beta-decay, and fission are crucial nucleosynthesis inputs, so we discuss these specially here. 

For neutron captures, very little experimental information exists beyond the stable species where the $s$ process operates. Additionally, the increasing complex structure of exotic species in the neutron-rich regions presents a challenge to reaction theory models, which leads to nucleosynthesis studies being dependent on Hauser-Feshbach (HF) models. Hauser-Feshbach utilizes the assumption that the reaction forms an intermediate compound nucleus that can be treated mostly independently of the initial state, as well as a statistical averaging to make use of a level density since modeling all level transitions and resonances in heavy systems can become infeasible. Although HF methods are known to have issues at closed shells and towards driplines where statistical averaging is no longer appropriate due to the level structure, at the present there are very few studies of alternative methods to calculate neutron capture rates globally 
%\cite{Rochman2017statisticalncap}.  
\cite{Rochman:17}.  

%Jutta-start:
The impact of optical-model uncertainties on calculated capture rates remains to be rigorously studied. This is particularly challenging for target nuclei far from stability.  New global optical models need to be developed, which requires microscopic predictions, experimental data, or a combination of both. 
Similarly, more work is required to obtain realistic gamma strength function (gSF) and level density (LD) models for unstable isotopes.
Studies have found that the gSF and LD models used in HF calculations can produce capture rates that differ by several factors to an order of magnitude or more, and these uncertainties propagate to the predictions for astrophysical abundances \cite{Nikas2020}. For more details on the present status of Hauser-Feshbach calculations, see Sec.~\ref{sec:reactions_statistical}.

Given the uncertainty in the calculated capture cross sections resulting from insufficiently-known level densities and gamma-ray strength functions, indirect methods will continue to play an important role for determining neutron capture rates.  It is via indirect approaches, such as the surrogate reactions method or Oslo-type measurements, that FRIB can make both immediate and long-term progress toward informing neutron capture in neutron-rich regions. The Oslo and beta-Oslo methods utilize charged-particle (transfer or scattering) reactions and beta-delayed gamma-emission measurements, respectively, to obtain information on the gamma-decay of the compound nucleus of interest to the neutron-capture reaction.  The experiments yield quantities that are a convolution of the LD and gSF; methods have been developed to extract both quantities and provide them for HF calculations; for a review, see 
%Ref.~\cite{Larsen:19}.
Ref.~\cite{Lar19a}.

The surrogate reaction method was originally developed to determine neutron-induced fission; for a review, see Ref.~\cite{Escher:12rmp}. In recent years, theory developments have made it possible to apply the method also to neutron-capture reactions~\cite{Escher:12rmp, Escher:16a, Potel:17}. Multiple applications have demonstrated that the method provides meaningful experimental constraints for cross sections calculations of neutron capture on isotopes near stability~\cite{Escher:18prl, Ratkiewicz:19prl, PerezSanchez:20}.  The approach does not rely on auxiliary quantities like s-wave resonance spacings ($D_0$) or average radiative widths ($\langle \Gamma_{\gamma} \rangle$), which are not available for unstable target isotopes, thus making it a promising method for exotic, neutron-rich or proton-rich, nuclei.
To obtain reaction rates relevant to astrophysical processes that involve isotopes well outside the valley of stability, it is necessary to further develop and validate the method beyond the range of current applications.  Both inelastic scattering and (d,p) transfer reactions are good candidates to reach a large number of exotic isotopes.  New developments are underway to explore these opportunities.
For more details, see the discussion in Section~\ref{sec:reactions_surrogate} and the references given there. In general, it is important to test the assumptions underlying indirect measurements and validate the resulting cross sections against additional measurements or theoretical predictions.  This will not only provide more reliable cross sections, but it will also yield insights into the underlying reaction mechanisms. 
%Jutta-end

Global models for beta-decay are of equal importance to those for capture rates. Here, more experimental information to benchmark models is available, but decisively disappears for many neutron-rich lanthanides, $N=126$ nuclei, and beyond. Additionally, the few relatively recent global beta-decay models have shown a significant difference between model predictions for rates near and beyond $N=126$, and have found that first-forbidden transitions play an important role here \cite{MarketinBeta,NeyBeta}. Nucleosynthesis studies have shown that the beta-decay model adopted can decisively change predictions for abundances and light curves (e.g. \cite{Famiano,Zhu2021Kilonova,Lund2023,Kullmann.Goriely.ea:2023}). Global beta-decay models typically make use of methods such as QRPA as well as Hauser-Feshbach in order to predict beta-delayed neutron emission probabilities~\cite{Minato.Marketin.Paar:2021}, but as is the case with neutron capture, how appropriate it is to apply approaches like HF should be an important focus of longer term FRIB investigations.

Fission in the neutron-rich regions is the ultimate unknown territory. With much of our experimental information quarantined to proton-rich actinides, on the neutron-rich side experiments have mostly probed species of relevance to reactor physics such as $^{239}$Pu, $^{235}$U, and $^{238}$U. Although such species are indeed produced in neutron-rich neutron capture nucleosynthesis such as the $r$ process (for instance $^{235}$U and $^{238}$U are present in stars and our Sun), fissioning species impacting nucleosynthesis calculations are predicted to be in regions that are much, much more neutron-rich \cite{Giuliani.Martinez-Pinedo.Robledo:2018,VasshJPhysGFiss,Kullmann.Goriely.ea:2023}. There are few models available which predict barrier heights globally for neutron-rich actinides, however already this limited set demonstrates a diversity of possible behavior. For instance, fission rates calculated using the barrier heights of the FRLDM model produce an $r$ process that terminates near $A\sim295$ (around $Z,N\sim94,200$, a few neutrons past the $N=184$ shell closure), whereas nucleosynthesis applications of rates using the HFB-14 fission barriers are capable of synthesizing nuclei much higher in the chart with an $A>300$ \cite{VasshJPhysGFiss}. Note that nucleosynthesis calculations reaching the neutron-rich actinides are very sensitive to the physics around $N=184$, which at the moment is purely a theoretically predicted shell closure, so searching for experimental hints of shell closures past $N=126$ in the future could directly impact nucleosynthesis studies of actinide production. 

Overall, fission can impact nucleosynthesis observables both by actively participating during the synthesis, introducing fission cycling into the dynamics and affecting the final abundances, and by impacting the electromagnetic afterglow (such as light curves and gamma rays) of the material present. Late-time fission effects on the electromagnetic emission depend on longer-lived fissioning species that lie closer to where current measurements have taken place, and so this will be discussed in the long-lived actinide section below as being potentially approachable by FRIB in upcoming years. However the fission cycling that influences abundances occurs at much earlier times when the nuclear flow lies in the very neutron-rich regions near $Z=94$, $N=184$. Studies have shown that when the synthesis reaches this deep into the actinides, the fission yields trends as a function of neutron number, along with which species the nuclear flow accesses, crucially impact the abundances around the second $r$-process peak at $A\sim130$ \cite{EichlerFissrproc,
%Giuliani.Martinez-Pinedo.ea:2020,
Giuliani2020,
VasshJPhysGFiss}. Furthermore depending on whether the fission deposition is dominantly symmetric or asymmetric, fission can also contribute significantly to the light precious metals region around $A\sim110$ \cite{VasshFissCoprod} as well as the lanthanide region around $A\sim150$ \cite{GorielyREP}. Nuclei which actively undergo fission as the abundances are being set lie well beyond the reaches of any current or planned experiments \cite{VasshJPhysGFiss,EichlerFissrproc,
%Giuliani.Martinez-Pinedo.ea:2020}. 
Giuliani2020}. 
Nevertheless, in the long-term future FRIB could potentially still provide some guidance to fission modelers by working to systematically extend the reach of probed nuclei in the neutron-rich actinides. Measurements of barrier heights, fission yields, and decay branchings could provide constraints for extrapolations and benchmarks for theoretical calculations. Additionally, the outcome of neutron-rich nucleosynthesis in the actinide region is determined by several fission processes (e.g. neutron-induced, beta-delayed and spontaneous fission) and alpha decays which could all play a role in setting the nucleosynthesis abundances. For all the fission processes it is important to investigate both rates and yields since both are dependent on the fission kinematics. Despite the challenges of probing the neutron-rich actinide region, it is important for experiments to push into this desolate frontier since it is connected to some of the biggest questions in nuclear astrophysics such as the ultimate nucleosynthesis reach of nature (i.e. how heavy of species can be synthesized) as well as whether or not the superheavy elements can be produced in astrophysical environments \cite{Petermann.Langanke.ea:2012,Goriely.Martinez-Pinedo:2015,MumpowerBDFrp}.

\subsubsection{Connecting FRIB to the synthesis of elements beyond tellurium and up to the actinides} 

\begin{figure}[H]
    \centering
    \includegraphics[width=17cm]
    {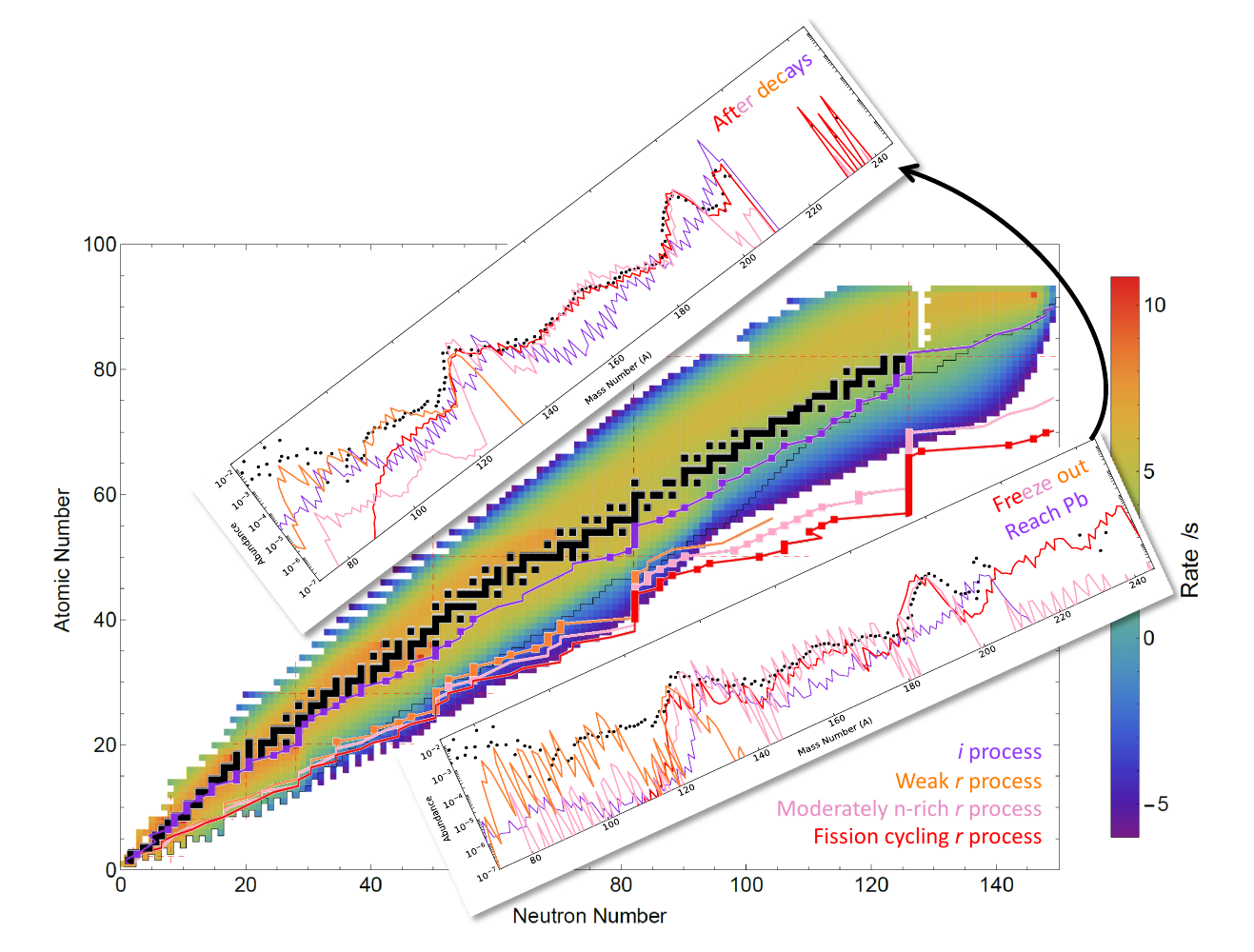}
   \caption{The FRIB production rates compared to the `paths' (most abundance species at a given proton number) for a `weak' $r$ process (orange), moderately neutron-rich $r$ process (pink) and a fission cycling $r$ process (red). The snapshot is taken at the moment of `freeze-out'  for each $r$-process case, and the abundances at this time are shown in the bottom plot as compared to the Solar $r$-process residuals from \cite{Sneden}. Also shown is the path of an $i$-process calculation (purple) at the moment that the path significantly populates lead species. The final abundances following the decay of the nuclei to stable species is shown in the top plot. For the $r$-process cases, the path is shown as a line when the abundance is more than $10^{-11}$ and as boxes connected by a line if the abundance exceeds $10^{-5}$. For the $i$-process case the abundance limits for the line only, boxes plus line is $10^{-15}$, $10^{-9}$ respectively. Note the significant number of species whose properties are important in shaping the final abundances that FRIB can reach between the black line of presently probed species and the paths.}
\label{fig:FRIBnucapnuc}%
\end{figure}

We first discuss the main $r$ process, which synthesizes elements beyond $\sim$Te all the way to the heaviest elements found in nature, though it may also be co-produced with some of the lighter elements. Astrophysical sites of the $r$ process eject material at a variety of neutron-richness, temperatures, and densities which can host nucleosynthesis. Figure~\ref{fig:FRIBnucapnuc} demonstrates the $r$-process path (most abundance species at a given proton number) for two distinct types of ejecta that both produce a main $r$ process, one being a hot, low entropy, low density condition of moderate neutron-richness ($Y_e = p/(n+p) = 0.2$ and another being a cold, low entropy, high density tidal tail condition which is extremely neutron-rich ($Y_e=0.01$). The figure shows the paths at the moment of `freeze-out', that is when the neutron-to-seed ratio goes to $1$ and the path begins to move back towards stability. During this stage late-time neutron captures are well as beta-decays shape the final abundances, which includes their influence on any fission fragments that have been deposited around or after freeze-out. As can be seen for comparing the location of the paths to the black line representing currently probed species, the $r$-process paths traverse over the presently unexplored territory outside the black line on their way to stability and over the course of this journey their abundances are altered from the bottom plot to that shown in the top. Therefore numerous opportunities to inform the evolution of the main $r$ process are possible for FRIB, from measurements of nuclei near $N=82$, to the lanthanide (a.k.a. rare-earth) species, as wells as near $N=126$ and beyond. We discuss each of these areas specifically here as well as the science questions that can be approached via dedicated studies of each region.

\noindent \textit{Nuclei near $N=82$ -} The Solar isotopic abundance pattern (see Fig.~\ref{fig:FRIBnucapnuc}) is clearly influenced by nuclear structure with ``peaks" of especially high abundance seen in the presence of magic numbers and enhanced stability, as is the case with the second $r$-process abundance peak at $A\sim130$. This enhancement in the abundances is known to come from a so-called ``pile-up" at neutron shell closure $N=82$, where the $r$ process must wait to beta-decay before capture can proceed beyond $N=82$ into the lanthanide region. Additionally processes with conditions that do not permit significant capture into the lanthanides, such as a so-called “weak” $r$ process (see Fig.~\ref{fig:FRIBnucapnuc}), can also reach $N=82$ nuclei. Therefore this peak in the Solar abundance distribution could be a convolution of multiple ejecta components from a single type of event or could in fact be shaped by different types of events. Disentangling these possibilities requires pinning down the predicted shape of this peak in the presence of different conditions. For instance if fissioning nuclei are produced and their fission yields are assumed to be symmetric, then their daughter products will deposit near $A\sim130$, influencing second peak abundances \cite{Shibagaki}. If instead the deposition is mostly asymmetric (e.g. \cite{MumpowerFRLDMYields, SadhukhanWitekFission}), the $N=82$ peak can be depleted relative to Solar in favor of enhancements in the abundances of lanthanides and light precious metals like silver \cite{VasshFissCoprod}.  Therefore the shape of the $A\sim130$ peak connected to $N=82$ hints at not only the involvement of different astrophysical sites but could also point to what specific types of conditions may be present at a given site. 

Thus, around $N=82$ is an interesting area to further pin down masses, beta-decay rates, and neutron emission probabilities ($P_n$ values) due to their influence on the population of key species. For instance the $^{129}$I abundance is connected the physics near $N=82$, which is of interest since the $^{129}$I/$^{247}$Cm abundance ratio in meteorites has been shown to be capable of peering directly into the conditions of the last $r$-process event that enriched our Solar System since this ratio does not change as a function of time due to the similar half-lives of these radioisotopes \cite{CoteICm}. Species near $N=82$ are also neutron-rich fission products, which can produce heat and gamma rays during the nucleosynthesis, affecting electromagnetic signals from events like mergers and their remnants \cite{Zhu2021Kilonova}. Thus pinning down nuclear structure at and near $N=82$ is tied to key astrophysical observables.

How nuclear structure evolves as one approaches the dripline is not only of interest for fundamental dripline studies, but also of relevance for nucleosynthesis predictions. For instance, astrophysical conditions with particularly high neutron densities can find the synthesis to occur very near the dripline (see for example the fission cycling $r$-process path in Fig.~\ref{fig:FRIBnucapnuc}). The first ever FRIB experiment explored isotopes approaching the neutron dripline near $N=28$ \cite{Crawford2022PRL}, which well demonstrates the special role FRIB can play in illuminating dripline physics in the neutron-rich regions. Note that the $r$-process paths of all $r$-process example cases in Fig.~\ref{fig:FRIBnucapnuc} show pile-up to occur at $N=82$ outside the black line of previously probed nuclei. Thus pressing $N=82$ measurements down towards the $N=82$ dripline would take full advantage of the wealth of neutron-rich species FRIB is capable of making for the first time and would be of great interest to nuclear astrophysics.

\noindent \textit{Lanthanide nuclei -} Neutron-rich lanthanide species (with $Z=57-71$, also called rare-earth nuclei) share a fortuitous overlap with the FRIB production reach. Measurements in this region have been extended and made more precise in recent years, for example at the CPT at CARIBU at Argonne \cite{OrfordVassh2018}, Jyvaskyla \cite{JYFLTRAP}, and BRIKEN \cite{BRIKENKiss2022}. Therefore, excellent benchmarks exist for future FRIB experiments to capitalize on when looking to further such explorations toward currently unprobed species. Thus this area provides a wealth of discovery opportunities while also capitalizing on more recent efforts.

Rare-earth nuclei are also connected to long standing questions of heavy element production as well as new observational capabilities. For instance, the Solar isotopic pattern hints of an enhanced stability of neutron-rich lanthanide species due to a peak found near $A\sim164$ which is not connected to a closed shell \cite{MattREP,GorielyREP}.  Since nuclear masses reflect underlying structure, here mass measurements are especially important. Mass measurements will also propagate into the nucleosynthesis predictions through separation energies determining neutron capture and photodissociation rates, as well as beta-decay half-lives and $P_n$ value predictions. In particular, the properties of $Z=58-62$ nuclei with neutron numbers $N\sim104-110$ have been found to be especially connected to whether simulations can reproduce the rare-earth peak \cite{VasshMCMC}. Given the overlapping FRIB reach in this area, the prospect of resolving the longstanding question of rare-earth peak origins within the next decade is promising.

Another important open question in heavy element nucleosynthesis is the origin of the so-called “robustness” or “universality” of lanthanide abundances in the abundance patterns of metal-poor stars rich in $r$-process elements. That is, here metal-poor stars have shown a remarkable consistency with each other as well as the Solar abundance pattern, when it is common for studies to find the abundance pattern to be sensitive to the astrophysical conditions present at the nucleosynthesis site \cite{Sneden,OlegRobustness}.  Could this be hinting that the astrophysical site of $r$-process production that enriched our Sun is the same as the source that enriched these stars? Does this mean that the $r$ process occurs in a similar way throughout the galaxy? Or does this hint that there are nuclear physics processes at play such as fission cycling that could wash away any initial differences in the astrophysical conditions? The dependence on the properties of local neutron-rich lanthanides (such as masses, beta-decay rates, and $P_n$ values) as rare-earth abundances are being finalized during the late phases of the $r$ process is crucial to address this question.

Other key investigations over the last few years also bolster the community’s interest in the lanthanides, such as modeling kilonova light curves from neutron star mergers as well as a budding interest in the $i$ process. Although the $i$ process was first proposed decades ago \cite{1977CowanRoseiprocess}, it has gained traction int recent years through indications that some metal-poor star patterns are best fit by something with a neutron density between that of an $s$ process and $r$ process \cite{Benoitiprocess,Hampel2019}. Additionally, recent hydrodynamic simulations of rapidly accreting white dwarfs and certain types of AGB stars have shown the neutron densities needed for an $i$ process to be possible in these environments \cite{DenissenkovRAWD,Choplin2022}. This process has been predicted to take place only a few neutrons away from stability where neutron capture timescales can exceed those for beta-decay (see Fig.~\ref{fig:FRIBnucapnuc}). Thus studies have found that the $i$ process is most sensitive to neutron capture rates and would greatly benefit from constraints on these models \cite{DenissenkovBatoW} (from either surrogate measurements or methods such as Beta-Olso informing the strength functions and level densities that enter neutron capture predictions). The lanthanides are a particularly interesting region to study from an $i$-process perspective, with elemental abundance ratios in stars such as [Ba/Eu] being used to differentiate $i$-process enrichment from $s$ process and $r$ process \cite{JINAbase}. As noted above another important aspect of the need to improve lanthanide abundance predictions lies in the sensitivity of kilonovae to the overall lanthanide mass fraction present in the ejecta. The high atomic opacities of lanthanide species give them the ability to push the light curve to longer timescales and towards the infrared, thereby providing the opportunity for conclusive identification of such species in the ejecta (as was the case with the first ever observed neutron star merger GW170817). New insights into mergers from the kilonova of this event are still being discussed and future gravitational wave detection prospects further present an urgency to refine our predictions of the evolution of spectral features of kilonova~\cite{Sneppen.Watson.ea:2024} so that more conclusive statements can be gleaned about the amount of mass ejected~\cite{BarnesKN2020}, the geometry of the ejecta~\cite{Collins.Shingles.ea:2024,Sneppen.Watson.ea:2023nat}, and the value of the Hubble constant~\cite{Sneppen.Watson.ea:2023}.

\noindent \textit{Nuclei near $N=126$ -} The solar abundance distribution also shows a peak at $A\sim195$ associated with nucleosynthesis undergoing pile-up at $N=126$. Studying this peak is directly connected to illuminating the production of platinum and gold as their stable isotopes can be found at $A=195-196$, $198$ and $A=197$ respectively. Here FRIB has the ability to inform modeling of reaction rates, beta-decays and masses greatly since at the present there are no measurements past $^{206}$Hg ($Z=80$) on the $N=126$ isotonic chain. Even though the FRIB reach will be unable to explore the dripline at $N=126$, pushing measurements down this chain toward the dripline is crucial for an understanding of the strength of this shell closure. Current mass models predict very different strengths of this shell closure (i.e. how big of a drop there is in the one neutron separation energy at $N=126$) which leads to different predictions for how long species are held up here before the synthesis can proceed to the actinides. Thus the $N=126$ shell closure has been referred to as the “gateway to the actinides” and studies have shown that predicting the abundances of long lived-actinides such as $^{232}$Th and $^{235,238}$U are sensitive to the mass trends~\cite{Mendoza-Temis.Wu.ea:2015,Holmbeck2019,MariusActinideBoost} and to beta-decay rates~\cite{EichlerFissrproc} around $N=126$. Additionally, for some nuclear structure models the ability to approach the heaviest regions of the nuclear chart may only be currently possible at or near closed shells like $N=126$ given the simpler valance structure here compared to that needed to be considered in the deformed rare-earths. Therefore this area offers a nice overlap between astrophysical interest and approachability by both nuclear structure theory and experiment. 

In addition to the influence that the masses have on the nuclear flow at $N=126$ (e.g. \cite{AlmudenaWitekDFTPRL,TrevorDFT}), beta-decay rates around this closed shell have been shown to also greatly affect abundance predictions. Global rate calculations have found that including first-forbidden transitions changes predictions significantly \cite{MarketinBeta,NeyBeta} which can lead to faster rates to the right versus left of $N=126$. Therefore studying beta-decay strength functions and rates around $N=126$ is impactful for global structure models and astrophysics. Currently, the nuclear physics uncertainties in the region stymie the ability to conclusively link unique features of the associated $A\sim195$ abundance peak, such as its overall width, with particular types of astrophysical conditions. For instance, it had previously been discussed that potentially neutron star mergers could not be the source of Solar System enrichment due to the narrow third peak predicted when these very neutron-rich environments see significant late time neutron capture. However studies later showed that using different beta-decay treatments can change the third peak abundance predictions toward something more in line with the Solar width \cite{Caballerorprocess}.  Another interesting link between third peak abundances and astrophysical conditions lies in considering `slow' versus `fast' ejecta components, since studies have shown that fast (i.e. rapidly expanding) ejecta can be tied to a shift in the third peak toward higher mass numbers \cite{MetzgerPrecursor,MendozaTemis,JustWind,BarnesKN2016,XiluSpallation,Placco2020}. In addition to clear connections between $N=126$ and the astrophysical environment, unlike $N=82$ and rare-earth regions, there are no potential contributions from fission deposition here, making $N=126$ abundances less convoluted than other parts of the pattern. Therefore measurements which put $N=126$ structure on more firm footing are directly connected to diagnosing whether or not abundance predictions for a particular astrophysical environment align with observations.

\noindent \textit{Long-lived actinides -} Information on actinide species past $N=126$ and $^{208}$Pb have been disproportionately mapped out in neutron-deficient regions over neutron-rich regions. For instance, although it is an important channel during the decay of heavy $r$-process species, beta-delayed fission has never been experimentally observed on the neutron-rich side, so confirmation alone would be important progress. Thus exploratory missions to push the boundaries on neutron-rich actinide measurements would take incredibly important steps of great benefit to those modeling neutron-rich actinide properties. This is also a regime that could shed light on nuclear structure given the speckled branching ratios between beta-decay, alpha-decay, and spontaneous fission seen here. Additionally, several species in this region are known to have anomalously long half-lives on the order of days, years, or longer which can have important consequences for observations of electromagnetic signals from astrophysical events such as mergers (light curves are sensitive to decays that take place on the order of the expansion timescale). 

For instance it has been discussed since the early days of nuclear astrophysics that $^{254}$Cf with its $\sim$60 day spontaneous fission half-life could greatly affect kilonovae \cite{B2FH} due to its high Q-value introducing heat into the system, as well as the high thermalization efficiency of fission fragments \cite{BarnesKN2016}. Note however that the effect $^{254}$Cf could have specifically on kilonovae has only been discussed in the last few years following GW170817 \cite{Cfpaper,Wu.Barnes.ea:2019} $^{254}$Cf serves as an excellent example of the key pieces of nuclear data lying just outside current measurements since its direct beta-feeder $^{254}$Bk, which provides the only pathway to populate $^{254}$Cf, has unknown branching ratios. Thus, should $^{254}$Bk be found to undergo alpha-decay 100\% of the time, the long-standing question of $^{254}$Cf’s potential impact of light curves could be definitively answered. It is not just long-lived $^{254}$Cf that makes neutron-rich actinide branching ratios important to probe. Since neutron-rich actinides are largely unprobed territory there lies the opportunity to find others like $^{254}$Cf. For instance, studies have found a few neutron-rich rutherfordium isotopes with theoretical spontaneous fission half-lives on the order of days to also be able to impact light curves greatly and allow for identification of superheavy elements in the ejecta \cite{Zhu2021Kilonova,Holmbeck2023Superheavy}. Additionally, studies of MeV gamma-ray emission from real-time $r$-process events have found gammas with energies $> 3.5$ MeV to be unique to fission since on the timescales of days the most neutron-rich species have already undergo beta-decay and so emission at such higher energies requires either the prompt gammas from the fission itself or the delayed gammas from the beta-decay of the associated neutron-rich fission daughters \cite{WangVasshMeVgammafiss}. Further probing the gamma spectrum from neutron-rich fission species would directly inform such studies and potentially change the expected contribution of prompt fission gammas relative to those from beta-decays. Note that the prospects of observing real-time MeV gamma rays from a merger are limited to either galactic events or events in nearby galaxies such as those within the Local Group (e.g. the Large Magellanic Cloud (LMC)), and further face the challenge of the relative rarity of merger events (estimates range from $\sim$20 to$\sim$400 mergers every Myr in the Milky Way, which is $\sim$100 to 1000 times less frequent than core-collapse supernovae) \cite{VasshWangTl208}. Nevertheless, since gamma spectral features provide a type of fingerprint for specific isotopes, they present the important prospect of detailed information on the ejecta composition \cite{Hotokezaka2016, OlegGammas}. Additionally, gamma-ray observations offer unique opportunities to identify the remnants of the Milky Way's last neutron star mergers~\cite{Wu.Banerjee.ea:2019,OlegGammas}.

Spontaneous fission is not the only process of interest since alpha particles have also been shown to efficiently thermalize in merger ejecta thus long-lived alpha-decays can play a dominant role in late-time kilonova heating when there are no species undergoing spontaneous fission to compete with \cite{BarnesKN2016,VasshJPhysGFiss,KentaKilonova}. Alpha-decays on the order of hours, days, or years can also lead to distinct MeV gamma ray emission from mergers, as is highlighted by the recently introduced prospect of utilizing the $2.6$ MeV gamma-ray line of $^{208}$Tl as a beacon of in situ neutron capture nucleosynthesis \cite{VasshWangTl208}. Although the role it can serve to identify the isotopic composition in astrophysical environments has only recently been highlighted, this particular emission line of $^{208}$Tl is well-known in numerous other fields of science, including clinical imagining studies utilizing $^{224}$Ra \cite{medimaging2022}, geological surveys estimating thorium concentrations \cite{watersoil1971,snowwatercontent1980}, nuclear safeguard measures looking for highly enriched $^{232}$U \cite{wastemat2009}, and as a well-established background and calibration point for experiments \cite{SNO1999,MajoranaExp}. The $^{208}$Tl emission line could be visible during the electromagnetic follow-up of a neutron star merger event, or could be observed locally from an AGB star or rapidly accreting white dwarf actively synthesizing the heaviest elements, and would correspond to an unambiguous signature of the production of $^{208}$Pb and thus elements lighter than lead such as gold could be conclusively stated to be present. The $^{208}$Tl $2.6$ MeV gamma-ray line shows itself on observable timescales due to the alpha decays feeding this species on the order of hours ($^{212}$Bi), days ($^{224}$Ra) and years ($^{228}$Ra). Since the observable timescales of kilonovae and MeV gamma emission align with the decay timescales of long-lived actinides, pushing the boundaries of known half-lives and branching ratios in the neutron-rich actinides has clear and direct connections to deciphering the electromagnetic emission from multi-messenger events. Thus should developments at FRIB permit studies of this area, this region offers high impact overlap between measurement opportunities and nucleosynthesis needs.

\subsubsection{Connecting FRIB to the synthesis of elements between iron and tellurium}

In astrophysical environments such as neutron star mergers and collapsars there is the possibility to form a disk around the central remnant (e.g. \cite{Siegel,Radice18}) and the ejecta from this disk can have conditions which support $r$-process nucleosynthesis. Although previous studies suggested disk material to be more processed and thus generally less neutron-rich than that which is dynamically ejected earlier during the merger \cite{JustWind,MillerWind}, more recent studies which run their hydrodynamic simulations out to longer times ($> 1$ sec) have changed this picture \cite{FahlmanWind,FernandezJustWind,SprouseWind,Just.Vijayan.ea:2023}. Thus both merger dynamical ejecta and post-merger accretion disks are comprised of a distribution of ejecta types (e.g. \cite{Wanajo,Foucart}), correspondingly producing of a wide range of elements all the way from the $r$-process first peak at $A\sim80$ to beyond the third peak at $A\sim195$. This is achieved via the total ejecta hosting a range of neutron-richness $Y_e$, as well as a variance in the manner that the temperature and density evolves. We refer to the components of the ejecta with synthesis taking place between $N=50$ and $N=82$, and forming elements with mass numbers between $A\sim80$ and $A\sim130$ such as strontium and silver, as undergoing a `weak' $r$ process (shown in Fig.~\ref{fig:FRIBnucapnuc}). For weak $r$ processes in both mergers and supernovae, ($\alpha$,n) reactions can also play an important role \cite{Bliss2017an}. The prediction of both weak and main $r$-process production in mergers is supported by the observations surrounding GW170817, where first a lanthanide-free blue component (such as a weak $r$ process) was first observed followed by a red component containing lanthanide elements \cite{Cowperthwaite2017,Villar}, and further bolstered by the identification of Sr in the absorption spectrum \cite{WatsonNatureSr}. 

Thus, in the weak $r$-process regime between $N=50$ and $N=82$, there are opportunities to impact $r$-process studies through measurements of masses, beta-delayed neutron emission probabilities $P_n$, and ($\alpha$,n) reactions. Additionally, surrogate measurements to inform neutron capture have begun to penetrate the neutron-rich regions in the $N=50$ to $82$ regime relevant for $i$-process abundance predictions \cite{DenissenkovN50}. Such measurements would not only inform $i$-process predictions, but serve to constrain and inform the neutron capture models relevant for the $r$ process by providing systematics that can be applied further along the isotopic chain. Studying the Solar System composition of the elements from Sr to Sn is therefore dependent on pinning down the nuclear physics properties of relevance for weak $r$ process and weak $i$ process so that their relative contributions, and how these compare to contributions from supernovae (e.g.\cite{Nishimura17,Mosta, ReichertArcones21}) and other events capable of forming the lighter heavy elements between iron and tellurium, can be better constrained. 

An additional recent development that makes this region of interest is the prospect of identifying the signature of fission fragment deposition here. Recent works have demonstrated that asymmetric fission taking place late during the $r$ process can inject neutron-rich daughter products which stabilize the abundances of light precious metals such as silver and palladium relative to the lanthanides \cite{VasshFissCoprod}. This introduces a new type of ``universality” of lanthanide abundances relative to light precious metals which can be verified through comparison with trends in stellar abundance ratios  \cite{VasshFissCoprod,RoedererVasshScience}. Thus constraining the nuclear properties along the decay path of fission daughter products would help to constrain whether fission deposition signatures have been hiding in metal poor stars, which would connect FRIB to approved future observations with the Hubble Space Telescope aimed at measuring the currently scarce cadmium ($Z=48$) abundances of relevance to this question.

Weak processes and in particular (anti)neutrino absorption reactions play a fundamental role in determining the proton-to-nucleon ratio both in core-collapse supernova and neutron-star mergers.  Those reactions are expected to operate on nucleons as nuclei typically form once the neutrino fluxes have substantially decreases as the ejecta moves away from the neutrino source. However, it is interesting to consider what will happen if neutrino fluxes remain large once nuclei are formed. Under these conditions, and assuming moderate neutron-rich ejecta, a kind of $r$ process may operate where the increase in element number is due to charged-current neutrino reactions instead of beta-decays. This process has been named $\nu r$ process~\cite{Xiong.Martinez-Pinedo.ea:2023} and is expected to contribute to the production of neutron deficient nuclei, the so-called p-nuclei, particularly $^{92,94}$Mo and $^{96,98}$Ru. Compared with other alternative excenarios like the $\nu p$ process, it has the advantage that can produce the long lived $^{92}$Nb that has been observed in the early solar system. At the moment, it is unclear the astrophysical scenario in which the $\nu r$ process may operate. Most likely involves magnetically driven outflows like those found in magneto-rotational supernova or collapsars. From the nuclear physics side, it requires neutrino-nucleus cross sections for nuclei to both sides of the stability valley. They require knowledge of both Gamow-Teller and forbidden strength distributions that can be determined by charge-exchange reactions.

\subsection{Nucleosynthesis on the proton-rich side of stability} 

\subsubsection{FRIB Experiments addressing stellar nucleosynthesis}
Stellar nucleosynthesis produces the majority of elements in nature, and proceeds mostly along the valley of stability through reactions on stable nuclei. This is a consequence of the long burning timescales that allow ample time for the majority of rare isotopes produced in a nuclear reaction to decay back to stability before the next nuclear reaction occurs. However, during the explosive burning of oxygen and silicon during a core collapse supernovae marking the endpoint of the stellar nucleosynthesis sequence for massive stars, timescales are much shorter and reactions on rare isotopes become important. This stellar burning stage is important to explain the origin of isotopes and elements in the Si-V range, and also contributes, together with thermonuclear supernovae, to the iron peak elements (e.g. \cite{ woosleyEvolutionExplosionMassive2002}). In addition, explosive silicon and oxygen burning is responsible for the synthesis of a broad range of $\gamma$-ray emitters that are sufficiently long-lived to be ejected by the supernovae to be observable with space or balloon based gamma-ray observatories \cite{AndrewsNucleosyntheticYieldsCorecollapse2020, hermansenReactionRateSensitivity2020, timmesCatchingElementFormation2019}. The production of the longest-lived gamma-ray emitters are also important to explain supernova light curves, meteoritic signatures of early solar system radioactivity (e.g. $^{53}$Mn), isotopic signatures in pre-solar grains (e.g. $^{44}$Ti) or the composition of cosmic rays (e.g. $^{44}$Ti, $^{49}$V, $^{51}$Cr, $^{55}$Fe, $^{57}$Co, and $^{59}$Ni) (see \cite{ hermansenReactionRateSensitivity2020} and references therein).

During explosive oxygen and silicon burning, the stable $N=Z$ products of previous stellar burning, predominantly $^{16}$O, $^{20}$Ne, $^{32}$S, and $^{28}$Si, are rapidly driven towards nuclear statistical equilibrium by a wide range of fusion reactions as well as $\alpha$, p, and n induced reactions. As weak interactions are not in equilibrium, the neutron to proton ratio remains roughly equal and nucleosynthesis proceeds mainly along the $N=Z$ line towards the iron region, thus involving neutron deficient rare isotopes. However, the region of nucleosynthesis is broad, and in a few cases, driven by captures of neutrons released in the nuclear reactions, extends a few mass units out on the neutron rich side, e.g. to $^{43}$K, $^{47}$Sc, or $^{59}$Fe.  

In principle, equilibrium nucleosynthesis is expected to be relatively independent of detailed nuclear reaction rates. However, as temperatures are dropping during nucleosynthesis, equilibrium breaks up into an increasing number of equilibrium clusters connected through slow reactions that are critical for determining the final nucleosynthesis outcome. The important reactions for the synthesis of isotopes of interest have been identified in sensitivity studies \cite{theNuclearReactionsGoverning1998, magkotsiosTRENDS44Ti56Ni2010, hermansenReactionRateSensitivity2020, subediSensitivity44Ti56Ni2020}  and include a large number of $\alpha$, p, and n induced reactions on stable and radioactive targets up to Ni. It will be feasible at FRIB to produce reaccelerated beam for many of the radioactive cases of interest, as the relevant nucleosynthesis temperatures are relatively high (typically $1-3$ GK but in some cases reaching up to $5-6$ GK) resulting in relatively larger cross sections, and as the target nuclei are relatively close to stability facilitating beam production. Proton and $\alpha$-induced reactions could then be readily measured directly with the SECAR recoil separator, the JENSA gas jet target, the AT-TPC active target time projection chamber, or the MUSIC multi-sampling ionization chamber detector. For (n,p) and (n,$\alpha$) reactions, probing the reverse reaction may provide valuable constraints. It will be important to carry out a parallel effort at stable beam and neutron beam facilities to determine or constrain the reactions on stable targets. Only a comprehensive program will ultimately advance our understanding of supernova physics and the origin of the elements. 

\subsubsection{FRIB Experiments addressing the p process}
The $p$ process is a classical nucleosynthesis process defined as being responsible for the synthesis of the rare neutron deficient stable isotopes between $^{74}$Se and $^{196}$Hg that cannot be produced by neutron capture processes, the so called p-nuclei. Nowadays a $\gamma$ process is thought to be responsible for the synthesis of the p-nuclei \cite{pignatariProductionProtonrichIsotopes2016a}. In the $\gamma$ process an initial distribution of heavy seed nuclei, for example produced by a previous $s$ process in a previous generation of stars, is exposed to sudden heating that induces ($\gamma$,n) photodisintegration reactions that drive the composition to neutron deficient unstable isotopes. The composition is further modified by subsequent ($\gamma$,p) and ($\gamma$,$\alpha$) reactions, as well as the inverse capture reactions capturing the released particles. Proposed sites are the explosive O/Ne burning in core collapse supernovae, or the outer layers of thermonuclear supernovae. 

A long-standing challenge has been the synthesis of the p-nuclei $^{92,94}$Mo and $^{96,98}$Ru that are particularly abundant in the solar system, and cannot be produced with sufficient abundance in standard $\gamma$-process scenarios. Accurate nuclear physics is needed to test various proposed model solutions to this puzzle, and to ultimately use $p$-process nucleosynthesis to probe the physics of core collapse supernova shock fronts, the physics of thermonuclear supernovae, and the physics of the supernova progenitor stars that potentially impacted the distribution of heavy seed nuclei for the $\gamma$ process. The most important reactions have been identified in sensitivity studies \cite{rappSensitivityPProcessNucleosynthesis2006a, rauscherUncertaintiesProductionNuclei2016, nishimuraUncertaintiesProductionNuclides2018} and include a mixture of reactions on stable and unstable neutron deficient nuclei in the $A=71$ to $A=196$ mass range. While the reactions on stable nuclei have been measured in many cases, information on the reactions on unstable nuclei is extremely limited, with first limited experiments only being performed very recently \cite{williamsCrossSections83Rb2023}. There is an opportunity to address this lack of data at FRIB. In particular the important branchings between ($\gamma$,n) and ($\gamma$,p) or ($\gamma$,$\alpha$) are mostly located a few mass units away from stability and can be probed by measuring directly the inverse (p,$\gamma$) and ($\alpha$,$\gamma$) reactions. As level densities tend to be high, important information can already be obtained with limited early FRIB beam intensities by measuring excitation functions at the upper end or above the Gamow-window. Such measurements can already provide constraints for statistical model inputs, especially the relatively uncertain $\alpha$-optical potential. Experiments can be carried out using gamma detection arrays such as SuN\cite{tsantiriCrosssectionMeasurement822023}, which can be combined with SECAR for increased sensitivity. 

\subsubsection{FRIB Experiments addressing the \texorpdfstring{$\nu p$} process}

Neutrino-driven outflows as part of core-collapse supernova ejecta have long been discussed as site for heavy element nucleosynthesis, especially for the $r$ process. There has been a tremendous progress in simulations of core-collapse supernova allowing for fully self-consistent three-dimensional explosions. These calculations seem to indicate that strong neutrino fluxes irradiating the innermost supernova ejecta drive the composition mostly to proton-rich conditions, though small pockets of neutron rich regions may still produce a weak $r$ process. Under such proton-rich conditions the $\nu p$ process is expected to operate. It is a sequence of proton captures and $(n,p)$ reactions producing neutron-deficient nuclei like $^{92,94}$Mo and $^{96,98}$Ru. Proton capture reactions operate closer to the stability than for the $rp$ process and may be more amenable to a statistical treatment. $(n,p)$ reactions are particularly important and several key reactions have been identified~\cite{Nishimura.Rauscher.ea:2019}. (n,p) reactions on unstable targets can be constrained at FRIB by directly measuring the inverse (p,n) reaction in inverse kinematics by using reaccelerated beams with for example the SECAR recoil separator in conjunction with a neutron detection array. Proton capture rates on unstable nuclei can be measured using the same techniques discussed for the $p$ process. 

\subsubsection{FRIB Experiments addressing the \texorpdfstring{$rp$} process in X-ray bursts}
X-ray bursts are thermonuclear explosions on the surface of neutron stars that accrete matter from a binary companion. In systems where the accreted material is hydrogen rich, and where the CNO cycle does not burn all the hydrogen prior to the ignition of the burst, bursts are powered by the rapid proton capture process ($rp$ process). The $rp$ process synthesizes heavy elements up to $A \approx 110$ via a proton captures and beta-decays proceeding on the proton rich side of the valley of stability \cite{schatzXrayBinaries2006b,fiskerExplosiveHydrogenBurning2008b,parikhNucleosynthesisTypeXray2013a}. The possible contribution of the rp-process to galactic nucleosynthesis is an open question. While models predict the ejection of some material during a burst \cite{weinbergExposingNuclearBurning2006a, herreraMasslossCompositionWind2023a} the historical frequency of events is likely insufficient for a significant contribution. However, accurate predictions of the nucleosynthesis are needed to predict possible spectral signatures of the ejected material that could be searched for with future X-ray observatories, and to predict the composition of the neutron star crust that is determined by the majority of burst ashes that remains on the neutron star \cite{meiselNuclearPhysicsOuter2018b}. The latter is essential for modeling crust processes and neutron star cooling observables in transiently accreting systems (section \ref{Sec:AccretedCrusts}). 

Furthermore, due to the slow beta-decays in the reaction sequence, the $rp$ process process is relatively slow and significantly extends and shapes the burst light curve. With accurate nuclear physics, burst light curve observations in systems with stable bursting behavior, where multiple bursts can be averaged to obtain a precise light curve, can be compared with models to extract the surface red shift, which provides a constraint on the mass-radius relationship of the neutron star and the nuclear equation of state \cite{meiselInfluenceNuclearReaction2019}. 

The important nuclear reactions that affect observables in X-ray bursts have been identified in sensitivity studies \cite{cyburtDependenceXRayBurst2016a, meiselInfluenceNuclearReaction2019} and involve mostly unstable, neutron deficient target nuclei. In the vast majority of cases, reaction rates have not been measured directly. With reaccelerated beams at FRIB direct measurements of (p,$\gamma$) reactions will be possible with the SECAR recoil separator once beam intensities reach $10^6$~pps or more. ($\alpha$,p) reactions can already be probed at somewhat lower beam intensities using either the JENSA gas jet target \cite{jensacollaborationFirstDirectMeasurement2023}, the AT-TPC active target time projection chamber \cite{randhawaFirstDirectMeasurement2020}, or the MUSIC multi-sampling ionization chamber \cite{jayatissaStudy22Mathrm2023}. As many of the important reactions are relatively far from stability, level densities are low and reaction rates tend to be dominated by single, isolated, narrow resonances. In these cases, indirect experimental approaches can already dramatically reduce reaction rate uncertainties. These include beta-delayed particle emission, for example with the GADGET active target system \cite{budnerConstraining30312022}, and (d,n) transfer reactions using the GRETINA gamma detection array with the LENDA neutron-detection array \cite{wolfConstrainingNeutronStar2019b}. Both techniques are particularly well suited for early FRIB experiments, the former because of the relatively low beam intensity requirements, and the latter due to the fact that the experiments can be performed with fast beams avoiding the beam losses from stopping and reaccelerating. (d,n) transfer reactions can also be measured with the SOLARIS spectrometer taking advantage of reaccelerated beams from ReA6 for measurements at lower energies. The future ISLA spectrometer will also open up opportunities for such indirect reaction measurements.

\subsection{Accreted neutron star crusts} \label{Sec:AccretedCrusts}
Transiently accreting neutron stars offer an observational window into the physics of the neutron star crust \cite{meiselNuclearPhysicsOuter2018b}. During an accreting outburst, the crust is heated by various types of thermonuclear X-ray bursts on the surface, as well as by nuclear reactions throughout the outer and inner crust, induced by the continuously increasing local pressure due to the ongoing accretion. During the subsequent quiescent phase that can last in quasi persistent transients for years, the cooling of the crust can be observed through repeated X-ray observations of the decreasing surface temperature. Such cooling curves provide insights into the structure of the crust, neutron superfluidity, nuclear pasta, and the nature of the neutron star core. The nuclear reactions may also induce density jumps that may lead to gravitational wave emission as the neutron star is rapidly spinning \cite{hutchinsGravitationalRadiationThermal2023a}. Interpretation of these observables requires accurate nuclear physics of nuclei up to mass 110 from stability to the neutron drip line. While the majority of these nuclei are expected to be within reach at FRIB in the future, including the FRIB400 upgrade, early experiments can already extend the boundary if known masses as well as electron capture and beta decay transition strengths that are important for heating and cooling, through neutrino losses, of the crust. The FDSi FRIB decay station initiator will be able to measure ground state to ground state beta-decay transition strengths by measuring beta-delayed neutrons and gamma rays via the MTAS or SuN total absorption spectrometers 
%\cite{ongDecay61VIts2020}. 
\cite{ong2020}. 
LEBIT Penning trap mass measurements and time-of-flight mass measurements can extend knowledge of the mass surface significantly, including the critical $^{40}$Mg region. New machine learning techniques, for example using Bayesian mass model averaging, are being developed to extend the impact of new measurements to short distance extrapolations 
%\cite{neufcourtNeutronDripLine2019, 
\cite{neufcourt2019neutron, 
saitoUncertaintyQuantificationMass2023a}. 

Charge exchange measurements on neutron rich nuclei in the EC/$\beta+$ direction provide unique experimental information on the Gamow-Teller strength in
this direction \cite{Langanke2021}. Experiments can be carried out on selected nuclei that can be used to evaluate theoretical models that are be used for the entire ranges of nuclei required for astrophysical modeling. 
Collaboration with theory is needed to help choose the optimal targets.
The (d,$^2$He) reaction in inverse kinematics with the AT-TPC active target time projection chamber has recently been developed as a tool \cite{PhysRevLett.130.232301}.

%\cite{giraudGamowTellerStrengthsUnstable2023}. 
%______________%

%______________%
\section{Fundamental symmetries\label{sec:FundamentalSymmetries}}

Despite its great success, the Standard Model (SM) of particle physics falls short to explain many important observed phenomena in cosmology, such as dark energy, dark matter, the baryon asymmetry of the universe and the neutrino mass, which calls for the search of physics beyond the Standard Model (BSM).
The test of various fundamental symmetries of the SM, such as C, P, T-symmetry, lepton and baryon number conservation, existence of Lorentz-invariant currents beyond the $V-A$ structure and the flavor-universality in weak interactions, constitutes a major component in the search of BSM physics at the ``precision frontier'', with the hope to unveil small differences between experimental results and SM predictions. Famous examples of such experiments include neutrinoless double beta decay ($0\nu\beta\beta$), searches for permanent electric dipole moments (EDMs), parity-violating electron scattering (PVES), muon $g-2$, and precision beta decays. Their importance is well articulated, for example, in the ``Fundamental Symmetries, Neutrons, and Neutrinos'' (FSNN) whitepaper for the 2023 Nuclear Science Advisory Committee (NSAC) long range plan~\cite{acharya2023fundamental}.

Many such experiments are carried out in hadron/nuclear systems. Apart from requirements of experimental precision, a proper understanding of the relevant hadronic or nuclear physics is necessary to (1) disentangle the SM background from possible BSM signals, and (2) properly translate the experimental bounds to constraints on BSM parameters. This is extremely challenging due to the large QCD uncertainties. Fortunately,  the rapid development of lattice QCD, effective field theory and nuclear ab-initio methods in recent years have made first-principles calculations with fully-controlled theory uncertainties possible. At the same time, a data-driven analysis based on dispersion relations allows us to relate the required hadron/nuclear theory inputs to experimental observables. In this section we discuss examples of such developments in both theory and experiment that lead to new breakthroughs at the precision frontier and open new research opportunities relevant to FRIB.

\subsection{Nuclear beta decays and CKM unitarity\label{sec:FS_CKM}}

\begin{figure}
    \centering
    \includegraphics[width=0.6\columnwidth]{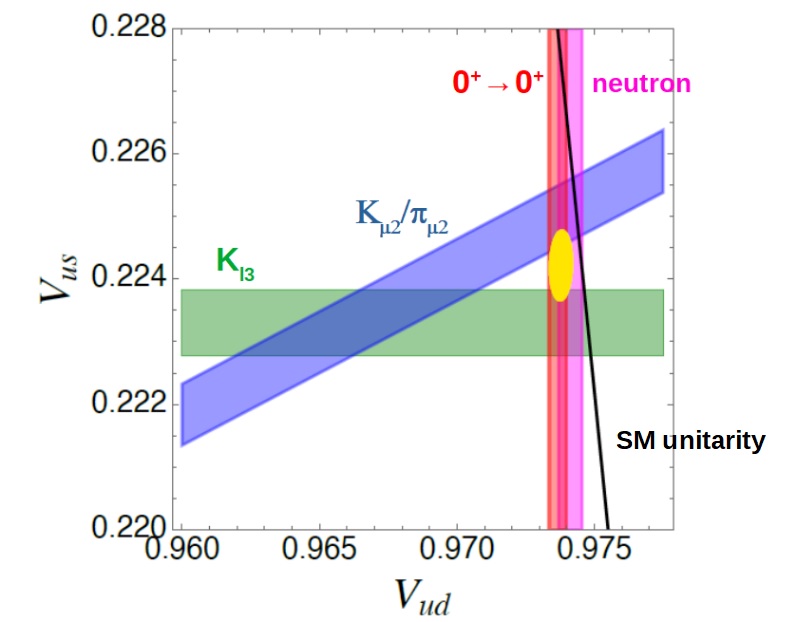}
    \caption{The CKM matrix elements $V_{ud}$ and $V_{us}$ measured from different charged weak decay processes: Superallowed nuclear beta decays ($0^+\rightarrow 0^+$), neutron beta decay, semiletonic kaon decay ($K_{\ell 3}$) and leptonic kaon/pion decays ($K_{\mu 2}/\pi_{\mu 2}$). Figure reproduced from Ref.\cite{cirigliano2023scrutinizing} with permission.}
    \label{fig:VudVus}
\end{figure}

Low-energy processes mediated by the charged-current weak interaction provide promising ways to test the SM and search for new physics.  
Within the SM, the strength of the charged current reactions is determined by elements of the Cabibbo-Kobayashi-Maskawa (CKM) matrix~\cite{cabibbo1963unitary,kobayashi1973cp}. Decays of pions, neutrons, nuclei, kaons, taus, and hyperons allow one to determine two CKM matrix elements, namely, $V_{ud}$ and $V_{us}$, respectively. The SM predicts that these parameters respect the so-called unitarity relation, $|V_{ud}|^2+|V_{us}|^2+|V_{ub}|^2=1$, where $V_{ub}$ is so small that it can be neglected at the current level of precision.
In recent years, there has been a resurgence of interest in the precise determination of these matrix elements due to the emergence of a number of intriguing tensions at the 3$\sigma$ level. Mutually-inconsistent results are observed in the determination of $V_{ud}$ from neutron and nuclear beta decays, and $V_{us}$ from leptonic and semileptonic kaon decays, and a global fitting also suggests a $\sim 3\sigma$ deficit in the unitarity relation (see Fig.\ref{fig:VudVus}) known as the ``Cabibbo-angle anomaly'', which provides a tantalizing hint for the existence of BSM physics (see, e.g., \cite{cirigliano2023scrutinizing} and references therein).

For a precise determination of $V_{ud}$ useful to test the unitarity relation, a 0.01\% precision is needed for both experiment and theory for nuclear beta decays. We will discuss how FRIB can play an important role, to provide both the direct experimental inputs (such as the decay half-lives), and the indirect ones necessary to narrow down the SM theory errors.

\subsubsection{Tree-level nuclear structure effects and relations to nuclear charge radii}

Superallowed $0^+\rightarrow 0^+$ beta decays of $T=1$ nuclei provide currently the best determination of $V_{ud}$ because it is a pure Fermi transition of which theory uncertainties are under better control. The direct experimental measurables are the so-called $ft$-values, where $t$ is the partial half-life and $f$ is the statistical rate function that depends on $Q_\text{EC}$, the mass difference between the initial and final atomic nuclei. 

An important tree-level nuclear matrix element in the superallowed nuclear beta decay is the weak transition form factor. At zero momentum transfer, it gives the well-known Fermi matrix element $M_F$; at finite momentum transfer it probes the actual distribution of the active nucleons eligible for charged weak transitions, which we denote as $\rho_\text{cw}$. Both quantities are crucial in the precise determination of the beta decay rate, but both are plagued with nuclear theory uncertainties:
\begin{itemize}
    \item For superallowed beta decays, we have $M_F\rightarrow M_F^0=\sqrt{2}$ in the isospin limit; but isospin-symmetry-breaking (ISB) interactions, predominantly the Coulomb repulsion between protons, introduces a correction $M_F^2=(M_F^{0})^2(1-\delta_\text{C})$. In the past 6 decades~\cite{macdonald1958coulomb} the quantity $\delta_\text{C}$ was computed in various nuclear models, but the results show no sign of convergence (see, e.g. \cite{Towner2010} and references therein). 
    \item The charged weak distribution $\rho_\text{cw}$ enters the statistical rate function $f$ which, in many existing reviews, was assumed to be very precisely determined. However, the current understanding of this distribution is based on traditional shell model calculations~\cite{hardy2005superallowed}, of which theory uncertainties are not properly quantified. 
\end{itemize}
Recent studies show that these quantities can be obtained, in a largely model-independent way, from experimental measurements of nuclear charge distributions which can be performed at, say, the BECOLA facility at FRIB. First, for superallowed beta decays of $T=1$ nuclei, isospin symmetry relates $\rho_\text{cw}$ to the charge distributions of the nuclei within the same isotriplet~\cite{seng2023model}; the latter can be inferred from the nuclear charge radii. At the same time, the isospin-mixing effects that generate $\delta_\text{C}$ are the same as those inducing ISB corrections to nuclear charge radii, so by precise measurements of the latter one can also probe the relevant nuclear matrix elements responsible for $\delta_\text{C}$~\cite{seng2023electroweak,seng2024}. Examples of isotopes whose radii are relevant inputs and accessible at FRIB are $^{14}$O, $^{26}$Si, $^{42}$Ti and $^{50}$Fe. 

\subsubsection{\label{sec:FS_Radiative}Radiative corrections and ab-initio calculations}

%CYS: Nuclear beta decays as a probe of new physics; precise determination of Vud and exotic couplings; nuclear structure effects in beta decay (decay form factor, radiative corrections, ISB corrections); Role of nuclear charge radii 

\begin{figure}
    \centering
    \includegraphics[width=0.6\columnwidth]{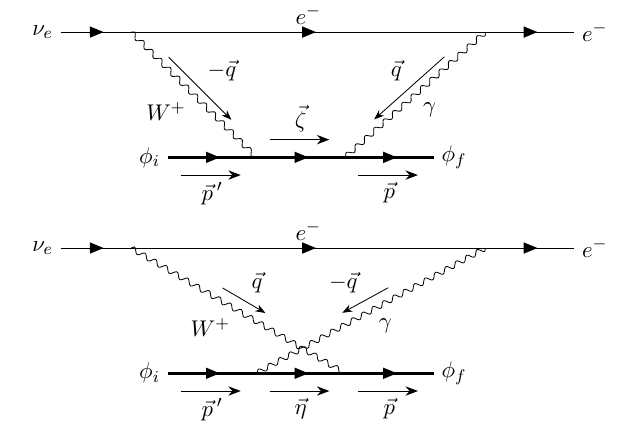}
    \caption{Diagrammatic description of the $\gamma W$-box, with top and bottom corresponding to time-reversed processes.}
    \label{fig:gW}
 \end{figure}

%Petr Navratil: NCSM calculations of the nuclear structure corrections needed to extract the Vud from Fermi transitions. NCSM calculations of the recoil and shape corrections to the electron spectrum in GT and unique first-forbidden transitions. Calculations of parity violating moments in light nuclei.  
To extract $V_{ud}$ from the measured Fermi superallowed nuclear beta decay $ft$ values, one needs to apply two nuclear structure dependent corrections. The first is the ISB correction $\delta_\text{C}$ discussed in the previous subsection, and the second is the $\delta_\text{NS}$ correction that corresponds to the modification of the single-nucleon axial $\gamma W$-box diagram (see Fig.~\ref{fig:gW}) due to nuclear structure effects. These corrections are known with a rather large uncertainty especially the latter one. In fact, the current uncertainty in $V_{ud}$ extracted from Fermi transitions is dominated by theory.

With advances in {\it ab-initio} nuclear theory, it now becomes feasible to calculate these corrections from first principles and with a quantifiable theoretical error. For the $p$- and light $sd$-shell nuclei, the quasi-exact no-core shell model (NCSM)~\cite{Barrett2013} and its extension the NCSM with continuum (NCSMC)~\cite{baroni13_753,Baroni2013,
%Navratil2016} 
navr2016} 
methods are applicable. Chiral two- and three-nucleon interactions serve as the input for these calculations.

The NCSM was applied to calculate the $\delta_\text{C}$ correction for the $^{10}$C$\rightarrow ^{10}$B Fermi transition in the past~\cite{PhysRevC.66.024314}. It was found that this correction is very slowly converging with the size of the NCSM  harmonic-oscillator (HO) basis due to the sensitivity of the correction to nuclear radii and to the admixture of the 1-particle-1-hole states in the $0^+$ ground- and the isospin analog final states by the Coulomb interaction. These issues should be resolved within the NCSMC that describes both bound and unbound states in atomic nuclei in a unified way and provides proper wave function tails for the bound states.  
%This approach describes the nuclear system using a basis expansion with two key components: one describing all nucleons close together, forming the composite nucleus, e.g., $^{10}$C ($^{10}$B), and a second one describing separated sub-clusters. The former part utilizes a square-integrable NCSM HO basis expansion treating all 10 nucleons on the same footing. The latter part factorizes the wave function into products of, e.g., $^9$B${+}p$ ($^9$Be${+}p$ and $^9$B${+}n$) components and their relative motion with proper scattering boundary conditions. The sub-clusters ($^9$B, $^9$Be) are also described within the NCSM. 
By applying the recently developed formalism for calculating beta decay transitions within the NCSMC~\cite{atkinson:2022}, the $\delta_\text{C}$ correction for the $^{10}$C$\rightarrow ^{10}$B Fermi transition are in progress.

The formalism for the calculation of the nuclear structure correction $\delta_\text{NS}$  has been developed in Ref.~\cite{PhysRevC.107.035503}. On the nuclear many-body side, one needs to evaluate matrix elements of operators consisting of a sequence of an energy-dependent and momentum-transfer-dependent electroweak current, the many-body nuclear propagator (Green's function), and an electroweak current again (see the diagrams in Fig.~\ref{fig:gW}) between the initial and final states, e.g., the $0^+$ ground state of $^{10}$C and the final $J^P(T)=0^+ (1)$ excited state of $^{10}$B. The computation of the many-body Green's function poses a challenge as it in principle involves the summation of an infinite number of intermediate eigenstates. However, it turns out that the calculation of the Green's function can be facilitated by the application of the continued fraction Lanczos algorithm~\cite{Haydock_1974,Marchisio2003} that has been implemented and applied within the NCSM formalism before, e.g., for calculations of anapole and electric dipole moments~\cite{Hao20, PhysRevC.104.025502}. Ongoing $\delta_\text{NS}$  calculations suggest that the convergence with the NCSM basis size is satisfactory, unlike in the $\delta_\text{C}$ case, and there is no need to apply the much more involved NCSMC approach. Still, the $\delta_\text{NS}$ calculations are quite challenging due to the momentum and energy transfer dependence of the electroweak operators, and integrations involving poles and residua. The first complete {\it ab-initio} calculation~\cite{gennari2024textit} agreed well with phenomenological calculations used so far in the $V_{ud}$ evaluations~\cite{hardy2020superallowed}, but with a significantly-reduced theory uncertainty which indicates the power of the new methodology.

While we presently focus on calculations for the $^{10}$C$\rightarrow ^{10}$B Fermi decay, the NCSM and NCSMC formalism would be immediately applicable to the $^{14}$O$\rightarrow ^{14}$N decay, and with some development to the $^{18}$Ne$\rightarrow ^{18}$F and  $^{22}$Mg$\rightarrow ^{22}$Na decays as well. Beyond contributing to  $V_{ud}$ extraction, studying the $^{10}$C$\rightarrow ^{10}$B itself is already quite interesting as it provides the most stringent bounds on scalar currents, and any deviation from the SM expectation would be strongly indicative of new physics.

With the {\it ab-initio} results for the nuclear structure corrections within reach, we call for new measurements of the $^{10}$C$\rightarrow ^{10}$B Fermi decay to reduce the experimental uncertainties in particular of the beta decay branching ratio.

%G.H. Sargsyan
\subsection{Tensor and scalar currents in electroweak interaction}

The “vector-minus-axial vector” ($V-A$) structure of the charged-current electroweak interaction in SM has been established thanks to detailed measurements of angular correlations in nuclear beta decay. In addition to the tests of CKM unitarity, beta decay efforts are at the front line in searches for evidence of additional Lorentz-invariant scalar (S) and tensor (T) interactions that arise in SM extensions. In the recent years, experimental developments coupled with state-of-the-art nuclear-theory calculations have resulted in a new generation of beta decay studies that continue to achieve unprecedented precision (see Ref. \cite{acharya2023fundamental} for a more comprehensive review).  
In particular, atom-traps and ion-traps have been used to accumulate and suspend samples of beta-decaying isotopes in vacuum, allowing for the measurements of the low-energy nuclear recoils, from which the neutrino momentum can be deduced. Experiments have reached increasingly precise results for the beta-neutrino angular correlations in $^{8}$Li~\cite{sternberg2015limit,burkey2022improved} and $^{8}$B~\cite{gallant2023angular, longfellow2023determination,Longfellow2024PRL}. Atom traps have been used to determine this correlation in $^{6}$He~\cite{muller2022beta} and to polarize $^{37}$K atoms to measure the beta asymmetry~\cite{fenker2016precision,fenker2018precision}. These experiments have achieved very high precision ($\sim 0.3\%$), placing stringent limits on the possible existence of tensor interactions and right-handed currents. There are well-defined paths to further improve this precision. In particular, with the help of \emph{ab-initio} calculations of beta decay observables, several independent studies have provided precision input for the interpretation of beta decays in $^{6}$He~\cite{King2023, Glick-Magid2022} and $^{8}$Li
%~\cite{Sargsyan2022}, 
~\cite{sarg2022}, 
which greatly improves the previous state-of-the-art results.

%%% SRS -- I can add some text about beta decay of Sn100, though this has already been approved by the PAC.

\subsection{Two-body axial vector currents}
Unlike the vector current, the weak axial current is not conserved in strong interactions, leading to non-negligible two-body currents which play a significant role in the quenching of GT decay rates~\cite{brow1985p347,mart1996,gysb2019}.
An appealing feature of chiral effective field theory is that it provides a framework for systematically constructing electroweak currents which are consistent with the nuclear force.

While this is a SM process, it also factors into searches for new physics.
For example, it has been an open question for decades whether the GT component of the $0\nu\beta\beta$ transition operator should use a quenched axial coupling constant $g_A$~\cite{Engel2017}.
The same axial current enters into the calculation of structure factors for direct detection of dark matter under the weakly interacting massive particle (WIMP) hypothesis~\cite{Engel1992,Menendez2012}, and the $\gamma W$-box correction for superallowed Fermi decays~\cite{Towner2002} (see section~\ref{sec:FS_Radiative}).

It is therefore of great interest to validate chiral EFT's systematic approach to the weak axial current.
Arguably, the cleanest test of the approach is not the few-nucleon system, but $^{100}$Sn.
This is because the presence of many particles enhances the effect of two-body currents to $\sim 30\%$ compared with the $\sim 2\%$ effect seen in the decay of $^{3}$H~\cite{Gazit2009,Gazit2019}.
In addition, the doubly-magic nature of $^{100}$Sn leads to the transition being dominated by a single configuration, so that the many-body calculation does not need to handle fine cancellations among equal-sized terms. 

\begin{figure}
    \centering
    \includegraphics[width=0.8\textwidth]{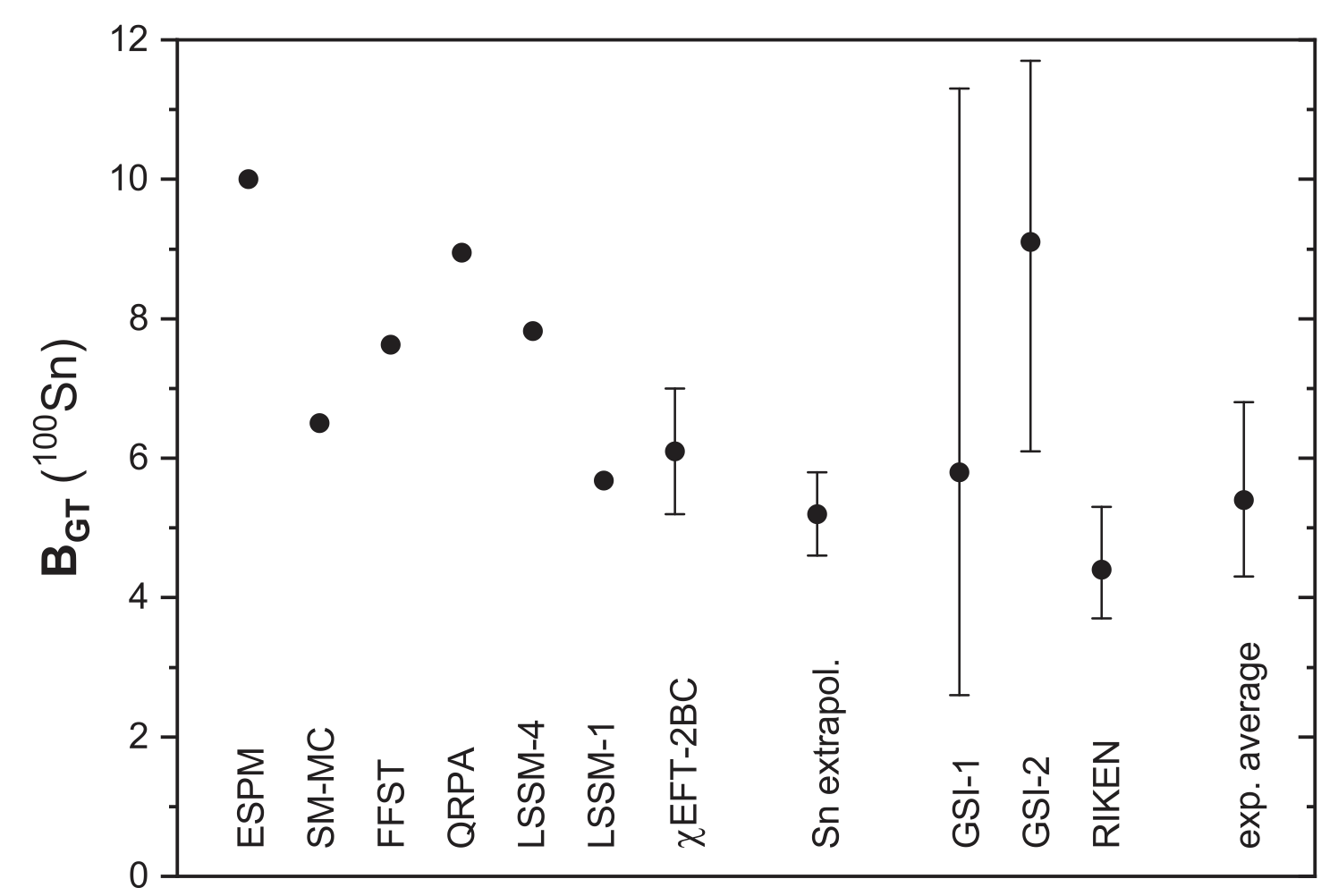}
    \caption{Reduced GT transition probability $B(GT)$ for the decay of $^{100}$Sn. The points labeled GSI-1, GSI-2 and RIKEN are experimental, while the other points are theory. Note that the ESPM point has been multiplied by a quenching factor $0.75^2$. Taken from Ref.~\cite{lubo2019}.}
    \label{fig:Sn100BGT}
\end{figure}

Currently, a significant limiting factor in utilizing the decay of $^{100}$Sn is the large experimental uncertainty, as illustrated in Fig.~\ref{fig:Sn100BGT}, from Ref.~\cite{lubo2019}.
Another measurement with similar precision to the RIKEN result would help clarify the situation.

%\paragraph{$CP$ violation}

\subsection{Heavy element isotopes for tests of $P$ and $T$ violation}

One of the major outstanding puzzles in physics is the predominance of matter over antimatter in the universe.
Explaining this asymmetry requires, among other things, violation of symmetry under $CP$, equivalent to $T$ violation.
While the Standard Model does contain sources of $CP$ violation, these are insufficient to explain the observed baryon density, indicating the need for new physics beyond the Standard Model.
Many new physics scenarios would manifest as a permanent electric dipole moment (EDM) of the electron, nucleons, or nuclei~\cite{Engel2013}, which can be searched for experimentally with very high precision.
An especially promising system for searching for EDMs is in nuclei with static octupole deformation, which leads to parity doubling and consequentially a dramatically enhanced EDM.
Recently it has been proposed that further significant enhancement in sensitivity could be obtained with radioactive molecules~\cite{arro2023}. Consequently, the light actinides and especially Ra and Rn have attracted a lot of interest in recent years because their odd-$A$ isotopes could be octupole deformed~\cite{Aue96a, Spe97a, Fla08a, Bis16a, Chu19a, Sin19a, Fla20a, Gar20a,Udrescu2024}.

A recent collaboration between FRIB, Harvard, and MIT has been formed to develop an experimental apparatus for searches for CP violation with radium-containing molecules, such as RaF and RaOH. Precision measurements of these systems are expected to provide insights into the $P$, $T$-violating nuclear Schiff moment of \(^{225}\mathrm{Ra}\). Nuclear theory will play an essential role in connecting Schiff moment measurements with the underlying sources of CP violation.

Recent Coulomb-excitation (CoulEx) experiments established that $^{222-226}$Ra are very likely statically octupole deformed in their ground states, while the even-even Rn isotopes are not\,\cite{Gaf13a, But19a, But20a}. Direct measurements of the relevant experimental observables in the odd-$A$ isotopes which would be used in EDM searches are still missing, leading to large theoretical uncertainties for the expected enhancement~\cite{Dob18a}.
These properties can be measured with laser spectroscopy on ions and molecules, with an early focus on thorium isotopes.
FRIB has developed $^{232}$Th$^+$, $^{232}$ThO$^+$ and $^{232}$ThF$^+$ beams from the BMIS, which will be used for preparatory studies for experiments on short-lived Th isotopes.

In addition to ground-state properties, sub-barrier Coulomb excitation with secondary beams of these odd-$A$ isotopes will become possible at FRIB. Especially the expected beam rate for $^{229}$Pa at 20-kW power \cite{frib-rates} would make such an experiment possible once the secondary beam is developed and GRETA \cite{greta, greta_fdr} is available at ReA. $^{229}$Pa could possibly have the closest-lying parity doublet \cite{Ahm82a}, though the experimental situation is still unclear~\cite{Ahm15a}. Anticipated early-FRIB beam rates for $^{225}$Ra are not sufficient to perform detailed $\gamma$-ray spectroscopy experiments after Coulomb excitation for which typically $10^5$ particles per second (pps) are needed.
CoulEx experiments could also test the predicted enhancement of the $B(E3;3^-_1 \rightarrow 0^+_1)$ strength in U and Th isotopes with mass $A \approx 224-228$ \cite{But16a, Rob12a}, which might prove even more octupole collective than $^{222-226}$Ra. Once the secondary beams are developed, these experiments would already be possible at 20-kW power. Either GRETA \cite{greta_fdr} with a Silicon detector array or CHICO2 \cite{Wu16a} for particle detection,  or the JANUS setup \cite{Lun18a} could be used at ReA.

The results from these experiments together with nuclear models
for pear-like nuclei will be essential for the extraction of new physics.
Understanding the nuclear structure of these elements is also important to benchmark theories to extend studies to even heavier elements \cite{dull2015, naza2018}.

Molecules can serve as highly sensitive probes for measuring hadronic P-violation, offering enhancements of over eleven orders of magnitude compared to atomic experiments \cite{Alt18}. Recent developments in the field are expected to be implemented at FRIB in the coming decade \cite{Kar24}. Advances in nuclear theory will be crucial for linking future precision studies of electroweak nuclear properties, such as anapole moments \cite{Hao20}, with tests of the Standard Model (SM) and constraints on new physics.

%______________%

%______________%
% \section{Emulators, Uncertainty Quantification, and Experimental Design}

\section{Experimental Design and Uncertainty Quantification\label{sec:ExpDesign}}
% * The manuscript is divided up by section, with each section placed in a folder. These folders contain a tex file and any other relevant files (e.g. figures).
% * When adding a block of text, please add a comment at the beginning of the block with your name so that others know who to contact for questions/discussions.
% * All references will be added to a single bibtex file “references.bib”.
%\label{sec:ExpDesign}

\begin{figure*}
    \centering{\includegraphics[width=1\textwidth]{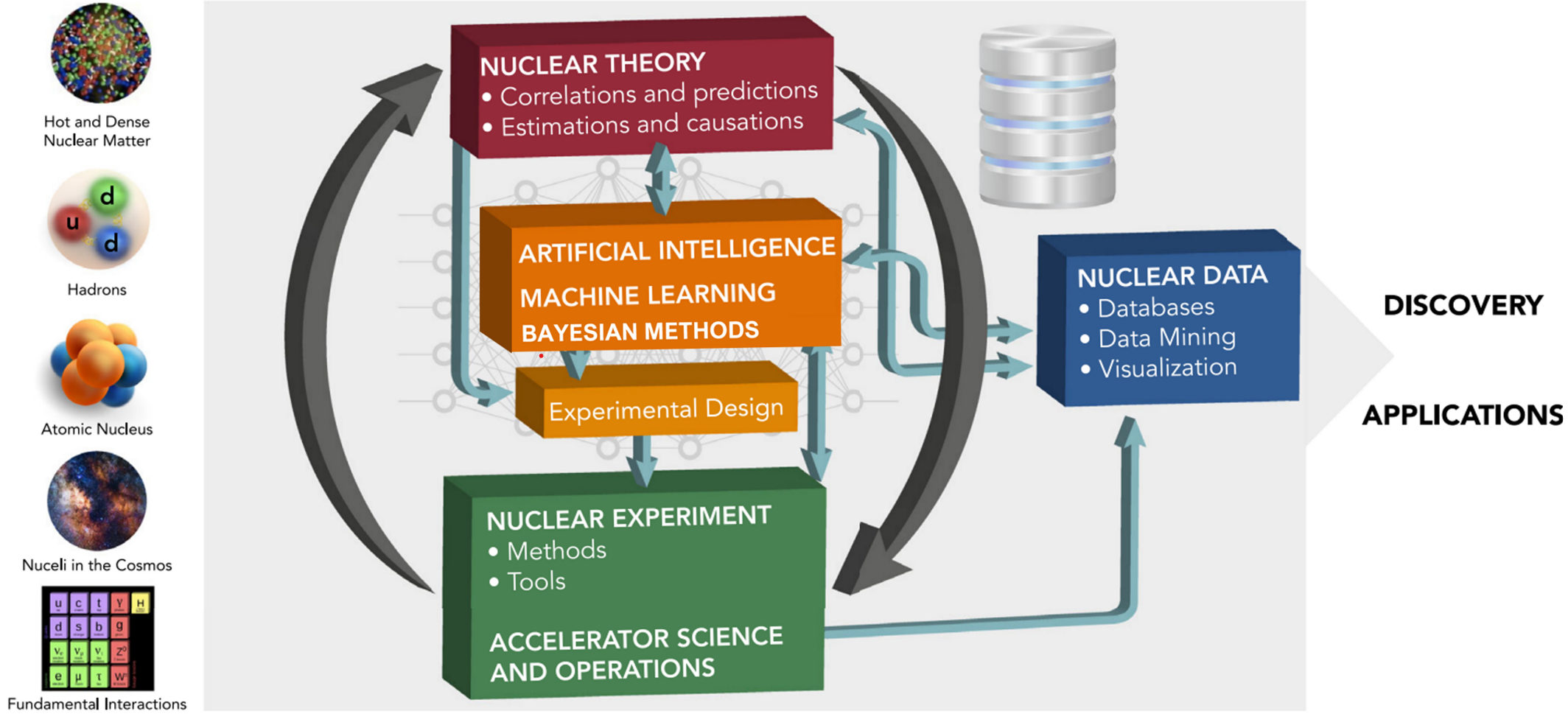}}
    \caption{Schematic representation of the theory-experiment cycle in nuclear physics, enabled by  Bayesian statistics and advanced scientific computing methods (Adapted from \emph{``Colloquium: Machine learning in nuclear physics (2022)"} \cite{Boehnlein2022}).}
      
    \label{fig:Exp_design}
\end{figure*}

All the topics we have explored cover immense ground in nuclear physics and closely related areas. We now close with a discussion on the role that scientific computing and computational statistics will play in tying everything together to foster the experiment-theory cycle for FRIB science. This cycle is represented schematically in Fig.~\ref{fig:Exp_design}.

Previous work on experimental design within nuclear physics has usually focused on developing a cost function to be optimized with respect to experimental choices, including the target selection. These approaches have tackled, for example, the selection of momentum transfer measurements for parity violating electron scattering experiments~\cite{piekarewicz2016power,giuliani2021noise}, the optimization of the angle and energy settings in proton Compton scattering~\cite{melendez2021designing}, and the selection isotopes candidates for mass measurements at FRIB \cite{farr2021decision}. The specific nuclear physics domain, as well as the details of the different experimental setup vary in a wide spectrum across these and other similar efforts, but in all of them a constant unifying theme is the focus on uncertainties, both from experimental and theoretical origin.
Bayesian statistics provides an ideal probabilistic framework to draw conclusions from data in the context of these uncertainties.
One main pillar of this framework lies in the explicit and clear statements of all assumptions about the various sources of uncertainties and their interactions. By making these assumptions transparent, the framework naturally lends itself to constant evaluation and refinement of the constructed statistical model. A second, equally important, pillar of Bayesian statistics resides in the philosophy of iteratively updating knowledge using new data. This iterative approach integrates expert knowledge and previous relevant data with new observations to refine our understanding as new information becomes available. This process is defined through Bayes's theorem 
\cite{gelman1995bayesian,BANDmanifesto}:
\begin{equation}\label{eq:post}
     P(\params|\obs) = \frac{P(\obs|\params) P (\params)}{P(\obs)}.
\end{equation}
The \emph{posterior} distribution $P(\params|\obs)$ contains the new information on the model parameters $\params$, or anything else we are interested in, conditional on the new data $\obs$. The model could be any of the theoretical frameworks described before, for example Density Functional Theory~\cite{mcdonnell2015uncertainty} or optical potentials~\cite{lovell2020quantifying}, or even data-driven approaches such as neural networks \cite{utama2016nuclear}, Gaussian processes  \cite{neufcourt2018bayesian}, or dynamic mode decomposition~\cite{brunton2022data}. The model parameters $\params$ could be those describing interaction strengths or nuclear potential sizes, for example, or hyperparameters controlling the data-driven model. The posterior distribution is built by the combination of the \emph{likelihood} $P(\obs|\params)$, which characterizes the process in which the observed data $\obs$ could correspond to a given parameter configuration $\params$, together with the prior $P (\params)$, which contains all the previous information, as well as expert knowledge, on the model parameters $\params$. The \emph{evidence} $P(\obs)$ serves to quantify how well the current overall framework, combining both the theoretical physical model and the statistical description of the uncertainties, accounts for the data $\obs$. See Figure~\ref{fig:MCMC} below as an example of a 2-D representation of a Bayesian posterior distribution for the parameters of a relativistic mean field model~\cite{giuliani2023bayes}.

\begin{figure*}
    \centering{\includegraphics[width=0.9\textwidth]{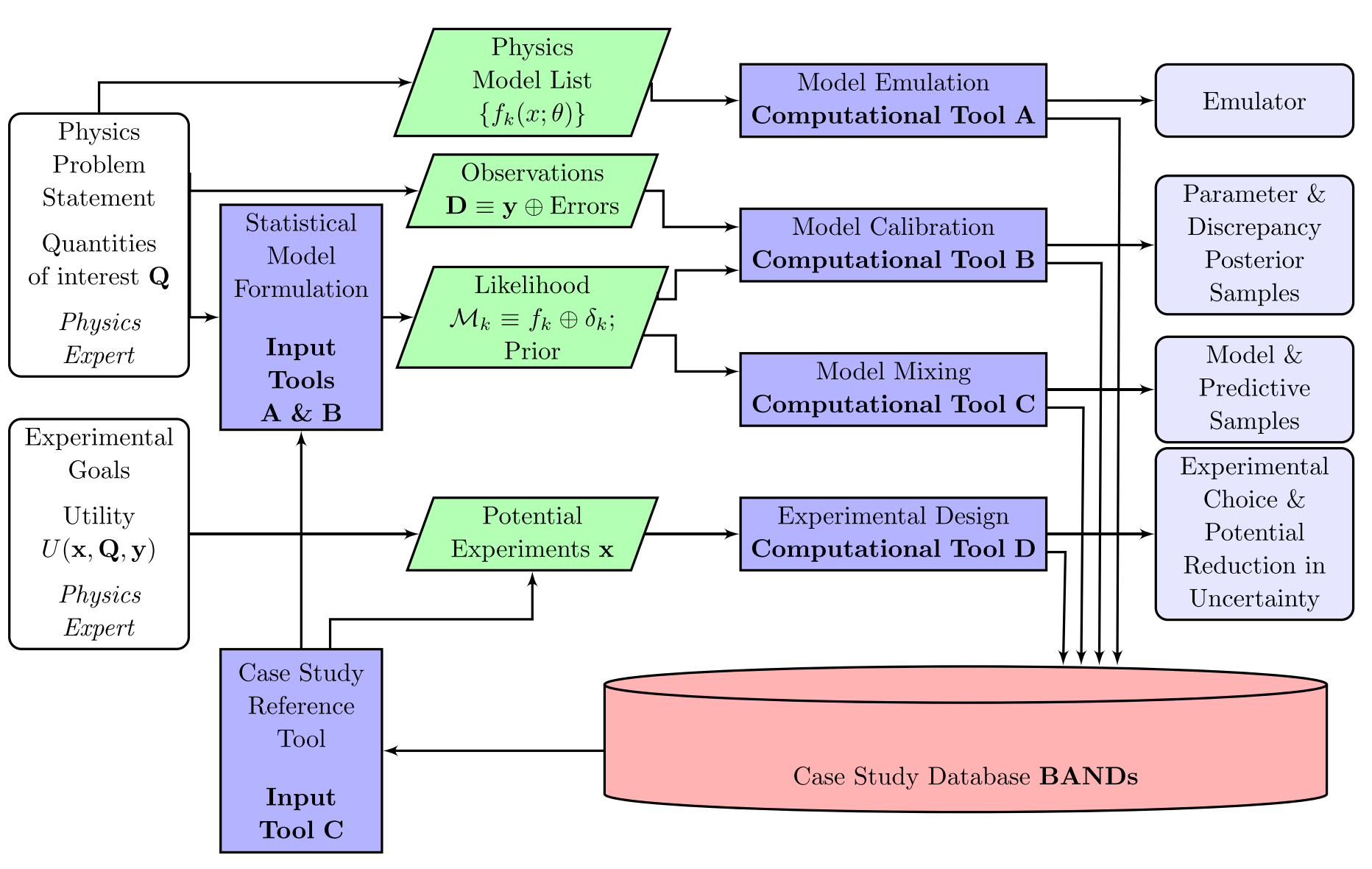}}
    \caption{Flowchart of the various tools that are part of the BAND framework \cite{bandframework} (Reproduced from \emph{``Get on the BAND Wagon: a Bayesian framework for quantifying model uncertainties in nuclear dynamics"} \cite{BANDmanifesto})}
      
    \label{fig:BAND_flowchart}
\end{figure*}

\begin{figure*}
    \centering{\includegraphics[width=0.95\textwidth]{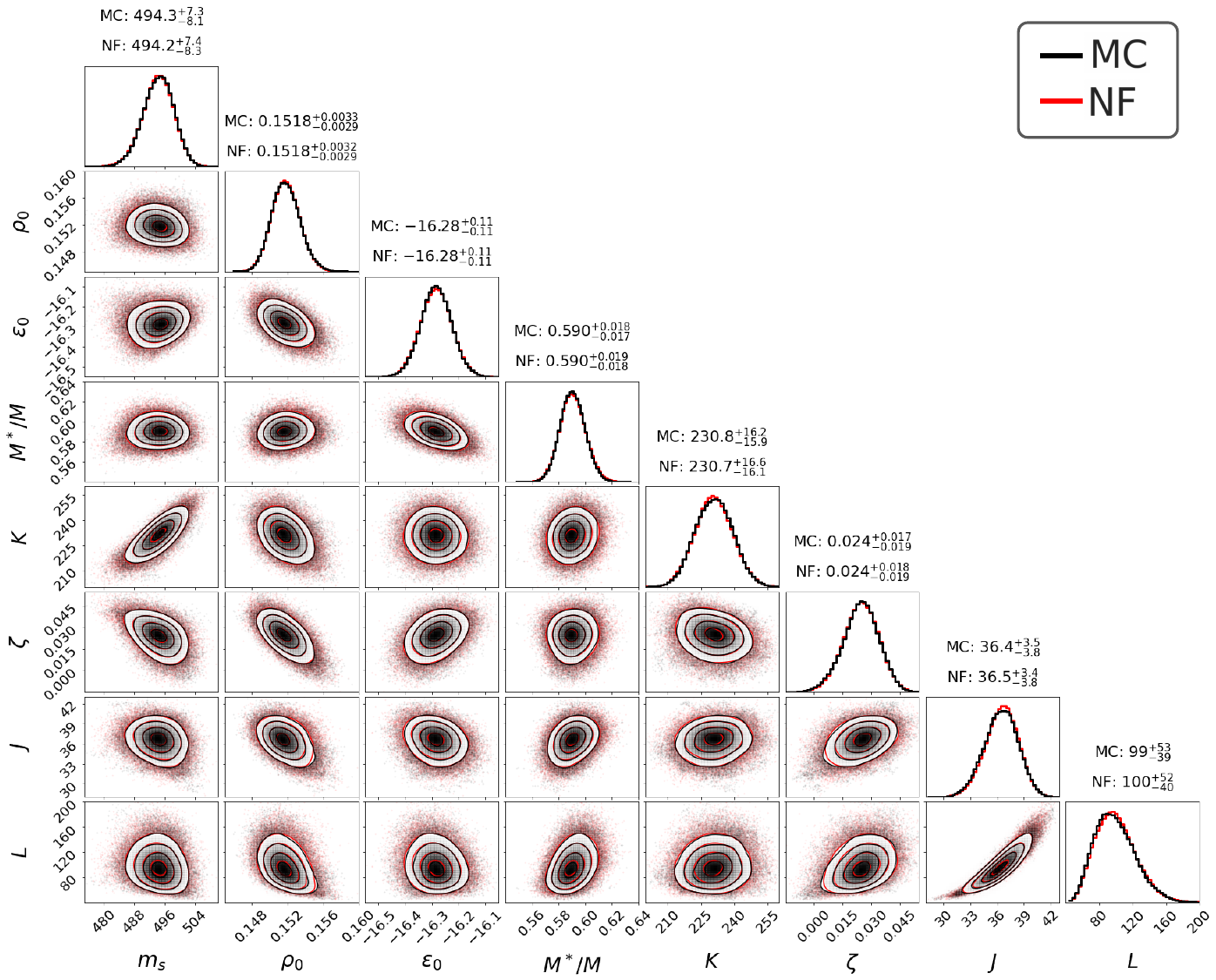}}
    \caption{Comparison between the samples obtained by Markov Chain Monte Carlo (black) and Normalizing Flows (red) for the relativistic mean field Bayesian calibration obtained in \cite{giuliani2023bayes}. The theoretical model has a total of 8 parameters that were calibrated to the masses and charge radii of 10 nuclei. The normalizing flow samples reproduce all the features of the corner plot of the original data, overlapping well enough to make it hard to distinguish between them. (Reproduced from \cite{yamauchi2023normalizing})}
      
    \label{fig:MCMC}
\end{figure*}

% In order to develop and asses the impact of experimental campaigns, especially within the context of our current theoretical understanding, the involved models need to provide quantified uncertainties with their respective predictions, a usually computationally challenging task. 
The process of constructing an adequate prior and likelihood form for a given situation can be a non-trivial statistical task, while the actual exploration and evaluation of the posterior distribution can be computationally challenging. %quickly become a serious computational challenge. 
The theory community has recognized \cite{join2011uncertainty,dobaczewski2014error} and supported a wide spectrum of efforts in these directions, including recurring meetings and workshops such as the Information and Statistics in Nuclear Experiment and Theory series \cite{ISNET2023}, and the FRIB-TA Summer School 2023 on Uncertainty Quantification and Emulator Development \cite{FRIBTASS2023,FRIBTASSVideos2023}, the creation of collaborations focused on Bayesian methods such as BAND~\cite{BAND} and BUQEYE~\cite{BUQEYE} and computational advancements such as NUCLEI~\cite{NUCLEI} , as well as explicit descriptions on the importance of these methods in the 2023 Long Range Plan for Nuclear Science~\cite{LRP2023}. Figure~\ref{fig:BAND_flowchart} illustrates the efforts pursued by the BAND collaboration to create useful software for the community in various aspects directly related to this section's discussion.  

The calibration and quantification of uncertainties of theoretical models through Eq.~\eqref{eq:post} is often done by Markov Chain Monte Carlo methods~\cite{gelman1995bayesian}, requiring hundreds, thousands, or even millions of evaluations of the model itself. To mitigate the computational burden that such task entails, a great effort has been done in recent years to develop and implement emulators~\cite{Duguet:2023wuh,frame2018eigenvector,furnstahl2020efficient,melendez2021fast,drischler2021toward,konig2020eigenvector,zhang2022fast,melendez2022model,drischler2023buqeye,garcia2023wave,sarkar2021convergence,sarkar2022self,demol2020improved,djarv2022bayesian,
%yapa2022volume,
Yapa:2022nnv,
franzke2022excited,bonilla2022training,anderson2022applications,giuliani2023bayes,bai2021generalizing,ekstrom2019global,surer2023sequential,liyanage2023bayesian,surer2022uncertainty,higdon2015bayesian,mcdonnell2015uncertainty,surmise2023,odell2024rose,drnuclear,Drischler:2022ipa,bakurov2024discovering,cook2024parametric,somasundaram2024emulators,reed2024towards,anderson2024universal}, which are algorithms or models that can mimic full theory calculations at a much faster pace with a negligible loss in accuracy. This is usually achieved by either finding and exploiting low-dimensional latent structures to build a reduced order model \cite{giuliani2023bayes}, or by finding alternative maps between inputs and outputs through universal approximants such as Gaussian processes~\cite{mcdonnell2015uncertainty} or neural networks~\cite{lay2023neural}. The construction of these emulators requires an appreciable computational cost up-front, but once deployed they allow nuclear theory to reach new calculational frontiers, with the inclusion of uncertainties. 

% Xilin and Pablo  
It is worth noting that these ideas and techniques are not unique to nuclear physics, and recent decades have seen tremendous developments across many other science disciplines and engineering~\cite{lassila2014model,quarteroni2015reduced,hesthaven2016certified}, especially with the rise and rapid development of machine learning \cite{brunton2022data,Boehnlein2022}. 
Even outside of the context of model calibration, surrogate models or emulators provide a more efficient way to perform large scale systematic studies or for exploring previously calibrated models for more expensive computational workloads~\cite{godbey2022}. They can also be used to access, via extrapolations, new computations that are otherwise infeasible.  Some examples include computing the properties of quantum systems with Monte Carlo sign problems using parameter extrapolation~\cite{frame2018eigenvector} and the properties of quantum states for different finite spatial volumes~\cite{Yapa:2022nnv} and as functions of energy in the complex plane~\cite{XilinZhangTRIUMF2023} (see  Sec.~\ref{sec:abinitio_reactions}). Alternatively, some of these methods can be used to directly create data-driven physical models \cite{brunton2022data,pan2022neural,brunton2016discovering,scheinker2021adaptive}, giving access to new paths for capitalizing on the available data through the power of machine learning. Further investments in developing and adapting these technologies for nuclear physics will be crucial in the immediate theory-experiment cycle, especially if they can also benefit the actual experimental data acquisition and control (see for example \cite{miskovich2022online} and \cite{blackbox}). Of particular importance will be the creation of user-focused open software tailored to the nuclear physics community as well as education and training activities\cite{surmise2023,odell2024rose,drnuclear,Drischler:2022ipa,FRIB2023Materials}.

Once various theoretical models have been calibrated and their uncertainties have been quantified, whenever possible, methods such as Bayesian Model Averaging~\cite{hoeting1999bayesian} and Bayesian Model Mixing~\cite{fernandez2002modelling} can help combine their predictions into a more precise and reliable overarching model within the Bayesian framework (see \cite{BANDmanifesto} for a detailed explanation of both methods within a nuclear physics context). These approaches 
%  done within the Bayesian framework, allowing 
compute the uncertainties associated with averaging or mixing models and help in identifying measured observables that are challenging to be reproduced by the collective of models (see for example \cite{hamaker2021precision}).  This helps to focus future community-wide experimental and theoretical efforts for resolving the discrepancies. In the absence of measured data to compare with, such approaches can help in identifying regions of interest for targeting future experiments where the variability of the collective wisdom of models is maximal \cite{neufcourt2019neutron}. Model extrapolations to uncharted territory is often aided by other machine learning and statistical tools \cite{neufcourt2018bayesian,lovell2022nuclear,lovell2020quantifying,utama2017refining,utama2016nuclear}. Similar to the development of emulators, Bayesian model averaging and mixing methods have received increasing attention by the nuclear physics community in recent years \cite{hamaker2021precision,semposki2022interpolating,neufcourt2019neutron,kejzlar2020statistical,
%neufcourt2020beyond,
Neufcourt2020b,
%neufcourt2020quantified,
Neufcourt2020a,
semposki2023mind,kejzlar2023black,kejzlar2023local,giuliani2024model,drischler2024bayesian}, as well as the creation of user-focused software \cite{SAMBA,Taweret}.

As new experimental data becomes available, the calibration and uncertainty quantification of models should be updated as quickly as possible. Without a swift update process, the combined theoretical knowledge will lag behind and will not be able to efficiently contribute to the theory-experiment cycle, leading to sub-optimal decisions and reducing the impact of new experimental campaigns. To address this issue, recent efforts have started for the construction of pipelines for the continuous calibration of theoretical models, exploiting modern machine learning tools for the sampling and storing of posterior distributions \cite{yamauchi2023normalizing}, as well as leveraging cloud computing infrastructures for the calculation, storage, and distribution, of data \cite{busk2024,bmex}. Figure~\ref{fig:MCMC} shows the Bayesian posterior parameter distribution from the calibration of a relativistic mean field model \cite{giuliani2023bayes} learned through the normalizing flow-neural network framework described in \cite{yamauchi2023normalizing}. Once the neural network is trained, it can be shared, deployed, and sampled from easily. Once new data becomes available, it can be used as a prior starting point for creating a new posterior, truly embracing the Bayesian philosophy of continuously updating knowledge. 

The construction of an online central resource for the storage and continuous update of these theoretical model calibrations and predictions, as well as other resources and tools such as the developed model emulators and updated experimental conditions, will play a critical role in community efforts. Facilitating interconnection through workshops~\cite{ExpTheory2023} and easy-to-use platforms \cite{ASCSN}, embracing the open source philosophy and leveraging cloud computing infrastructures, will help lower the entry barrier for researchers across a wide institutional spectrum, scientific backgrounds, and career stages.  This allows many people to add their efforts to the task of sustaining an efficient and effective theory-experiment cycle in FRIB science. These efforts will help to broaden participation in the scientific community in this new era of discovery.

% The construction of an online database for the storage and update of these theoretical model calibrations and predictions will play a critical role in the efforts to construct an efficient and effective theory-experiment cycle at FRIB. This central online platform could also host other resources and tools, such as the developed model emulators, for example. Embracing the open source philosophy and leveraging cloud computing infrastructures will help lower the entry barrier for researchers across a wide institution spectrum and career stages to add their efforts to fostering the theory-experiment cycle, further pushing for a more diverse and inclusive scientific community.

%______________%
%Pablo G: Something specific about exp design
%______________%

%\section{Other topics}
%\input{Other/other}
%%\input{Othertopcis/othertopics}

\section{Summary}

During the last decades, the nuclear theory community sketched a rough road map for a predictive theory of nuclei. Today, low-energy nuclear theory research has substantially matured due to new information on exotic nuclei from exotic beam facilities, novel theoretical methods resulting in quantified predictions, and high-performance computing speeding up the experiment-theory cycle. The important challenge for the field is to bridge different many-body approaches, describing the nucleus  at different resolution scales,  in the regions of the nuclear landscape where they overlap. 

%On the journey to a comprehensive nuclear theory, important milestones are marked by designer nuclei \cite{jone1010}  with characteristics adjusted to specific research needs. The designer nuclei are stepping stones of the FRIB research program.

A goal of research at FRIB is to understand the properties of nuclei
by identifying the relevant degrees of freedom, and by connecting these properties to the
underlying interaction between nucleons. The relevant degrees of freedom can be collective
modes like clustering, superfluidity, vibrations and deformation; they can be related to a
set of active nucleon orbitals defined by shell gaps; or they may be related to the proximity of the
particle continuum near the driplines. The main new information provided by FRIB will relate to how these features are intertwined and modified in nuclei near the fringes of the nuclear chart, and how these various degrees
of freedom interplay with the continuum. In Section \ref{sec:Structure}, we used several specific regions
of nuclei of importance for the evolution of nuclear structure to illustrate how neutron clustering may impact nuclear properties in \nuc{28}{O}, the interplay of quadrupole correlations and the neutron continuum in the region around \nuc{42}{Si}, the impact of pairing in the continuum on the neutron-rich end of the Ca isotopic chain, shell evolution leading to a 5$^{th}$ island of inversion around \nuc{78}{Ni}, and the proximity of self-conjugate, doubly-magic \nuc{100}{Sn} near the proton dripline. While these are specific examples, it is clear that the very concepts and interplays outlined there will repeat as new territory becomes available on the nuclear chart during FRIB's capabiliy ramp-up. Then we singled out
aspects of collective degrees of freedom, isospin symmetry, the spin-isospin response of nuclei, and pairing effects.

Near-threshold physics is poised to play an important role in the FRIB scientific program, first and foremost in the exploration of the drip lines, but also in understanding the structure of low-lying resonances of astrophysical interest involving, for instance, clustering. 
In addition, near-threshold physics strongly impact properties of many light nuclei, which remain the ideal testing ground for nuclear forces using quasi-exact \textit{ab initio} approaches. 

Several priorities related to near-threshold physics were identified to advance the FRIB scientific mission. 
One of the main challenges that both theorists and experimentalists will have to face in the exploration of the drip lines is the existence of interplays between standard emergent phenomena such as deformation, clustering, pairing, or collective motion, and near-threshold phenomena. 
In that regard, and considering FRIB technical capabilities, the region around the $N=20$ island of inversion near the neutron drip line, where deformation is known to emerge, presents a unique opportunity.
Specifically, the study of $p$-wave halo states in this region, in which $p$ waves are intruder states, could provide valuable insight on how weak-binding affects deformation and rotational motion. 
Similarly, neutron-rich fluorine isotopes recently appeared as an ideal place to study the effect of the intrusion of a low-lying unbound $p$ wave on nuclear structure in this region. Obtaining a detailed spectroscopy of $^{28-31}$F should be prioritized as well as it may provide important clues as to how far the drip line extends up to $N=28$. 
At $N=28$, the isotope $^{40}$Mg, which may or may not have a two-neutron halo ground state, continues to elude our current understanding of the region. 
Beta-decay studies in this region should be prioritized as well.
All these studies will provide stringent tests for nuclear models and help in understanding interplays involving near-threshold phenomena. 
At a theory level, continuum couplings need to be better accounted for both in effective and \textit{ab initio} approaches to deal with multi-nucleon resonances and clustered states. 

In a broader context, the FRIB scientific program should also seize opportunities to study exotic phenomena such as two-proton decay or near-threshold clustering, which provide unique windows on nuclear correlations and few-body dynamics in nuclei. 
Finally, while not the primary focus of FRIB, light nuclei must continue to be studied both theoretically and experimentally to improve the quality of first-principle calculations in regions of interest of FRIB and provides insight into complex near-threshold emergent phenomena. 

FRIB will bring unprecedented opportunities for exploring bulk behavior of nuclear matter, including carrying out comprehensive studies of the nuclear matter equation of state~(EOS) and its dependence on both density and isospin content. 
Experimental campaigns centered on constraining the EOS will lead to a better understanding of properties of heavy nuclei as well as elucidate the evolution and composition of astronomical objects characterized by extreme densities (such as supernovae, neutron stars, and merging binary neutron star systems) and provide robust constraints for \textit{ab initio} nuclear theory.
At and below the saturation density of symmetric nuclear matter, experimental opportunities to probe the effects of isospin on the EOS include measurements of giant monopole resonances, electric dipole polarizabilities, and scattering data for proton- and neutron-rich nuclei. 
Both well below and well above the saturation density, relativistic collisions of heavy nuclei (also referred to as central heavy-ion reactions) provide unparalleled opportunities to constrain the isospin-dependence of the EOS.
Importantly, extraction of the EOS from measurements relies on comparisons of experimental data to model predictions.
A particular challenge in constraining the isospin dependence of the EOS with experimental measurements lies in the fact that, compared to, e.g., neutron stars, the isospin excess in experimentally-available nuclei is relatively small %. 
%
%Moreover, the most promising experimental observables are often affected not only by the isospin dependence of the EOS, but also by other properties of nuclear matter, so that 
and 
the effects due to isospin are subtle.
Here, FRIB's unique capability to provide high-statistics beams of proton- and neutron-rich heavy nuclei will push the boundaries of accessible values of the isospin imbalance as well as lead to exceptional experimental precision. 
The proposed FRIB400 upgrade will further dramatically increase the yields of nuclei with large isospin asymmetry, allowing for measuring properties of extreme neutron-rich nuclei, such as neutron skins, as well as for studies of dense nuclear matter up to twice saturation density, critical for multi-messenger astrophysics.
On the theoretical side, this opportunity must be matched by development of comprehensive models characterized by high accuracy. 
Measurements constraining the EOS at and above the saturation density will lead to rigorous tests of \textit{ab initio} approaches such as chiral Effective Field Theory (see Sections~\ref{Microscopic_calculations_of_the_EOS_within_chiral_EFT} and~\ref{Isospin_dependence_of_the_EOS}). 
Observables measured in energetic heavy-ion collisions will not only enable studies of the isospin dependence of the EOS, but also of the momentum dependence of single-particle potentials, the dependence of particle cross sections on density, subthreshold particle production, and nuclear clusters and correlations; here, robust inferences are contingent on reliable inclusion of relevant physical processes in transport model simulations of heavy-ion collisions (see Section~\ref{EOS_and_in-medium_properties_of_nuclear_matter_from_heavy-ion_collisions}).
Explorations of the influence of isospin imbalance on nucleon elastic and inelastic scattering on medium-mass and heavy isotopes will lead to tight constraints on optical potentials and, more generally, nucleon self-energies, directly related to the EOS (see Section~\ref{EOS_OMP}).
With some of the relevant FRIB experimental programs underway (see Section~\ref{Upcoming_and_proposed_experiments_at_FRIB}), this is an exciting time for developing strong theoretical foundations for future interpretations of observables related to the EOS.

Nuclear astrophysics studies aim to illuminate the ultimate origin of the elements we observe in nature. From the early days of the field, studies have been centered around the intimate connection between nuclear physics properties and astrophysical observables (such as Solar and stellar element abundances) and thus have always had theory, experiment, and observation recognized to be of equal importance. Given that astrophysical environments produce exotic, unstable nuclei far from stability and beyond currently explored territory, FRIB experiments have the opportunity to shed light on key nuclear properties affecting the formation of elements. The information that FRIB can provide is of relevance to nucleosynthesis studies on both the neutron-rich and proton-rich side of stability, and would aid to understand x-ray bursts, accreting neutron stars, core-collapse supernovae, and neutron star mergers to name a few astrophysical sites. Thus opportunities for FRIB to revolutionize our understanding of element synthesis exist across the nuclear chart. Particularly exciting is the huge range of new isotopes FRIB will be able to produce of relevance for the rapid neutron capture process ($r$ process) which is believed to be responsible for making the heaviest naturally occurring nuclei. Such experimental efforts are quite timely given recent leaps forward in the observational community’s abilities in “multi-messenger science” via collecting numerous distinct observations from an astrophysical event (such as light curves, gamma rays, neutrinos, and gravitational waves). Such multi-messenger information combined with experimental constraints on nuclear properties is the holy grail of nuclear astrophysics, serving to inform models of astrophysical environments and the synthesis they are capable of. In this paper we discuss the various needs for nuclear astrophysics studies, including advancements of global models for nuclear masses, decays (e.g. beta and alpha decays), reactions (e.g. neutron and proton capture), and fission. We also highlight how probing specific regions of the nuclear chart and specific isotopes can help to address some of the most fundamental questions in the field of nuclear astrophysics. 

FRIB also offers many opportunities for precision tests of the SM predictions and probe new physics at low energies; the latter, among other things, is needed to address the various unexplained observations in cosmology. As an example, precise measurements of nuclear beta decay lifetimes and correlations (such as $\vec{p}_e\cdot\vec{p}_\nu$) allow us to test the CKM matrix unitarity and to constrain exotic electroweak interactions such as scalar and tensor currents; we have seen a tremendous progress in the development of  {\it ab-initio} methods to compute tree- and loop-level nuclear matrix elements in the decay processes. Furthermore, FRIB experiments provides invaluable inputs to pin down important nuclear effects in beta decays, for instance the measurements of nuclear charge radii help to determine the weak decay form factor and isospin-breaking corrections, and the measurement of GT decay rates help to valide chiral EFT's treatment of the weak axial current. Finally, future FRIB experiments on nuclear deformations and excitations will greatly propel the search for permanent EDMs that may provide sources of {\it CP} violation to explain the matter-antimatter asymmetry. 

In the experimental design and uncertainty quantification section we explore the key role that scientific computing and Bayesian statistics play in the theory-experiment cycle for FRIB science.  Within this context, we discuss the use of emulators for speeding up the necessary expensive computations, methods such as Bayesian Model Mixing for combining the predictions of different theoretical models, and the construction of cloud computing-based pipelines for the continuous calibration of these models as new data becomes available. We close the discussion by highlighting that these technologies not only can help in the design of future experiments at FRIB, but can also lower the barrier for new generations to join the scientific efforts, creating a more diverse and inclusive scientific community in this new era of discovery.

\section{Abbreviations}
%\section{Abbreviations}
\phantom{x}

%\squeezetable
%\begin{table}
%\caption{Abbreviations (A-E)}
\topcaption{Abbreviations \label{tab:Abbreviations}}
\tablefirsthead{ \hline }
%\tablefirsthead{\multicolumn{2}{c}{Abbreviations } \\ \hline }
\tablehead{\multicolumn{2}{c}{Abbreviations (cont.)} \\ \hline }
%\centering
\begin{center}
%\begin{longtable}{|l|l|}
%\begin{tabular}{|l|l}
\begin{xtabular}{|l|l|}
%\hline
ADWA   &  Adiabatic Distorted Wave Approximation \\
ANC    &  Asymptotic Normalization Coefficient \\
AT-TPC &  Active Target Time Projection Chamber \\ 
\hline
BAND   &  Bayesian Analysis of Nuclei Dynamics \\
BBN    &  Big Bang Nucleosynthesis \\
BCS    &  Bardeen Cooper Schrieffer Theory of Superconductivity \\
BEC    &  Bose-Einstein Condensation \\
BECOLA & BEam COoling and LAser spectroscopy \\
BMIS   &  Batch Mode Ion Source \\ 
BSM    &  Beyond the Standard Model \\ 
\hline
CAESAR  &  CAESium-iodide scintillator ARray \\
CC      & Coupled Channels \\
CKM     & Cabibbo Kobayashi Maskawa \\
CI      & Configuration-Interaction  \\
CDCC    & Continuum-Discretized Coupled Channels \\ 
CLS     & Collinear Laser Spectroscopy \\
CN      & Compound Nucleus \\
CREX    & $^{48}$Ca Radius Experiment (JLAB) \\
\hline
DFT     & Density Functional Theory  \\
DT      & deuteron-tritium \\
DWBA    & Distorted-Wave Born Approximation \\ 
\hline
EDF     & Energy-Density Functional \\
EDM     & Electric Dipole Moments \\
EFT     & Effective Field Theory  \\
EOS     & Equation Of State \\
ESPE    & Effective Single-Particle Energy \\
\hline
FDSi    & FRIB Decay Station Initiator \\
FSNN    & Fundamental Symmetries, Neutrons, and Neutrinos \\
FRIB    & Facility for Rare Isotopes \\
FRIB400 & FRIB 400 MeV/u energy upgrade \\
FRIB-TA & FRIB Theory Alliance \\
\hline
G-DMRG  & Gamow Density Matrix Renormalization Group \\
$\gamma$SF     & Gamma-ray Strength Function \\
\hline
GRETA   & Gamma-Ray Energy Tracking Array \\  
GRETINA & Gamma-Ray Energy Tracking In-beam Nuclear Array \\
\hline
HF      & Hauser Feshbach  \\
HFB      & Hartree-Fock Boguliubov  \\
HIC     & Heavy-Ion Collision \\
HPGe    & High Purity Germaniun (detector) \\
HRS     & High Resolution Spectrometer \\
\hline
ISB     & Isospin Symmetry Breaking \\
IMME    & Isobaric Multiplet Mass Equation \\ 
ISOL    & Isotope Separator On Line \\
ITER    & International Thermonuclear Experimental Reactor \\
\hline
LEBIT   & Low-Energy Beam Ion Trap \\
LEC     & Low-Energy Coupling Constant \\
LENDA   & Low-Energy Neutron Detector Array \\
LD      & Level Density \\
LISA    & Large Multi-institutional Scintillator Array \\
\hline
JENSA   & Jet Experiments in Nuclear Structure and Astrophysics \\
JLAB    & Jefferson Laboratory \\
\hline
MED     & Mirror Energy Difference \\
MONA    & Modular Neutron Array \\
MTAS    & Modular Total Absorption Spectrometer \\
MUSIC   & Multi-Sampling Ionization Chamber Detector \\  
\hline
NCSM    & No-Core Shell Model \\
NEXTi   & Neutron (Xn) Tracking initiator \\
NIF     & National Ignition Facility \\
NCGSM   & No-Core Gamow Shell Model \\
NNDC    & National Nuclear Data Center \\
NUCLEI  & Nuclear Computational Low-Energy Initiative \\
NSAC    & Nuclear Science Advisory Committee \\
NSCL    & National Superconducting Cyclotron Laboratory \\
\hline
OMP     & Optical Model Potential \\
\hline
PDR     & Pygmy-Dipole Resonance \\
PREX    & $^{208}$Pb Radius Experiment (JLAB) \\
\hline
QCD     & Quantum Chromodynamics \\
QRPA    & Quasiparticle Random-Phase Approximation \\
\hline
RGM     & Resonating Group Method \\
RIB     & Radioactive Ion Beam  \\
RIBF    & Radioactive Ion Beam Factory \\ 
RIKEN   & Institute of Physical and Chemical Research (Japan) \\
\hline
S800    & FRIB's high-resolution spectrometer \\
SA      & Symmetry Adapted (basis) \\
SA-NCSM & Symmetry Adapted No-Core Shell Model \\
SAMURAI & Superconducting Analyzer for Multi-particles from Radioisotope beams \\
SECAR   & Separator for Capture Reactions \\ 
SeGA    & Segmented Germanium Array \\
SM      & Standard Model \\
S$\pi$RIT & SAMURAI Pion-Reconstruction and Ion-Tracker \\
SOLARIS  & SOLenoid spectrometer Apparatus for ReactIon Studies \\
SuN     & Summing NaI(Tl) (gamma-ray detector) \\
\hline
TED     & Triple Energy Difference \\
TOF     & Time Of Flight \\
TPC     & Time Projection Chamber \\ 
\hline
VANDLE  & Versatile Array of Neutron Detectors at Low Energy \\  
WIMP   & Weakly Interacting Massive Particle \\
\hline
\end{xtabular}
%\end{tabular}
%\end{longtable}
\end{center}

\section{Appendix}
%%% Any appendices we may want
\appendix

\section{Orbitals and magic numbers\label{app:magic}}

\begin{figure}
\includegraphics[scale=0.8]{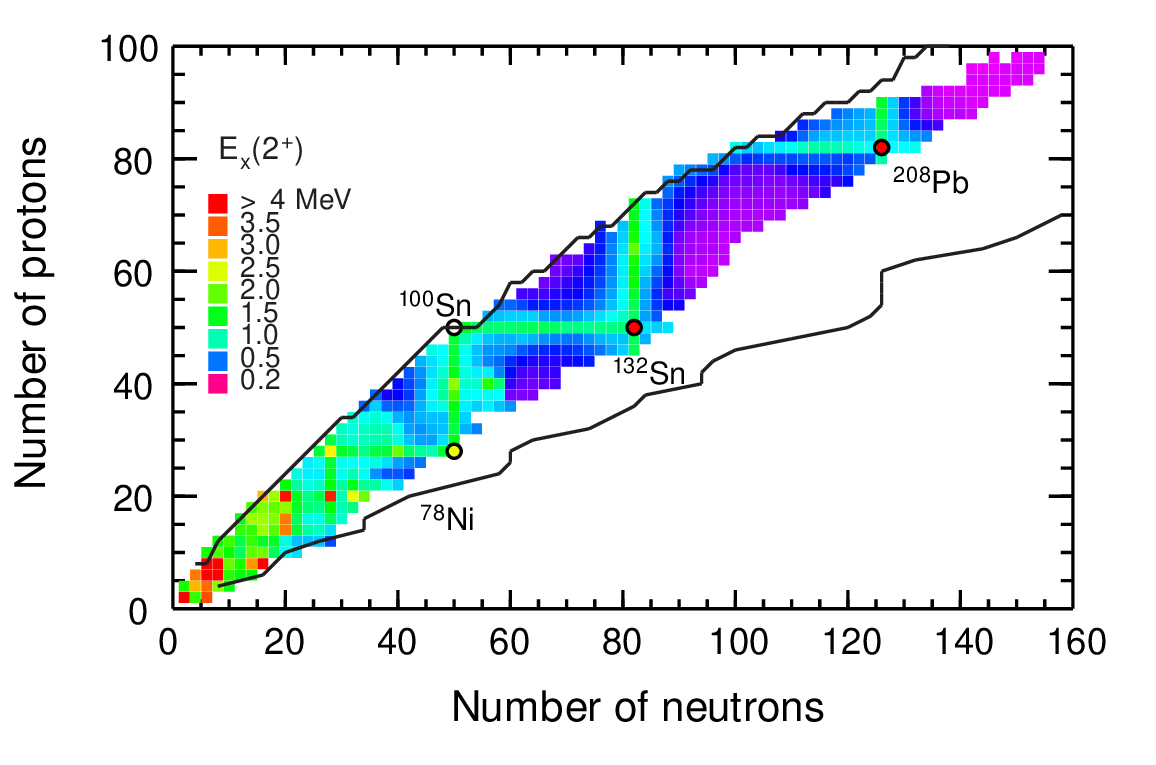}
\caption[]{\small
The nuclear chart showing the observed excitation energies of the 2$^{ + }_{1}$ states.
The black lines show
where the two-proton (upper) and two-neutron (lower) separation energies obtained with
the UNEDF1 energy-density functional (EDF) \cite{UNEDF2013} cross one MeV.
}
\label{[ex2]}
\end{figure}

The nuclear shell model starts with the association of nuclear structure
properties with those expected from the motion of a single neutron or proton
in the average potential of all other nucleons.
The gaps in the spacings of the single-particle energies
give rise to "magic numbers" for the number of protons $  Z  $
and the number of neutrons $  N  $.
The magic numbers are associated with those nuclei that have
a relatively high energy for the
2$^{ + }_{1}$ state. Experimental results shown in Fig. \ref{[ex2]}.

\begin{figure}
\includegraphics[scale=0.8]{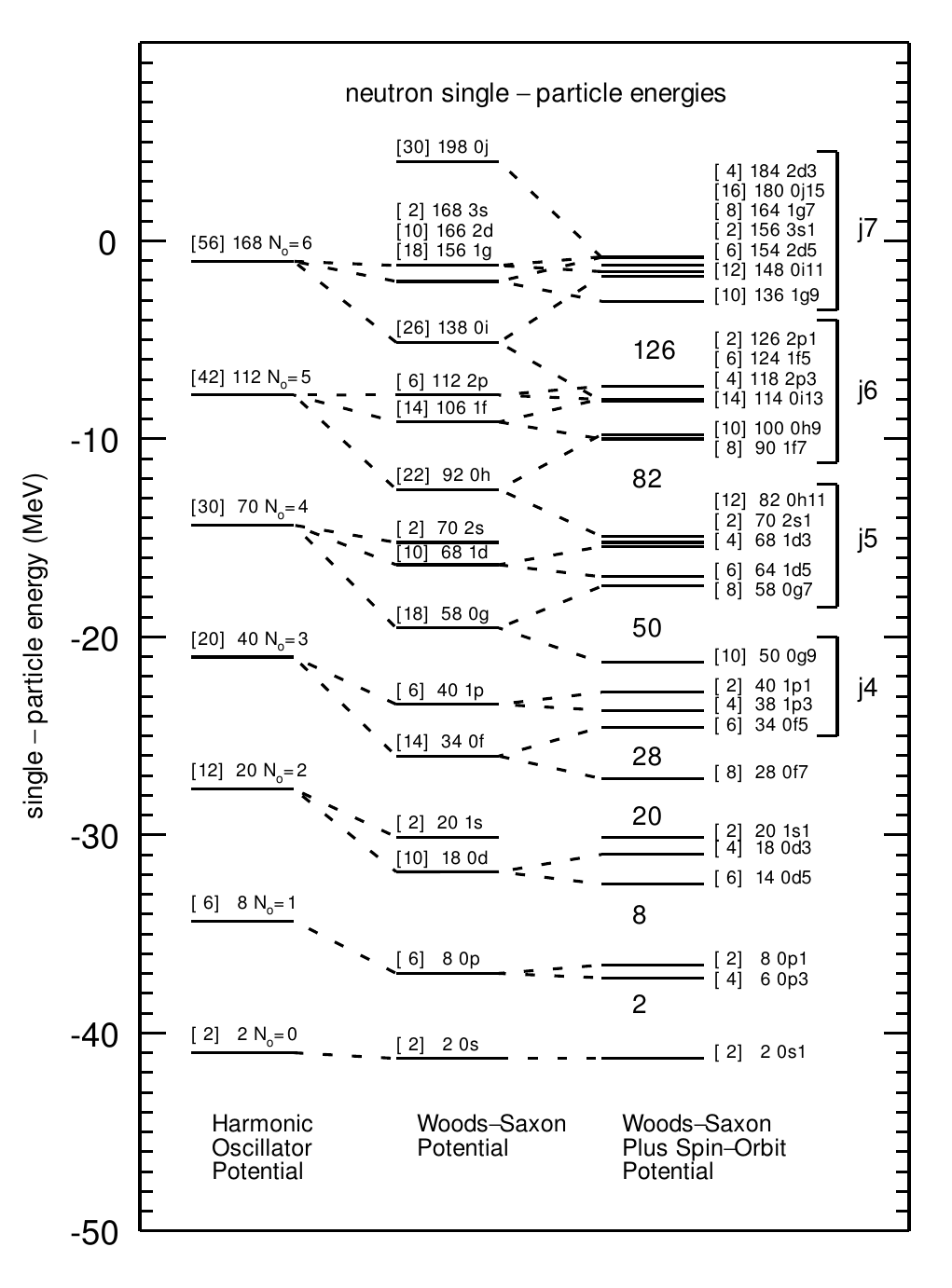}
\caption[]{\small Neutron single-particle states in $^{208}$Pb with
three potential models, harmonic oscillator (left), Woods-Saxon without
spin-orbit (middle) and Woods-Saxon with spin orbit (right). The numbers
in square brackets are the maximum number of neutrons in that each level
can contain, the following number is a running sum of the total. In
addition the harmonic oscillator is labeled by the major quantum number
$  N=2n+\ell   $, the Woods Saxon is labeled by $  (n,\ell )  $ and the
Woods-Saxon with spin-orbit is labeled by $  (n,\ell ,2j)  $.
}
\label{[pbe]}
\end{figure}

An example for the neutron
single-particle energies (SPE) for $^{208}$Pb is shown in Fig. \ref{[pbe]}.
These are eigenstates in a spherical potential with
a Woods-Saxon shape.
The Woods-Saxon
parameters were adjusted to qualitatively reproduce observed
structure properties. Similar SPE spacings
are obtained for protons. The neutron number for the energy gaps
in the SPE shown in Fig. \ref{[pbe]} are
the shell-model magic numbers; 2, 8, 20, 28, 50, 82 and 126.
The SPE evolve with
the number of protons and neutrons. The detailed ordering
and the associated magic numbers can change.

\begin{figure}
\includegraphics[scale=0.8]{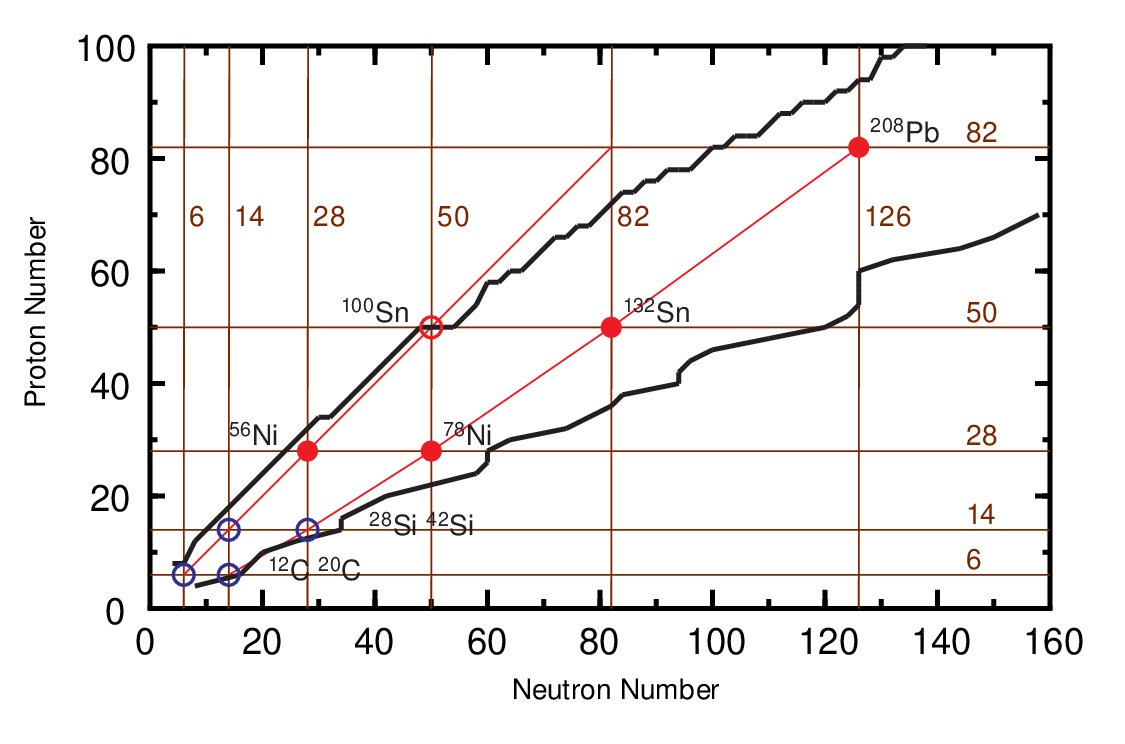}
\caption[]{\small
The nuclear chart showing $  jj  $ magic numbers. The black lines show
where the two-proton (upper) and two-neutron (lower) separation energies obtained with
the UNEDF1 EDF \cite{UNEDF2013} cross one MeV. The filled red circles shows the locations of doubly-$ 
 jj  $
magic nuclei established from experiment. The open red circle for $^{100}$Sn indicates
it is probably
doubly-magic, but that the 2$^{ + }_{1}$ energy is not yet measured.
The blue circles in the
bottom left-hand side are nuclei in the $  jj  $ doubly-magic number sequence that are oblate
deformed.
}
\label{[jj]}
\end{figure}

\begin{figure}
\includegraphics[scale=0.8]{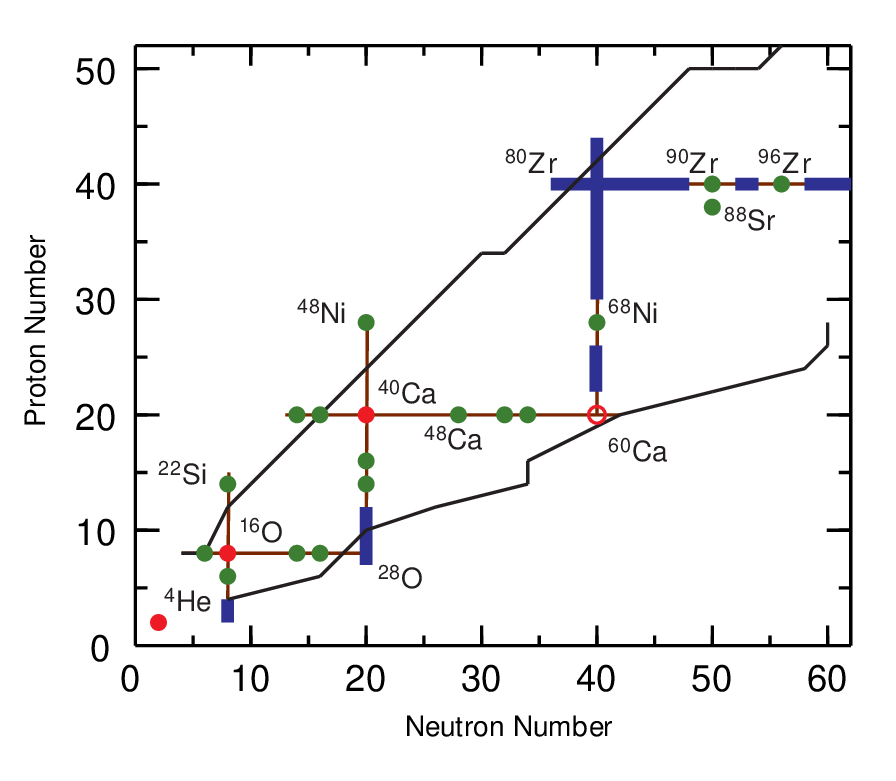}
\caption[]{\small The nuclear chart showing the locations of the
$  ls  $ magic numbers and the associated doubly-magic nuclei.
Lower mass region of the nuclear chart showing $  ls  $ magic numbers, 8, 20 and
40. The black lines show where the two-proton (upper) and two-neutron (lower) separation
energies obtained with the UNEDF1 EDF \cite{UNEDF2013} functional cross one MeV. The filled red 
circles
show the doubly-magic nuclei $^{4}$He $^{16}$O and $^{40}$Ca. The open red circle for $^{60}$Ca 
indicates
a possible doubly-magic nucleus, but the 2$^{ + }_{1}$ energy is not yet measured. The green
circles are doubly-magic nuclei associated with the $  j  $-orbital fillings. The blue lines 
indicate
isotopes or isotones where the $  ls  $ magic number is observed to be broken.
}
\label{[ls]}
\end{figure}

The magic numbers 28, 50, 82 and 126, referred to as
$  jj  $ magic numbers, are associated with
the gaps formed from the lowering of the $  j=\ell +1/2  $ orbitals
due to the spin-orbit single-particle potential.
The nuclei associated with $  jj  $ magic numbers are shown in nuclei Fig. \ref{[mjj]}.
Doubly-magic nuclei are those with magic numbers for both protons and neutrons.
The $  jj  $ doubly-magic nuclei are shown in Fig. \ref{[mjj]}.
The proton-neutron tensor interaction contained
in the Hamiltonians used in CI calculations
modifies the effective spin-orbit splittings for protons/neutrons
depending on the filling of the  neutron/proton spin-orbit doublets
\cite{otsu2005p232502}, \cite{otsu2020}.

For light nuclei, the magic numbers 2, 8 and 20
are observed. These are referred to as
$  ls  $ magic numbers, and are associated with the filling of a
major harmonic-oscillator shell with $  N_{o}=(2n+\ell )  $ where both members of the
spin-orbit pair $  j=\ell  \pm s  $ ($  s  $=1/2) are filled (except for
$  N_{o}=0  $ that contains only the $  0s_{1/2}  $ orbital).
The nuclei associated with $  ls  $ magic numbers are shown in Fig. \ref{[mls]}.
These include three nuclei with $  ls  $ doubly-magic numbers, $^{4}$He, $^{16}$O
and $^{40}$Ca. There are many doubly-magic nuclei (shown with green circles)
associated with nuclei
where one kind  of nucleon (proton or neutron) has an $  ls  $ magic number and the other
kind of nucleon fills the $  j  $ levels in the order shown at the bottom right-hand side
of Fig. \ref{[pbe]}.

\section{Model spaces\label{app:modelspaces}}

\begin{figure}
\includegraphics[scale=0.8]{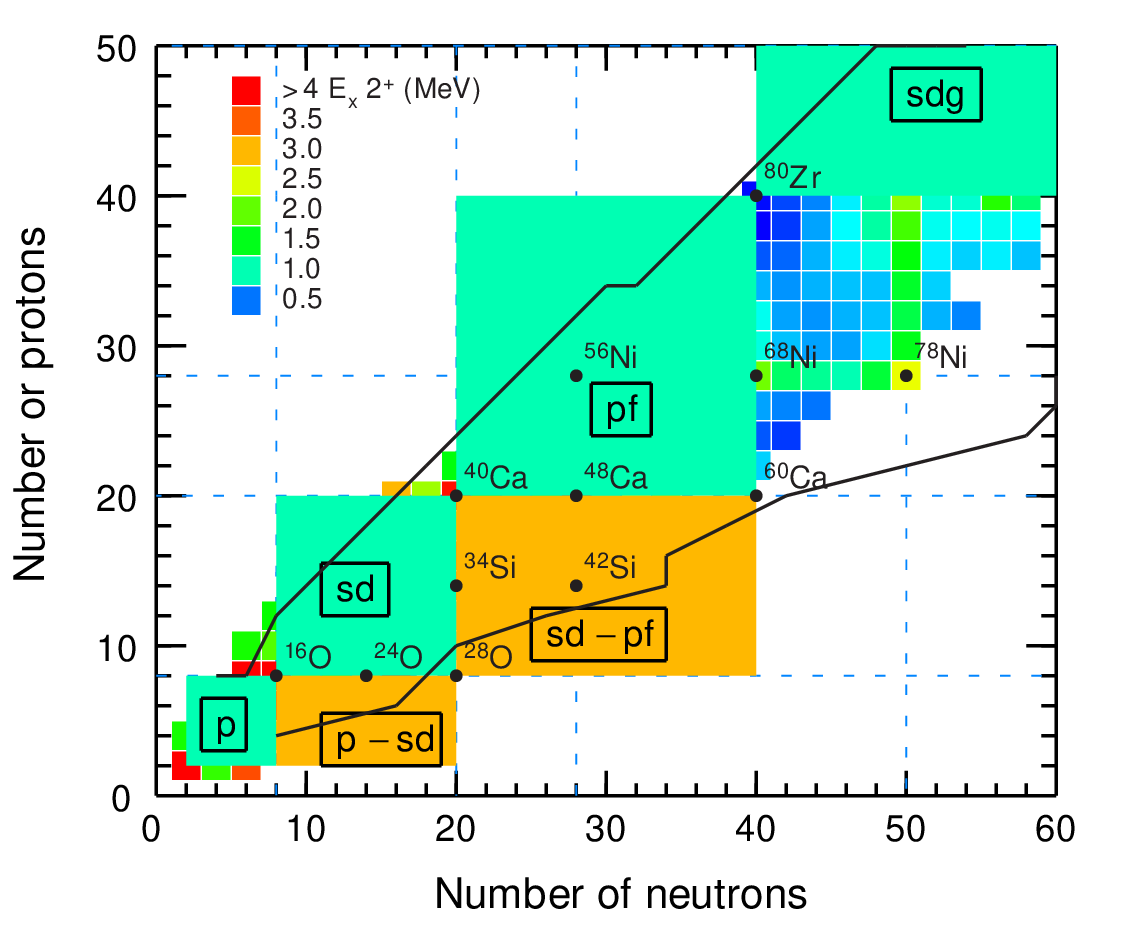}
\caption[]{\small
The nuclear chart showing the regions of nuclei associated with
the harmonic-oscillator $  ls  $ model spaces.
}
\label{[mls]}
\end{figure}

In this appendix, we describe various model spaces relevant for interacting shell model calculations.
The magic numbers described in appendix~\ref{app:magic} serve to define model spaces that are used to
truncate the number of configurations that are used for shell-model
calculations. The traditional names for model spaces based on the
$  ls  $ magic numbers for light nuclei are shown in Fig. \ref{[mls]}.
For example, $  sd  $ is stands for the model space with
active nucleons in the $  (0d_{5/2},0d_{3/2},1s_{1/2})  $ set of
orbitals. The model spaces $  s  $, $  p  $, $  sd  $, and $  pf  $
are those for the major oscillator quantum numbers
with $  N_{o}  $=0, 1, 2, and 3, respectively, in Fig. \ref{[pbe]}.
Near $  N=Z  $, protons and neutrons occupy the same
major (oscillator) shell.  Configurations for neutron-rich can
involve protons and neutrons in two different major shells.
For example, in the $  sd-pf  $ region of nuclei,
protons occupy the $  sd  $ major shell and neutrons occupy the
$  fp  $ major shell (the region of mirror nuclei, not shown,
would be called $  fp-sd  $).

\begin{figure}
\includegraphics[scale=0.8]{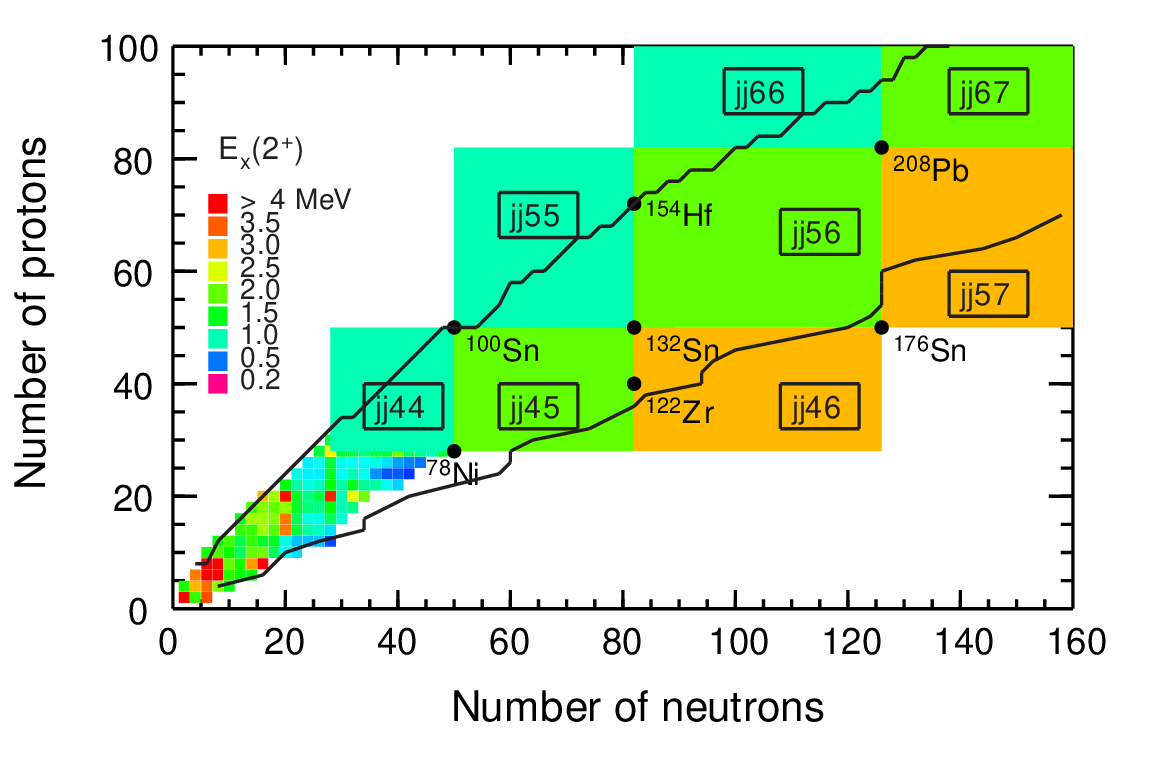}
\caption[]{\small
The nuclear chart showing the regions of nuclei associated with
the $  jj  $ model spaces.
}
\label{[mjj]}
\end{figure}

In the harmonic-oscillator basis, all of these $  ls  $
model spaces involve nucleons in the lowest possible
oscillator-energy configuration,
often called the 0$\hbar\omega$ configurations.
%For 0$\hbar\omega$ configurations, the nuclear center of mass
%must be in the lowest energy $  0s_{1/2}  $ configuration,
%and this motion must be removed from the
%calculated observables.

Model spaces can involve more complex configuations.
In the oscillator-basis, excited states involve
$  n\hbar \omega   $ configurations. For example, starting with the $  sd  $ model space,
a 1$\hbar\omega$ basis involves one nucleon excited from $  p  $ to $  sd  $
or one nucleon excited from $  sd  $ to $  pf  $.
%The $  n\hbar \omega   $ basis
%contains linear combinations of configurations that
%correspond to spurious motion of the nuclear center of mass.
%These spurious states must be removed, for example, using the
%Gloeckner-Lawson method of adding a term to the Hamiltonian
%that moves up the energy of spurious states \cite{gloe1974}.

\begin{figure}
\includegraphics[scale=0.8]{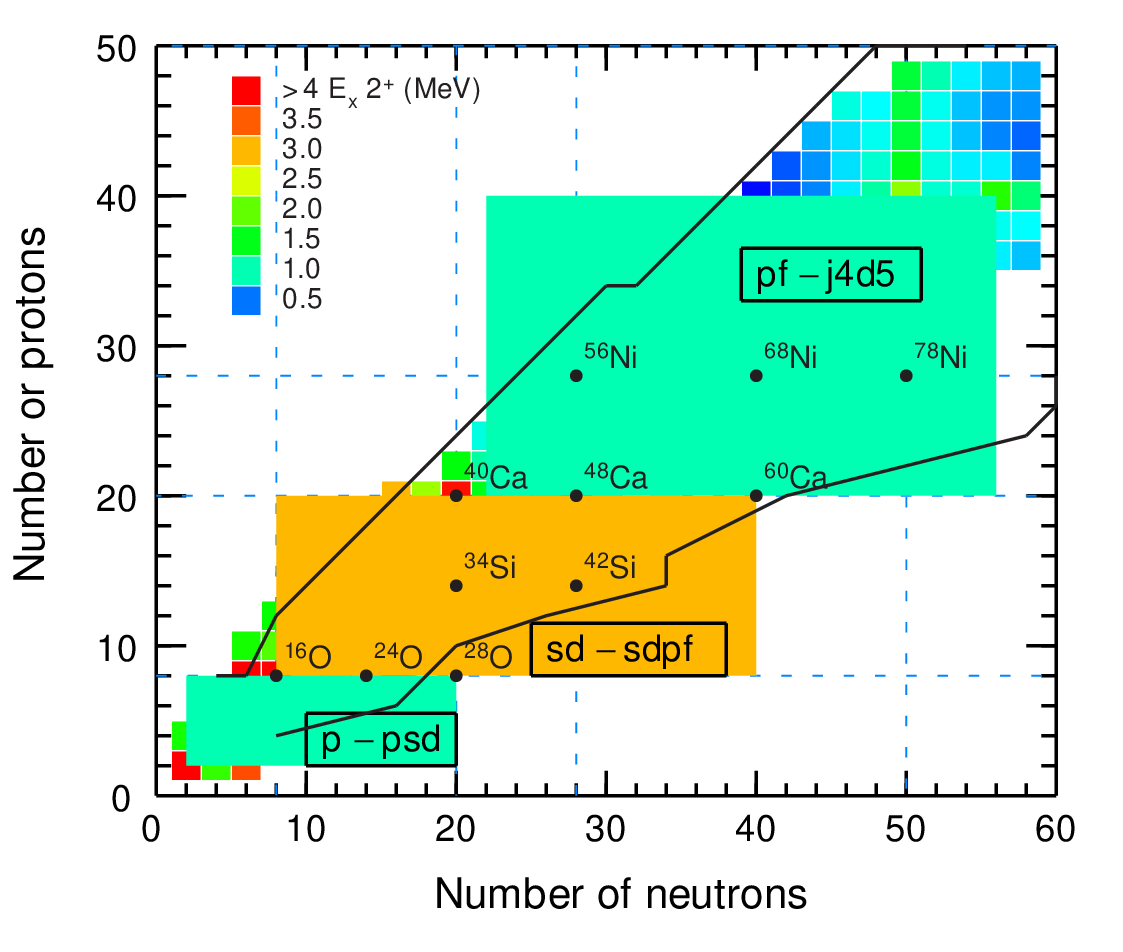}
\caption[]{\small
The nuclear chart showing the regions of nuclei associated with
some larger model spaces.
}
\label{[mex]}
\end{figure}

In the $  N=8  $, 20 and 40 neutron-rich islands of inversion, the model space
for neutrons must involve two major shells.
For example, for the $  N=20  $ island of inversion, the
model space for neutrons must contain both $  sd  $
and $  pf  $ major shells as shown in Fig. \ref{[mex]}.
%Spurious states involve
%both protons and neutrons excited across major shells.
%For neutron-rich nuclei, low-lying states
%are often dominated by just neutrons excited across
%major shells, and the spurious state contribution is small.

Model spaces for heavy nuclei are shown in Fig.~\ref{[mjj]}, labeled with the naming convention used in the NuShellX code~\cite{brow2014p115}
%The model space names used in the NuShellX code \cite{brow2014p115}
%for heavy nuclei
%are shown in Fig. \ref{[mjj]}.
Groups of orbitals are labeled by $  jn  $
as shown by the labels on the right-hand side of Fig. \ref{[pbe]}.
The notation $  jjnm  $ indicates that protons
occupy the $  jn  $ set of orbitals and neutrons
occupy the $  jm  $ set of orbitals.

For a given calculation, the choice of model space is based on
several interrelated factors.

\begin{enumerate}
\item One considers the
limitation on the basis dimension connected to
the computational methods and capabilities,
as well as the practical consideration
how many different nuclei and states
need to be considered for a given experimental situation.
%With conventional shell-model codes calculations for
%basis dimensions up to the order of 10$^{11}$ can be
%considered.

\item One considers to what extent the desired spectroscopic
properties are
contained within the model space. For example,
an $\ell$=3 transfer reaction cannot be described
within the sd model space.
Likewise, quadrupole 
deformation is connected to the mixing of orbitals
with $\Delta\ell$=2 and $\Delta  j  $=2, and so it is preferable to have both
type of orbitals contained in the model space if one is interested in collective properties
%\cite{caur2005p427}, 
\cite{Caurier2005}, 
\cite{nowa2021}.

%(3) One considers to what extent the collective properties
%are contained in the model space. Quadrupole nuclear
%deformation is connected to the mixing of orbitals
%with $\Delta\ell$=2 and $\Delta  j  $=2. It is preferable to have both
%type of orbitals contained in the model space 
%%\cite{caur2005p427}, 
%\cite{Caurier2005}, 
%\cite{nowa2021}.
%For example, for the $  fp-j4d5  $ model space shown in Fig. \ref{[mex]}
%the neutrons orbitals involve both $  0g_{9/2}  $
%and $  1d_{5/2}  $ to allow for the collectivity involved in the
%$  N=40  $ island of inversion below $^{68}$Ni.

\end{enumerate}

In ``full-space'' calculations all possible configuration within
a given model space are included. This is currently possible
for matrix dimensions up to about 10$^{11}$.
To reach beyond this,
there are several types of truncations that can be made.
For light nuclei it is common to consider truncations based on
a selected set of oscillator excitations. The WBP
and WBT Hamiltonians obtained in \cite{warb1992} were designed for $  n\hbar \omega   $ 
configurations
within the $  0s-0p-1s0d-1p0f  $ model space. For example, the 1$\hbar\omega$ basis can be
used for giant dipole excitations \cite{orc2023}. One can also consider a fixed number
of excitations across one of the oscillator magic numbers. The
FSU Hamiltonian was obtained from data related to 1$\hbar\omega$ excitations,
and then used to successfully describe the energies of states
that are considered to have two nucleons excited from $  sd  $ to $  pf  $ \cite{lubn2020}.

%It is well known that low-lying states of many nuclei can be described by
%collective models.
In many cases, low-lying states obtained from large-basis calculations
can also be related to a selective set of group theoretical based configurations
or by the Bohr-Mottelson collective model for even-even nuclei
and the Nillson model for odd-even nuclei.
%This observation motivates truncations based on these considerations.
For light nuclei
the SU(3)-adapted basis has been used since the 1950's (see \cite{arim1999} and
references therein).
%In a no-core
%basis one can use a Symmetry-Adapted No-Core Shell Model framework
%based on the Sp(3,R)  basis 
%%\cite{laun2021} 
%\cite{Launey:21} 
%which allows for
%configurations that include quadrupole excitations across two major
%oscillator shells (e.g., $  p  $ to $  fp  $).
%Effective charges for $  E2  $ transitions used for
%calculations within a  single-oscillator model space such as $  sd  $ are ascribed to
%the absence of 2$\hbar\omega$ excitations in this basis. In the  Sp(3,R)  basis
%these excitations are explicitly contained in the wavefunctions.
For heavier nuclei, one can utilize methods like the generator-coordinate method~\cite{otsu2020,shim2021,dao2022,lian2024} or density matrix renormalization group~\cite{Tichai2024} to approximately diagonalize within the model space.
%A variety of methods have been proposed
%to truncate the very large basis dimensions associated
%with many nucleons in model spaces for heavy nuclei
%to obtain collective spectra with traditional Hamiltonians
%used for nuclei near the closed shells \cite{otsu2020,shim2021,dao2022,lian2024}.
%The model space can be truncated by first carrying out
%deformed Hartree-Fock calculations in the ($\beta$,$\gamma$) collective space.
%Then one obtains a  non-orthoginal set of basis states from
%configurations in the energy minima that can be used within the
%spherical shell-model basis. The discrete nonorthogonal shell model (DNO-SM)
%basis developed in \cite{dao2022} has been used to obtained
%abd $\beta$-$\gamma$ energy surface for $^{254}$No within the jj67 basis above $^{208}$Pb.
%The  Monte Carlo Shell Model (MCSM) method \cite{otsu2020}
%that uses selected linear combinations of basis states.
%The resulting wavefunctions describe spherical and deformed
%states with basis configurations related to probabilities
%on the $\beta$-$\gamma$ energy surface \cite{otsu2020}.
%In large-basis calculations, states above the low-lying collective bands
%are formed from more complex non-collective configurations \cite{long2020}.

%\input{Othertopcis/othertopics}

\begin{acknowledgments}

Danial Bazin, Alexandra Gade, Hiro Iwasaki, Sean Liddick, and Rebeka Lubna: Supported by the U.S. Department of Energy, Office of Science, Office of Nuclear Physics award DE-SC0023633 (FRIB).

Kei Minamisono and Ryan Ringle: Supported by NSF grant PHY-2111185.

B. Alex Brown; Supported from NSF grant PHY-2110365.

Kyle W. Brown: Supported by the National Science Foundation under Grant No. PHY-2309923.

Mark A.~Caprio: Supported by the U.S.~Department of Energy, Office of Science,
under Award No.~DE-FG02-95ER40934.

Heather Crawford: Supported by the U.S. Department of Energy, Office of Science, Office of Nuclear Physics under Contract No. DE-AC02-05CH11231 (LBNL).

Pawel Danielewicz: Supported by the U.S. Department of Energy Office of Science Grant No. DE-SC0019209.

Chrisitan Drischler: Supported by the U.S. Department of Energy, Office of Science, Office of Nuclear Physics, under the FRIB Theory Alliance award DE-SC0013617 and under the STREAMLINE collaboration award DE-SC0024233, and by the National Science Foundation under PHY~2339043.

Jutta Escher: Supported under the auspices of the U.S. Department of Energy by Lawrence Livermore National Laboratory under Contract DE-AC52-07NA27344, with support from LDRD projects 21-ERD-006 and 24-ERD-023.

Kevin Fossez: Supported by the U.S.\ NSF (PHY-2238752) and the U.S.\ DOE, Office of Science, Office of Nuclear Physics, under the FRIB Theory Alliance, award DE-SC0013617. 

Pablo Giuliani: Supported by the National Science Foundation
CSSI program under award number 2004601 (BAND collaboration)

Calem Hoffman: Supported by the U.S. Department of Energy under Awards No. DEFG02-95ER40934 (ND), DE-AC02-06CH11357 (Argonne), DE-SC0009883 (FSU), DESC0023633 (FRIB), and DE-SC0013365 (MSU) through the Office of Science, Office of Nuclear Physics,

Ronald F. Garcia Ruiz: Supported by by the U.S. Department of Energy under Awards No.DE-SC0021176 and DE-SC0021179.

Kyle Godbey: Supported by the department of Energy under Award Numbers DOE-DE-NA0004074 (NNSA, the Stewardship Science Academic Alliances program), DE-SC0023175 (Office of Science, NUCLEI SciDAC-5 collaboration), and DE-SC0023688; by the National Science Foundation CSSI program under award number 2004601 (BAND collaboration).

Robert Grzywacz: Supported in part by the Ofﬁce of Nuclear Physics, U.S. 
Department of Energy under award No. DE-FG02-96ER40983, and by the 
National Nuclear Security Administration under the Stewardship Science 
Academic Alliances program through DOE Awards No. DE-NA0004068.

Linda Hlophe: Supported by the U.S. Department of Energy, Office of Science, Office of Nuclear Physics, under Work Proposal No. SCW0498 and LLNL LDRD project No. 22-LW-003, and U.S. Department of Energy, Office of Science, Office of Nuclear Physics, under the FRIB Theory Alliance award DE-SC0013617.

Jeremy Holt: Supported by the U.S. NSF under grant PHY-2209318.

Dean Lee: Supported by DOE grants DE-SC0013365, DE-SC0023175, and DE-SC0024586.

Augusto O. Macchiavelli: Support from the Laboratory Directed Research and Development Program of Oak Ridge National Laboratory, managed by UT-Battelle, LLC, for the U. S. Department of Energy.

Gabriel Martinez-Pinedo: Supported by the European Research Council
(ERC) under the European Union’s Horizon 2020 research
and innovation programme (ERC Advanced Grant
KILONOVA No. 885281) and the Deutsche
Forschungsgemeinschaft (DFG, German Research
Foundation)—Project-ID 279384907—SFB 1245, and
MA 4248/3-1. 

Anna McCoy: Supported by the U.S.Department of Energy, Office of Science, under Award Numbers DE-SC0021027 and DE-SC0013617 (FRIB Theory Alliance).

Kei Minamisono: Support from NSF grant PHY-2111185.

Alexis Mercenne: Supported by the U.S. NSF grant PHY-2327385.

Belen Monteagudo: Supported by the National Science Foundation under Grant No. PHY-2209138 and Grant No. G0002815

Petr Navratil: Support from NSERC Grant SAPIN-2022-00019. TRIUMF receives federal funding via a contribution agreement with the National Research Council of Canada. Computing support is acknowledged from an INCITE Award on the Summit and Frontier supercomputer of the OLCF at ORNL.

Witek Nazarewicz:  Supported by grants DE-SC0013365 and DE-SC0023688 (MSU) through the Office of Science, Office of Nuclear Physics, and DE-NA0004074  (NNSA, the Stewardship Science Academic Alliances program). 

Ryan Ringle: Support from NSF grant PHY-2111185.

Grigor Sargsyan: Supported by the U.S. Department of Energy, Office of Science, Office of Nuclear Physics, under the FRIB Theory Alliance award DE-SC0013617, and the U.S. Department of Energy, Office of Science, Office of Nuclear Physics, under the FRIB Theory Alliance award DE-SC0013617.

Hendrik Schatz: Supported by NSF award PHY-2209429. The work also benefited from interdisciplinary discussions facilitated by CeNAM DOE award DE-SC0023128 and IReNA NSF award OISE-1927130.

Chien-Yeah Seng: Supported in part by the U.S. DOE, Office of Science, Office of Nuclear Physics, under the FRIB Theory Alliance award DE-SC0013617, and by the DOE grant DE-FG02-97ER41014.

Agnieszka Sorensen: Support by the U.S.\ Department of Energy, Office of Science, Office of Nuclear Physics, under Grant No.\ DE-FG02-00ER41132. 

Mark-Christoph Spieker: Supported by U.S. National Science Foundation (NSF) under Grant No. PHY-2012522, (WoU-MMA: Studies of Nuclear Structure and Nuclear Astrophysics).

Nicole Vassh: Supported by the Natural Sciences and Engineering Research Council of Canada (NSERC).

Alexander Volya: Supported by the U.S. Department of Energy Office of Science, Office of Nuclear Physics under Award No. DE-SC0009883.

Remco Zegers and Hendrik Schatz: Supported by US National Science Foundation No. PHY-2209429, “Windows on the Universe: Nuclear Astrophysics at FRIB.

Xilin Zhang: Supported by the U.S. Department of Energy, Office of Science, Office of Nuclear Physics, under the FRIB Theory Alliance Award No. DE-SC0013617 and under the STREAMLINE Collaboration Award No. DE-SC0024586. 

\end{acknowledgments}

%\appendix

%%%%%%%%%%%%%%%%%%%%%%%%%%%%%%%%%%%%%%%%%%%%%%%%%%%%%%%%%%%%%%%%%%%%%%%%%%
%%% Who commented this out and why? -- SRS
%% At the workshop we agreed to have one bib file, so I am confused why there are two now. -- GHS
%\bibliography{references}%,totaln}% Produces the bibliography via BibTeX.

%KF: I had to break down the file totaln.bib because it was too large for the Overleaf editor to open and contained errors. It seems that the limit is around 22000 lines.
%\bibliography{references,totaln_1,totaln_2,totaln_3,totaln_4,totaln_5,totaln_6,continuum,reactions,babadd,astro}

\bibliography{final.bib,Reactions/reactions_20240811.bib}

\end{document}